\documentclass{aastex61}

\usepackage{amsmath}
\usepackage{multirow}
\usepackage{ulem}
\usepackage{epsfig}

\newcolumntype{Q}{>{\centering\arraybackslash}m{2.2cm}}
\newcolumntype{P}{>{\centering\arraybackslash}m{2.7cm}}
\newcolumntype{M}{>{\centering\arraybackslash}m{1.5cm}}
\newcolumntype{N}{>{\centering\arraybackslash}m{1.25cm}}
\newcolumntype{O}{>{\centering\arraybackslash}m{0.9cm}}

\newcommand{\veritas}{VERITAS}
\newcommand{\magic}{MAGIC}
\newcommand{\hess}{H.E.S.S.}
\newcommand{\fermi}{\textit{Fermi}}
\newcommand{\hawc}{HAWC}
\newcommand{\cta}{CTA}
\newcommand{\LAT}{\textit{Fermi}-LAT}
\newcommand{\milagro}{Milagro}
\newcommand{\nustar}{\textit{NuSTAR}}
\newcommand{\xmm}{\textit{XMM-Newton}}

\newcommand{\hegra}{HEGRA}
\newcommand{\argo}{ARGO-YBJ}
\newcommand{\whipple}{Whipple}
\newcommand{\rosat}{\textit{ROSAT}}
\newcommand{\spitzer}{\textit{Spitzer}}
\newcommand{\suzaku}{\textit{Suzaku}}

\newcommand{\cisne}{VER J2019\allowbreak+368}
\newcommand{\TeVJ}{VER J2031\allowbreak+415}
\newcommand{\GCyg}{VER J2019\allowbreak+407}
\newcommand{\CTB}{VER J2016\allowbreak+371}
\newcommand{\Sh}{\textit{Sh} 2-104}
\newcommand{\MGROcisne}{MGRO J2019\allowbreak+37}

\newcommand{\cisneA}{VER J2018\allowbreak+367*}
\newcommand{\cisneB}{VER J2020\allowbreak+368*}

\newcommand{\fermipy}{\emph{fermipy}}
\newcommand{\GR}{$\gamma$-ray}
\newcommand{\plm}{$\pm$}

\newcommand{\GeV}{\mathrm{GeV}}
\newcommand{\cm}{\mathrm{cm}}
\newcommand{\s}{\mathrm{s}}

\newcommand{\point}{\textit{Point}}
\newcommand{\ext}{\textit{Extended}}
\newcommand{\on}{\texttt{ON}}
\newcommand{\off}{\texttt{OFF}}

\newcommand{\No}{$N_0$}
\newcommand{\Eo}{$E_0$}

\newcommand{\ra}{$\alpha_{J2000}$}
\newcommand{\dec}{$\delta_{J2000}$}
\newcommand{\hm}[2]{$#1^\mathrm{h}\allowbreak#2^\mathrm{m}$}
\newcommand{\hms}[3]{$#1^\mathrm{h}\allowbreak#2^\mathrm{m}\allowbreak#3^\mathrm{s}$}
\newcommand{\hmsee}[5]{$#1^\mathrm{h}\allowbreak#2^\mathrm{m}\allowbreak#3^\mathrm{s}\allowbreak\pm#4^\mathrm{s}_{stat}\allowbreak\pm#5^\mathrm{s}_{sys}$}
\newcommand{\dm}[2]{$#1\arcdeg\allowbreak#2\arcmin$}
\newcommand{\dms}[3]{$#1\arcdeg\allowbreak#2\arcmin\allowbreak#3\arcsec$}
\newcommand{\dmsee}[5]{$#1\arcdeg\allowbreak#2\arcmin\allowbreak#3\arcsec\allowbreak\pm#4''_{stat}\allowbreak\pm#5''_{sys}$}

\newcommand{\dstat}[2]{$#1\arcdeg\allowbreak\pm#2^\circ_{stat}$}
\newcommand{\dee}[3]{$#1\arcdeg\allowbreak\pm#2^\circ_{stat}\allowbreak\pm#3^\circ_{sys}$}

\newcommand{\nstat}[2]{$#1\allowbreak\pm#2_{stat}$}
\newcommand{\nee}[3]{$#1\allowbreak\pm#2_{stat}\allowbreak\pm#3_{sys}$}

\newcommand{\fstat}[3]{$(#1 \pm #2_{stat})\allowbreak \times 10^{#3} \, \allowbreak \GeV^{-1}\cm^{-2}\s^{-1}$}
\newcommand{\fee}[4]{$(#1\allowbreak\pm#2_{stat}\allowbreak\pm#3_{sys} )\allowbreak \times  10^{#4}\allowbreak \, \mathrm{GeV}^{-1}\allowbreak  \mathrm{cm}^{-2}\allowbreak  \mathrm{s}^{-1}$}

\newcommand{\pee}[4]{$(#1\allowbreak\pm#2_{stat}\allowbreak\pm#3_{sys} )\allowbreak \mathrm{E}{#4}$}

\received{September 09, 2017}
\revised{March 28, 2018}
\accepted{May 11, 2018}

\submitjournal{ApJ}

\shorttitle{VERITAS Cygnus Region Survey}
\shortauthors{Abeysekara et al.}

\begin{document}

\title{A Very High Energy $\gamma$-Ray Survey towards the Cygnus Region of the Galaxy.}

\correspondingauthor{R.~Bird, M.~Krause, A.~Popkow}
\email{ralphbird@astro.ucla.edu, maria.krause@desy.de, apopkow@ucla.edu}

\author{A.~U.~Abeysekara} 
\affiliation{Department of Physics and Astronomy, University of Utah, Salt Lake City, UT 84112, USA}
\author{A.~Archer} 
\affiliation{Department of Physics, Washington University, St. Louis, MO 63130, USA}
\author{T.~Aune} 
\affiliation{Department of Physics and Astronomy, University of California, Los Angeles, CA 90095, USA}
\affiliation{The Climate Corporation, San Francisco, CA 94103, USA}
\author{W.~Benbow} 
\affiliation{Fred Lawrence Whipple Observatory, Harvard-Smithsonian Center for Astrophysics, Amado, AZ 85645, USA}
\author{R.~Bird} 
\affiliation{Department of Physics and Astronomy, University of California, Los Angeles, CA 90095, USA}
\author{R.~Brose} 
\affiliation{Institute of Physics and Astronomy, University of Potsdam, 14476 Potsdam-Golm, Germany}\affiliation{DESY, Platanenallee 6, 15738 Zeuthen, Germany}
\author{M.~Buchovecky} 
\affiliation{Department of Physics and Astronomy, University of California, Los Angeles, CA 90095, USA}
\author{V.~Bugaev} 
\affiliation{Department of Physics, Washington University, St. Louis, MO 63130, USA}
\author{W.~Cui} 
\affiliation{Department of Physics and Astronomy, Purdue University, West Lafayette, IN 47907, USA}\affiliation{Department of Physics and Center for Astrophysics, Tsinghua University, Beijing 100084, China.}
\author{M.~K.~Daniel} 
\affiliation{Fred Lawrence Whipple Observatory, Harvard-Smithsonian Center for Astrophysics, Amado, AZ 85645, USA}
\author{A.~Falcone} 
\affiliation{Department of Astronomy and Astrophysics, 525 Davey Lab, Pennsylvania State University, University Park, PA 16802, USA}
\author{Q.~Feng} 
\affiliation{Physics Department, McGill University, Montreal, QC H3A 2T8, Canada}
\author{J.~P.~Finley} 
\affiliation{Department of Physics and Astronomy, Purdue University, West Lafayette, IN 47907, USA}
\author{H.~Fleischhack} 
\affiliation{DESY, Platanenallee 6, 15738 Zeuthen, Germany}
\author{A.~Flinders} 
\affiliation{Department of Physics and Astronomy, University of Utah, Salt Lake City, UT 84112, USA}
\author{L.~Fortson} 
\affiliation{School of Physics and Astronomy, University of Minnesota, Minneapolis, MN 55455, USA}
\author{A.~Furniss} 
\affiliation{Department of Physics, California State University - East Bay, Hayward, CA 94542, USA}
\author{E.~V.~Gotthelf} 
\affiliation{Columbia Astrophysics Laboratory, Columbia University, New York, NY 10027, USA}
\author{J.~Grube} 
\affiliation{Department of Physics, Stevens Institute of Technology, Hoboken, NJ 07030, USA}
\author{D.~Hanna} 
\affiliation{Physics Department, McGill University, Montreal, QC H3A 2T8, Canada}
\author{O.~Hervet} 
\affiliation{Santa Cruz Institute for Particle Physics and Department of Physics, University of California, Santa Cruz, CA 95064, USA}
\author{J.~Holder} 
\affiliation{Department of Physics and Astronomy and the Bartol Research Institute, University of Delaware, Newark, DE 19716, USA}
\author{K.~Huang} 
\affiliation{Van Nuys High School, Van Nuys, CA 91411, USA}\affiliation{Department of Physics and Astronomy, University of California, Los Angeles, CA 90095, USA}
\author{G.~Hughes} 
\affiliation{Fred Lawrence Whipple Observatory, Harvard-Smithsonian Center for Astrophysics, Amado, AZ 85645, USA}
\author{T.~B.~Humensky} 
\affiliation{Physics Department, Columbia University, New York, NY 10027, USA}
\author{M.~H\"utten} 
\affiliation{DESY, Platanenallee 6, 15738 Zeuthen, Germany}
\author{C.~A.~Johnson} 
\affiliation{Santa Cruz Institute for Particle Physics and Department of Physics, University of California, Santa Cruz, CA 95064, USA}
\author{P.~Kaaret} 
\affiliation{Department of Physics and Astronomy, University of Iowa, Van Allen Hall, Iowa City, IA 52242, USA}
\author{P.~Kar} 
\affiliation{Department of Physics and Astronomy, University of Utah, Salt Lake City, UT 84112, USA}
\author{N.~Kelley-Hoskins} 
\affiliation{DESY, Platanenallee 6, 15738 Zeuthen, Germany}
\author{M.~Kertzman} 
\affiliation{Department of Physics and Astronomy, DePauw University, Greencastle, IN 46135-0037, USA}
\author{D.~Kieda} 
\affiliation{Department of Physics and Astronomy, University of Utah, Salt Lake City, UT 84112, USA}
\author{M.~Krause} 
\affiliation{DESY, Platanenallee 6, 15738 Zeuthen, Germany}
\author{S.~Kumar} 
\affiliation{Department of Physics and Astronomy and the Bartol Research Institute, University of Delaware, Newark, DE 19716, USA}
\author{M.~J.~Lang} 
\affiliation{School of Physics, National University of Ireland Galway, University Road, Galway, Ireland}
\author{T.~T.~Y.~Lin} 
\affiliation{Physics Department, McGill University, Montreal, QC H3A 2T8, Canada}
\author{G.~Maier} 
\affiliation{DESY, Platanenallee 6, 15738 Zeuthen, Germany}
\author{S.~McArthur} 
\affiliation{Department of Physics and Astronomy, Purdue University, West Lafayette, IN 47907, USA}
\author{P.~Moriarty} 
\affiliation{School of Physics, National University of Ireland Galway, University Road, Galway, Ireland}
\author{R.~Mukherjee} 
\affiliation{Department of Physics and Astronomy, Barnard College, Columbia University, NY 10027, USA}
\author{S.~O'Brien} 
\affiliation{School of Physics, University College Dublin, Belfield, Dublin 4, Ireland}
\author{R.~A.~Ong} 
\affiliation{Department of Physics and Astronomy, University of California, Los Angeles, CA 90095, USA}
\author{A.~N.~Otte} 
\affiliation{School of Physics and Center for Relativistic Astrophysics, Georgia Institute of Technology, 837 State Street NW, Atlanta, GA 30332-0430}
\author{D.~Pandel} 
\affiliation{Department of Physics, Grand Valley State University, Allendale, MI 49401, USA}
\author{N.~Park} 
\affiliation{Enrico Fermi Institute, University of Chicago, Chicago, IL 60637, USA}
\author{A.~Petrashyk} 
\affiliation{Physics Department, Columbia University, New York, NY 10027, USA}
\author{M.~Pohl} 
\affiliation{Institute of Physics and Astronomy, University of Potsdam, 14476 Potsdam-Golm, Germany}\affiliation{DESY, Platanenallee 6, 15738 Zeuthen, Germany}
\author{A.~Popkow} 
\affiliation{Department of Physics and Astronomy, University of California, Los Angeles, CA 90095, USA}\affiliation{Department of Physics and Astronomy, University of Hawaii at Manoa, Honolulu, HI 96822, USA}
\author{E.~Pueschel} 
\affiliation{DESY, Platanenallee 6, 15738 Zeuthen, Germany}
\author{J.~Quinn} 
\affiliation{School of Physics, University College Dublin, Belfield, Dublin 4, Ireland}
\author{K.~Ragan} 
\affiliation{Physics Department, McGill University, Montreal, QC H3A 2T8, Canada}
\author{P.~T.~Reynolds} 
\affiliation{Department of Physical Sciences, Cork Institute of Technology, Bishopstown, Cork, Ireland}
\author{G.~T.~Richards} 
\affiliation{School of Physics and Center for Relativistic Astrophysics, Georgia Institute of Technology, 837 State Street NW, Atlanta, GA 30332-0430}
\author{E.~Roache} 
\affiliation{Fred Lawrence Whipple Observatory, Harvard-Smithsonian Center for Astrophysics, Amado, AZ 85645, USA}
\author{J.~Rousselle} 
\affiliation{Department of Physics and Astronomy, University of California, Los Angeles, CA 90095, USA}
\author{C.~Rulten} 
\affiliation{School of Physics and Astronomy, University of Minnesota, Minneapolis, MN 55455, USA}
\author{I.~Sadeh} 
\affiliation{DESY, Platanenallee 6, 15738 Zeuthen, Germany}
\author{M.~Santander} 
\affiliation{Department of Physics and Astronomy, Barnard College, Columbia University, NY 10027, USA}
\author{G.~H.~Sembroski} 
\affiliation{Department of Physics and Astronomy, Purdue University, West Lafayette, IN 47907, USA}
\author{K.~Shahinyan} 
\affiliation{School of Physics and Astronomy, University of Minnesota, Minneapolis, MN 55455, USA}
\author{J.~Tyler} 
\affiliation{Physics Department, McGill University, Montreal, QC H3A 2T8, Canada}
\author{V.~V.~Vassiliev} 
\affiliation{Department of Physics and Astronomy, University of California, Los Angeles, CA 90095, USA}
\author{S.~P.~Wakely} 
\affiliation{Enrico Fermi Institute, University of Chicago, Chicago, IL 60637, USA}
\author{J.~E.~Ward} 
\affiliation{Institut de Fisica d’Altes Energies (IFAE), The Barcelona Institute of Science and Technology, Campus UAB, 08193 Bellaterra (Barcelona), Spain}
\author{A.~Weinstein} 
\affiliation{Department of Physics and Astronomy, Iowa State University, Ames, IA 50011, USA}
\author{R.~M.~Wells} 
\affiliation{Department of Physics and Astronomy, Iowa State University, Ames, IA 50011, USA}
\author{P.~Wilcox} 
\affiliation{Department of Physics and Astronomy, University of Iowa, Van Allen Hall, Iowa City, IA 52242, USA}
\author{A.~Wilhelm} 
\affiliation{Institute of Physics and Astronomy, University of Potsdam, 14476 Potsdam-Golm, Germany}\affiliation{DESY, Platanenallee 6, 15738 Zeuthen, Germany}
\author{D.~A.~Williams} 
\affiliation{Santa Cruz Institute for Particle Physics and Department of Physics, University of California, Santa Cruz, CA 95064, USA}
\author{B.~Zitzer} 
\affiliation{Physics Department, McGill University, Montreal, QC H3A 2T8, Canada}

\begin{abstract}

We present results from deep observations towards the Cygnus region using 300 hours of very-high-energy (VHE) \GR\ data taken with the \veritas\ Cherenkov telescope array and over seven years of high-energy \GR\ data taken with the \fermi\ satellite at an energy above 1~GeV.
As the brightest region of diffuse \GR\ emission in the northern sky, the Cygnus region provides a promising area to probe the origins of cosmic rays. 
We report the identification of a potential \LAT\ counterpart to \TeVJ\ (TeV J2032+4130), and resolve the extended VHE source \cisne\ into two source candidates (\cisneA\ and \cisneB ) and characterize their energy spectra.
The \LAT\ morphology of 3FGL J2021.0+4031e (the Gamma-Cygni supernova remnant) was examined and a region of enhanced emission coincident with \GCyg\ was identified and jointly fit with the \veritas\ data.
By modeling 3FGL J2015.6+3709 as two sources, one located at the location of the pulsar wind nebula CTB 87 and one at the quasar QSO J2015+371, a continuous spectrum from 1 GeV to 10 TeV was extracted for \CTB\ (CTB 87).
An additional 71 locations coincident with \LAT\ sources and other potential objects of interest were tested for VHE \GR\ emission, with no emission detected and upper limits on the differential flux placed at an average of 2.3\% of the Crab Nebula flux. 
We interpret these observations in a multiwavelength context and present the most detailed \GR\ view of the region to date.
\end{abstract}


\section{Introduction}
\label{sec:1}
Very-high-energy (VHE, E $>$ 100 GeV ($10^{11}$ eV)) \GR s provide insights into the most extreme environments in our local Universe. 
Produced by the interactions of relativistic particles, this radiation enables the study of the non-thermal astrophysical processes by which these particles are accelerated. 

The Cygnus region is the brightest region of diffuse high-energy (HE, 0.1 GeV ($10^{8}$ eV) $<$ E $<$ 100 GeV (10$^{11}$ eV)) \GR s in the northern sky.
Seen as a small-scale version of a whole galaxy, the Cygnus region harbors a wealth of objects including over ten supernova remnants (SNRs) \citep{2014BASI...42...47G}, more than fourteen pulsars \citep{Manchester:2004bp}, nine OB associations, as well as numerous pulsar wind nebulae (PWNe), HII regions, Wolf-Rayet binary systems,  microquasars, dense molecular clouds, and a superbubble. 
Cygnus-X, a large, diffuse region (roughly 5\arcdeg~$\times$~5\arcdeg\ centered at \hm{20}{31}, \dm{40}{20}) of bright radio emission, is one of the richest known regions of star formation in the Galaxy, with OB associations that have a total stellar mass as high as $10^6$ M$_{\odot}$ \citep{Reipurth:2008uh} and a total mechanical stellar wind energy output of $\geq 10^{39}$ erg s$^{-1}$ \citep{Lozinskaya:2002rc}, corresponding to several percent of the energy output by supernova remnants in the entire Galaxy \citep{Verschuur:1988}. 
This makes the Cygnus region the largest known star-forming region in the Galaxy outside the Galactic center and this, combined with its proximity, is thought to be the reason for its brightness in HE \GR s. 

The Cygnus region has already been observed by various instruments at different wavelengths, including radio observations by the Canadian Galactic Plane Survey (CGPS) \citep{Taylor:2003cp} and the Giant Metrewave Radio Telescope \citep{Paredes:2009vi}, infrared observations by the \spitzer\ space telescope \citep{Benjamin:2003rm, Rieke:2004sp, Churchwell:2009sp},  and X-ray observations by \emph{Chandra} \citep{Butt:2003xc, Butt:2005dx}, \xmm\ \citep{Horns:2007xq} and \suzaku\ \citep{Murakami:2011de,2017arXiv170502733M}. 
These observations have highlighted the variety of objects and processes within the region and firmly established it as a key region for understanding our Galaxy.

In the HE waveband, 36 \GR\ sources have been detected by the \LAT\ in the region covered by this analysis (described in \autoref{sec:VERObs}) and published in the \fermi\ Large Area Telescope Third Source Catalog (3FGL, \citealt{TheFermi-LAT:2015hja}) of which seven are pulsars and one is the large star forming region, the Cygnus cocoon.
Of these sources, twelve appear in the 3FHL: The Third Catalog of Hard Fermi-LAT Sources (3FHL, \citealt{2017ApJS..232...18A}), a catalog of sources above 10~GeV, of which five are pulsars, and three in the Second \LAT\ Catalog of High-Energy Sources (2FHL, \citealt{2015arXiv150804449T}), a catalog above 50~GeV.

The Cygnus region has been observed by several VHE \GR\ instruments including \hegra\ \citep{Aharonian:2005ak}, \milagro\ \citep{Abdo:2007ad}, \argo\ \citep{Bartoli:2012tj}, \magic\ \citep{2007ApJ...665L..51A,2010ApJ...721..843A}, \veritas\ \citep{2013ApJ...779..150A,aliu_discovery_2013,aliu_observations_2014,aliu_spatially_2014} and \hawc\ \citep{2017arXiv170202992A}.
In the region defined in \autoref{sec:VERObs}, seven VHE sources have already been detected.
TeV J2032+4130 is an unidentified VHE emitter which lies within the extended Cygnus Cocoon.
\GCyg\ is also located within the Cygnus Cocoon and is associated with the Gamma-Cygni SNR (G78.2+2.1). 
The large, bright, unidentified \milagro\ source \MGROcisne\ has since been resolved into two sources after observations by \veritas: \CTB\ is associated with the SNR CTB 87, and \cisne\ is a spatially extended source whose origin has yet to be identified.
\hawc\ has recently published the 2HWC catalog \citep{2017arXiv170202992A}, their first catalog with the completed detector.
The catalog was produced using 507 days of data and identified three new sources in the survey region: 2HWC~J1953+294, which lies at the edge of the survey region; 2HWC~J2006+341; and 2HWC~J2024+417* (the * signifies that the source is separated from neighboring sources by a $\Delta(\sqrt{Test\; Statistic})$ of between 1 and 2).

\begin{figure*}[htb!]
\centering
\plotone{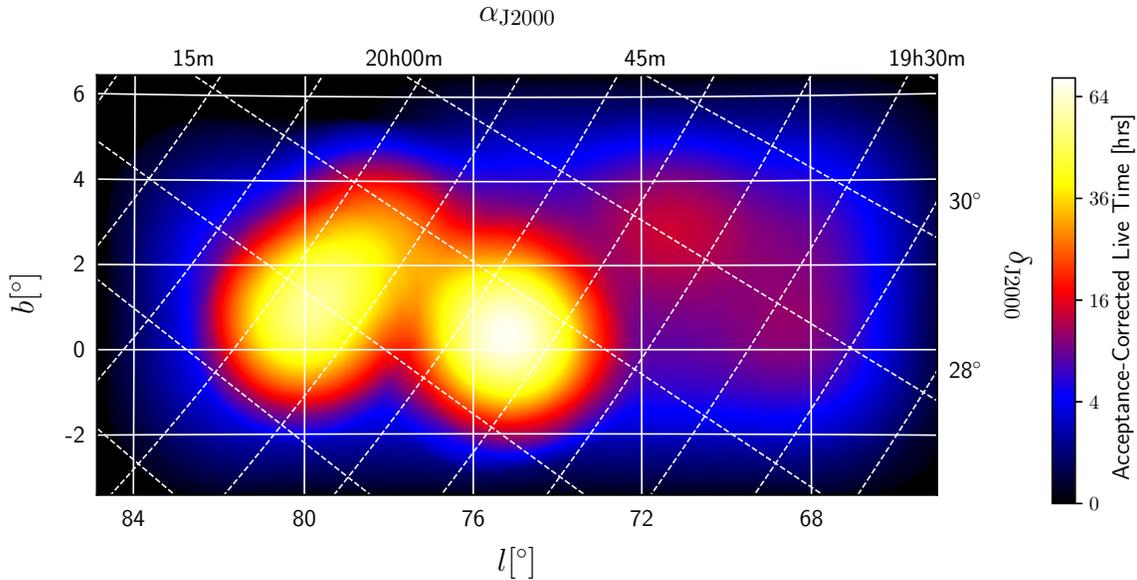}
\caption{The \veritas\ exposure map of the observations used in this analysis, in Galactic (longitude, $l$ , and latitude, $b$) and equatorial (right ascension, $\alpha_{J2000}$, and declination, $\delta_{J2000}$) coordinates, solid and dashed lines respectively. The color scale represents the acceptance-corrected live time in hours. \label{fig:Exposure} The bright regions correspond to the primary targets within the region, \MGROcisne, TeV~J2032+4130 and the Gamma-Cygni supernova remnant.}
\end{figure*}

In this paper we present a survey over a 15\arcdeg\ by 5\arcdeg\ portion of the Cygnus region centered on Galactic longitude ($l$) 74.5\arcdeg\ and Galactic latitude ($b$) 1.5\arcdeg\ conducted by \veritas\ between 2007 April and 2008 December with a total observing time of 135 h (120 h live time). 
We also include targeted and follow-up observations of 174~h (151~h live time) made by \veritas\ between 2008 November and 2012 June, for a total observing time of 309 h (271 h live time) distributed as shown in \autoref{fig:Exposure}. 
Initial results from these observations, including the deeper study of three regions have already been published by the \veritas\ collaboration \citep{aliu_discovery_2013, aliu_observations_2014, aliu_spatially_2014}. 
Here we report the results from the combined sample of survey and follow-up observations from \veritas .

After a description of \veritas\ , its dataset and analysis methods are presented in \autoref{sec:VERObs}. This is followed by a description of the \LAT\ and its data analysis in \autoref{sec:FermiAnalysis}. 
Descriptions of the \veritas\ results for the whole region are given in \autoref{sec:VERRes} and the \LAT\ results in \autoref{sec:FermiRes}.
This is followed by the detailed study of the individual VHE sources previously detected by \veritas\ in \autoref{sec:KnownSources}.

\section{\veritas\ Observations and Data Analysis}
\label{sec:VERObs}
\label{sec:VERAnalysis}
The Very Energetic Radiation Imaging Telescope Array System (\veritas) is an array of four imaging atmospheric Cherenkov telescopes (IACTs), located at the Fred Lawrence Whipple Observatory in southern Arizona (31\arcdeg 40\arcmin~N, 110\arcdeg 57\arcmin~W, 1.3~km a.s.l.). 
Each telescope is of Davies-Cotton design, with a 12-m diameter reflector comprised of 350 hexagonal mirror facets \citep{Holder:2006T1}. The focal length of each telescope is 12~m and each telescope is equipped with a camera consisting of 499 photomultiplier tube (PMT) ``pixels'' at the focus.
The angular spacing between PMTs is 0.15\arcdeg, which yields a total field of view (FoV) of 3.5\arcdeg.
Full array operations began in 2007 and, in the Summer of 2009, the first telescope was relocated to increase the sensitivity of the array \citep{T1}.
The trigger was upgraded in the winter of 2012. 
In summer 2012 the cameras in each telescope were replaced with new, high quantum efficiency PMTs which has resulted in a decrease of the array energy threshold to about 85~GeV \citep{Park:2015vts}. 
All the data presented in this work were taken prior to the 2012 camera upgrade using standard \veritas\ operational procedures. 

From 2007 April through 2008 December, \veritas\ undertook a survey of a 15\arcdeg\ by 5\arcdeg\ area of the Cygnus region, covering Galactic longitude ($l$) from 67\arcdeg\ to 82\arcdeg\ and Galactic latitude ($b$) from -1\arcdeg\ to 4\arcdeg. 
With a 1.75\arcdeg\ radius FoV, the \veritas\ exposure extends out to cover approximately a 19\arcdeg\ by 9\arcdeg\ area (65.5\arcdeg\ $<$ $l$ $<$ 83.5\arcdeg, -2.5\arcdeg\ $<$ $b$ $<$ 5.5\arcdeg).
The survey consisted of nearly 135 h (120 h live time) of observations, reaching an average point-source sensitivity of better than 4\% of the Crab Nebula flux  above 200~GeV. 
Runs were taken at pointed positions with a spacing of 0.8\arcdeg\ in the longitudinal direction and 1\arcdeg\ in the latitudinal direction. 
An exposure time of approximately one hour per grid point was taken, with that hour broken into three 20-minute observation runs, giving the survey region a relatively uniform exposure of about 6~h. 
Observations were scheduled to keep the average zenith angle close to 20\arcdeg\ to avoid the higher energy threshold associated with larger zenith angle observations. 

In addition to this survey, additional observations totaling 174 h (151 h live time) were conducted in this region, targeting objects of interest and following up on hot spots identified in the initial survey.
These observations were typically conducted in wobble mode \citep{1991ICRC....1..468F}, where the center of the FoV is offset from the target in one of the cardinal directions, allowing for simultaneous background estimation.

After removing periods affected by bad weather or hardware issues, 309 hours of data were taken.
Accounting for dead time results in a dataset totaling 271 hours of quality-selected live time at an average zenith angle of 20\arcdeg.
The acceptance-corrected exposure is depicted in \autoref{fig:Exposure}.

The results presented here were generated using one of the standard \veritas\ event reconstruction packages as described in \citet{Daniel:2007da}. 
The air shower images are parameterized using the Hillas moment analysis \citep{1985ICRC....3..445H} following calibration and image cleaning. 
Four analyses were conducted with four sets of selection criteria (``cuts'') used to identify good quality images and to discriminate between \GR s and the cosmic-ray background, these cuts consist of two different image intensity cuts and two different integration region sizes.
The two different image intensity cuts that were applied as part of the selection of good quality images were at a  medium threshold (about 70 photoelectrons) and at a high threshold (about 130 photoelectrons), giving minimum energy thresholds for these observations of around 200 and 400 GeV, respectively.
The minimum energy threshold depends on the observing setup used for that point in the sky (energy threshold is defined here as the peak in the differential counting rate, the effective area multiplied by a power law spectrum of -2).
For both image intensity cuts, a reconstructed image was required in a minimum of three telescopes and a constant image intensity cut was used across the survey region.
After selecting good quality images, selection cuts were applied on the reconstructed images to select \GR\ like events before two source searches were conducted for each of the image intensity cuts.
For a point source search, we used an integration radius of $\theta_{int}$ = 0.1\arcdeg\ (\point), and for an extended source search $\theta_{int}$ = 0.23\arcdeg\ (\ext).
In addition to selecting these cuts for optimum sensitivity to typical Galactic sources, they were chosen to reduce the impact of the numerous background stars in the region.
With the higher energy threshold cuts applied, the point spread function (PSF; 68\% containment radius) at 1~TeV was 0.1\arcdeg.

To produce a sky map of the region, the survey region was divided into a grid of trial source positions with grid points spaced by 0.025\arcdeg, a value well below the PSF of the instrument. 
For each of these points, for both the \point\ and \ext\ integration radii, a ring background model (RBM) analysis \citep{2007A&A...466.1219B} was conducted and the significance of the deviation from a background-only model was calculated using Equation 17 of \citet{LiMa:1983}. 
Spectral analysis was conducted using the reflected region (RR) method \citep{2007A&A...466.1219B}. 
In both cases regions around known sources and optically bright stars (magnitude brighter than 6) which cause a reduction in the local rate of events \citep{2007A&A...466.1219B} are excluded from the background regions.

The source extension was determined by fitting a sky map of the excess events with a two-dimensional Gaussian distribution convolved with the \veritas\ PSF.  
The PSF was modeled as a King function and fit to observations of the Crab Nebula, with a correction applied for the difference between the spectral indices of the examined source and the Crab Nebula. 
To estimate the systematic uncertainty, the source extension was also determined using different models of the PSF.
In addition the model generated using the Crab Nebula models were also generated using Markarian 421 observations and the PSF was also modeled as the sum of two symmetric Gaussian functions, giving four different models of the PSF.
The standard deviation of these four results is quoted as the systematic error.

For all sources except for \CTB, to conduct a spectral analysis, only data taken with four telescopes in operation and with a pointing offset of less than one degree were used, to reduce systematic errors on the energy reconstruction.
In the case of \CTB, due to the larger pointing offset of the majority of the data, the offset requirement was relaxed to observations within two degrees.
Due to the nature of the IACT technique, though the FoV is only 1.75\arcdeg\ radius, large images can be reconstructed with origins that lie outside of the FoV.  
This allows for the offset requirement to be larger than the FoV.
We found, through observations of the Crab Nebula and through simulations, that increasing the maximum offset up to two degrees increased the energy resolution and introduced a small ($<$ 10\%) bias in reconstructed energies.  To account for these effects, we have increased the systematic error on \CTB to $\pm0.4$ on the spectral index and $\pm40$\% in the flux normalization from the estimate of the error of $\pm0.2$ on the spectral index and $\pm20$\% for the rest of these observations.
All spectral points with a significance of at least 1$\sigma$ were fit with one of two spectral types, a power law (PL, \autoref{eq:PL}) or a log parabola (LP, \autoref{eq:LP}), also known as a curved power law, with an F-test conducted to determine whether there was significant evidence of curvature.

\begin{align}
F = \frac{dN}{dE} &= N_0  \left( \frac{E}{E_0} \right) ^{-\gamma} \label{eq:PL} \\
F = \frac{dN}{dE} &= N_0  \left( \frac{E}{E_b} \right) ^{-\gamma -\beta\ln(E/E_b)} \label{eq:LP}
\end{align}

Here, $F$ is the differential flux, $-\gamma$ is the spectral index, $\beta$ the curvature of the spectral index, $N_{0}$ the normalization, $E_{0}$ the energy normalization and $E_{b}$ the energy of the spectral break.
For fits to curved spectra we insist upon concavity, that is for the LP $\beta > 0$. 
Upper limits were calculated using the method of Rolke \citep{2005NIMPA.551..493R} at the 95\% level (statistical uncertainty only) and with an assumed spectral index of -2.5.

\section{\LAT\ Analysis}
\label{sec:FermiAnalysis}

The \textit{Fermi Gamma-ray Space Telescope} has been operating since its launch in 2008 with two \GR\ instruments: the Large Area Telescope (LAT) and the Gamma-ray Burst Monitor (GBM). 
The LAT \citep{2009ApJ...697.1071A}, \fermi 's primary instrument, is a pair conversion \GR\ detector that is sensitive to \GR s with energies from 20~MeV to greater than 500~GeV. 
The effective collecting area is approximately 6500~cm$^{2}$ at 1~GeV, and the angular resolution is strongly energy-dependent, with a 68\% containment radius of about 0.8\arcdeg\ at 1~GeV. 
Tables describing the energy resolution, effective area, and angular resolution are provided with the publicly available analysis tools.\footnote{\url{https://www.slac.stanford.edu/exp/glast/groups/canda/lat_Performance.htm}}

We have undertaken an analysis of the Cygnus region using over 7 years (2008 August - 2016 January) of \LAT\ Pass 8 data \citep{atwood_pass_2013} using \LAT\ science tools v10r0p5\footnote{\url{http://fermi.gsfc.nasa.gov/ssc/data/analysis/}} and the \fermipy\ tools v0.13.5 \citep{2017arXiv170709551W}.
In order to reduce the contribution of the Galactic diffuse emission and for improved angular resolution, ``FRONT+BACK'' events were selected in the energy range from 1 to 500~GeV. 
They were selected within a 30\arcdeg\ radius region centered at ($l$, $b$) = (74.5\arcdeg, 1.5\arcdeg). 
The region of interest was taken to be 65.5\arcdeg\ $<$ $l$ $<$ 83.5\arcdeg, -2.5\arcdeg\ $<$ $b$ $<$ 5.5\arcdeg\ to match the \veritas\ data. 
``SOURCE'' class photons were selected to maximize the effective area, while the corresponding IRF ``P8R2\_SOURCE\_V6'' was used with a maximum zenith angle of 90\arcdeg\ and only using good time intervals.
We used the binned analysis technique \citep{2009ApJS..183...46A},  implemented in the \emph{fermipy} routine \emph{optimize} to conduct an iterative fit to all of the sources and optimize the model parameters. 
It was then possible to calculate the significance of the source detection, the flux of the source, and the spectrum for sources in the region of interest.
The base model was derived from the third \LAT\ source catalog (3FGL) \citep{TheFermi-LAT:2015hja} using the provided templates for extended sources.\footnote{\url{http://fermi.gsfc.nasa.gov/ssc/data/access/lat/4yr\_catalog/LAT\_extended\_sources\_v15.tgz}} 
It was confirmed that all sources from the earlier \LAT\ high energy catalogs (1FHL \citep{collaboration_first_2013} and 2FHL \cite{2015arXiv150804449T}) in the region are also 3FGL sources (the 3FHL was published after this analysis was conducted and a comparison with the results is presented in \autoref{sec:FermiRes}).
For this analysis the Galactic diffuse emission model ``gll\_iem\_v06.fits'' and isotropic diffuse model ``iso\_P8R2\_SOURCE\_V6\_v06.txt'' provided by the \LAT\ collaboration were used.\footnote{\url{http://fermi.gsfc.nasa.gov/ssc/data/access/lat/BackgroundModels.html}} 
After an initial fit, sources with a Test Statistic (TS) less than 16.0 were removed from the subsequent fits. 
The \fermipy\ tool \emph{find\_sources} was used to identify new sources in the region of interest with a TS of at least 20, which were then added to the model. 
The region was then refit with sources below a minimum TS (which was incrementally increased to 25 over a number of fits) removed. 

As observed by the \LAT, up to a few 10s of GeV, by far the brightest sources in the region are the pulsars, but they all exhibit a spectral break at around a few GeV \citep{2013ApJS..208...17A} with an extrapolated flux that at 1000~GeV exceeds the \veritas\ sensitivity by several orders of magnitude.
The pulsed emission is therefore not expected to contribute at very high energies. 
In contrast, PWNe are very common VHE emitters \citep{2018A&A...612A...2H}, and nebulae associated with these bright pulsars might be detectable by \veritas. 
To search for potential PWN emission associated with the observed \veritas\ sources, a technique was employed to remove (or at least significantly reduce) the pulsar emission. 
To do this, a ``gated'' analysis was conducted, where a time cut was applied to the pulsar phase to remove the \on-pulse contribution. 
We used tempo2 with the \textit{Fermi}-LAT plugin \citep{2006MNRAS.369..655H,Ray:2011psr} to assign pulsar phases to the photons in the region for the pulsars using the timing models from \citet{kerr_timing_2015} for two pulsars that lie close to detected \veritas\ sources (3FGL J2021.1+3651, and 3FGL J2032.2+4126), and the resulting phaseograms were checked against the published results. 
The \on-pulse region was defined using these phaseograms to cover any \on-pulse and bridge emission, with the region defined conservatively to minimise the contamination of the pulsar flux into the \off-pulse region.
Then, using only data from the time periods covered by the pulsar ephemeris, we looked for steady emission from a putative PWN in the \off-pulse intervals.
In addition to this standard gated analysis, which suffers from reduced sensitivity due to the reduced exposure time after phase cuts and the limited time range of the publicly available ephemerides, we performed an additional analysis.
The aim of this analysis was to increase the exposure available, increasing the sensitivity and reducing the statistical errors.
This was performed by conducting an \on-pulse analysis on the data covered by the available ephemerides to determine the spectral parameters of the pulsar emission.
The parameters of the pulsar in the full model were then fixed to these parameters and the whole dataset (covering the full time range) was refit.
Nebula emission should then be apparent as a positive residual. 
For this method to work, the \on-pulse flux needs to be steady over time, thus we checked the catalog light curves and also produced new light curves of each of these objects to check for flux variability. 
The analysis method was not applied to 3FGL J2021.5+4026 (the Gamma-Cygni pulsar) due to its large flux variability over time and the lack of a clearly definable \off-pulse \citep{2013ApJ...777L...2A}. 
This did not unduly impact the analysis of the region since the Gamma-Cygni SNR was already detected as an extended object in the \LAT\ data \citep{Lande:2012fe} and is clearly distinguishable from the pulsar emission.  
Any sources found in this way were added to the model, the pulsar spectral parameters freed, and a new fit was conducted.
Provided the new source has a TS of at least 25 after being refit with the pulsar parameters free, it was kept in the model.

The  \LAT\ sources were fit with one of three spectral types: a power law (PL, \autoref{eq:PL}), a log parabola (LP, \autoref{eq:LP}), or (in the case of identified pulsars as in the 3FGL) a power law with exponential cutoff (PLEC, \autoref{eq:PLEC}).

\begin{equation}
 F = \frac{dN}{dE} = N_0 \left( \frac{E}{E_0} \right) ^{-\gamma} \exp \left( \frac{-E}{E_c} \right) \label{eq:PLEC}
\end{equation}
where $E_{c}$ the energy of the spectral cutoff.

A number of sources which were fit with a LP in the 3FGL were fit with a PL in this analysis because $\beta$ was either negative or consistent with zero within errors, which is likely due to the higher energy threshold in this analysis.
The decorrelation energy of each source with a power law spectrum was then calculated and the spectrum refit with the pivot energy set to the decorrelation energy.
This produced the base model from which all analyses were conducted.

For \LAT\ sources associated with \veritas\ sources, a $\chi^2$ fit was conducted to the spectral points for both the \veritas\ and \LAT\ data, considering statistical errors only.  Three different spectral models were tested, PL (\autoref{eq:PL}), LP (\autoref{eq:LP}) and broken power law (BPL, \autoref{eq:BPL}).

\begin{equation}
F = \frac{dN}{dE} =
  \begin{cases}
    N_0  \left( \frac{E}{E_b} \right) ^{-\gamma_1}  & \quad \text{if } E \leqslant E_b \\
    N_0  \left( \frac{E}{E_b} \right) ^{-\gamma_2}  & \quad \text{if } E >E_b . \\
  \end{cases} \label{eq:BPL} \\
\end{equation}
As mentioned previously, for fits to curved spectra we insist upon concavity, that is for the LP $\beta > 0$ and for the BPL $-\gamma_1 > -\gamma_2$.

\section{\veritas\ Results}
\label{sec:VERRes}
\begin{figure*}[htbp]
\centering
\gridline{
	\fig{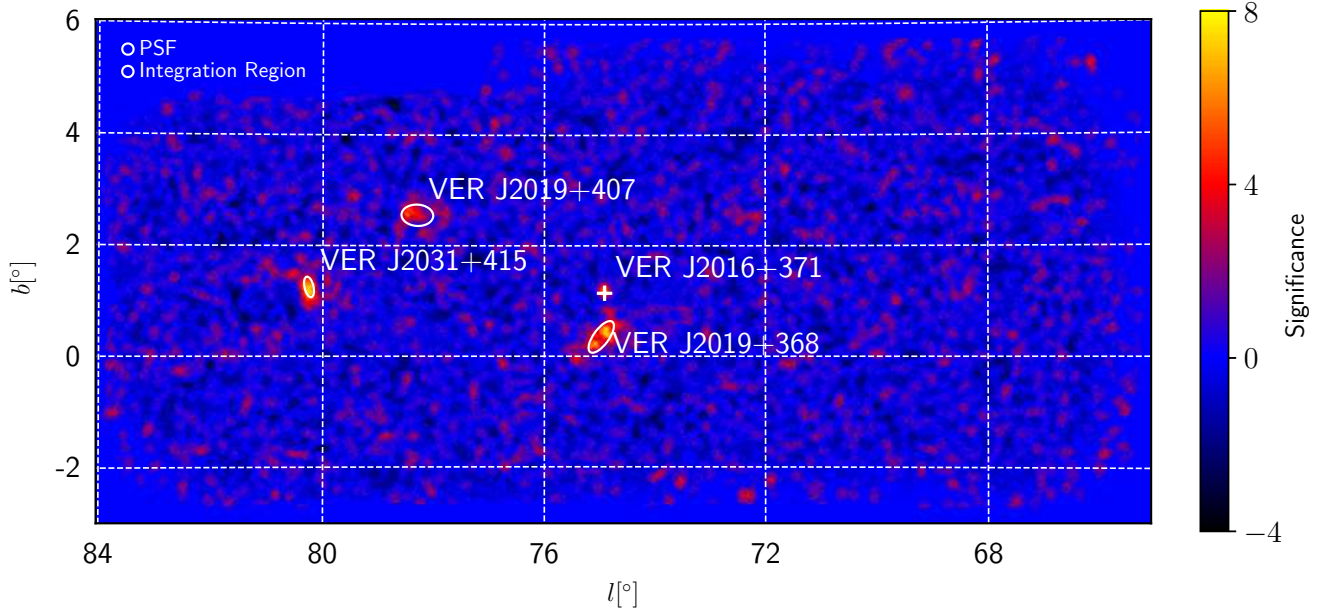}
	{0.98\textwidth}
	{(a) \point\ integration region (radius = 0.1\arcdeg)}}
\gridline{
	\fig{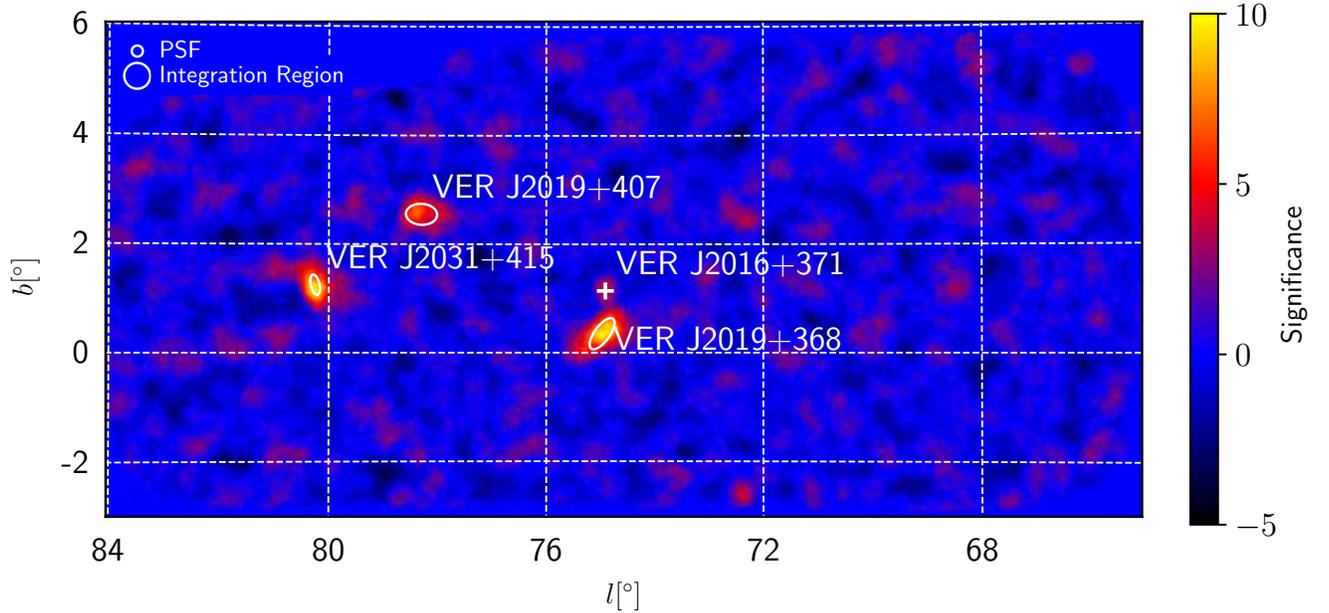}
	{0.98\textwidth}
	{(b) \ext\ integration region (radius = 0.23\arcdeg)}}
\caption{Significance sky maps above 400~GeV of the entire region using the \point\ (0.1\arcdeg) and \ext\ (0.23\arcdeg) integration radii.  
Significances were calculated using Equation 17 of \citet{LiMa:1983} and the ring background method.  Areas around known sources and bright stars were excluded from background regions.  Overlaid are the 1$\sigma$ ellipses for the source extension fits with an asymmetric Gaussian function for the three extended sources (\GCyg, \TeVJ, \cisne) and the position for \CTB\ (cross). }
\label{fig:VERSkymaps}
\end{figure*}

We used the ring background method (\autoref{sec:VERAnalysis}) over a grid of points at a spacing of 0.025\arcdeg\ and the higher energy image intensity cut to produce a sky map of the significances for the observed region.
This resulted in an average energy threshold of about 400~GeV which is roughly uniform across the survey region.
Examination of these significance sky maps (\autoref{fig:VERSkymaps}) produced with both integration region radii show the four known VHE sources as regions of high significance.
The extended sources \TeVJ, \GCyg, and \cisne\ were detected at 10.1, 7.6 and 10.3$\sigma$ respectively, using the \ext\ integration region and are described in more detail in Sections \ref{sec:TeVJ}, \ref{sec:GammaCyg} and \ref{sec:CisneRes}.
The point source \CTB\ was detected at  6.2$\sigma$ using the \point\ integration region and is described in \autoref{sec:CTB}.
For this analysis, the statistical significance is quoted at the pre-defined locations of these sources without accounting for statistical trials. 
Full descriptions of the analyses leading to the detection of each of these objects, and the statistical trials involved, are given in the relevant discovery papers. 

\begin{figure}[htb!]
\gridline{
	\fig{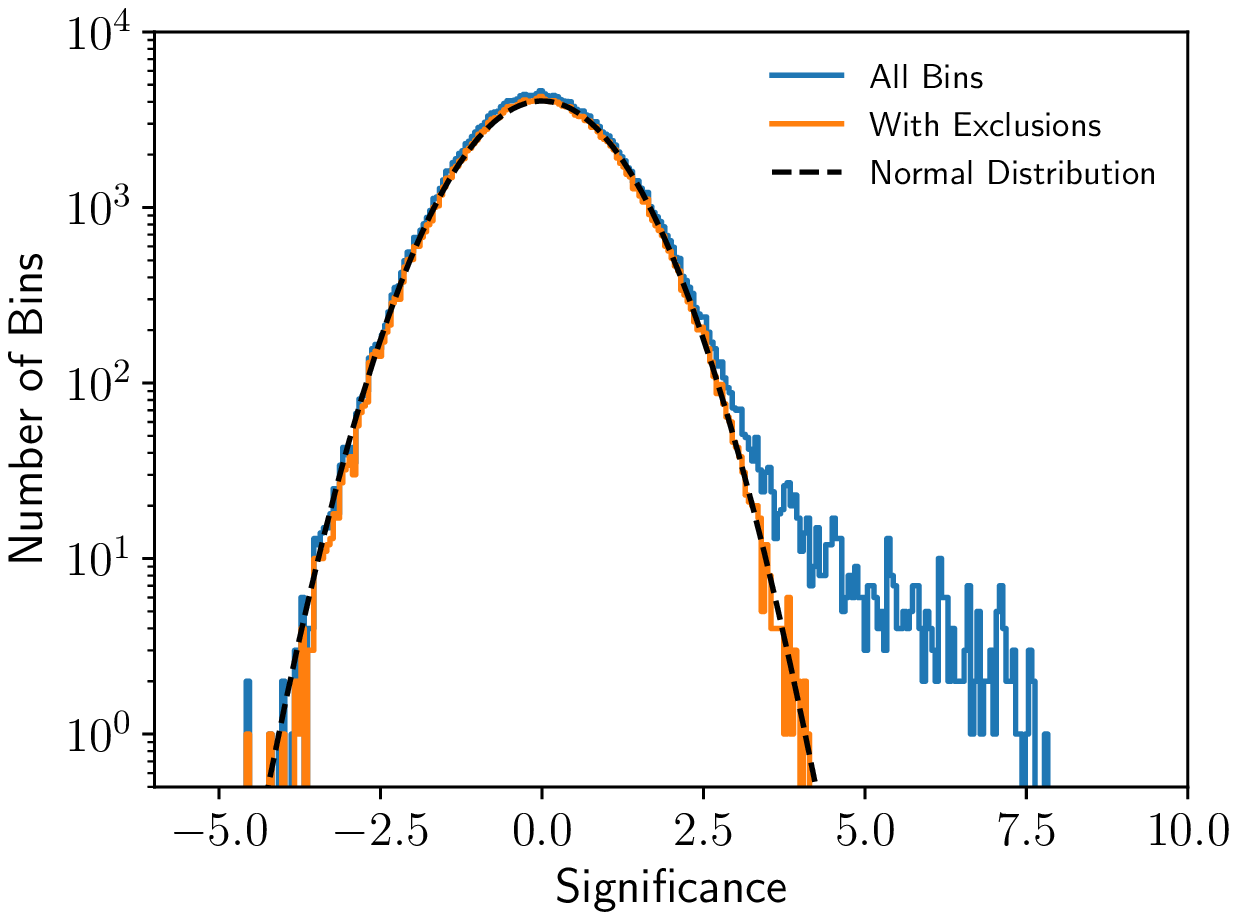}
	{0.48\textwidth}
 	{(a) \point\ integration region (radius = 0.1\arcdeg). Mean = 0.03, Standard Deviation = 1.00.}
	\fig{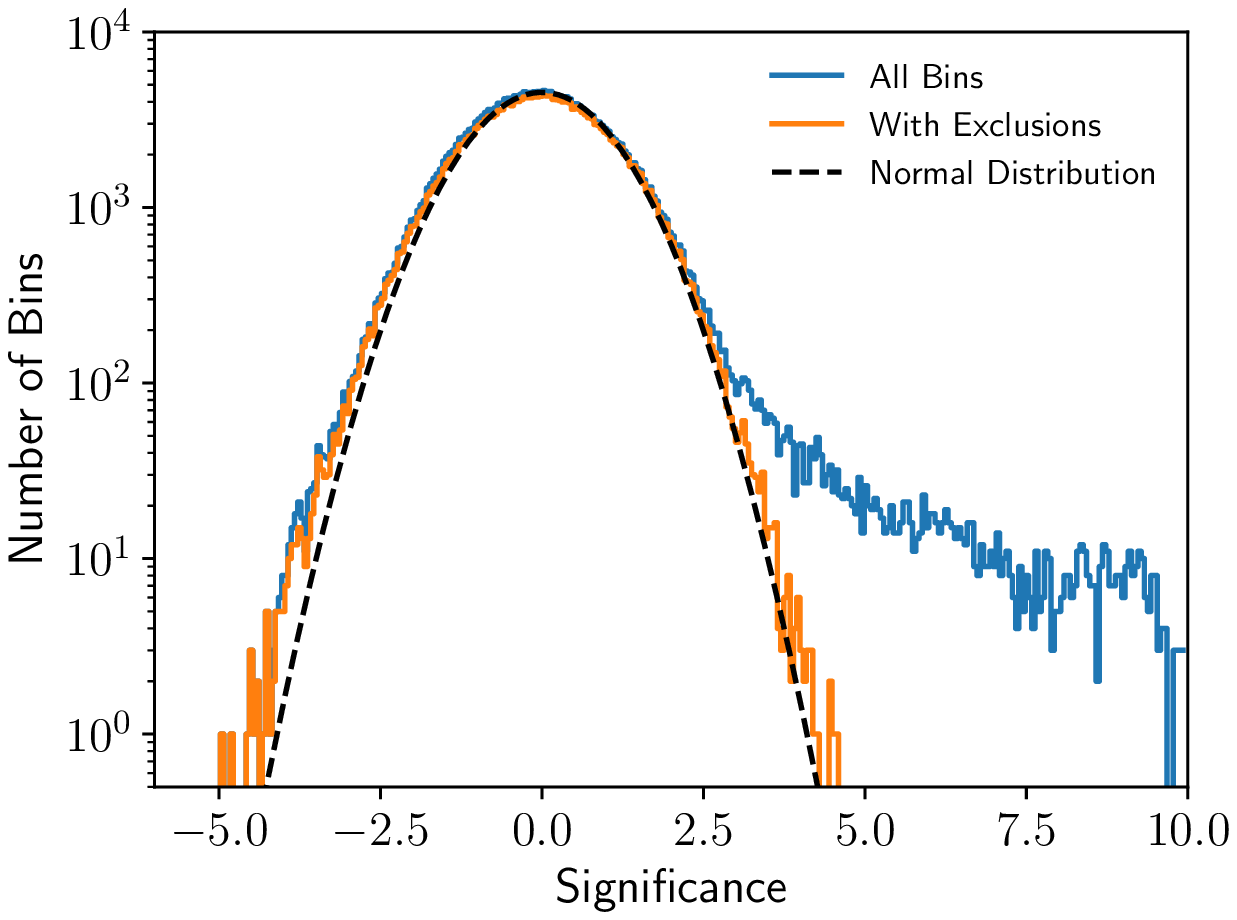}
	{0.48\textwidth}
 	{(b) \ext\ integration region (radius = 0.23\arcdeg). Mean = 0.03, Standard Deviation = 1.18.}
}
\caption{Histograms of the significances of the bins in \veritas\ sky maps (\autoref{fig:VERSkymaps}).  
Blue includes all the bins, orange all the bins that are not excluded due to their proximity to a source or a bright star, and black is a normal distribution (with mean = 0 and standard deviation = 1.0). The mean and standard deviation indicated for each figure correspond to the orange histogram. \label{fig:VEGASSigDist}}
 \end{figure}

The presence of \GR\ sources in these sky maps can be determined by examining a histogram of all of the bins in the significance sky map.
If the null hypothesis is true (that there is a uniform background and that the camera acceptance is well modeled), then the only variation in the significances of the bins should be statistical and they should be normally distributed.
In this case, we know that there are both \GR\ sources and bright stars which invalidate the null hypothesis, thus we only expect the significances to be normally distributed in the bins which lie away from these regions.  
We have produced two histograms, one with all of the bins in the sky map included, which should show significant variation from a normal distribution due to the stars and the \GR\ sources; and one with \GR\ sources and bright stars excluded, which should be normally distributed.
Examining these histograms (\autoref{fig:VEGASSigDist}) shows the presence of the known sources prior to their exclusion (blue line) but no additional sources after the known sources have been excluded (orange line).
The distribution with both the sources and bright stars excluded is close to a normal distribution (black line) with the mean approximately equal to zero and standard deviation approximately equal to one.
The wider distribution for the \ext\ analysis reflects the greater correlation between the bins and is common in such analyses.
With no clear positive excess there is no evidence for additional sources and the good fit to a normal distribution shows that the background and acceptance functions are well modeled and understood. 

To check for weak sources whose presence could be masked by the number of bins in the distribution, the survey region was broken up into smaller, overlapping $6\arcdeg \times 6\arcdeg$  tiles and the significance distributions from those regions were also examined.
Again, no evidence for any sources beyond the four previously detected sources was found, and the background was found to be well modeled for all of the regions.

\subsection{Upper Limits of Undetected Sources}
\label{sec:RestOfSurvey}
In addition to the four VHE sources previously detected, there are a number of other objects within the Cygnus region that are potential VHE \GR\ emitters.
These are objects that show significant, non-thermal emission in other wavebands, or belong to source classes which are either known to be or thought to be VHE \GR\ emitters.
Foremost amongst these objects are the two bright X-ray sources, Cygnus X-1 and X-3.

Cygnus X-1 is a high mass X-ray binary (HMXB) which consists of a (14.8\plm1.0)~M$_{\sun}$ black hole orbiting around a O9.7Iab companion of (19.2\plm1.9)~M$_{\sun}$. 
The binary system is located at a distance of 1.86$^{+0.12}_{-0.11}$~kpc \citep{2011ApJ...742...83R,2011ApJ...738...78X} in a circular orbit of 5.6 days and at an inclination of 27.1\plm0.8\arcdeg.

There has been no detection of steady VHE emission from Cygnus X-1 although \magic\ observations showed evidence of variable emission at the 4.9$\sigma$ level (4.1$\sigma$ after accounting for the number of statistical trials) for emission during 79 minutes effective on time between 2006 September 24 22h17 and 23h41 UTC \citep{2007ApJ...665L..51A}.
Further observations by \magic\ are reported in \citet{2017MNRAS.472.3474A} where they set integral upper limits at 95\% confidence level for energies above 200~GeV at 2.6$\times10^{-12}$ and 1.0$\times10^{-11}$ photons cm$^{−2}$ s$^{-1}$ for the hard and soft states respectively.
\veritas\ observations were presented in \citet{2009arXiv0908.0714G} giving a 99\% upper limit on the flux above 400~GeV of 1.05$\times 10^{−12}$ cm$^{-2}$~s$^{-1}$ was presented.
Evidence for HE emission was first found by \citet{2013MNRAS.434.2380M} at the 4$\sigma$ level using Pass 7 \LAT\ data.  This was for emission in the hard spectral state when the lower energy thermal component is weak and the emission is dominated by high energy X-rays.
Using Pass 8 data \citet{2016arXiv160505914Z} found emission correlated with the hard spectral state later on confirmed by \citet{Zdziarski2016}.

Cygnus X-3 was the first microquasar seen emitting in HE \GR s by AGILE \citep{Tavani:2009cx} and \LAT\ \citep{Fermi:2009cx} (although it does not appear in the 3FGL catalog due to flux variability, it is associated with 1FGL J2032.4+4057 and 2FGL J2032.1+4049). 
Unlike Cygnus X-1 where HE emission is associated with the hard spectral state and the presence of jets, in Cygnus X-3, HE emission is strongly anti-correlated with the hard X-ray emission, is emitted prior to major radio flares and is associated with discrete-blob jets \citep{2012MNRAS.421.2947C,2012A&A...545A.110P}.
Cygnus X-3 is a HMXB where the compact object powering the system is either a neutron star in an unusual state of accretion, or a black hole of 10-20 M$_{\sun}$ orbiting around a Wolf-Rayet companion \citep{1992Natur.355..703V}.
Study of this system could shed light on the non-thermal processes associated with the formation of relativistic jets from accretion processes. 
Currently, the source is undetected at very high energies and flux upper limits have been set by \veritas\ and \magic, with the former reporting a limit of $0.7\times10^{-12}$ photons cm$^{-2}$ s$^{-1}$ above 263 GeV \citep{2013ApJ...779..150A} and the latter reporting an upper limit of $2.2\times 10^{-12}$ photons cm$^{-2}$ s$^{-1}$ above 250 GeV \citep{2010ApJ...721..843A}.
\veritas\ has also conducted observations during radio and \GR\ flaring episodes, placing a preliminary upper limit on the  flux at 3.11$\times 10^{-12}$ cm$^{-2}$~s$^{-1}$ above 500~GeV.
The results presented here for both Cygnus X-1 and X-3 are the average over all the observations, irrespective of the spectral state. 
Breakdowns by spectral state are beyond the scope of this work.

In addition to these two well known HMXBs there are other high and low mass X-ray binaries in the region along with a number of other potential sources of VHE emission including PWNe, SNRs, and colliding wind binaries (CWB).
In this paper we have generated a list of PWNe/SNRs using Green's catalog \citep{2014BASI...42...47G} and cross checked it with ``A census of high-energy observations of Galactic supernova remnants''\footnote{\url{http://www.physics.umanitoba.ca/snr/SNRcat}} \citep{2012AdSpR..49.1313F}.
We developed lists of binaries from the catalogs produced by \citet{2006A&A...455.1165L,2007A&A...469..807L} and \citet{2013A&A...558A..28D}.
The list of sources presented here is not intended to be exhaustive; rather, the sources have been selected by the authors of this paper as potential VHE emitters.
All of the sources from the \LAT\ analysis in this work (\autoref{sec:FermiRes}) are also included.
In addition, two sources from the 3FHL (3FHL J1950.5+3457 and 3FHL J2026.7+3449) that are not associated with sources in the 3FGL are included (3FHL J2015.9+3712 which lies within the \CTB\ region is discussed in \autoref{sec:CTB}) along with the three undetected \hawc\ sources.
If two upper limit locations lie within 0.05\arcdeg\ (typically because the source is also detected by the \LAT) then the more accurate radio/X-ray position is taken as the nominal position.

In total 71 locations were tested, the upper limits are presented in \autoref{tab:UpperLimits} and, for all of the \fermi\ sources, they are also plotted on the relevant \LAT\ SEDs in \autoref{fig:Fermi3FGLSpectra}.
In the majority of cases, the upper limits do not constrain an extrapolation of the \LAT\ flux with the remaining sources having upper limits that lie within approximately 1$\sigma$ of the extrapolated flux.
The mean upper limit was 2.3\% (2.9\%) of the Crab Nebula flux\footnote{Throughout this paper, the Crab Nebula flux is taken from \citet{2004ApJ...614..897A}.} for the \point\ (\ext) integration regions above 600~GeV.
The most significant location tested is G69.7+1.0, which has $\sigma_{local}$ of 3 for the \point\ integration region and 2.3 for the \ext\ integration region.
Since we have tested 71 locations for upper limits, a corresponding trials factor (calculated using \autoref{eq:Trials Factor}) needs to be imposed, after applying this correction we determined  $\sigma_{global}$ values of 1.3 and 0.6 respectively.

2HWC J1953+294 lies at the edge of the survey region, only covered by a few runs at a large offset in this analysis, with 109 minutes of live time.
At its location, the RBM significance is 0.75$\sigma$ (-0.12$\sigma$) for the \point\ (\ext) analysis.
\citet{2016arXiv160902881H} presented an analysis of 37 hours of \veritas\ data that covers this source (most of which is not included in this paper) and found a statistically significant \GR\ source located within the \hawc\ source contours. 
This emission is coincident with the PWN DA 495 (G65.7+1.2) and its central object, WGA J1952.2+2925.
Further work on this region will form part of a forthcoming publication.

2HWC J2006+341 was not detected in this analysis with the RBM analysis (0.93$\sigma$ and 1.61$\sigma$ for the \point\ and \ext\ analyses respectively) in 513 minutes of live time.
Locally the significance reached 2.73$\sigma$ (2.78$\sigma$) for the \point\ (\ext) analysis at ($l$, $b$) =  (71.07\arcdeg, 1.04\arcdeg) ((71.51\arcdeg, 1.24\arcdeg)) lying at a distance of 0.29\arcdeg\ (0.20\arcdeg) from the source.
At the location of 2HWC~J2024+417* an RBM analysis showed significances of 0.93$\sigma$  (1.61$\sigma$) for the \point\ (\ext) analysis with 2428 minutes of live time.

Upper limits on the fluxes from WR 146 and WR 147 were reported in \citet{2008ApJ...685L..71A} at $5.6\times 10 ^{-13}$ cm$^{-2}$s$^{-1}$ and $7.3\times 10 ^{-13}$ cm$^{-2}$s$^{-1}$ above 600~GeV.  The results presented here represent an improvement by approximately a factor of two over both of these results.

\begin{longrotatetable}
\begin{deluxetable}{ccccccccccccccccccc}
\tabletypesize{\scriptsize}
\tablecaption{\veritas\ upper limits of the integral and differential flux for a variety of locations that could emit VHE \GR s.  
The upper limits were calculated using the method of Rolke \citep{2005NIMPA.551..493R} at the 95\% level and with an assumed spectral index of -2.5.
They have a mean decorrelation energy of 980~GeV and a standard deviation on the decorrelation energy of 100~GeV.
Sources identified as simply FGL are found in the analysis of \LAT\ data performed for this work and are reported here (and in \autoref{tab:Fermi_NewPS_Results}) for the first time.
\on\ are the counts from within the integration region, \off\ are the counts from the background regions, $\alpha$ is the ratio of the number of signal to background regions. 
\textit{Sig.} is the significance calculated using the RR method (\autoref{sec:VERAnalysis}, in $\sigma$), \textit{Energy Thresh.} is the energy threshold (in GeV), \textit{Int. UL} is the integral upper limit in units of $\mathrm{cm}^{-2}\mathrm{s}^{-1}$ between the energy threshold and 3$\times 10^{4}$~GeV and \textit{Diff. UL} is the differential upper limits  in units of $\mathrm{cm}^{-2}\mathrm{s}^{-1}\mathrm{GeV}^{-1}$ and at an energy of 1000~GeV.
SNR = supernova remnant, HMXB = high mass X-ray binary, LMXB = low mass X-ray binary, CWB = particle-accelerating colliding-wind binary, BCU = active galaxy of uncertain type, U = unknown.
\label{tab:UpperLimits}}

\tablehead{
\multirow{3}{*}{Source}  & \multirow{3}{*}{$l$[\arcdeg]}  & 
\multirow{3}{*}{$b$[\arcdeg]}  & 
\multirow{3}{*}{Class}  &  \colhead{Live}  & 
\multicolumn{7}{c}{\point} & \multicolumn{7}{c}{\ext} \\
\colhead{} & \colhead{} & \colhead{} & 
\colhead{} & \colhead{Time} &
\multirow{2}{*}{\on} & \multirow{2}{*}{\off} & \multirow{2}{*}{$\alpha$} & 
\multirow{2}{*}{Sig.} &  \colhead{Energy}  &  \colhead{Int.} &  \colhead{Diff.} &
\multirow{2}{*}{\on} & \multirow{2}{*}{\off} & \multirow{2}{*}{$\alpha$} & 
\multirow{2}{*}{Sig.}   &  \colhead{Energy}  &  \colhead{Int.}  &  \colhead{Diff.} \\
\colhead{} & \colhead{} & \colhead{} & 
\colhead{} & \colhead{[min]} &
\colhead{} & \colhead{} & \colhead{} &
\colhead{} & \colhead{Thresh.} & \colhead{UL} & \colhead{UL} & \colhead{} &
\colhead{} & \colhead{} & \colhead{} &
\colhead{Thresh.} & \colhead{UL} & \colhead{UL} 
}

\startdata
 2HWC~J1953+294 & 65.85 & 1.07 & U & 109 & 4 & 50 & 0.027 & 1.8 & 790 & 3.16E-12 & 3.37E-15 & 10 & 98 & 0.073 & 0.8 & 790 & 2.6E-12 & 2.78E-15 \\
 2HWC~J2006+341 & 71.32 & 1.16 & U & 513 & 12 & 214 & 0.038 & 0.9 & 500 & 9.7E-13 & 5.16E-16 & 67 & 392 & 0.117 & 1.6 & 500 & 1.98E-12 & 1.06E-15 \\
 2HWC~J2024+417* & 79.75 & 2.21 & U & 2428 & 69 & 1503 & 0.042 & 0.8 & 600 & 3.65E-13 & 2.57E-16 & 280 & 1945 & 0.137 & 1.9 & 600 & 6.66E-13 & 4.68E-16 \\
 3A~1954+319 & 68.39 & 1.93 & LMXB & 702 & 22 & 372 & 0.037 & 1.6 & 660 & 9.35E-13 & 7.58E-16 & 57 & 648 & 0.096 & -0.8 & 550 & 6.57E-13 & 4.02E-16 \\
 3FGL~J1951.6+2926 & 65.67 & 1.32 & SPP & 109 & 2 & 53 & 0.025 & 0.5 & 790 & 2.2E-12 & 2.35E-15 & 8 & 109 & 0.068 & 0.2 & 870 & 1.89E-12 & 2.31E-15 \\
 3FGL~J1958.6+2845 & 65.88 & -0.35 & PSR & 89 & 2 & 66 & 0.029 & 0.1 & 660 & 2.28E-12 & 1.84E-15 & 7 & 131 & 0.091 & -1.4 & 660 & 1.48E-12 & 1.19E-15 \\
 3FGL~J2004.4+3338 & 70.67 & 1.19 & U & 504 & 7 & 215 & 0.037 & -0.6 & 550 & 4.39E-13 & 2.69E-16 & 60 & 387 & 0.109 & 1.6 & 500 & 1.98E-12 & 1.06E-15 \\
 3FGL~J2018.5+3851 & 76.59 & 1.66 & BCU & 2336 & 43 & 995 & 0.042 & 0.2 & 600 & 2.98E-13 & 2.1E-16 & 238 & 1784 & 0.125 & 0.3 & 600 & 4.63E-13 & 3.26E-16 \\
 3FGL~J2018.6+4213 & 79.4 & 3.53 & U & 1034 & 32 & 614 & 0.046 & 0.4 & 660 & 5.3E-13 & 4.28E-16 & 170 & 951 & 0.16 & 0.3 & 660 & 8.65E-13 & 7.02E-16 \\
 3FGL~J2021.5+4026 & 78.23 & 2.08 & PSR & 2129 & 69 & 1029 & 0.063 & 0.5 & 660 & 3.54E-13 & 2.85E-16 & 338 & 1896 & 0.18 & -0.2 & 660 & 3.9E-13 & 3.15E-16 \\
 3FGL~J2023.5+4126 & 79.25 & 2.34 & U & 3305 & 89 & 1952 & 0.039 & 0.9 & 600 & 3.9E-13 & 2.75E-16 & 484 & 3486 & 0.112 & 2.6 & 600 & 8.38E-13 & 5.9E-16 \\
 3FGL~J2025.2+3340 & 73.1 & -2.41 & BCU & 101 & 0 & 35 & 0.025 & 0 & 550 & 1.34E-12 & 8.24E-16 & 3 & 74 & 0.069 & -0.8 & 550 & 1.78E-12 & 1.09E-15 \\
 3FGL~J2028.5+4040c & 79.19 & 1.13 & U & 3785 & 89 & 1867 & 0.049 & 0.5 & 550 & 1.72E-13 & 1.05E-16 & 471 & 3023 & 0.151 & 1.2 & 550 & 2.35E-13 & 1.44E-16 \\
 3FGL~J2030.0+3642 & 76.13 & -1.43 & PSR & 974 & 26 & 439 & 0.045 & 1.4 & 600 & 6.93E-13 & 4.87E-16 & 105 & 777 & 0.149 & -0.1 & 600 & 4.23E-13 & 2.97E-16 \\
 3FGL~J2030.8+4416 & 82.35 & 2.89 & PSR & 251 & 8 & 201 & 0.047 & -1 & 660 & 6.7E-13 & 5.41E-16 & 39 & 355 & 0.1 & 0.5 & 550 & 3.06E-12 & 1.87E-15 \\
 3FGL~J2032.5+3921 & 78.57 & -0.27 & U & 652 & 18 & 550 & 0.039 & -0.8 & 550 & 5.14E-13 & 3.14E-16 & 104 & 938 & 0.12 & -1 & 550 & 6.47E-13 & 3.97E-16 \\
 3FGL~J2032.5+4032 & 79.51 & 0.44 & U & 3239 & 100 & 1466 & 0.057 & 1.9 & 550 & 4.75E-13 & 2.91E-16 & 415 & 2294 & 0.183 & 0.6 & 550 & 4.07E-13 & 2.49E-16 \\
 3FGL~J2034.4+3833c & 78.16 & -1.04 & U & 270 & 7 & 285 & 0.037 & -1.2 & 550 & 7.09E-13 & 4.34E-16 & 51 & 500 & 0.107 & -0.5 & 550 & 1.48E-12 & 9.02E-16 \\
 3FGL~J2034.6+4302 & 81.77 & 1.6 & U & 760 & 29 & 655 & 0.037 & 1.1 & 600 & 9.44E-13 & 6.68E-16 & 146 & 1339 & 0.107 & 0.7 & 600 & 1E-12 & 7.03E-16 \\
 3FGL~J2035.0+3634 & 76.63 & -2.32 & U & 54 & 0 & 49 & 0.032 & 0 & 720 & 1.1E-12 & 1.02E-15 & 6 & 90 & 0.111 & -1 & 720 & 2.24E-12 & 2.08E-15 \\
 3FGL~J2038.4+4212 & 81.53 & 0.54 & U & 1880 & 46 & 1139 & 0.04 & 0.7 & 600 & 2.94E-13 & 2.07E-16 & 190 & 2079 & 0.123 & -2.5 & 600 & 1.05E-13 & 7.41E-17 \\
 3FGL~J2039.4+4111 & 80.83 & -0.21 & U & 2015 & 40 & 1107 & 0.045 & -0.8 & 550 & 2.02E-13 & 1.24E-16 & 260 & 2090 & 0.14 & 0 & 550 & 2.76E-13 & 1.69E-16 \\
 3FGL~J2042.4+4209 & 81.93 & -0.07 & U & 462 & 17 & 383 & 0.039 & 0.1 & 660 & 8.21E-13 & 6.64E-16 & 77 & 774 & 0.11 & -0.6 & 660 & 6.17E-13 & 4.98E-16 \\
 3FHL~J1950.5+3457 & 70.3 & 4.3 & U & 282 & 3 & 102 & 0.041 & -0.7 & 790 & 4.33E-13 & 4.62E-16 & 27 & 182 & 0.124 & -0.3 & 790 & 7.82E-13 & 8.35E-16 \\
 3FHL~J2026.7+3449 & 74.21 & -1.99 & U & 152 & 0 & 85 & 0.038 & 0 & 660 & 4.11E-13 & 3.31E-16 & 17 & 143 & 0.128 & 0.3 & 600 & 1.35E-12 & 9.49E-16 \\
 Cygnus~X-1 & 71.33 & 3.07 & HMXB & 871 & 13 & 254 & 0.061 & -1.1 & 600 & 1.86E-13 & 1.3E-16 & 107 & 416 & 0.2 & 1 & 600 & 8.85E-13 & 6.21E-16 \\
 Cygnus~X-3 & 79.85 & 0.7 & HMXB & 1269 & 52 & 268 & 0.136 & 2.3 & 500 & 7.47E-13 & 3.98E-16 & 475 & 1551 & 0.301 & 1.4 & 550 & 6.69E-13 & 4.1E-16 \\
 EXO 2030+375 & 77.15 & -1.24 & HMXB & 359 & 12 & 263 & 0.042 & 0.2 & 600 & 1.06E-12 & 7.45E-16 & 70 & 468 & 0.137 & 0.2 & 660 & 1.44E-12 & 1.17E-15 \\
 G65.7+1.2 & 65.7 & 1.2 & SNR & 37 & 1 & 25 & 0.026 & 0.4 & 790 & 3.73E-12 & 3.98E-15 & 6 & 43 & 0.074 & 1.3 & 790 & 6.33E-12 & 6.75E-15 \\
 G65.8-0.5 & 65.8 & -0.5 & SNR & 89 & 1 & 75 & 0.028 & -0.8 & 660 & 1.52E-12 & 1.23E-15 & 7 & 152 & 0.078 & -1.5 & 660 & 1.55E-12 & 1.25E-15 \\
 G66.0-0.0 & 66 & -0 & SNR & 126 & 2 & 94 & 0.031 & -0.6 & 660 & 1.36E-12 & 1.1E-15 & 11 & 175 & 0.092 & -1.3 & 720 & 1.16E-12 & 1.08E-15 \\
 G67.6+0.9 & 67.6 & 0.9 & SNR & 624 & 14 & 326 & 0.038 & -0.4 & 660 & 4.37E-13 & 3.54E-16 & 61 & 638 & 0.095 & -0.6 & 660 & 6.1E-13 & 4.93E-16 \\
 G67.7+1.8 & 67.7 & 1.8 & SNR & 609 & 13 & 314 & 0.038 & 0.2 & 660 & 5.15E-13 & 4.17E-16 & 68 & 635 & 0.099 & 0.1 & 660 & 8.38E-13 & 6.75E-16 \\
 G67.8+0.5 & 67.8 & 0.5 & SNR & 537 & 18 & 238 & 0.05 & 1.3 & 600 & 8.93E-13 & 6.28E-16 & 95 & 443 & 0.152 & 2 & 600 & 1.8E-12 & 1.27E-15 \\
 G68.6-1.2 & 68.6 & -1.2 & SNR & 424 & 5 & 229 & 0.037 & -1.5 & 600 & 3.99E-13 & 2.8E-16 & 29 & 383 & 0.1 & -2 & 550 & 5.97E-13 & 3.66E-16 \\
 G69.0+2.7 & 69 & 2.7 & SNR & 582 & 11 & 279 & 0.036 & -0.1 & 550 & 5.88E-13 & 3.6E-16 & 59 & 496 & 0.099 & 0.6 & 550 & 1.26E-12 & 7.68E-16 \\
 G69.7+1.0 & 69.68 & 1.01 & SNR & 655 & 23 & 278 & 0.036 & 3 & 550 & 1.65E-12 & 1.02E-15 & 80 & 549 & 0.099 & 2.3 & 550 & 2.05E-12 & 1.26E-15 \\
 G73.9+0.9 & 73.9 & 0.9 & SNR & 2633 & 52 & 727 & 0.059 & 0.6 & 600 & 3.16E-13 & 2.23E-16 & 229 & 949 & 0.205 & 0.9 & 550 & 7.05E-13 & 4.34E-16 \\
 G76.9+1.0 & 76.89 & 0.97 & SNR & 2028 & 42 & 811 & 0.049 & 0.4 & 550 & 4.04E-13 & 2.48E-16 & 198 & 1136 & 0.191 & -0.2 & 600 & 3.07E-13 & 2.16E-16 \\
 G83.0-0.3 & 83 & -0.3 & SNR & 108 & 4 & 91 & 0.042 & 0 & 660 & 1.7E-12 & 1.38E-15 & 30 & 192 & 0.125 & 1 & 660 & 3.85E-12 & 3.11E-15 \\
 GS~2023+338 & 73.12 & -2.09 & LMXB & 101 & 1 & 38 & 0.032 & -0.2 & 550 & 1.64E-12 & 1.01E-15 & 10 & 82 & 0.09 & 0.8 & 550 & 3.39E-12 & 2.07E-15 \\
 HD~190603 & 69.49 & 0.39 & CWB & 471 & 14 & 209 & 0.044 & 1.2 & 550 & 1.09E-12 & 6.66E-16 & 44 & 436 & 0.118 & -1.7 & 550 & 4.39E-13 & 2.69E-16 \\
 KS~1947+300 & 66.01 & 2.08 & HMXB & 165 & 1 & 83 & 0.03 & -1 & 790 & 6.26E-13 & 6.67E-16 & 17 & 163 & 0.082 & 0.8 & 790 & 1.88E-12 & 2.01E-15 \\
 FGL~J1949.0+3412 & 69.49 & 4.21 & U & 338 & 7 & 144 & 0.036 & 0.6 & 660 & 9.5E-13 & 7.67E-16 & 29 & 241 & 0.091 & 0.9 & 660 & 1.74E-12 & 1.4E-15 \\
 FGL~J1955.0+3319 & 69.36 & 2.68 & U & 417 & 10 & 183 & 0.043 & 0.3 & 550 & 7.84E-13 & 4.8E-16 & 50 & 294 & 0.13 & 0.8 & 550 & 1.49E-12 & 9.11E-16 \\
 FGL~J2005.7+3417 & 71.36 & 1.3 & U & 644 & 14 & 272 & 0.036 & 0.9 & 500 & 8.73E-13 & 4.66E-16 & 78 & 519 & 0.11 & 1.4 & 500 & 1.76E-12 & 9.38E-16 \\
 FGL~J2009.9+3544 & 73.04 & 1.36 & U & 537 & 8 & 218 & 0.04 & -0.6 & 500 & 4.58E-13 & 2.44E-16 & 62 & 345 & 0.126 & 1.4 & 500 & 1.81E-12 & 9.65E-16 \\
 FGL~J2017.3+3526 & 73.62 & -0.07 & U & 2501 & 38 & 1004 & 0.044 & -0.8 & 600 & 1.41E-13 & 9.9E-17 & 211 & 1548 & 0.144 & -1.1 & 550 & 2.44E-13 & 1.49E-16 \\
 FGL~J2018.1+4111 & 78.47 & 3.03 & U & 2077 & 73 & 883 & 0.071 & 1.3 & 720 & 3.8E-13 & 3.52E-16 & 316 & 1305 & 0.245 & 0.3 & 660 & 4.2E-13 & 3.4E-16 \\
 FGL~J2022.6+3727 & 75.88 & 0.21 & U & 4606 & 91 & 1455 & 0.059 & -0.3 & 550 & 1.36E-13 & 8.34E-17 & 402 & 2182 & 0.171 & 0.7 & 550 & 3.27E-13 & 2E-16 \\
 FGL~J2024.4+3957 & 78.13 & 1.35 & U & 1443 & 36 & 597 & 0.057 & -0 & 600 & 2.91E-13 & 2.05E-16 & 183 & 804 & 0.208 & 0.9 & 550 & 7.69E-13 & 4.71E-16 \\
 FGL~J2025.9+3904 & 77.58 & 0.61 & U & 818 & 25 & 359 & 0.057 & 1.2 & 550 & 8.74E-13 & 5.35E-16 & 108 & 527 & 0.207 & 0.7 & 550 & 1.05E-12 & 6.43E-16 \\
 FGL~J2029.4+3940 & 78.46 & 0.41 & U & 1088 & 26 & 681 & 0.036 & 0.3 & 550 & 6.15E-13 & 3.77E-16 & 161 & 1248 & 0.111 & 1.8 & 550 & 1.67E-12 & 1.02E-15 \\
 FGL~J2031.3+3857 & 78.1 & -0.32 & U & 490 & 17 & 391 & 0.044 & -0.1 & 550 & 7.53E-13 & 4.62E-16 & 98 & 658 & 0.143 & 0.9 & 550 & 1.55E-12 & 9.47E-16 \\
 FGL~J2032.7+4333 & 81.96 & 2.19 & U & 540 & 14 & 498 & 0.039 & -1.2 & 600 & 4.59E-13 & 3.23E-16 & 83 & 941 & 0.115 & -1.9 & 600 & 4.53E-13 & 3.19E-16 \\
 FGL~J2032.9+3956 & 79.08 & 0.03 & U & 2788 & 49 & 1578 & 0.032 & -0.3 & 550 & 2.75E-13 & 1.68E-16 & 265 & 2710 & 0.096 & -0.2 & 550 & 4.37E-13 & 2.68E-16 \\
 FGL~J2034.3+4219 & 81.14 & 1.23 & U & 3284 & 74 & 1821 & 0.038 & 0.6 & 550 & 2.73E-13 & 1.67E-16 & 376 & 3214 & 0.108 & 0.8 & 550 & 4.32E-13 & 2.65E-16 \\
 FGL~J2036.9+4314 & 82.16 & 1.39 & U & 393 & 17 & 294 & 0.039 & 1.1 & 660 & 1.21E-12 & 9.77E-16 & 78 & 537 & 0.116 & 0.8 & 660 & 1.26E-12 & 1.02E-15 \\
 FGL~J2037.0+4005 & 79.67 & -0.53 & U & 1750 & 26 & 890 & 0.037 & -1.6 & 550 & 1.81E-13 & 1.11E-16 & 168 & 1509 & 0.124 & -1.9 & 550 & 2.52E-13 & 1.55E-16 \\
 FGL~J2037.6+4152 & 81.15 & 0.47 & U & 2253 & 79 & 1216 & 0.048 & 2.8 & 550 & 6.57E-13 & 4.02E-16 & 323 & 2159 & 0.16 & 1.1 & 550 & 2.37E-13 & 1.46E-16 \\
 FGL~J2038.8+4235 & 81.85 & 0.72 & U & 809 & 18 & 648 & 0.034 & -1 & 600 & 3.53E-13 & 2.48E-16 & 123 & 1301 & 0.1 & -0.2 & 600 & 4.97E-13 & 3.5E-16 \\
 FGL~J2040.1+4152 & 81.43 & 0.1 & U & 1898 & 46 & 1201 & 0.035 & 1.1 & 600 & 4.37E-13 & 3.08E-16 & 230 & 2391 & 0.103 & 0.8 & 550 & 3.62E-13 & 2.21E-16 \\
 FGL~J2054.6+4130 & 82.85 & -2.23 & U & 55 & 2 & 59 & 0.024 & 0.4 & 720 & 4.47E-12 & 4.15E-15 & 5 & 109 & 0.071 & -1 & 720 & 2.54E-12 & 2.36E-15 \\
 PSR~J1952+3252 & 68.76 & 2.82 & PSR & 526 & 16 & 245 & 0.037 & 1.7 & 550 & 1.31E-12 & 8.03E-16 & 37 & 450 & 0.091 & -1.1 & 550 & 6.04E-13 & 3.7E-16 \\
 PSR~J2006+3102 & 68.67 & -0.53 & PSR & 424 & 10 & 234 & 0.048 & -0.4 & 550 & 5.53E-13 & 3.38E-16 & 54 & 355 & 0.156 & -0.6 & 550 & 8.02E-13 & 4.92E-16 \\
 WR~133 & 72.65 & 2.07 & CWB & 535 & 9 & 237 & 0.04 & -0.2 & 500 & 6.46E-13 & 3.45E-16 & 48 & 414 & 0.124 & -0.8 & 500 & 7.18E-13 & 3.83E-16 \\
 WR~137 & 74.33 & 1.09 & CWB & 3119 & 64 & 882 & 0.055 & 1.8 & 600 & 4.37E-13 & 3.07E-16 & 231 & 1547 & 0.143 & -0.1 & 600 & 4.02E-13 & 2.83E-16 \\
 WR~140 & 80.93 & 4.18 & CWB & 110 & 5 & 99 & 0.03 & 1 & 600 & 3.77E-12 & 2.66E-15 & 17 & 210 & 0.087 & -0.3 & 550 & 3.13E-12 & 1.91E-15 \\
 WR~146 & 80.56 & 0.44 & CWB & 3273 & 92 & 1194 & 0.07 & 1.9 & 550 & 3.01E-13 & 1.84E-16 & 273 & 1839 & 0.145 & 1.7 & 550 & 4.4E-13 & 2.69E-16 \\
 WR~147 & 79.85 & -0.31 & CWB & 2135 & 48 & 1049 & 0.043 & 0.2 & 550 & 3.61E-13 & 2.21E-16 & 272 & 1780 & 0.145 & 0.4 & 550 & 5.81E-13 & 3.56E-16 \\
 XTE~J2012+381 & 75.39 & 2.25 & LMXB & 1342 & 38 & 597 & 0.047 & 0.8 & 660 & 5.15E-13 & 4.16E-16 & 153 & 807 & 0.193 & -1.7 & 600 & 2.73E-13 & 1.92E-16 \\
\enddata
\end{deluxetable}
\end{longrotatetable}

\begin{figure}[htb!]
\centering
\gridline{
	\fig{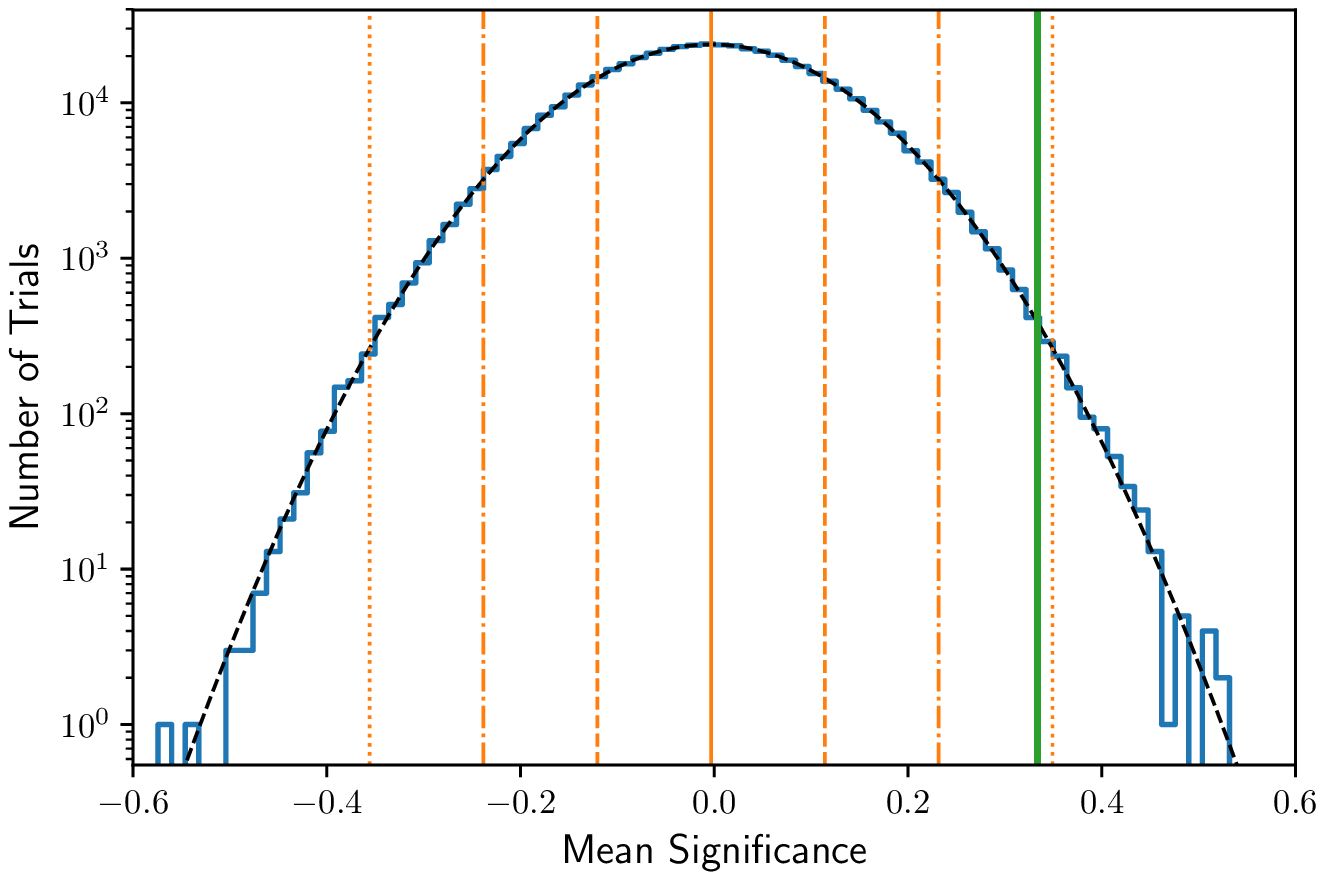}
	{0.4\textwidth}
	{(a) \point\ integration region}
	\fig{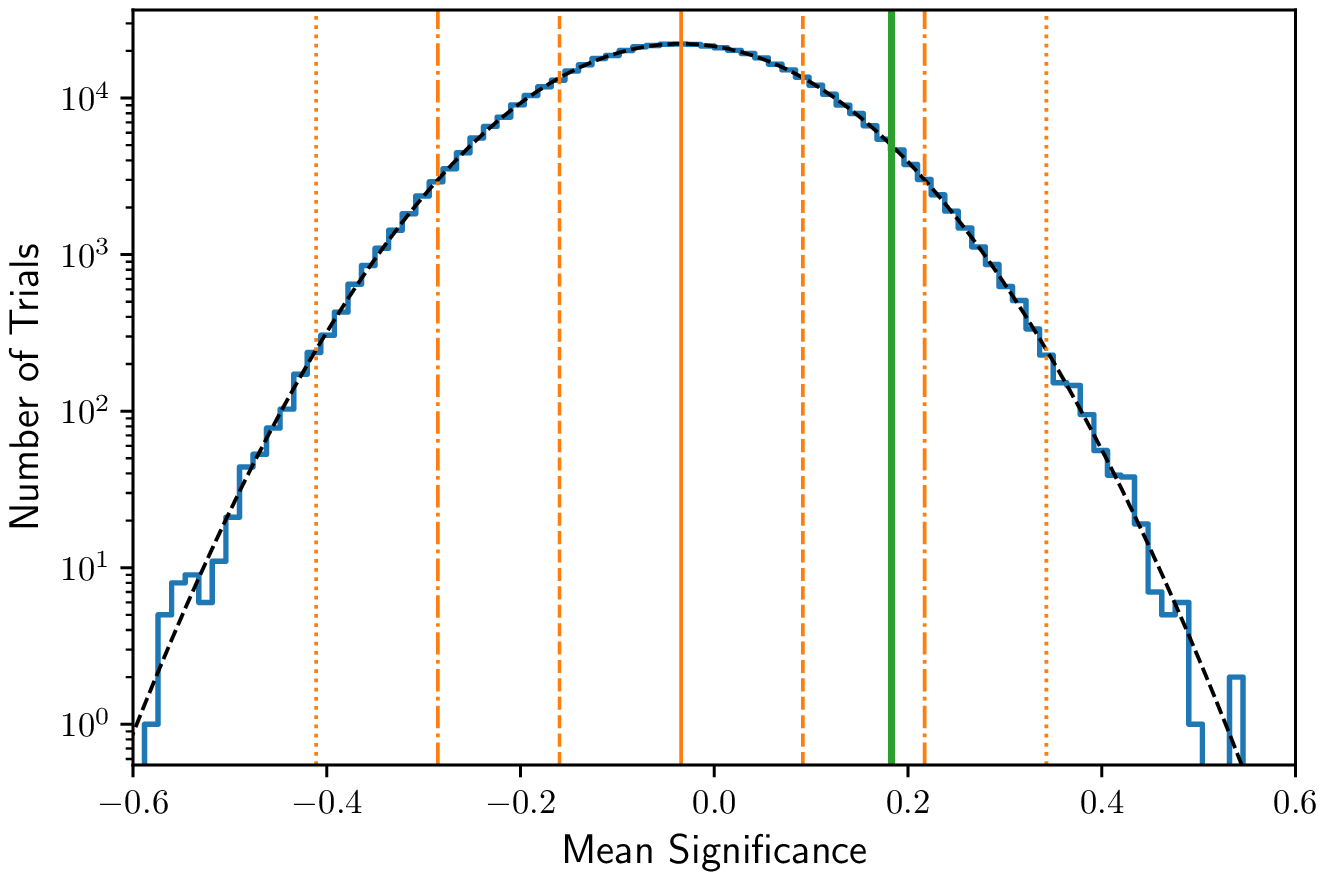}
	{0.4\textwidth}
	{(b) \ext\ integration region}
	\fig{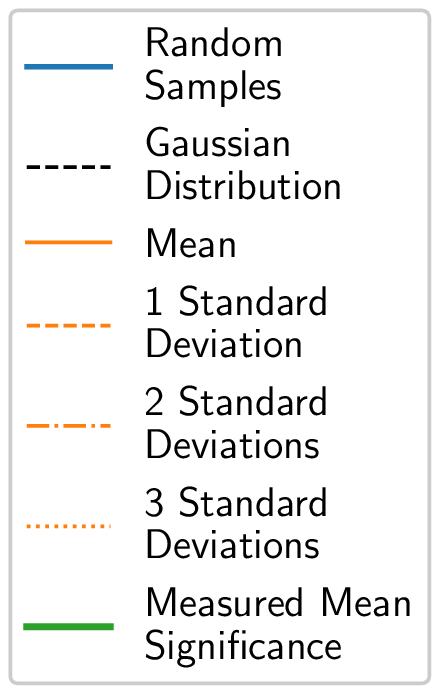}
	{0.16\textwidth}
	{}}
\caption{Histograms of mean significances calculated from 71 random locations outside the source exclusion regions in the significance sky map repeated 50,000 times for both the \point\ (a) and \ext\ (b) integration regions.
The blue line shows the results of the test; the black dashed line is the best fit Gaussian distribution.
The mean and the one, two, and three standard deviations of the distribution are shown in orange.
Note that the mean significance is slightly less than zero in both instances, reflecting the fact that the locations of bright stars have not been excluded from the possible locations of the randomly drawn locations.
The mean of the significance of the upper limit locations is marked by the green line and in both cases is clearly offset from the mean, corresponding to a 2.9$\sigma$ and 1.7$\sigma$ fluctuations for the \point\ and \ext\ integration regions respectively.
\label{fig:ULSimulation}}
\end{figure}

It is interesting to examine whether there is any evidence of emission that is not detectable for any of the locations individually.
If so, this would show up as a positive mean significance across all 71 results.
The mean significances for all of the upper limit positions are 0.33 and 0.18 for the \point\ and \ext\ integration regions respectively.
To test whether these positive mean significances are significant deviations from the expected mean significances of zero comparisons are made with the same number of significances drawn at random from the significance sky maps shown in \autoref{sec:VERRes} but with the VHE sources masked out. 
This was repeated 500,000 times using both the \point\ and \ext\ sky maps (\autoref{fig:ULSimulation}).
From the \point\ sky map a mean significance of at least 0.33$\sigma$ occurred 1029 times, giving a probability of occurrence of $2.1\times 10^{-3}$ (a 2.9$\sigma$ fluctuation), whereas for the \ext\ sky map a deflection of at least 0.18$\sigma$ occurred 21091 times with probability $4.22\times 10^{-2}$ (a 1.7$\sigma$ fluctuation).
There is no significant evidence for additional sources with fluxes that are just below the sensitivity of this work.

\begin{figure*}[htbp]
\centering
\gridline{
	\fig{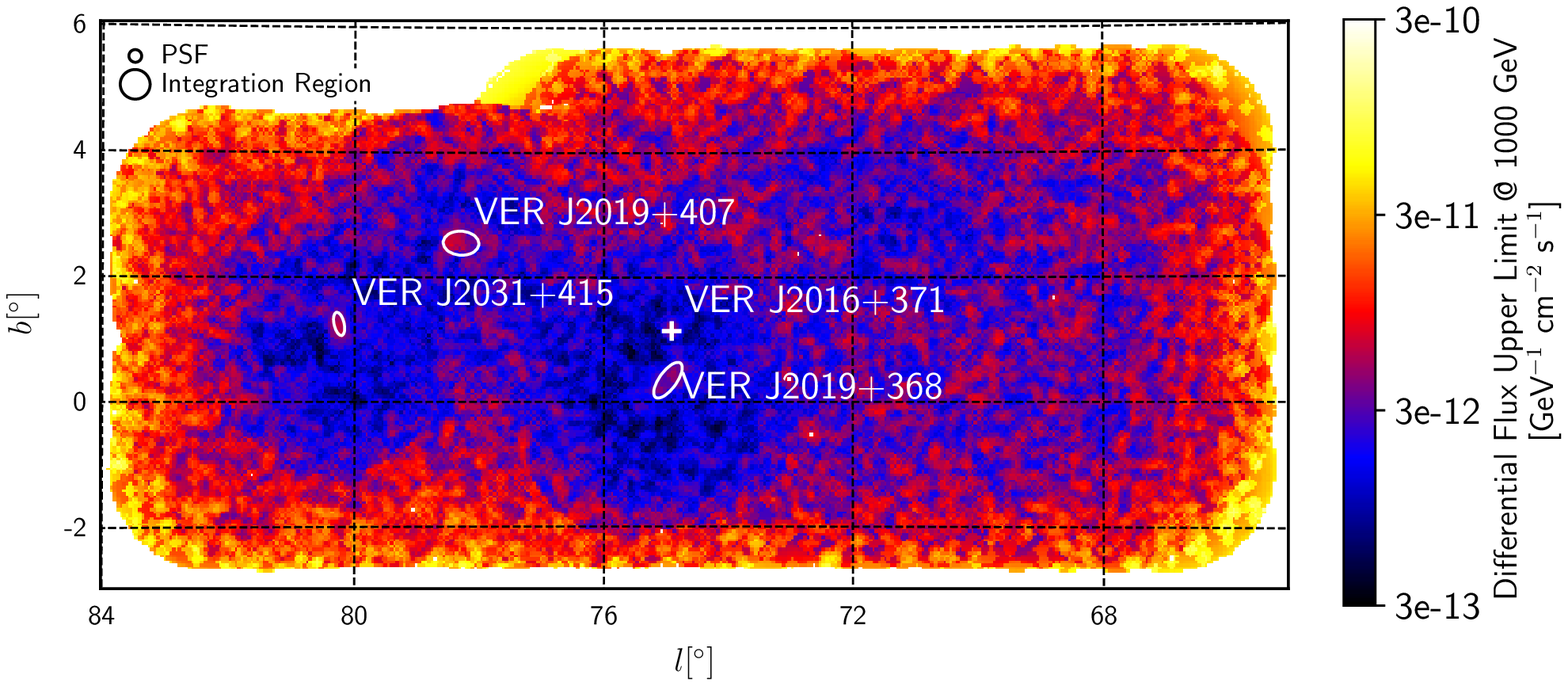}
	{0.98\textwidth}
	{(a) \point\ integration region (radius = 0.1\arcdeg)}}
\gridline{
	\fig{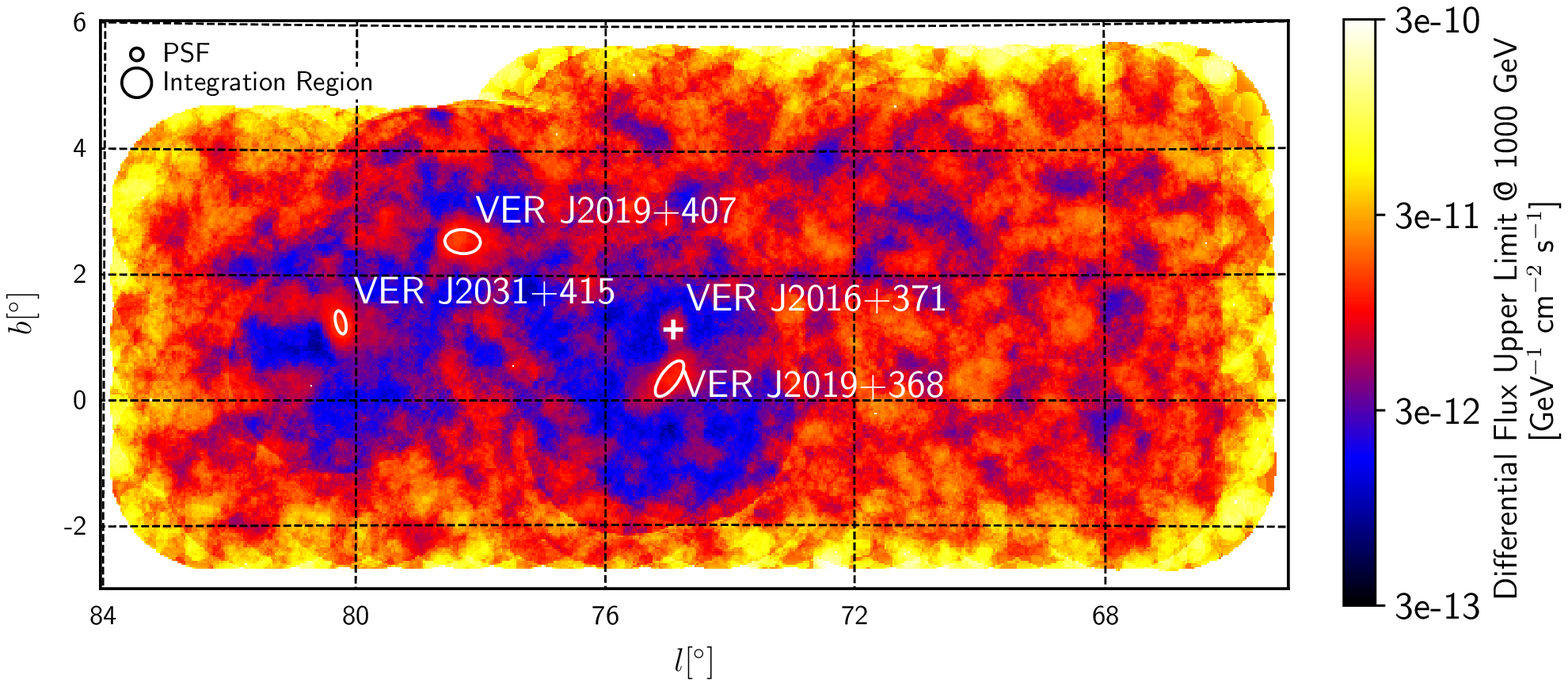}
	{0.98\textwidth}
	{(b) \ext\ integration region (radius = 0.23\arcdeg)}}
\caption{Map of the 95\% upper limits on the differential flux at 1000~GeV using the \point\ (0.1\arcdeg) and \ext\ (0.23\arcdeg) integration radii.  
Upper limits were calculated using the method of  Rolke \citep{2005NIMPA.551..493R} at the 95\% level and with an assumed spectral index of -2.5 and using the ring background method.  Areas around known sources and bright stars were excluded from background regions.  Overlaid are the 1$\sigma$ ellipses for the source extension fits with an asymmetric Gaussian function for the three extended sources (\GCyg, \TeVJ, \cisne) and the position for \CTB\ (cross). }
\label{fig:VERULSkymaps}
\end{figure*}

In addition to calculating upper limits at specific locations for the chosen sources we calculated an upper limit map for each of the integration regions. Upper limits were calculated using the method of  Rolke \citep{2005NIMPA.551..493R} at the 95\% level (statistical uncertainty only) and with an assumed spectral index of -2.5 and using the ring background method and are presented in \autoref{fig:VERULSkymaps}.

\section{\LAT\ Results}
\label{sec:FermiRes}

\begin{figure*}[htb!]
\centering
\plotone{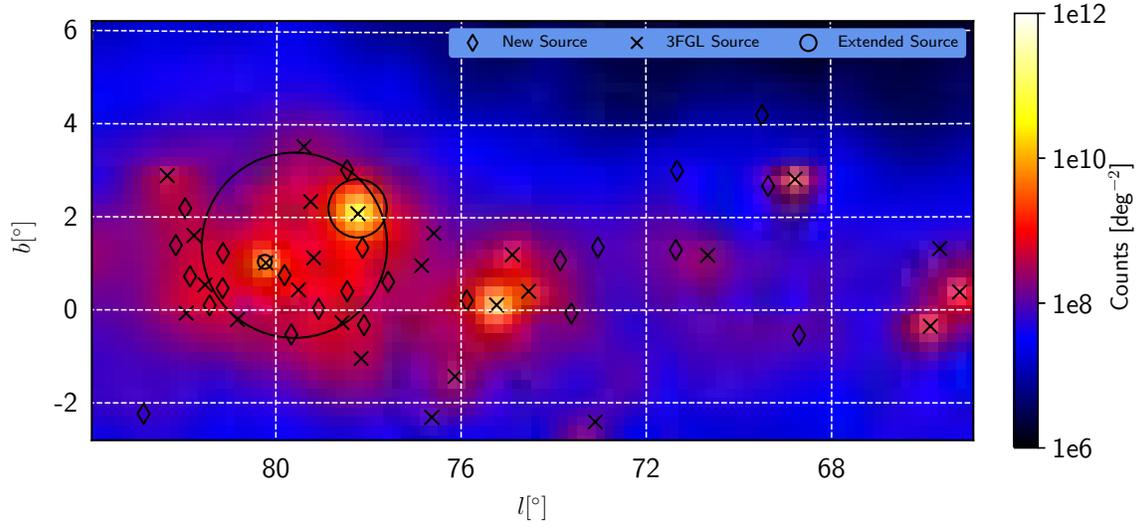}
\caption{$>$ 1 GeV counts map of the entire region obtained with \LAT. 
Point sources of the 3FGL catalog that are significant in this analysis are shown with small black crosses, the extended sources are shown with larger circles showing their characteristic extension.  New sources identified in this analysis are shown with black diamonds.}
\label{fig:FermiCounts}
\end{figure*}

The \LAT\ counts map of the survey region is shown in \autoref{fig:FermiCounts} together with the locations (and extensions for extended sources) of the 3FGL sources and the new sources identified in this analysis. 
In the survey region 27 3FGL catalog sources were identified which overlap with eight 1FHL sources, and four 2FHL sources.
The 3FGL sources in this region that were removed from the model due to their low TS ($<$ 25) are listed in \autoref{sec:3FGLremoved}. 
In addition, 25 new point sources were identified within the survey region.
This number of new sources is not unexpected due to two main factors, the increase in the exposure time (7.5 years rather than 4 years) and the increased sensitivity of Pass 8 in comparison to Pass 7 (the increase in the differential sensitivity is about a factor of 1.25 above 1~GeV \citep{2013arXiv1303.3514A}).
Furthermore nine of the new sources lie within the Cygnus Cocoon.
It is noted that in the residual map of this analysis a deficit was located in the region around ($l$, $b$) = (81\arcdeg, 2.5\arcdeg).
The low-TS sources around the edge of the Cygnus Cocoon along with this deficit suggest that the symmetrical Gaussian model of the cocoon is overly simplistic.
However, producing a more detailed analysis of the region with an improved model of the Cygnus Cocoon is beyond the scope of this work.
The results for individual \fermi\ catalog sources are presented in \autoref{tab:Fermi_Results}, with their SEDs shown in \autoref{fig:Fermi3FGLSpectra}.
The results for the new sources identified in this analysis are presented in \autoref{tab:Fermi_NewPS_Results} and \autoref{fig:Fermi3FGLSpectra}. 

There are two 3FHL catalog sources that were not detected in this analysis, 3FHL J1950.5+3457 and 3FHL J2026.7+3449.
Both of these sources are close to the 3FHL detection threshold (5.5$\sigma$ and 4.2$\sigma$ respectively) and are hard (spectral indices of -1.8 and -1.9). 
Combined with the higher energy threshold of the 3FHL (10~GeV cf. 1~GeV), this could explain why neither of these sources is detected in this analysis.

\begin{figure}[htb!]
\gridline{
\fig{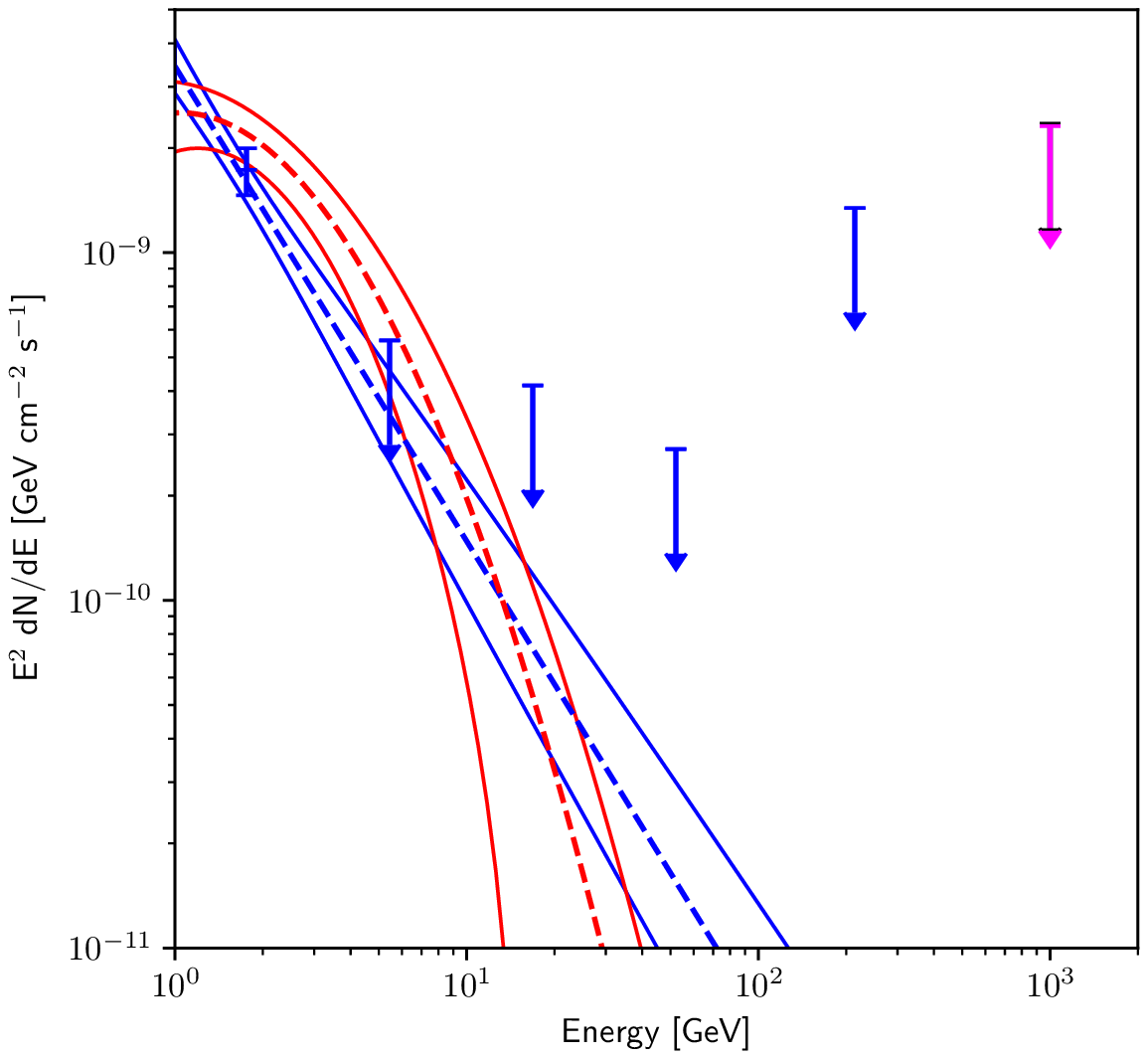}{0.21\textwidth}{(a) 3FGL~J1951.6+2926}
\fig{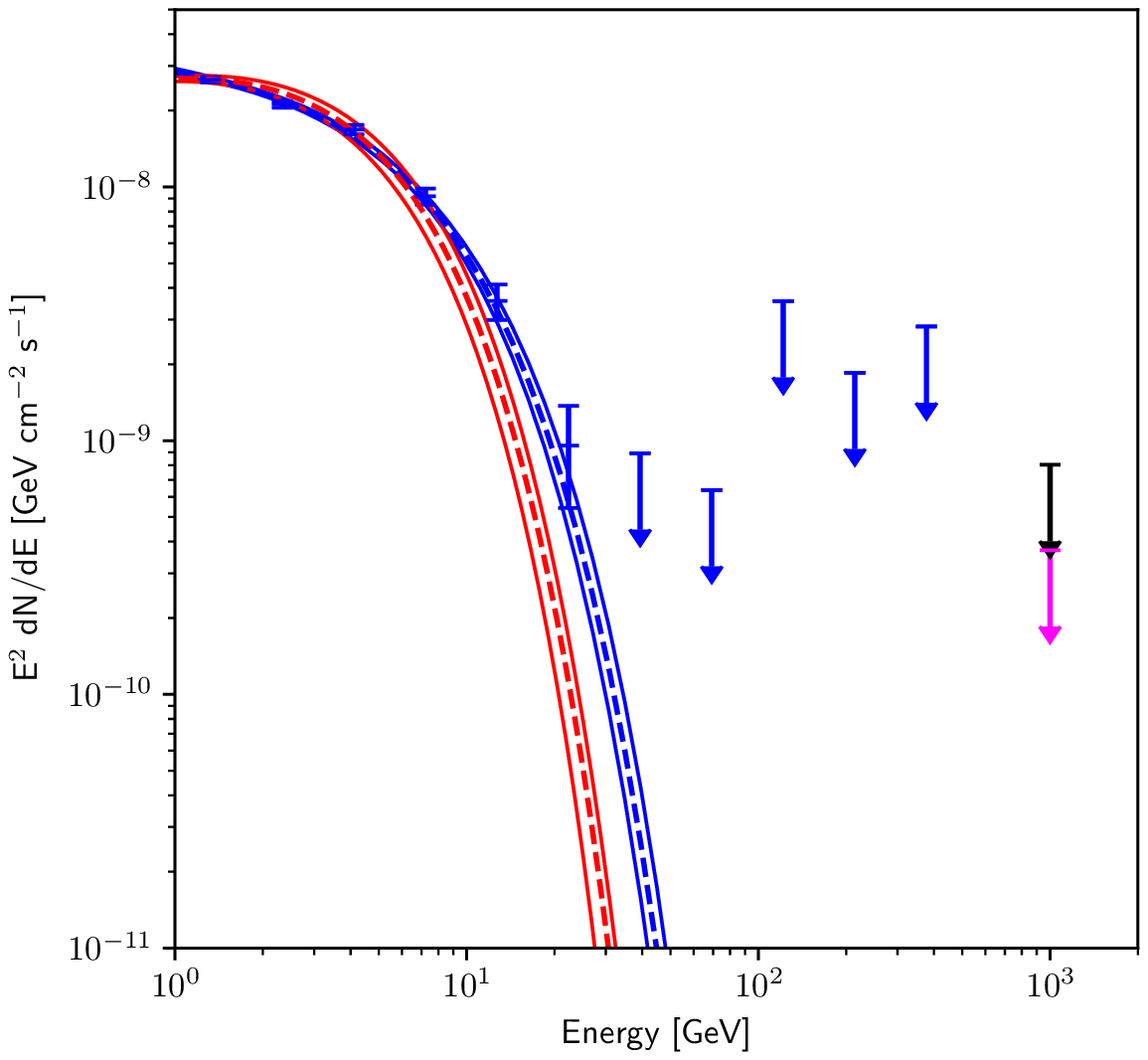}{0.21\textwidth}{(b) 3FGL~J1952.9+3253}
\fig{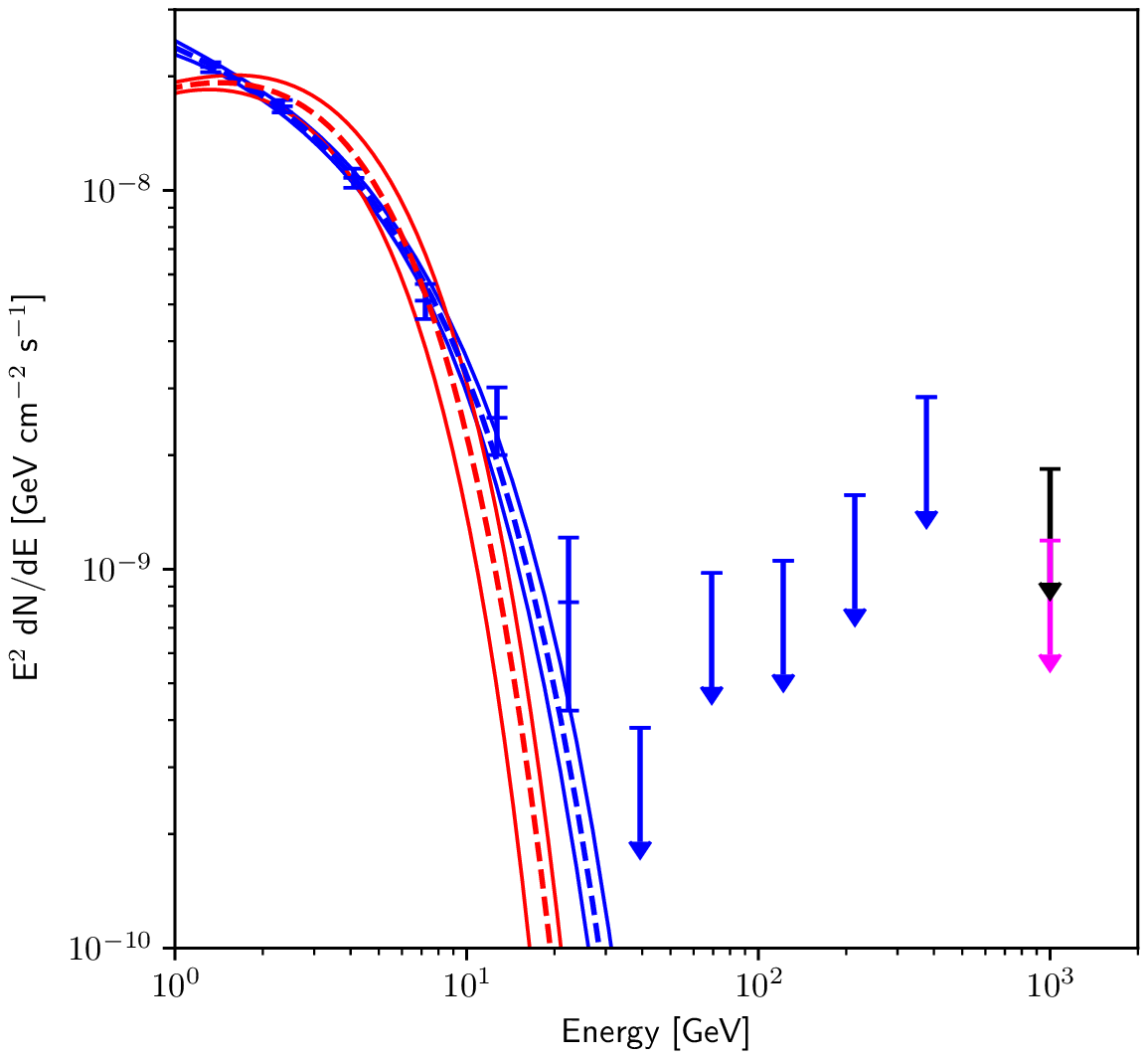}{0.21\textwidth}{(c) 3FGL~J1958.6+2845}
\fig{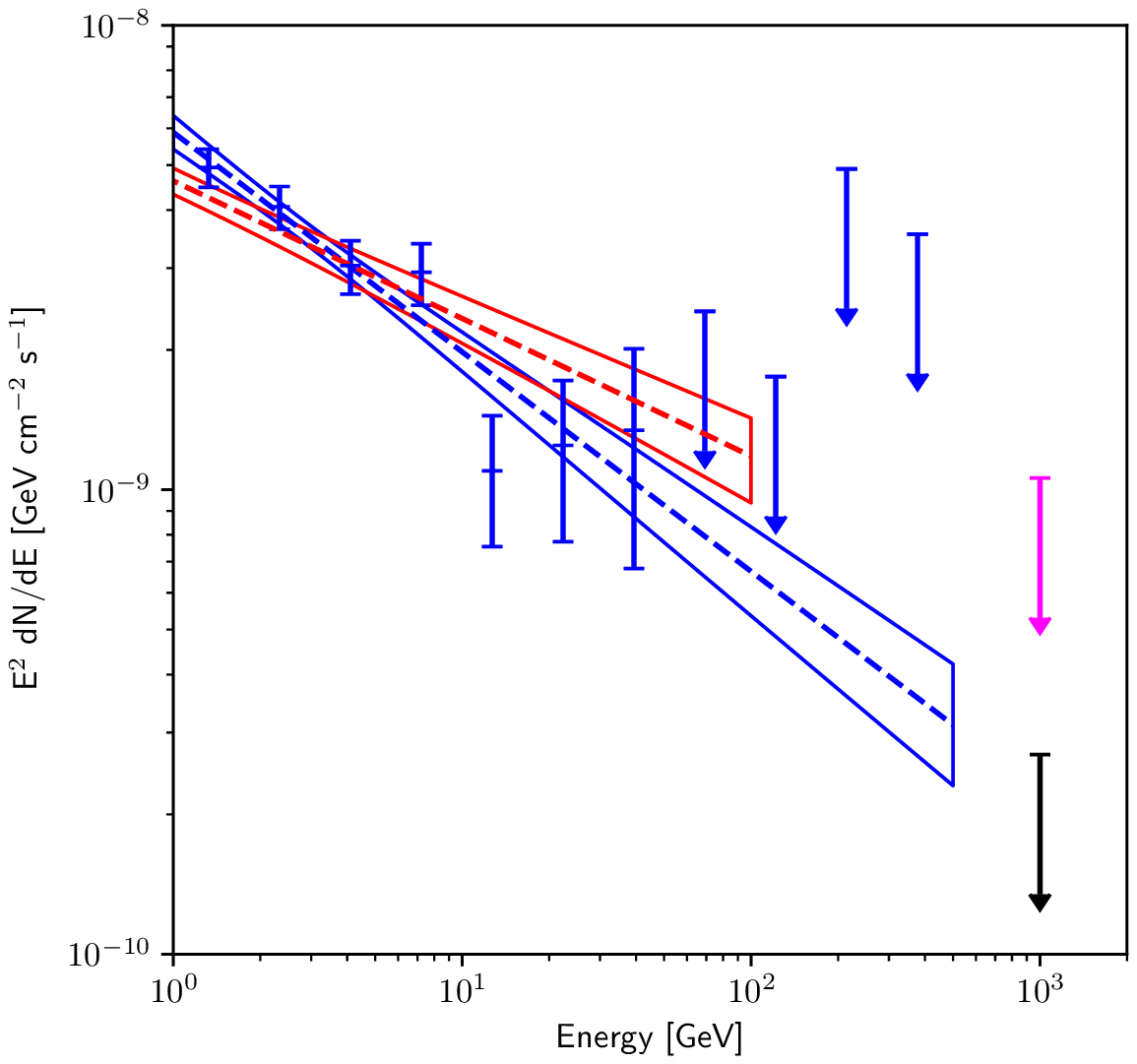}{0.21\textwidth}{(d) 3FGL~J2004.4+3338}
}
\gridline{
\fig{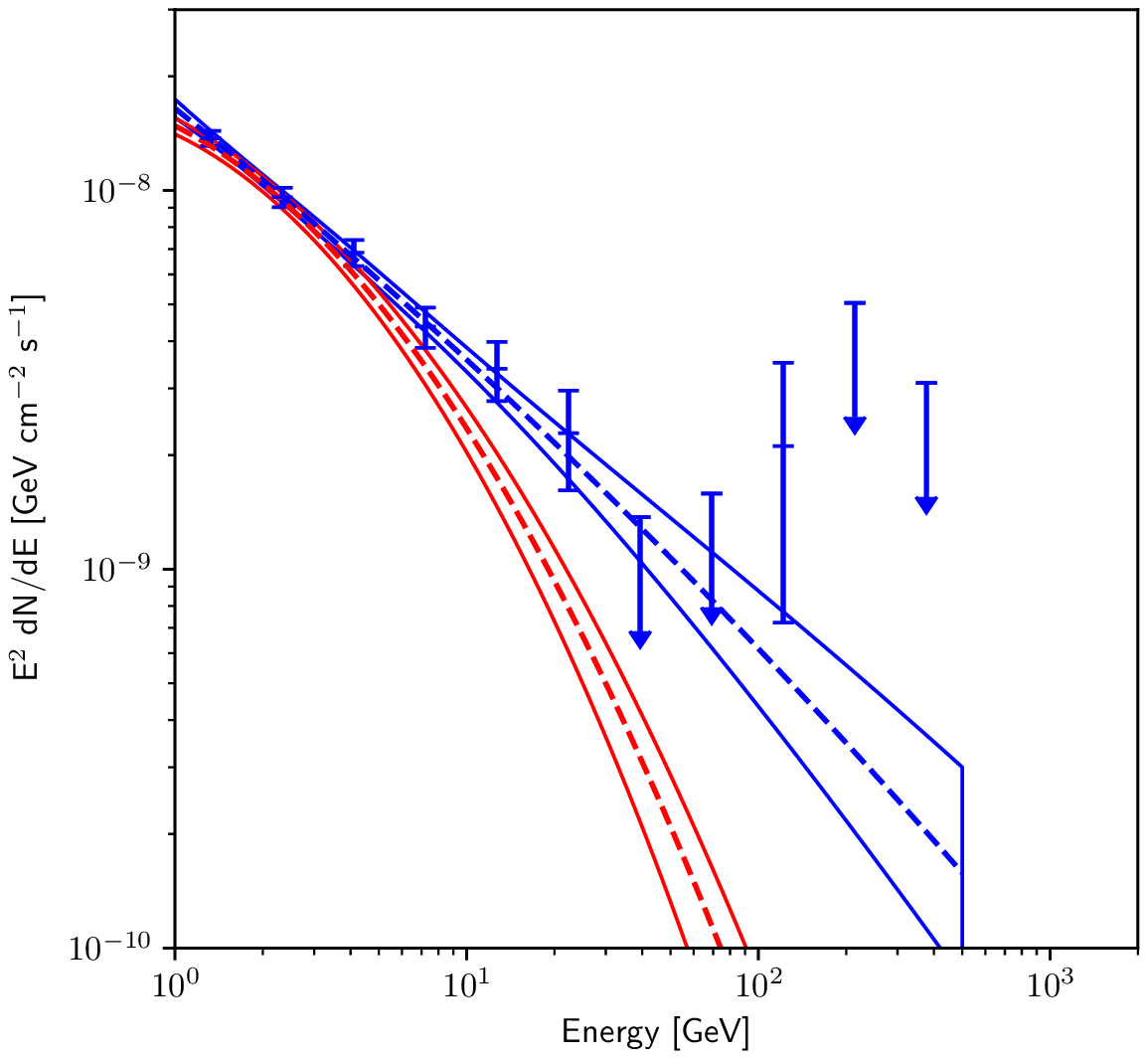}{0.21\textwidth}{(e) 3FGL~J2015.6+3709}
\fig{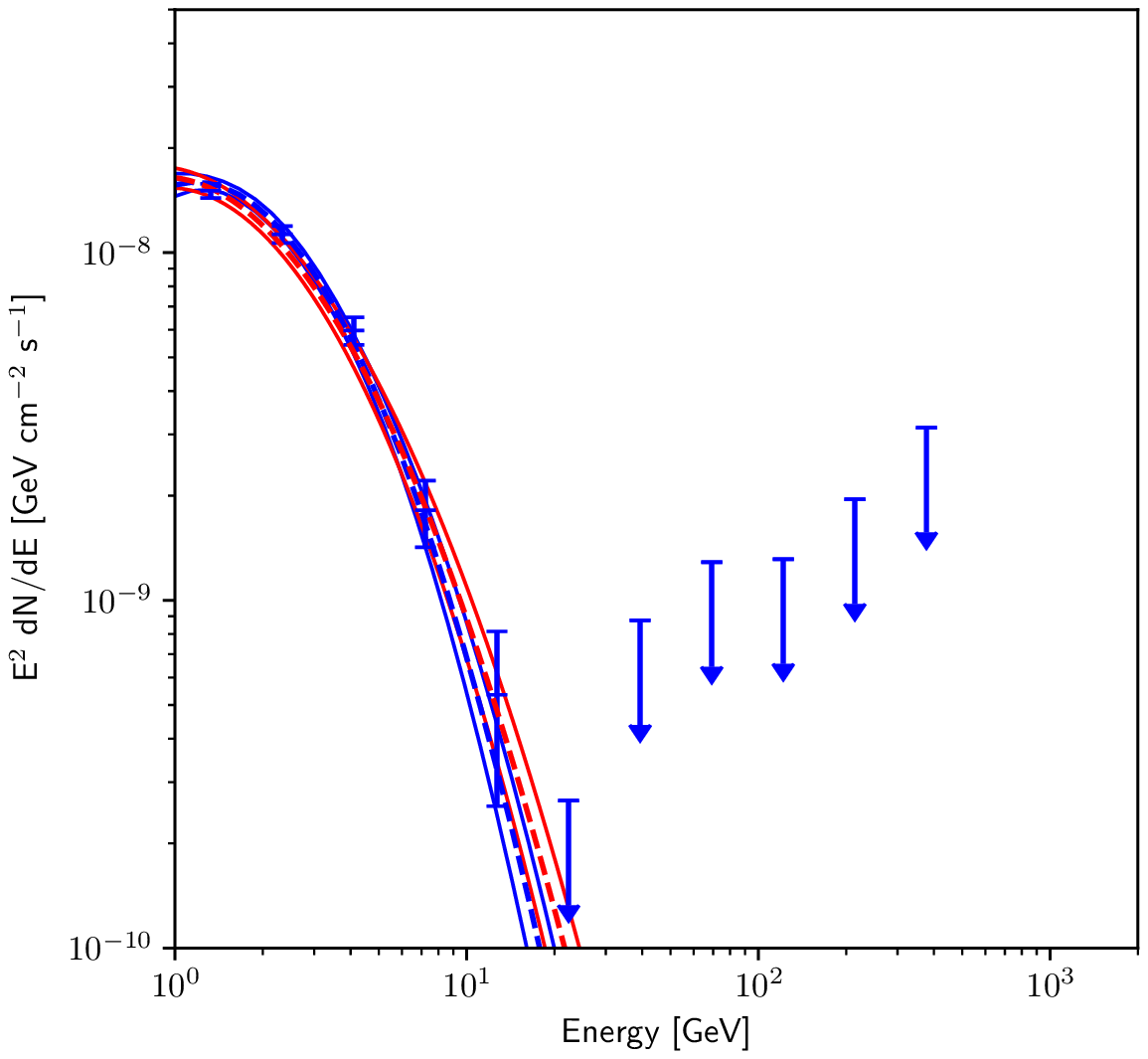}{0.21\textwidth}{(f) 3FGL~J2017.9+3627}
\fig{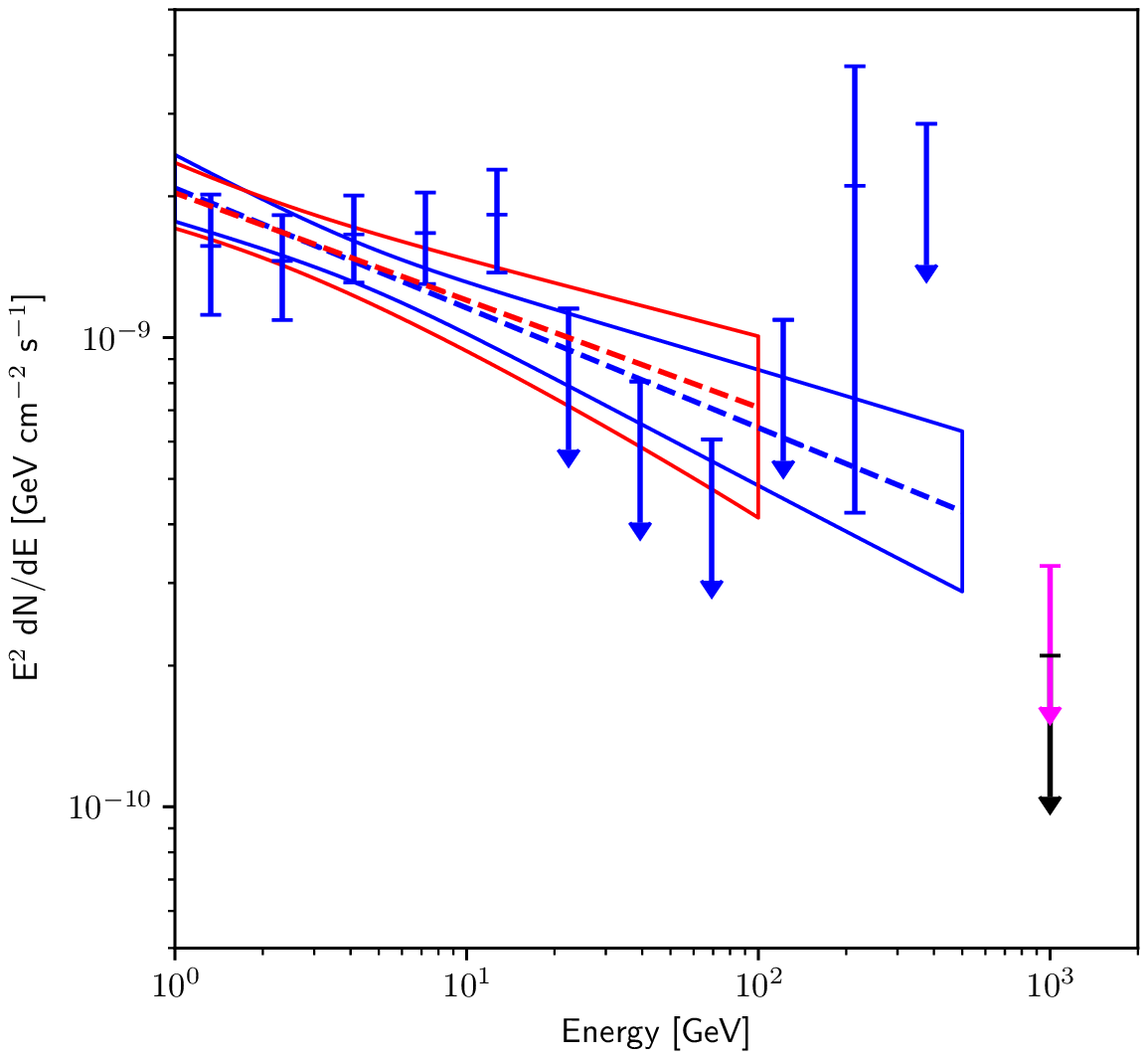}{0.21\textwidth}{(g) 3FGL~J2018.5+3851}
\fig{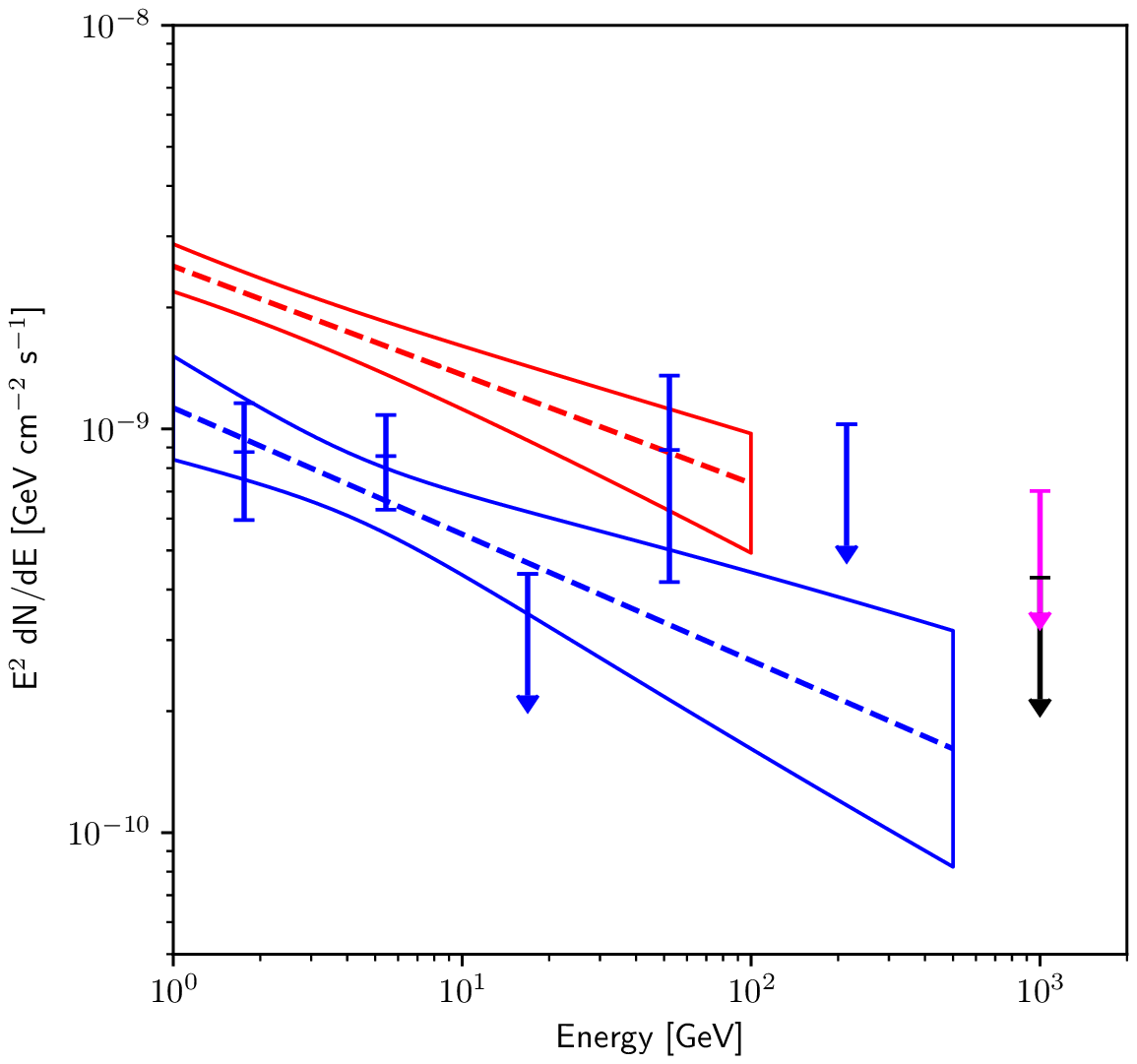}{0.21\textwidth}{(h) 3FGL~J2018.6+4213*}
}
\gridline{
\fig{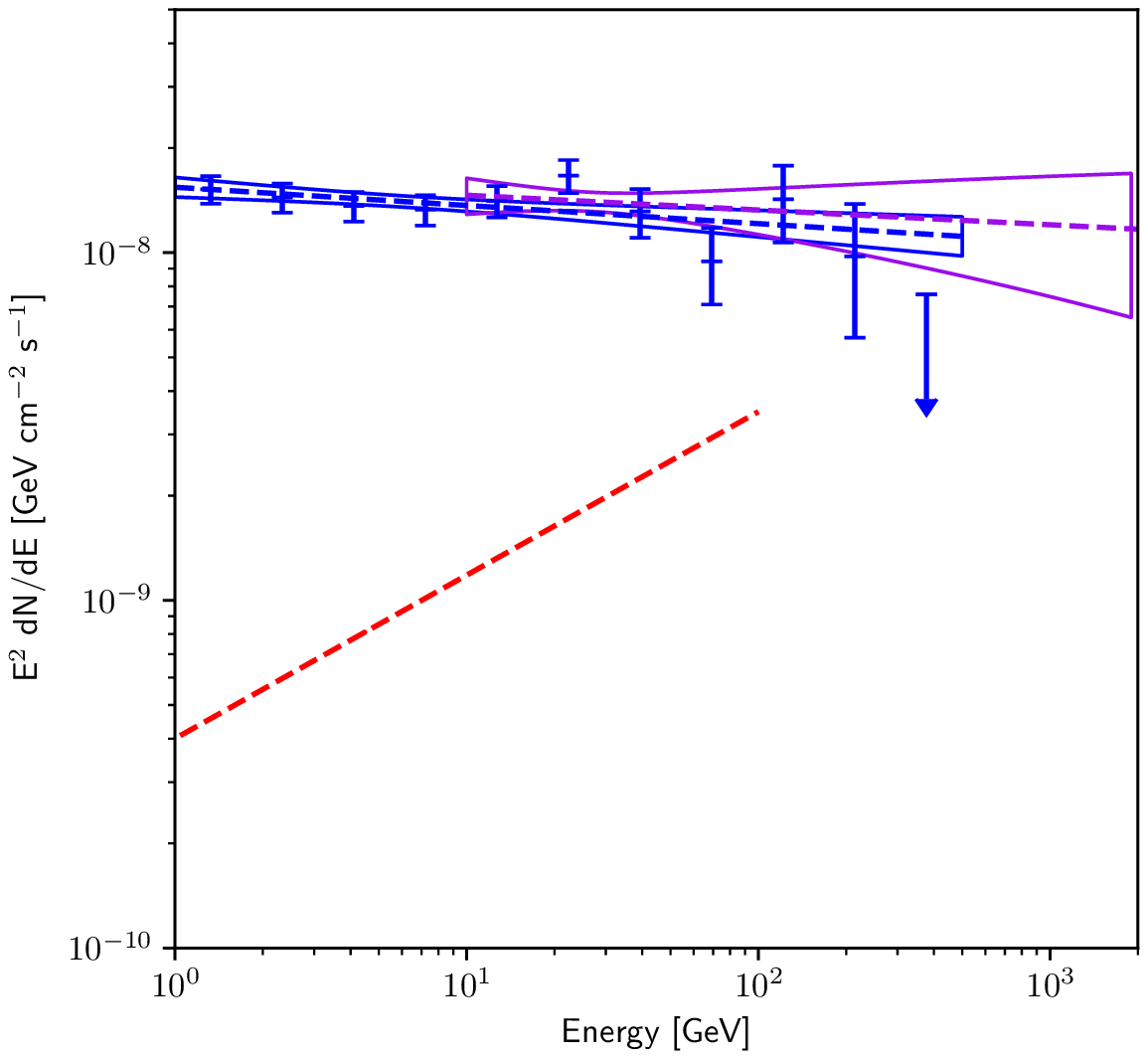}{0.21\textwidth}{(i) 3FGL~J2021.0+4031e}
\fig{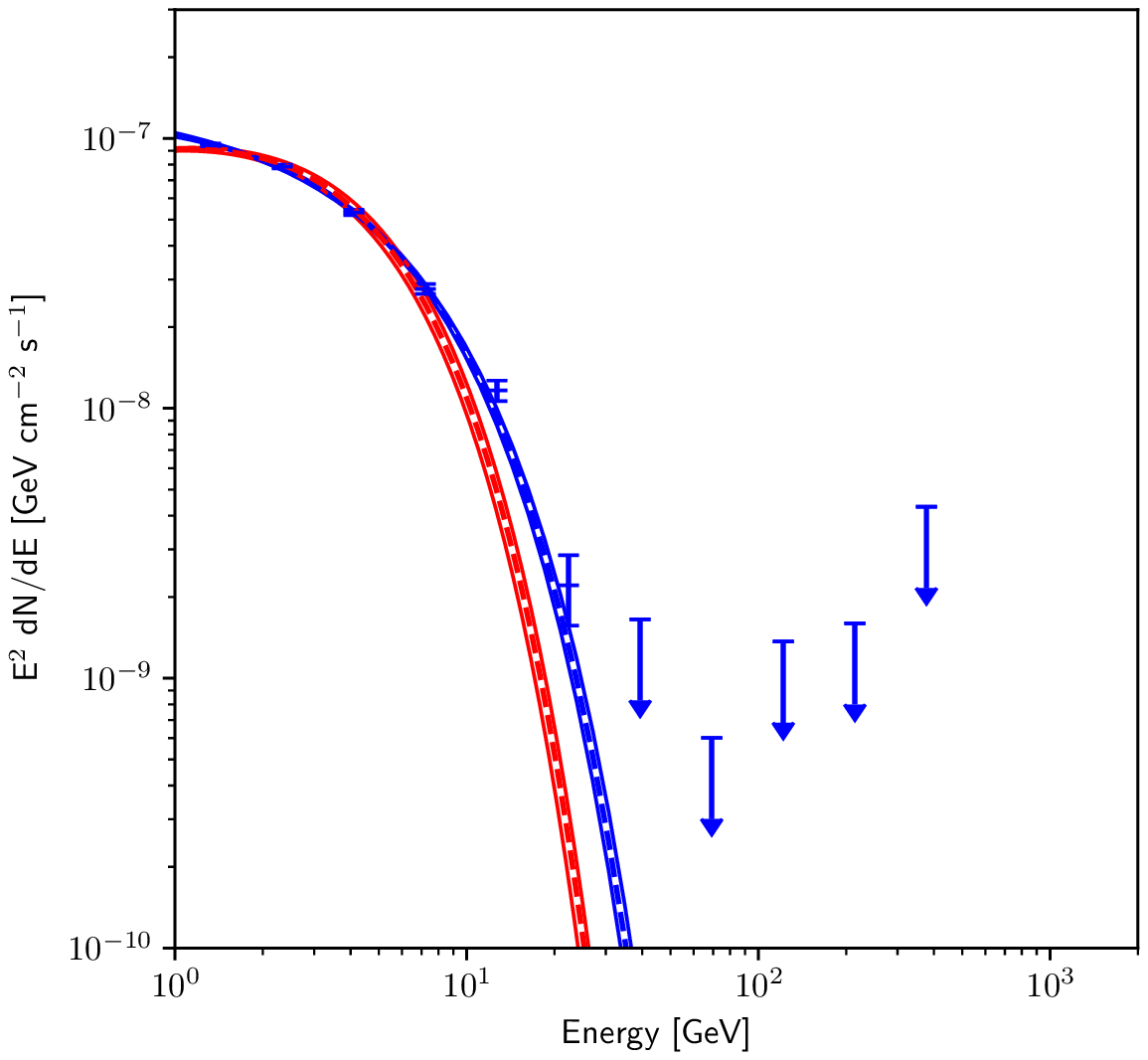}{0.21\textwidth}{(j) 3FGL~J2021.1+3651}
\fig{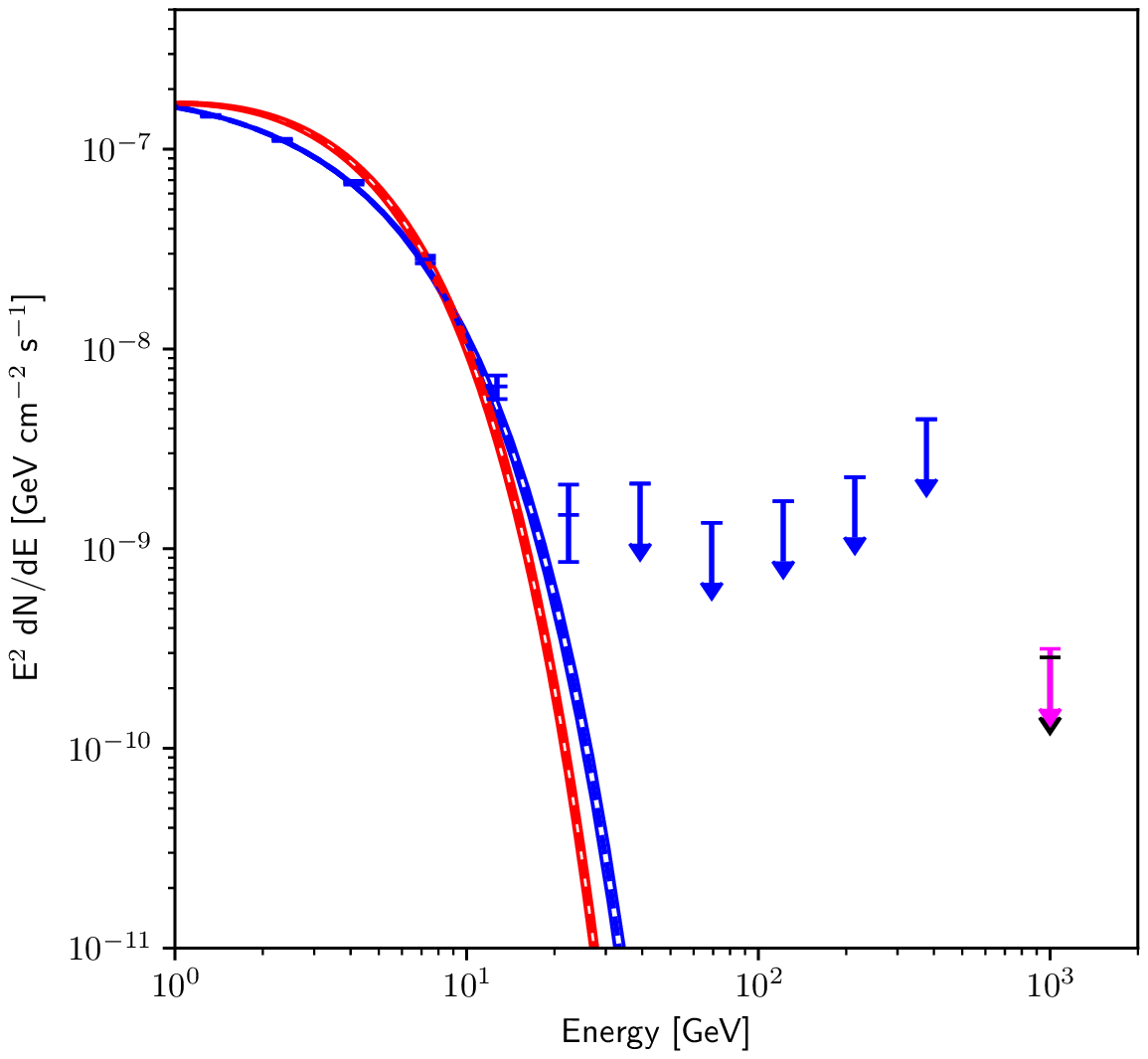}{0.21\textwidth}{(k) 3FGL~J2021.5+4026}
\fig{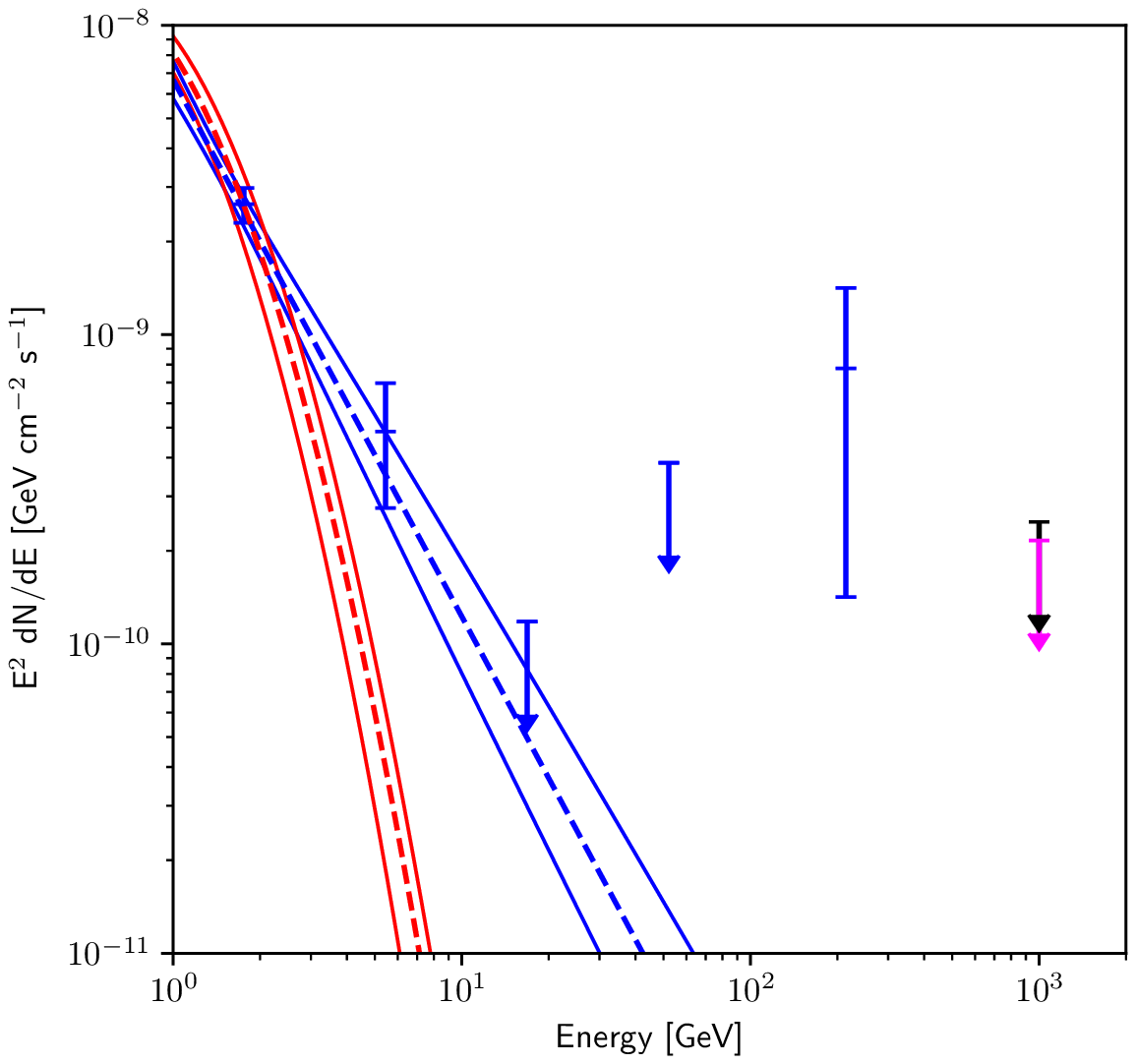}{0.21\textwidth}{(l) 3FGL~J2022.2+3840*}
}
\gridline{
\fig{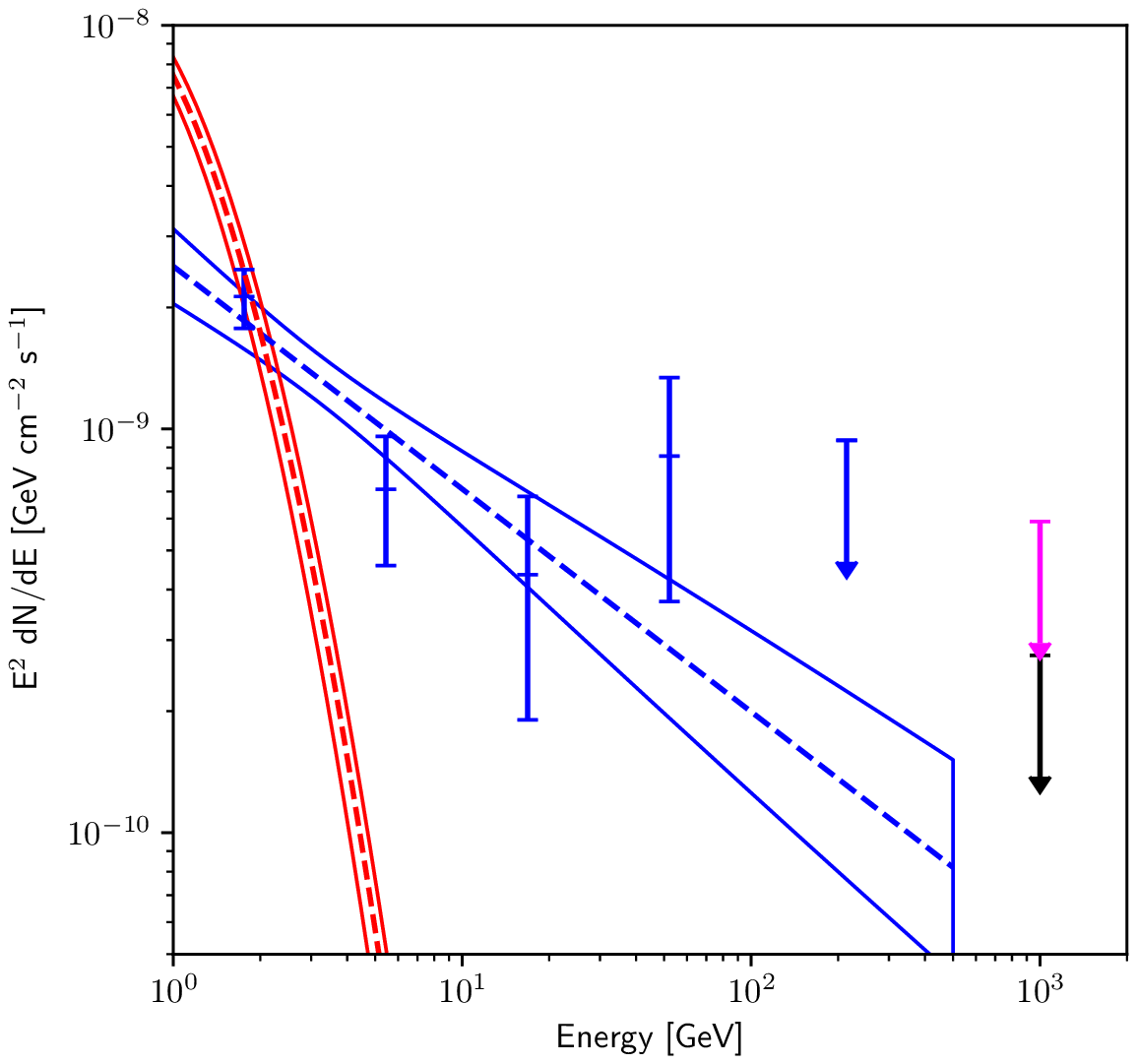}{0.21\textwidth}{(m) 3FGL~J2023.5+4126*}
\fig{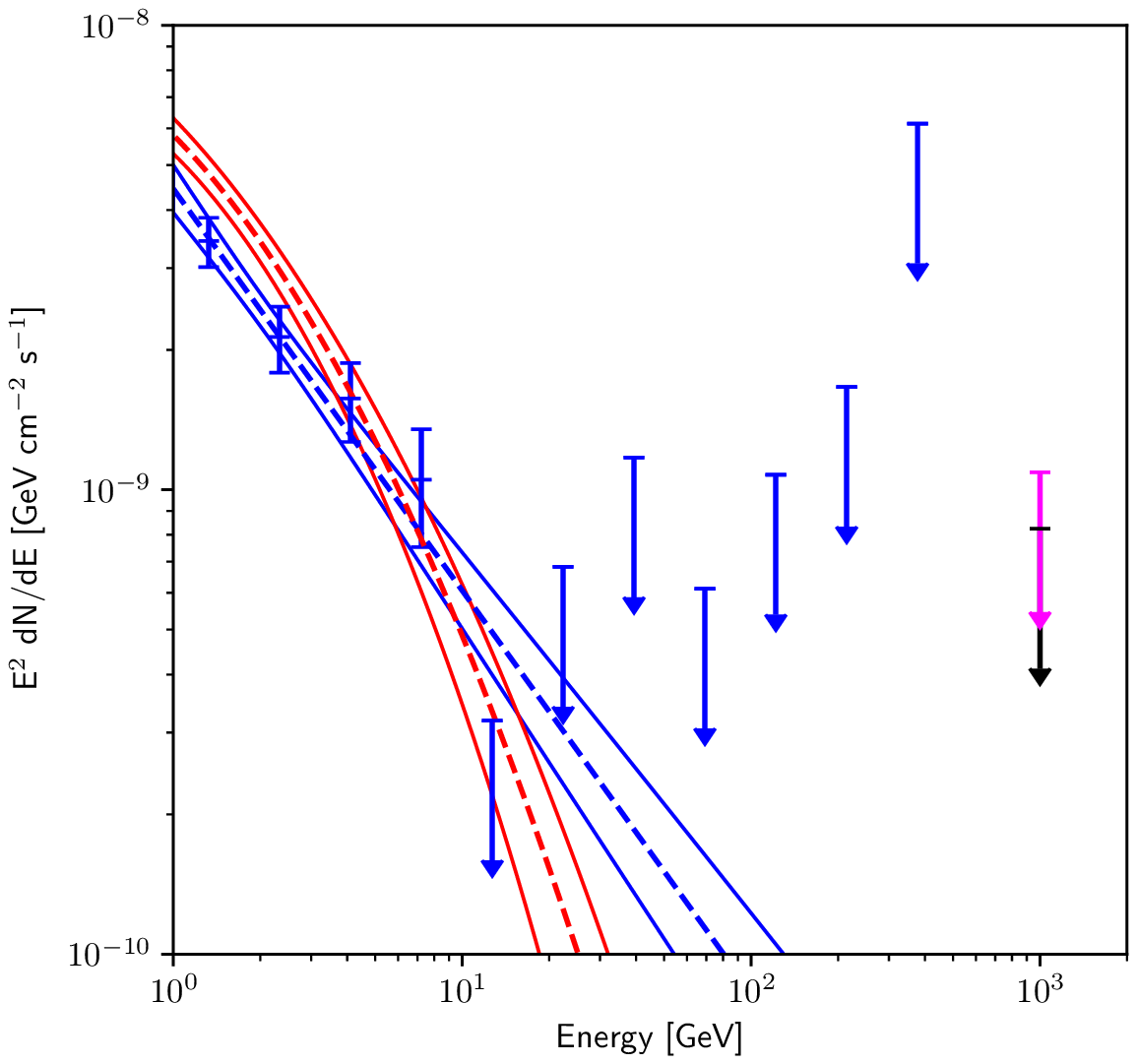}{0.21\textwidth}{(n) 3FGL~J2025.2+3340}
\fig{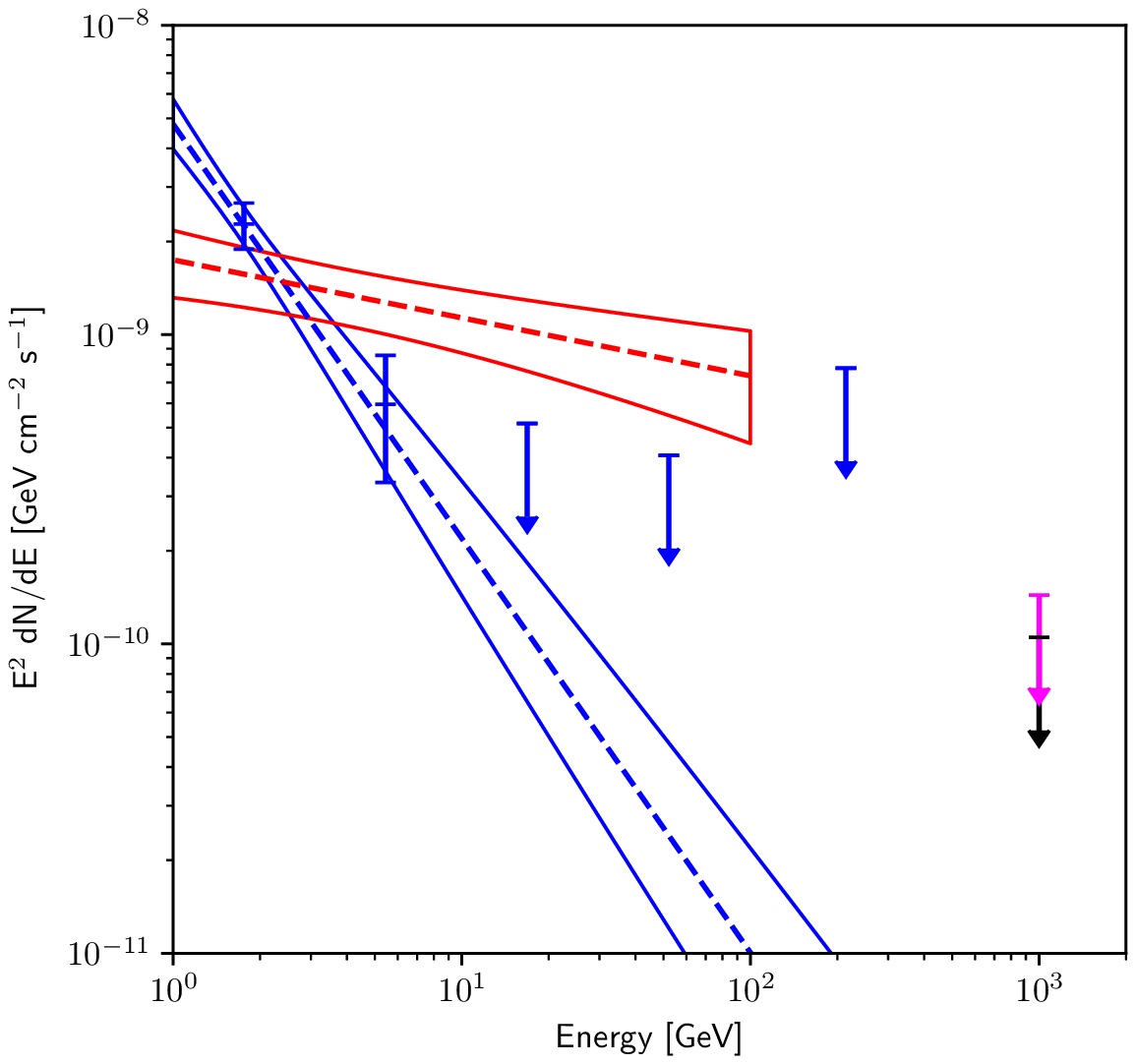}{0.21\textwidth}{(o) 3FGL~J2028.5+4040c*}
\fig{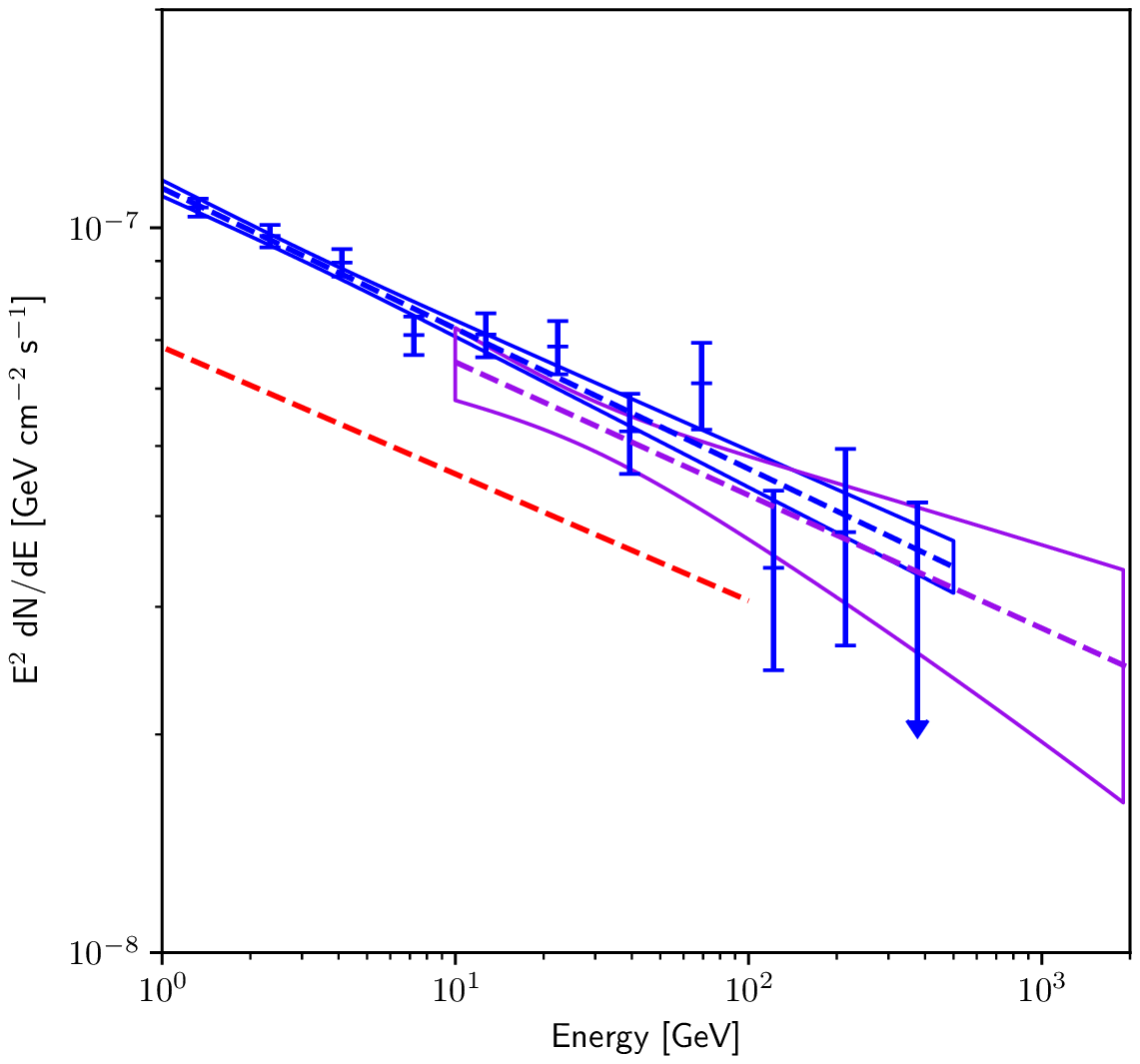}{0.21\textwidth}{(p) 3FGL~J2028.6+4110e}
}
\gridline{
\fig{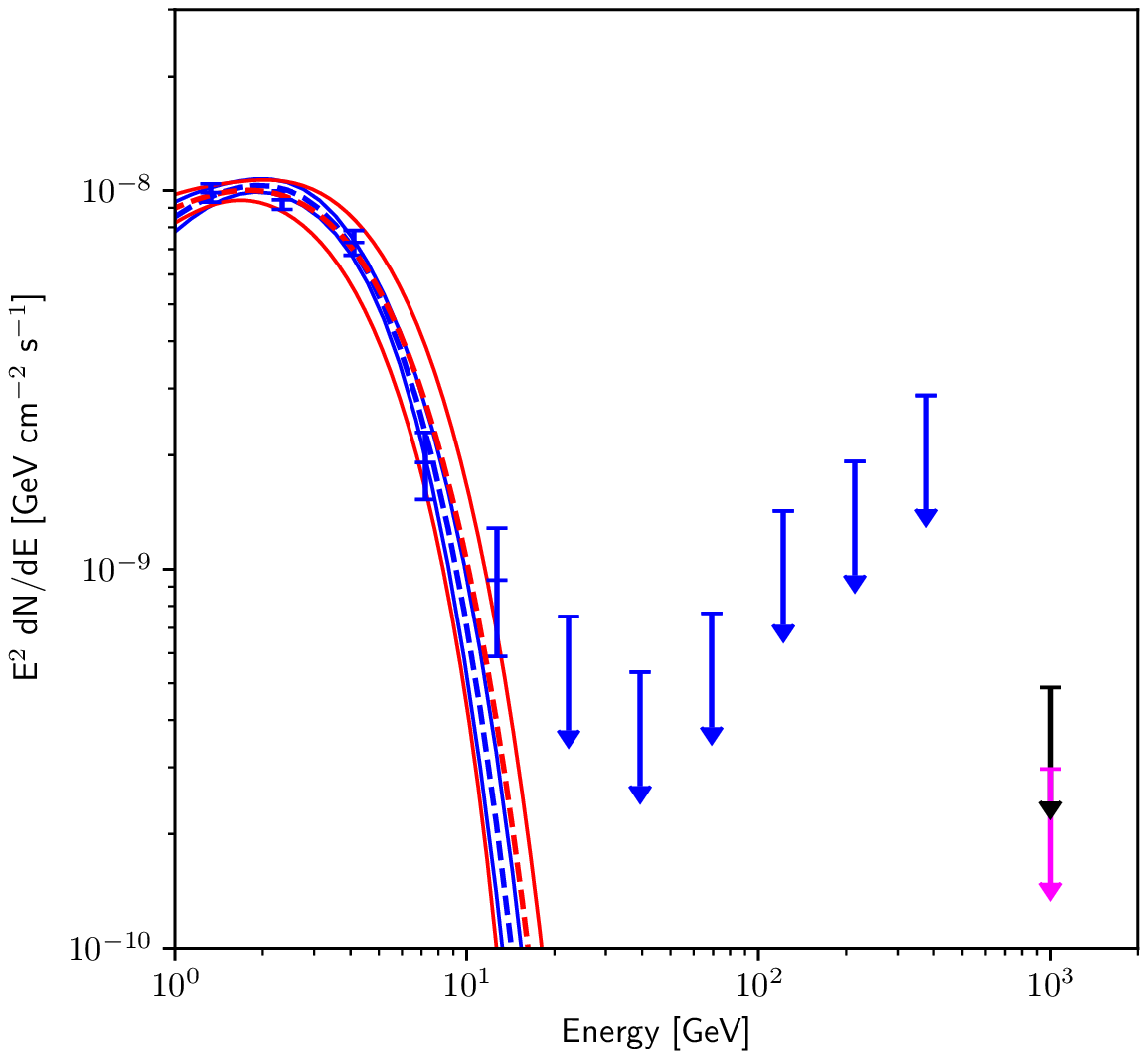}{0.21\textwidth}{(q) 3FGL~J2030.0+3642}
\fig{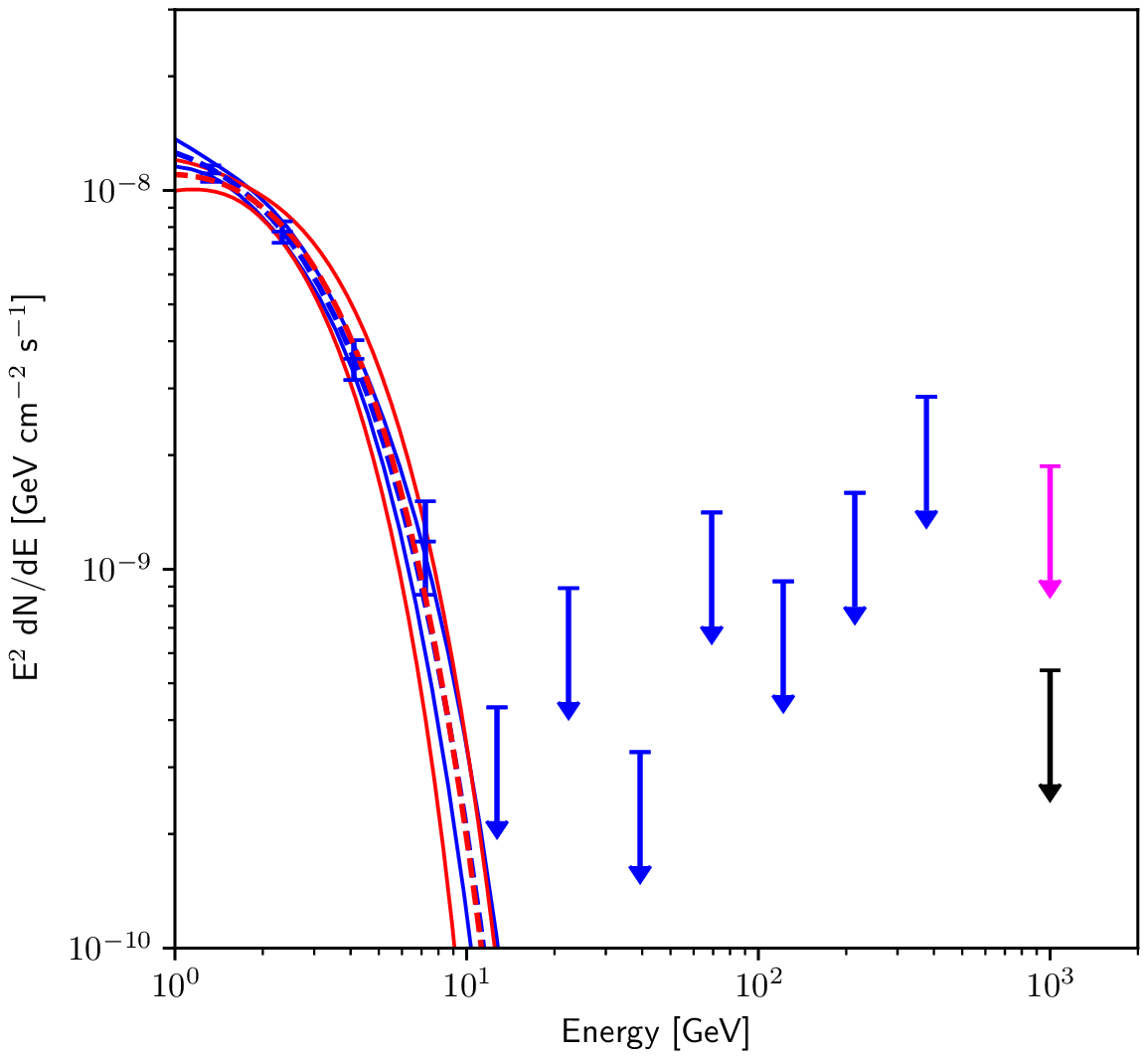}{0.21\textwidth}{(r) 3FGL~J2030.8+4416}
\fig{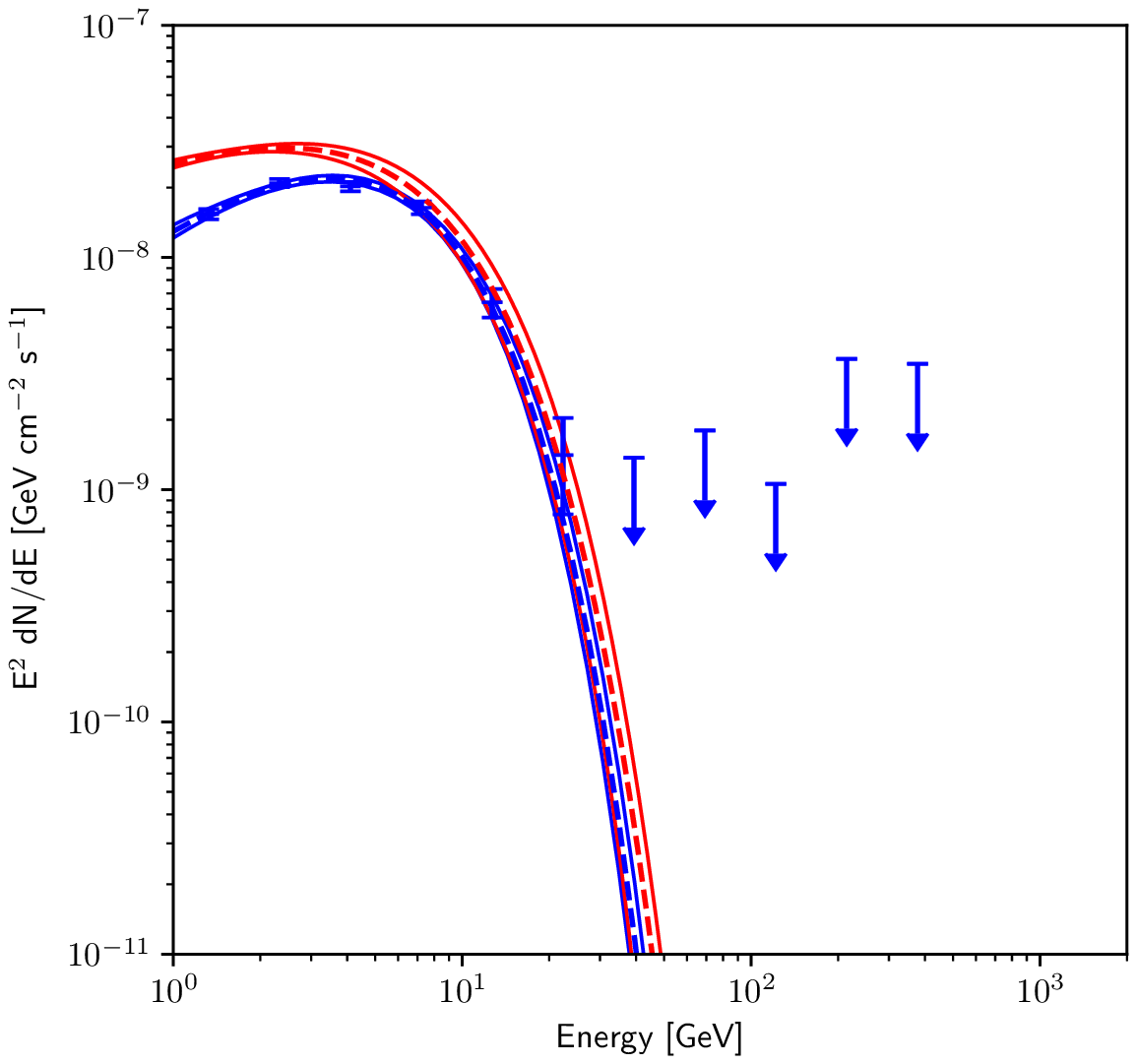}{0.21\textwidth}{(s) 3FGL~J2032.2+4126}
\fig{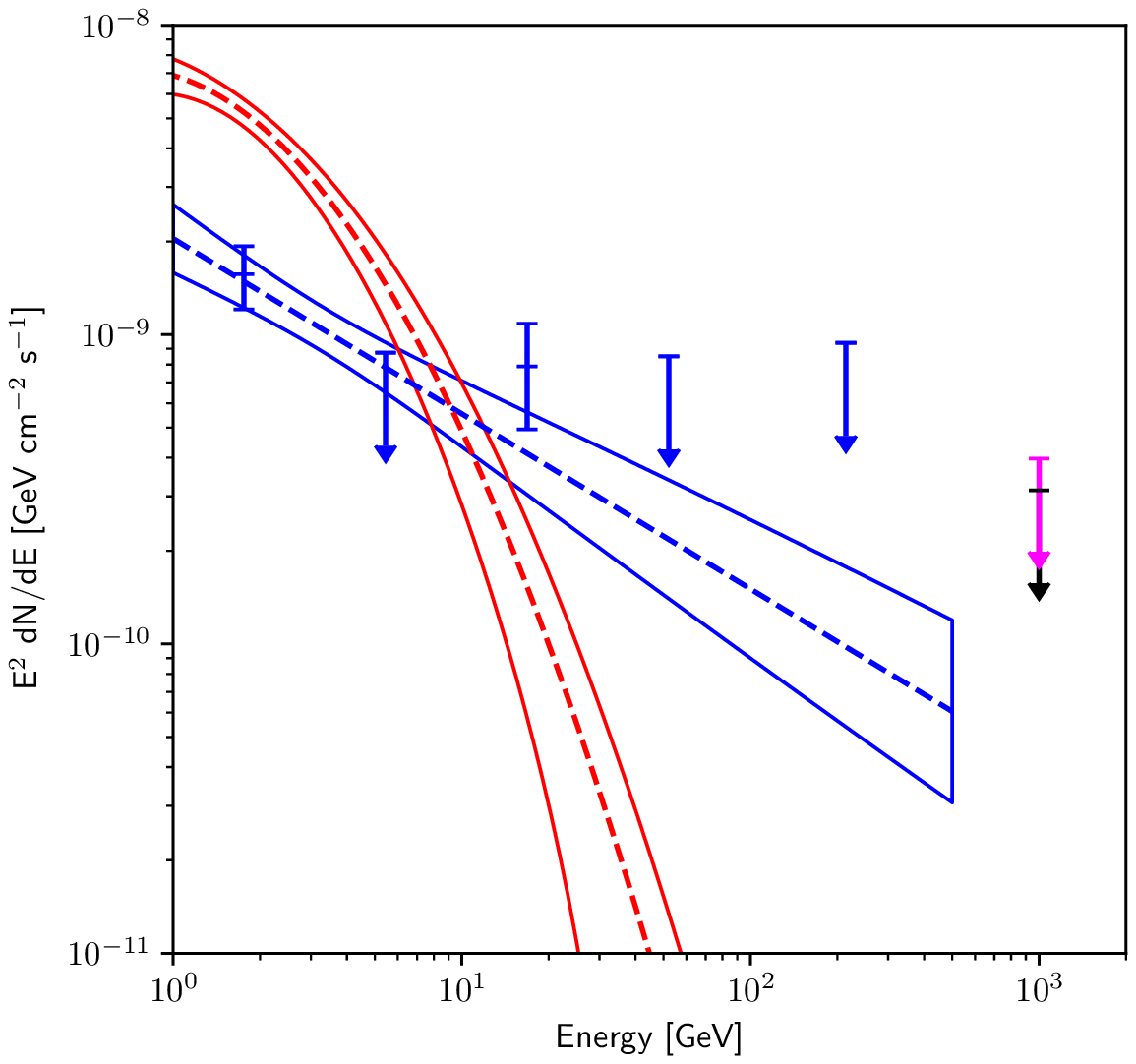}{0.21\textwidth}{(t) 3FGL~J2032.5+3921*}
}
\end{figure}

\begin{figure}[htb!]
\gridline{
\fig{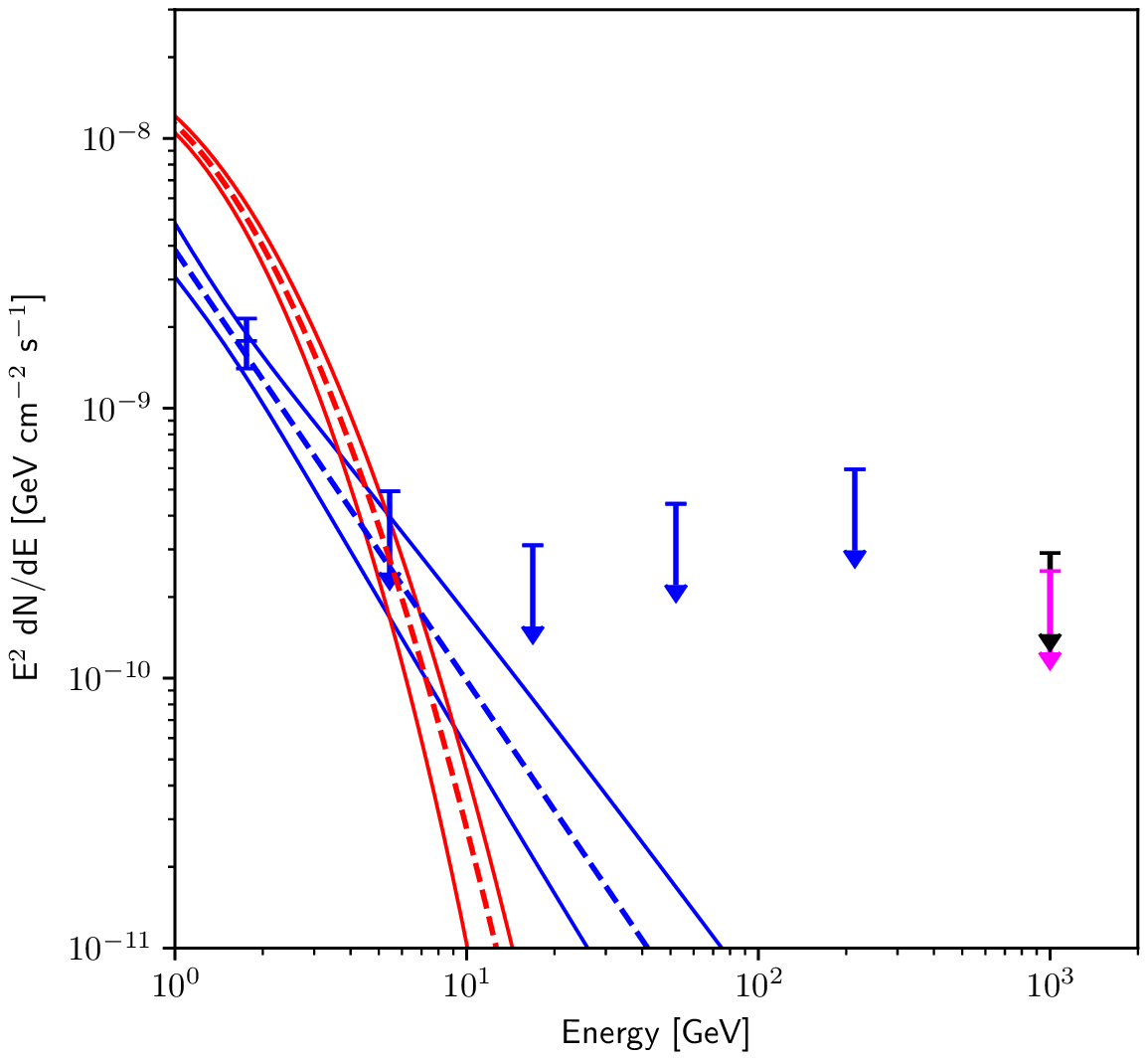}{0.21\textwidth}{(u) 3FGL~J2032.5+4032*}
\fig{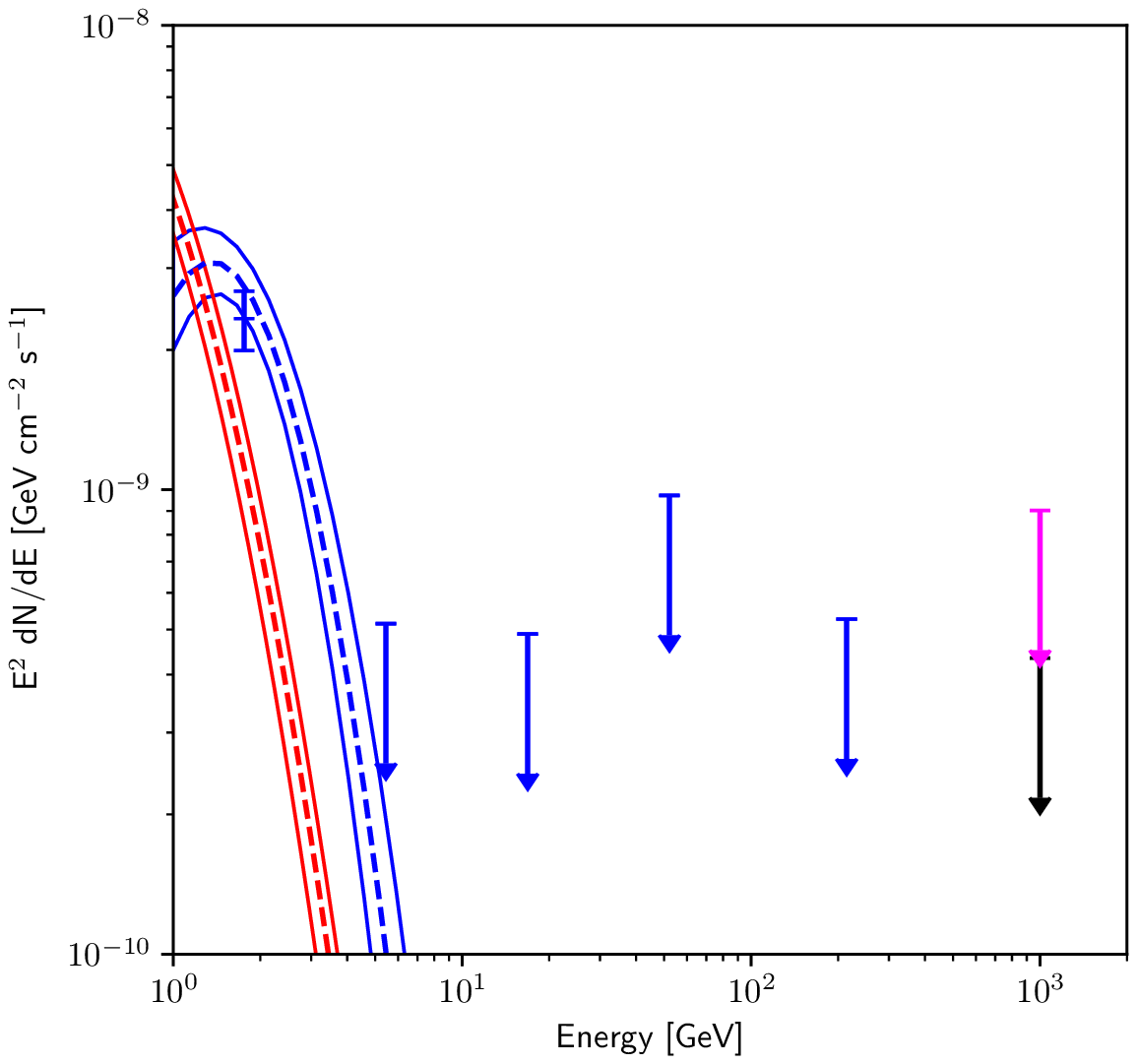}{0.21\textwidth}{(v) 3FGL~J2034.4+3833c*}
\fig{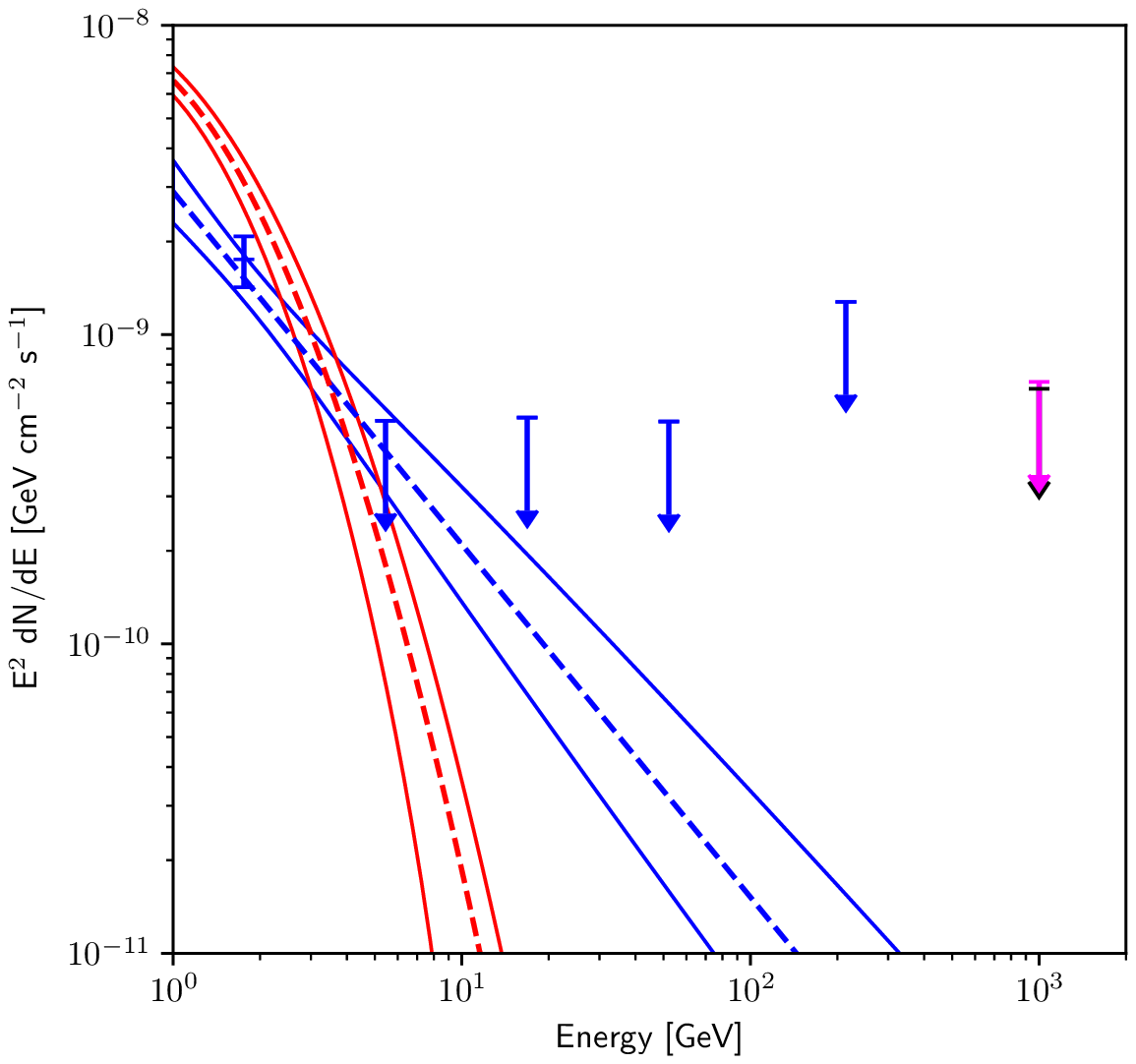}{0.21\textwidth}{(w) 3FGL~J2034.6+4302*}
\fig{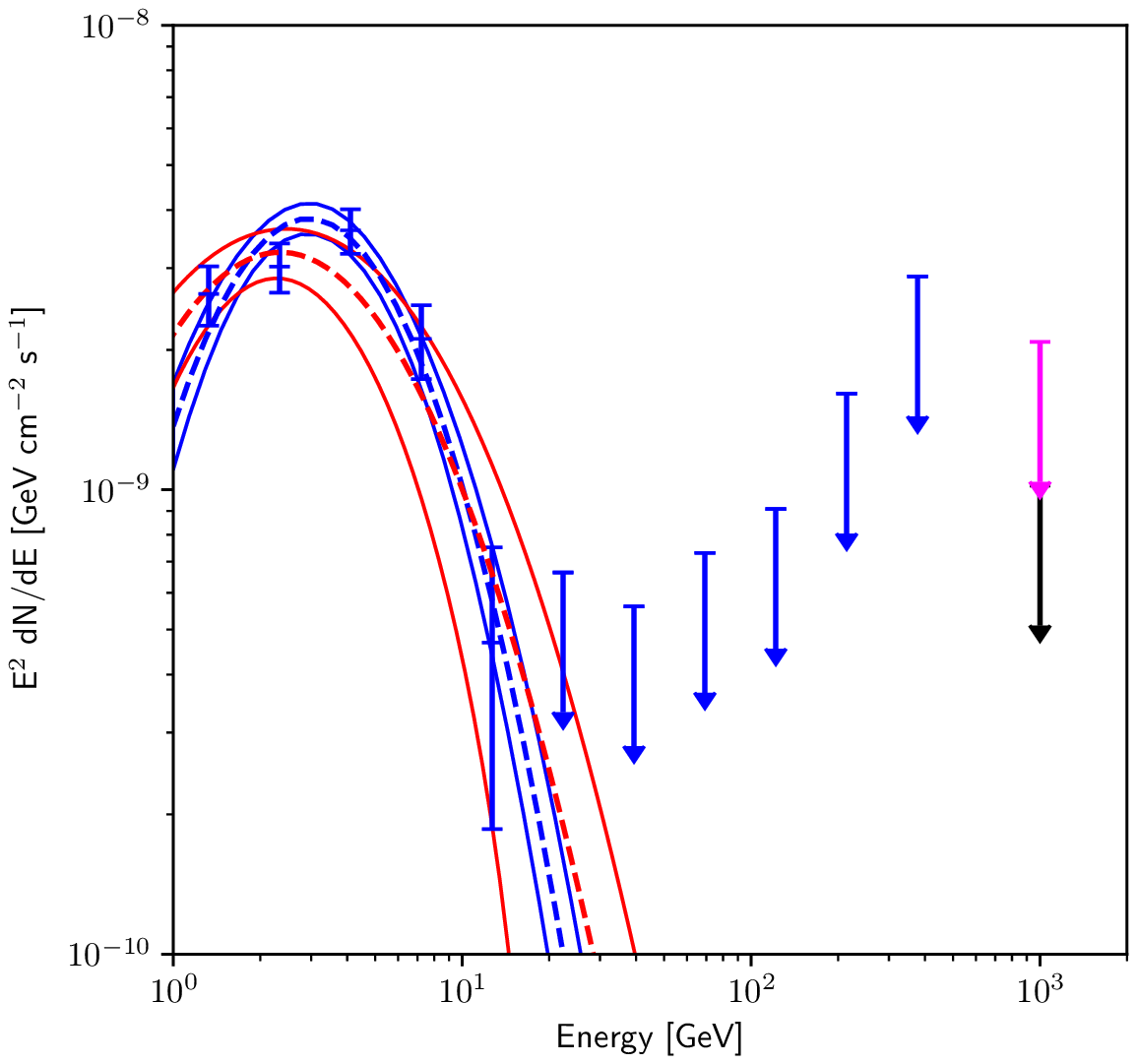}{0.21\textwidth}{(x) 3FGL~J2035.0+3634}
}
\gridline{
\fig{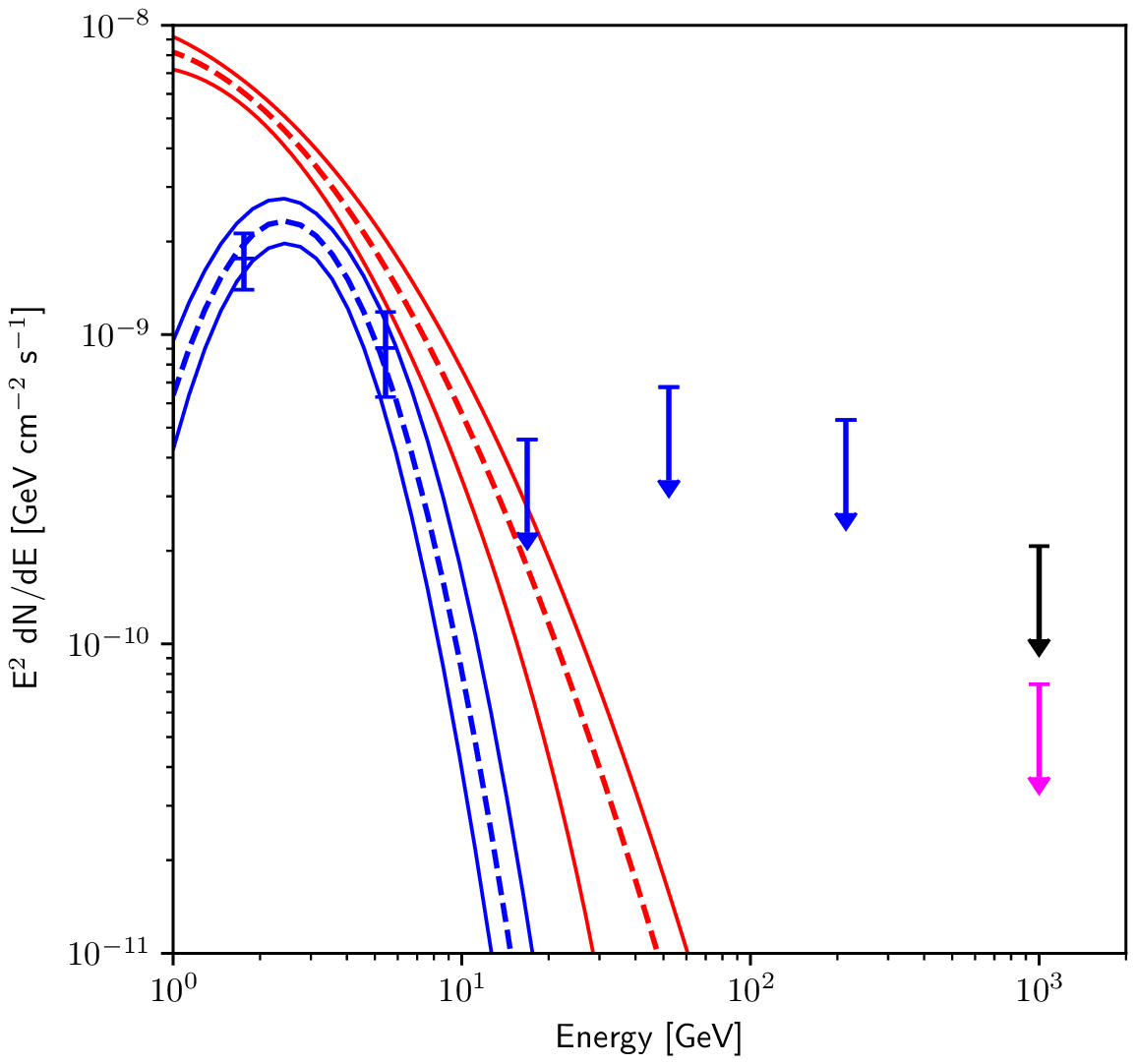}{0.21\textwidth}{(y) 3FGL~J2038.4+4212*}
\fig{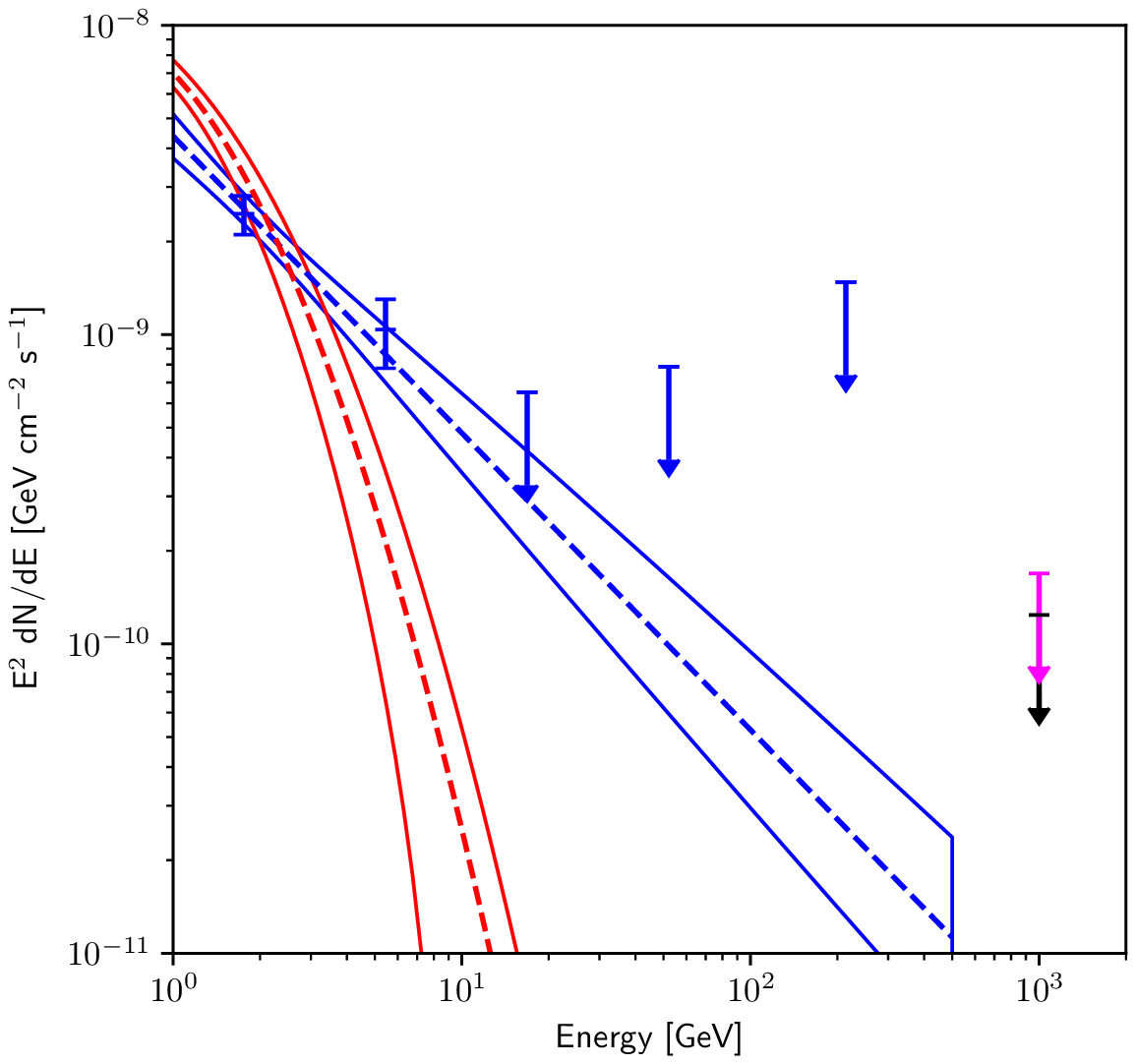}{0.21\textwidth}{(z) 3FGL~J2039.4+4111*}
\fig{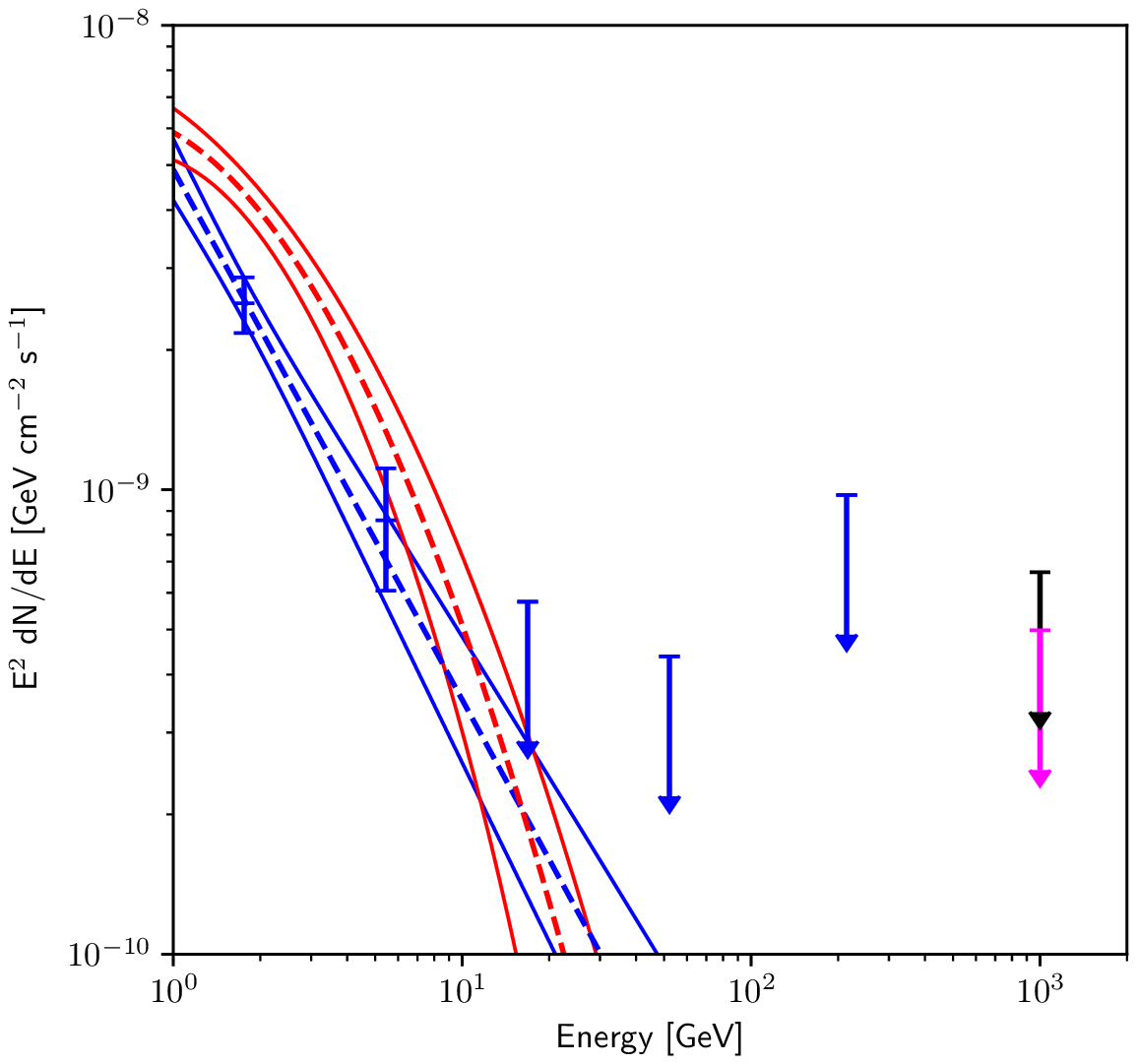}{0.21\textwidth}{(ab) 3FGL~J2042.4+4209*}
\fig{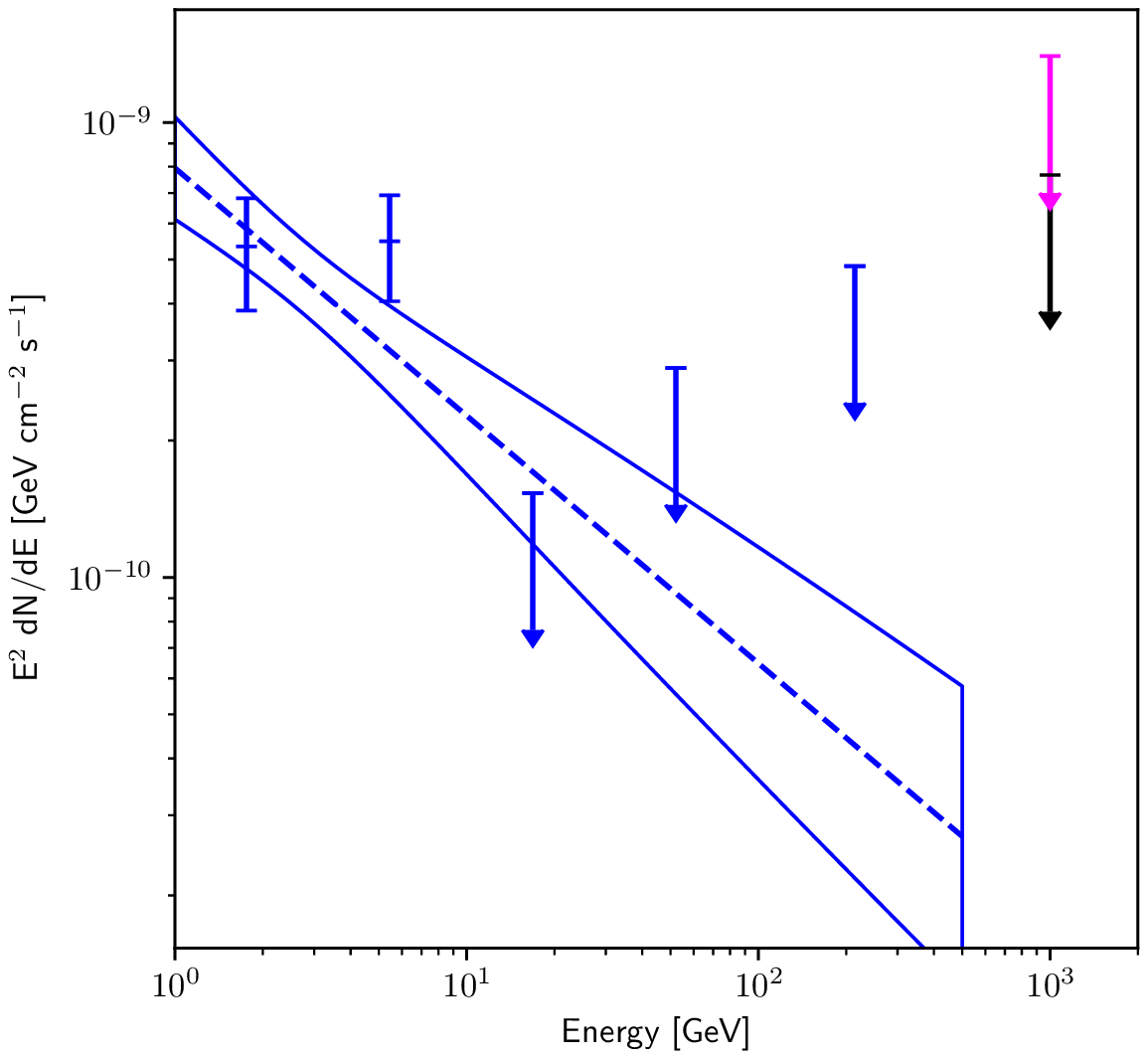}{0.21\textwidth}{(ac) FGL J1949.0+3412}
}
\gridline{
\fig{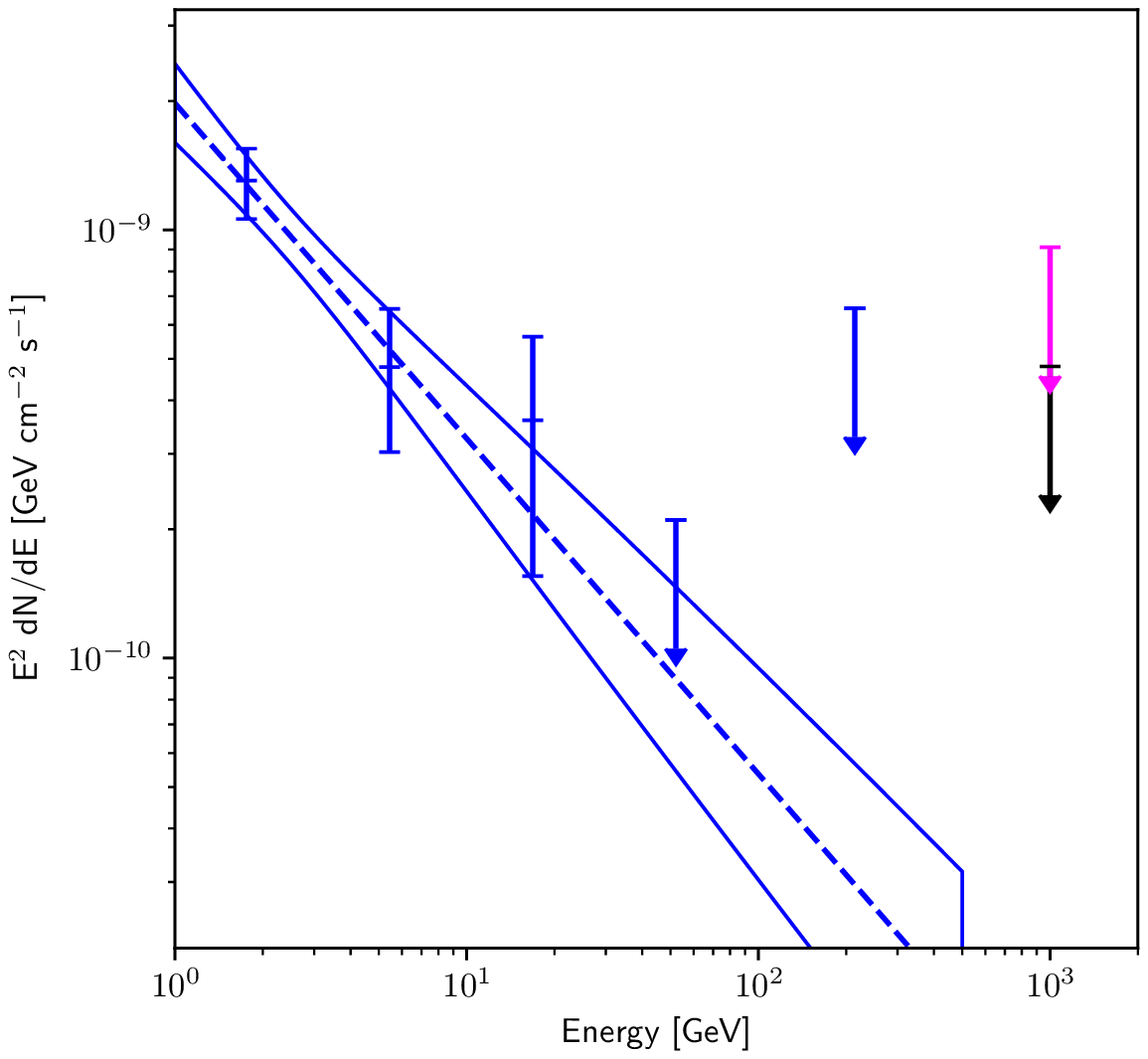}{0.21\textwidth}{(ad) FGL J1955.0+3319}
\fig{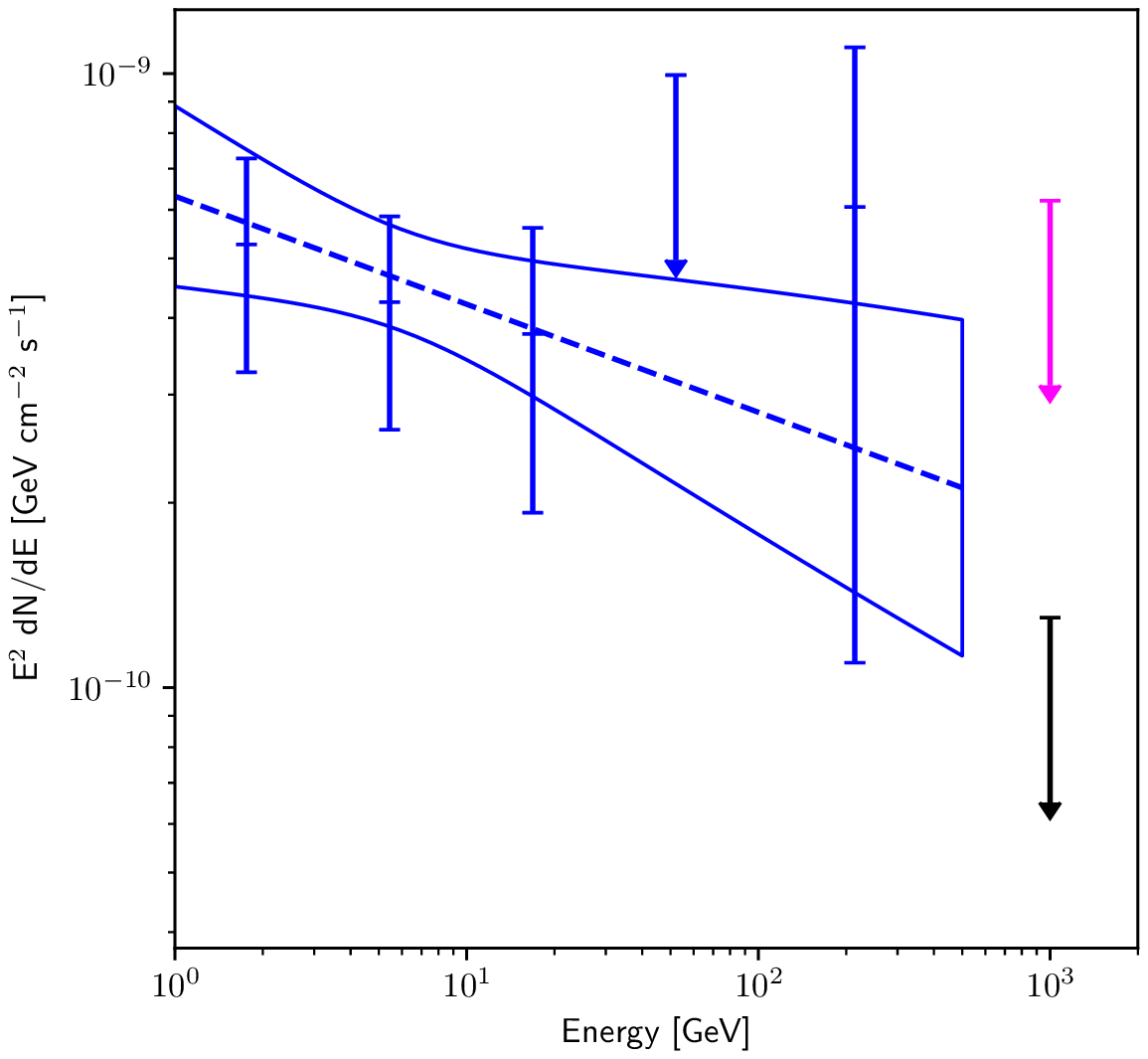}{0.21\textwidth}{(ae) FGL J1958.6+3510}
\fig{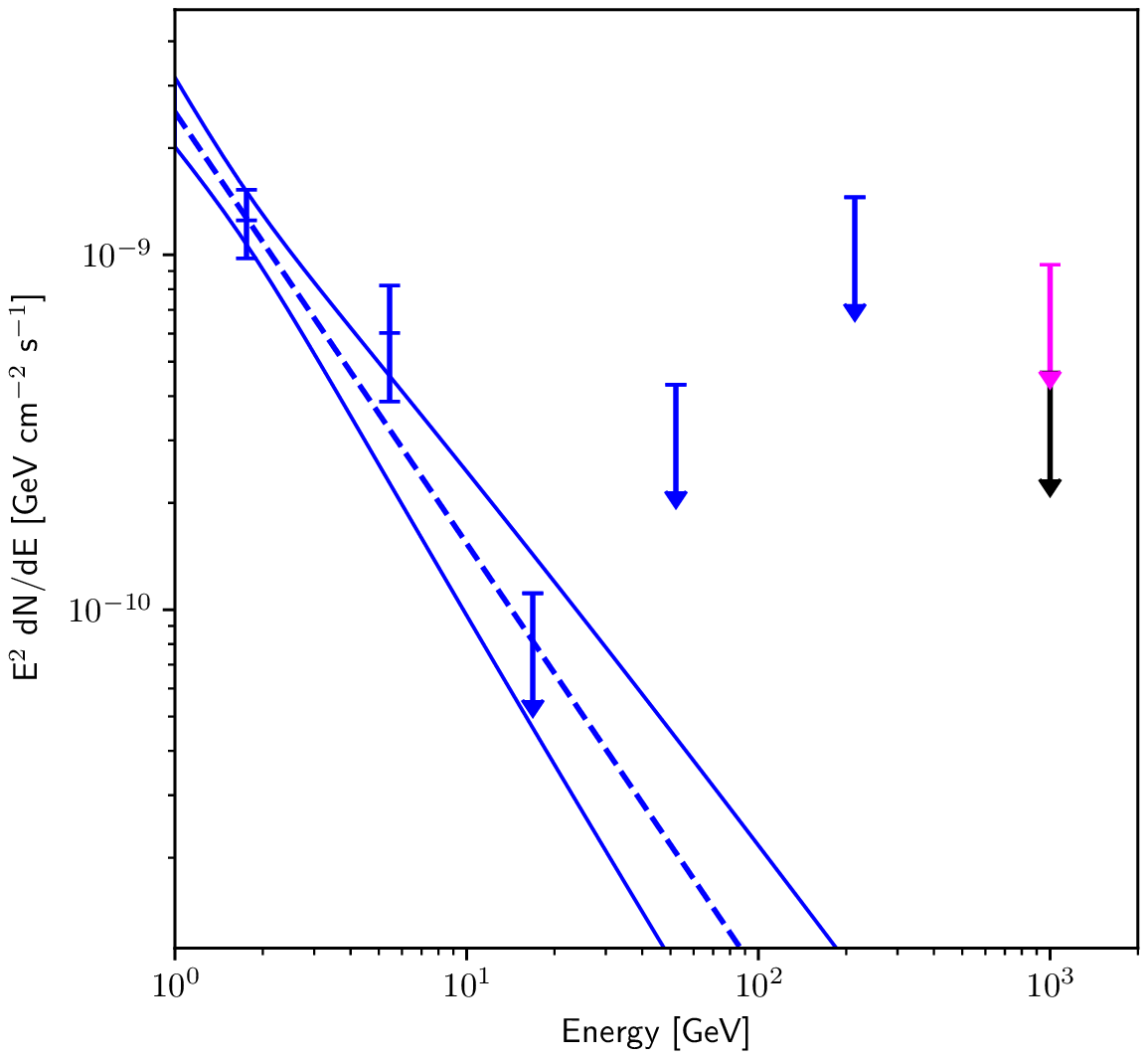}{0.21\textwidth}{(af) FGL J2005.7+3417}
\fig{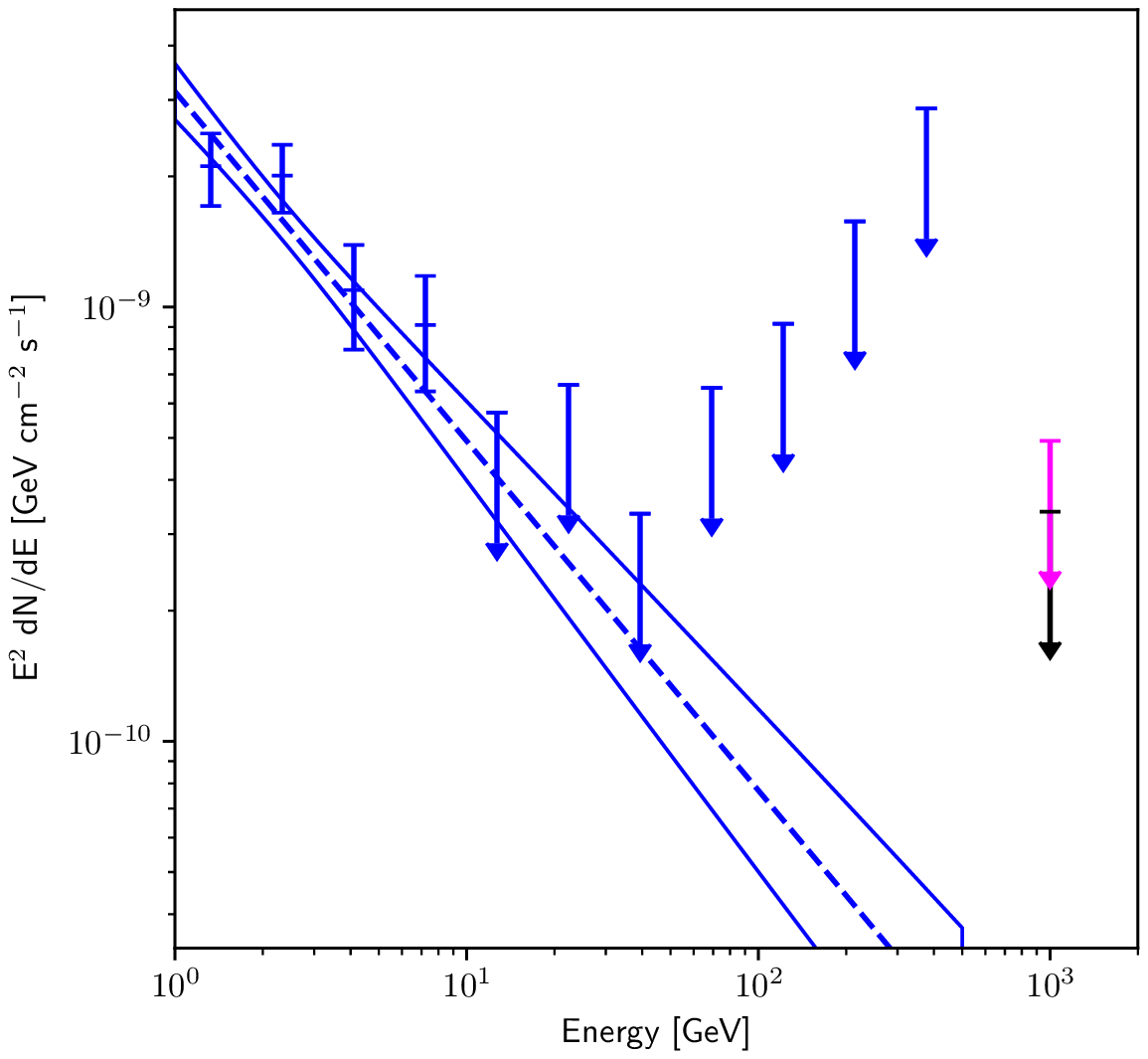}{0.21\textwidth}{(ag) FGL J2006.3+3103}
}
\gridline{
\fig{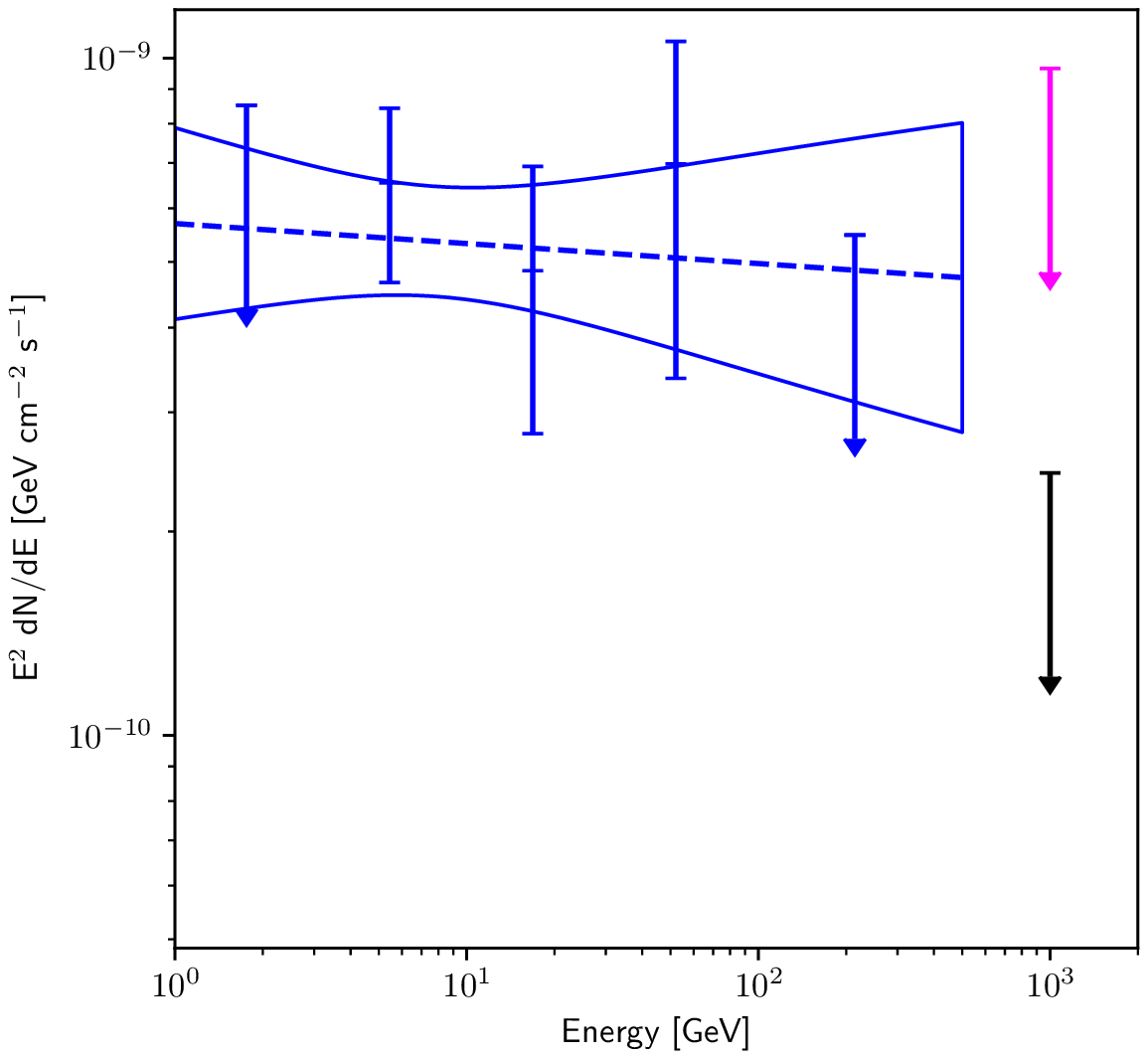}{0.21\textwidth}{(ah) FGL J2009.9+3544}
\fig{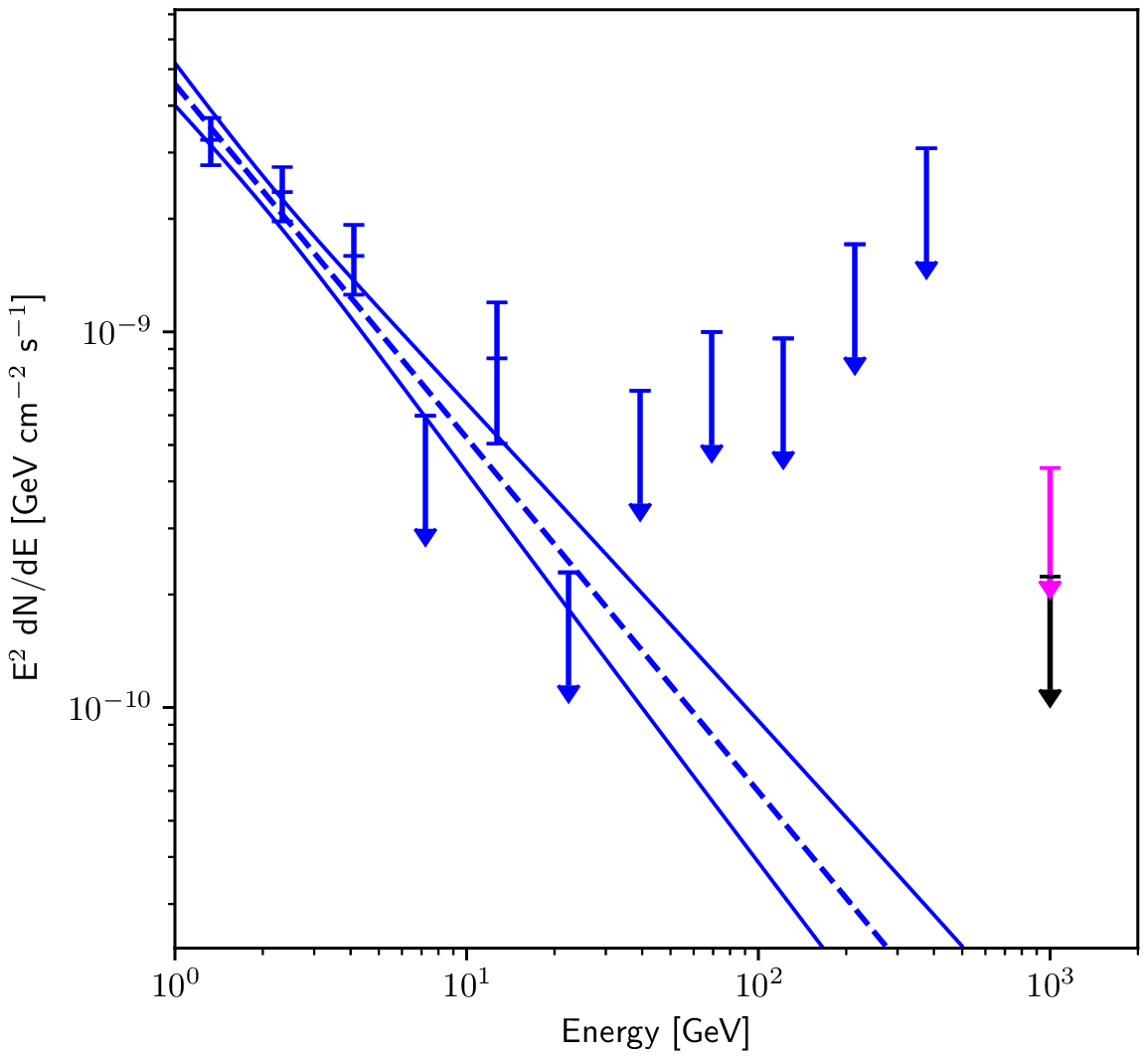}{0.21\textwidth}{(ai) FGL J2013.3+3616}
\fig{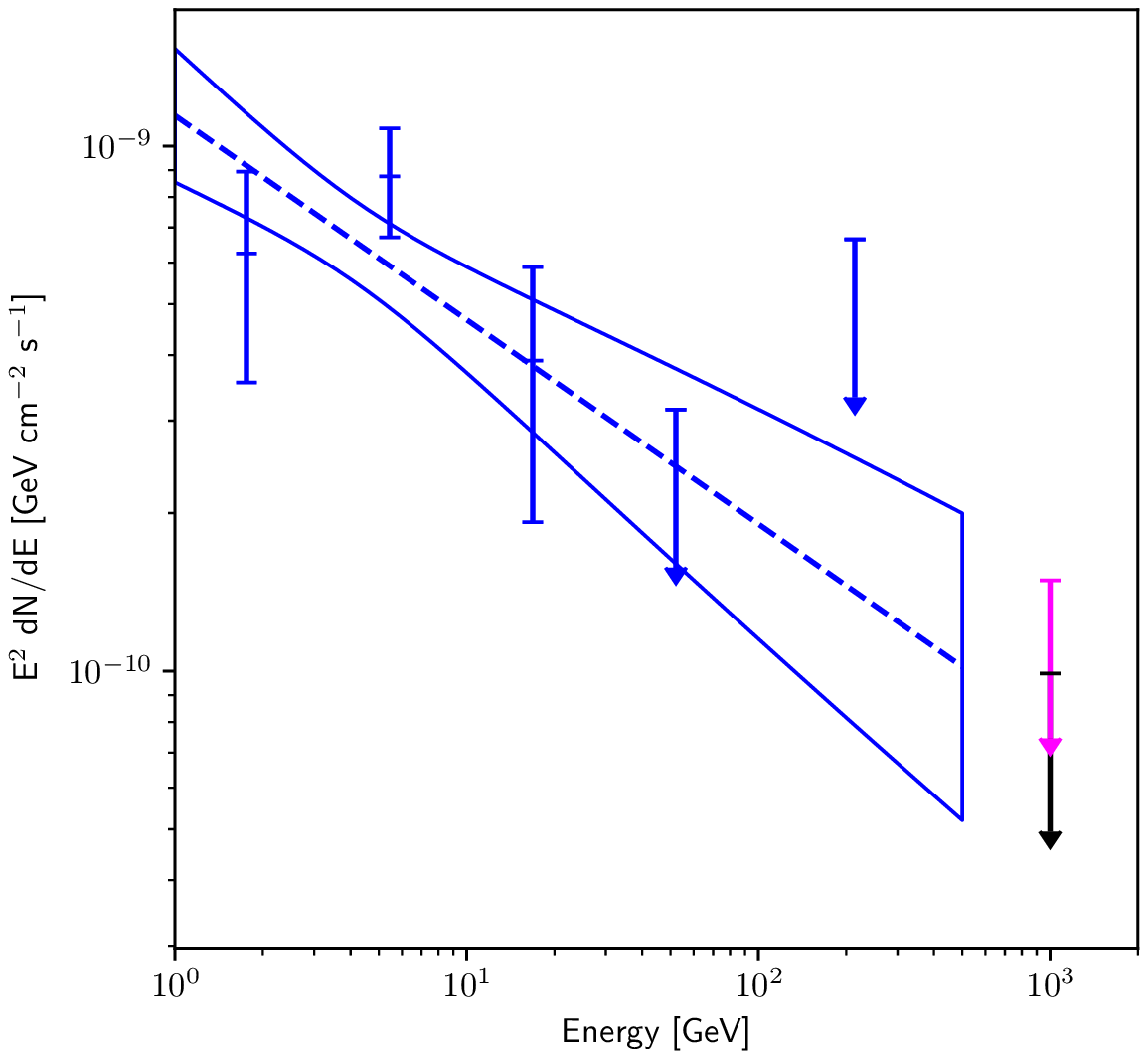}{0.21\textwidth}{(aj) FGL J2017.3+3526}
\fig{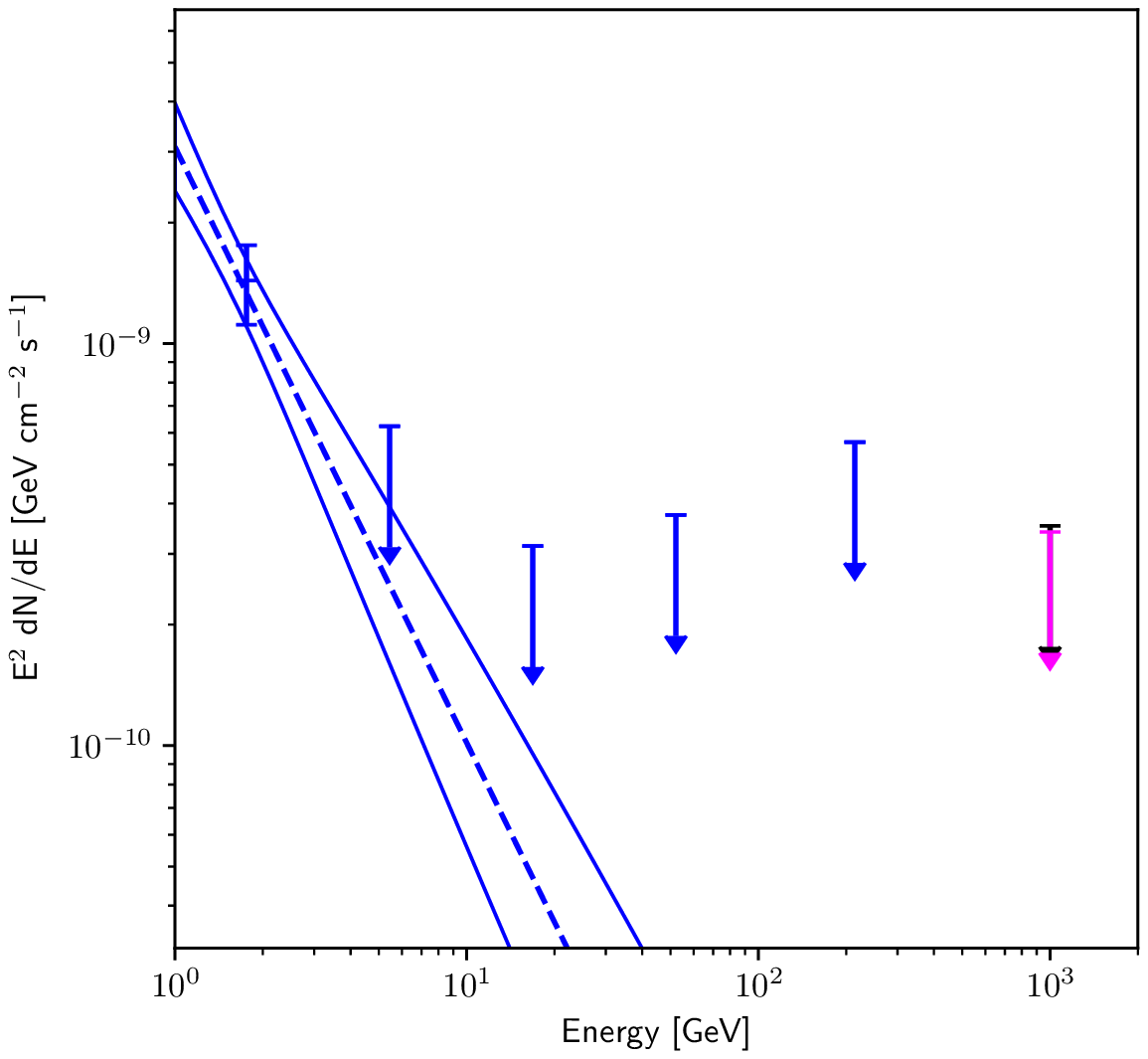}{0.21\textwidth}{(ak) FGL J2018.1+4111*}
}
\gridline{
\fig{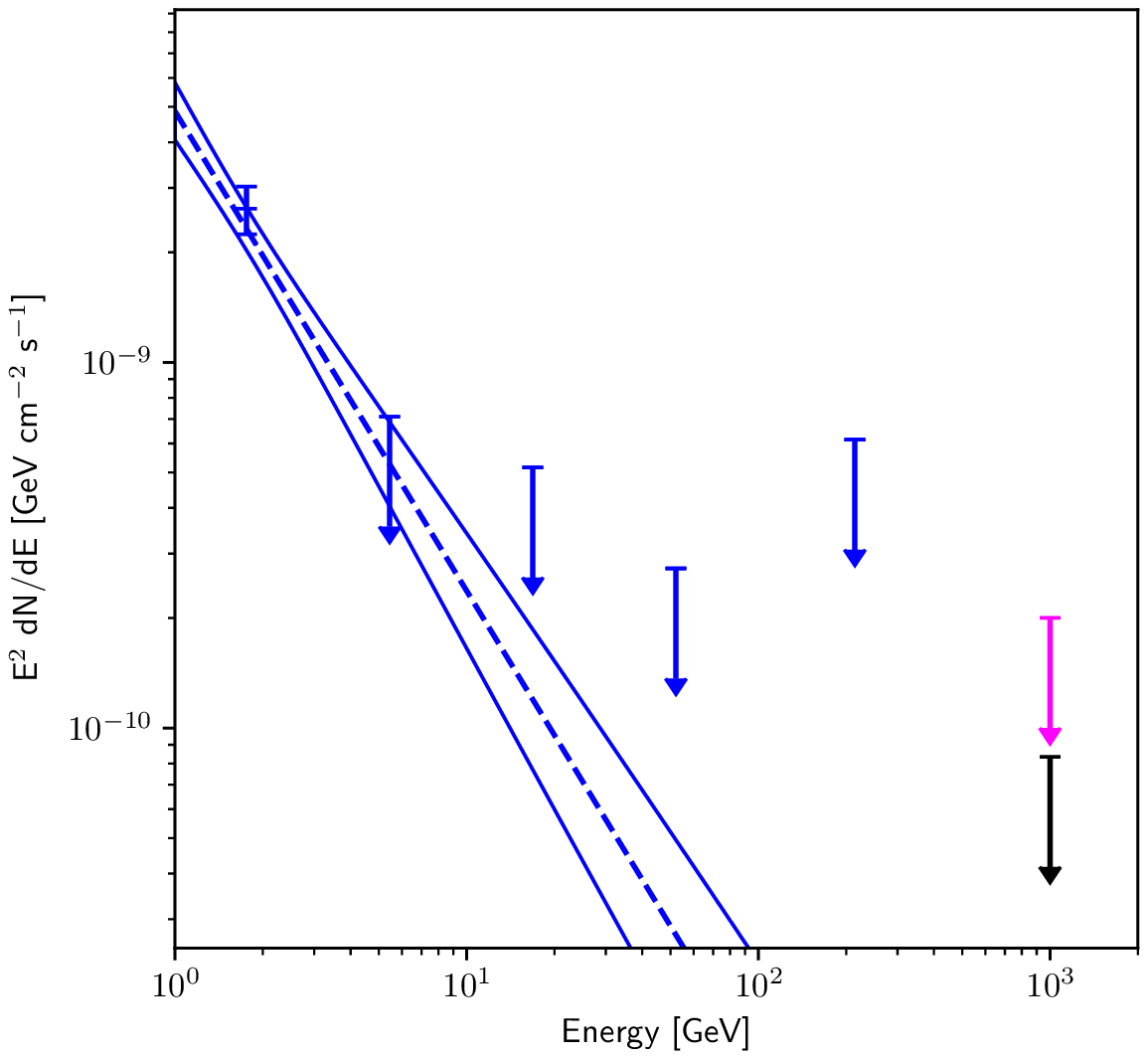}{0.21\textwidth}{(al) FGL J2022.6+3727}
\fig{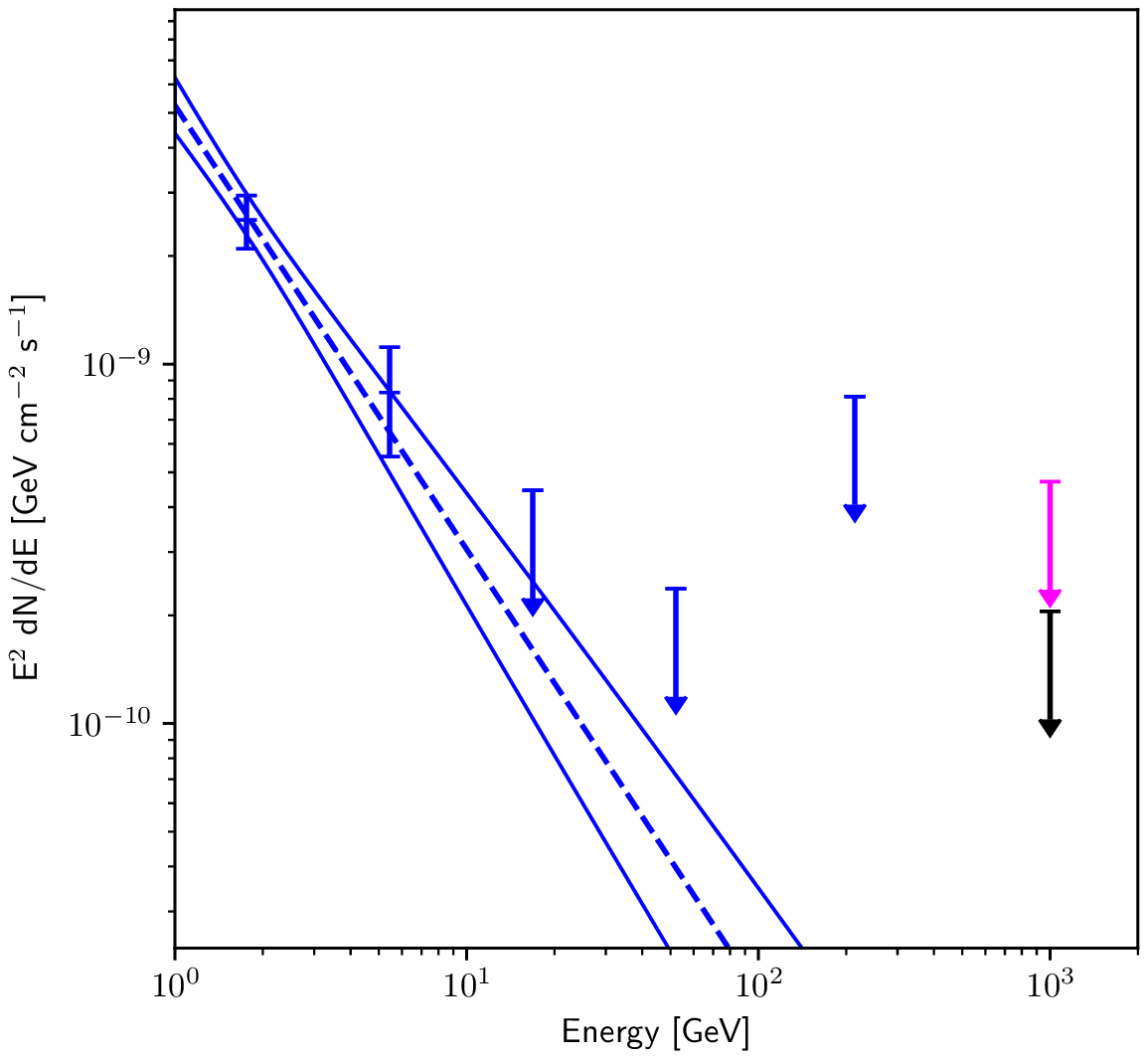}{0.21\textwidth}{(am) FGL J2024.4+3957*}
\fig{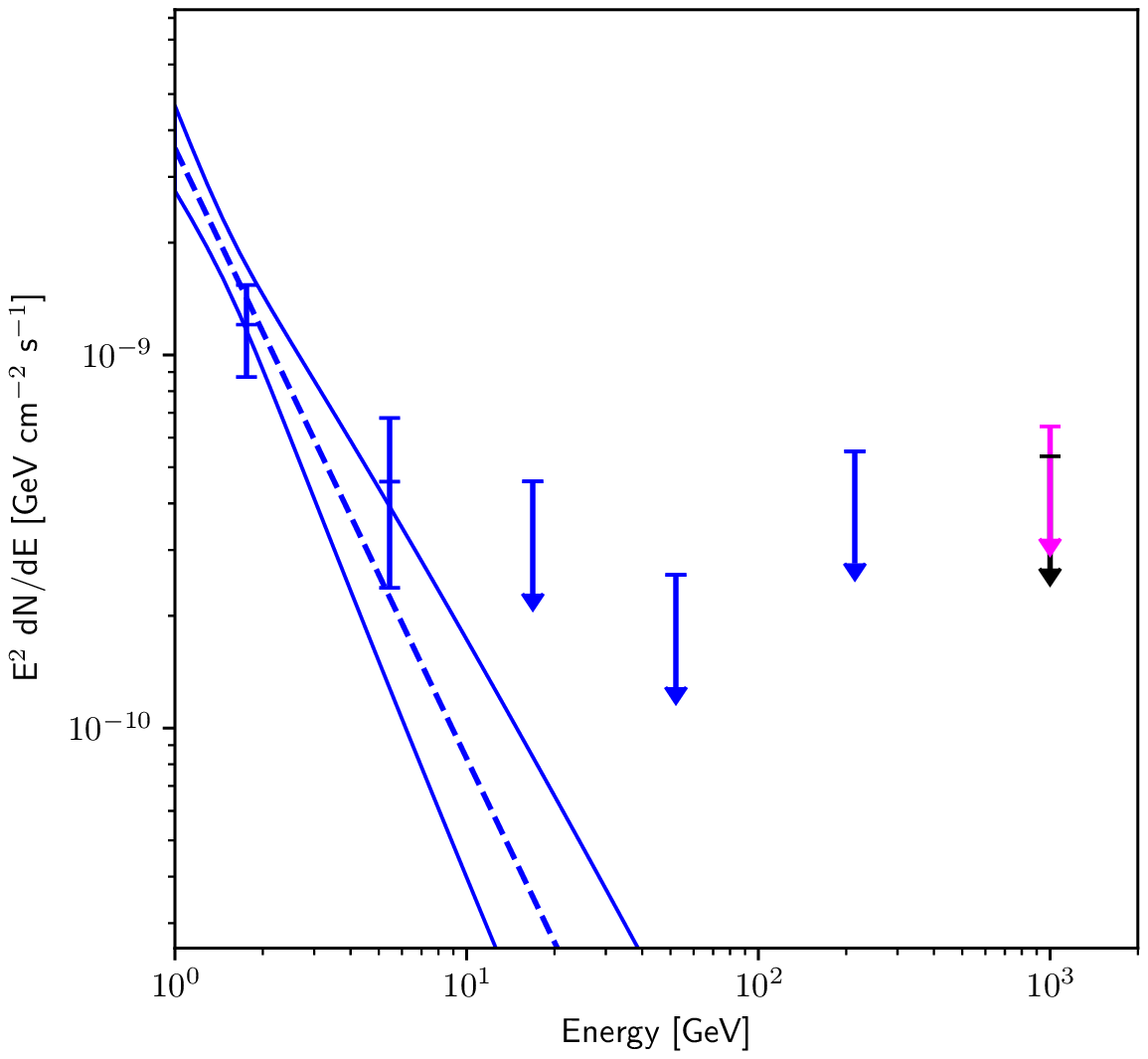}{0.21\textwidth}{(an) FGL J2025.9+3904}
\fig{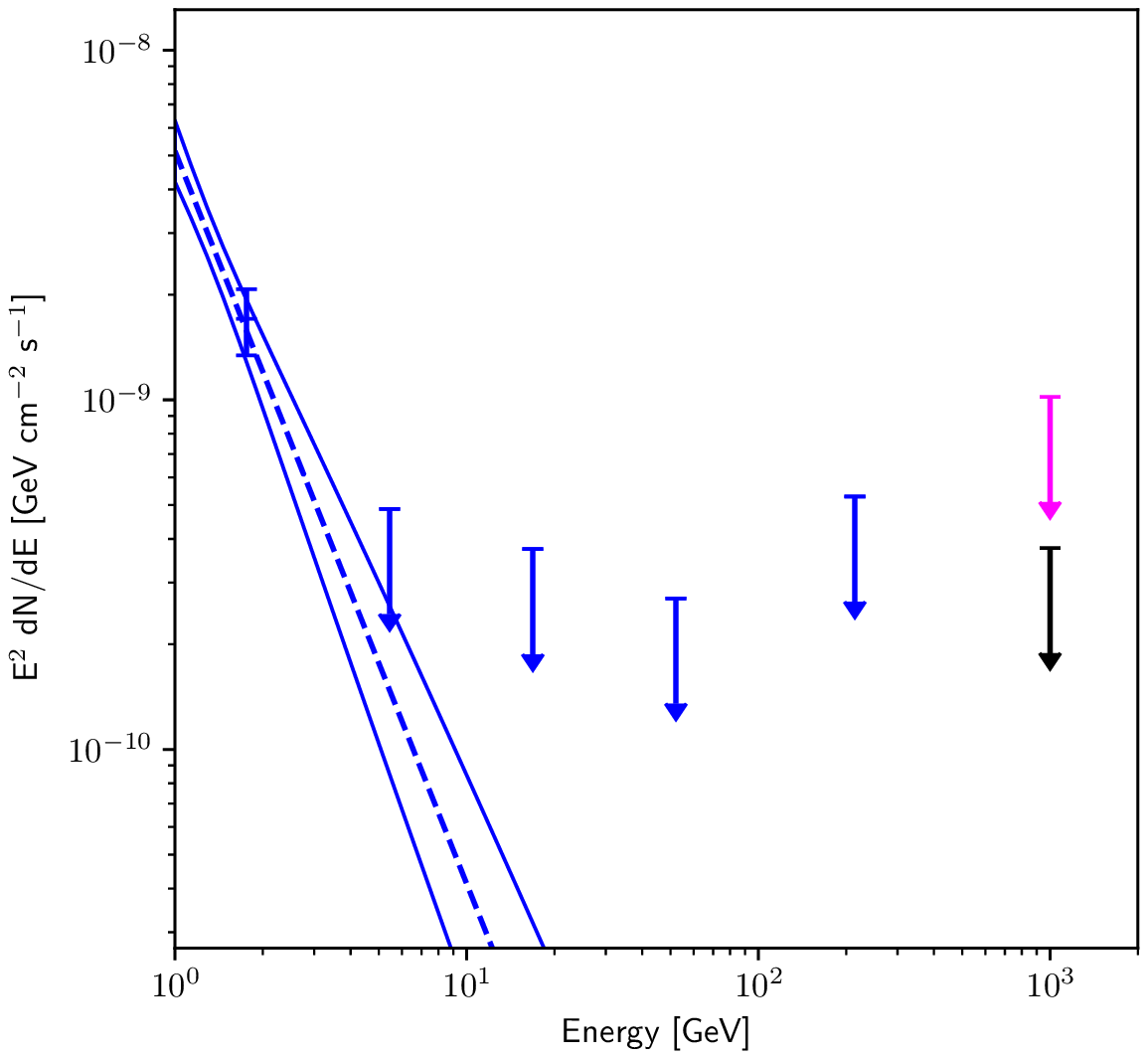}{0.21\textwidth}{(ao) FGL J2029.4+3940*}
}
\end{figure}

\setcounter{figure}{6}

\begin{figure}[htb!]
\gridline{
\fig{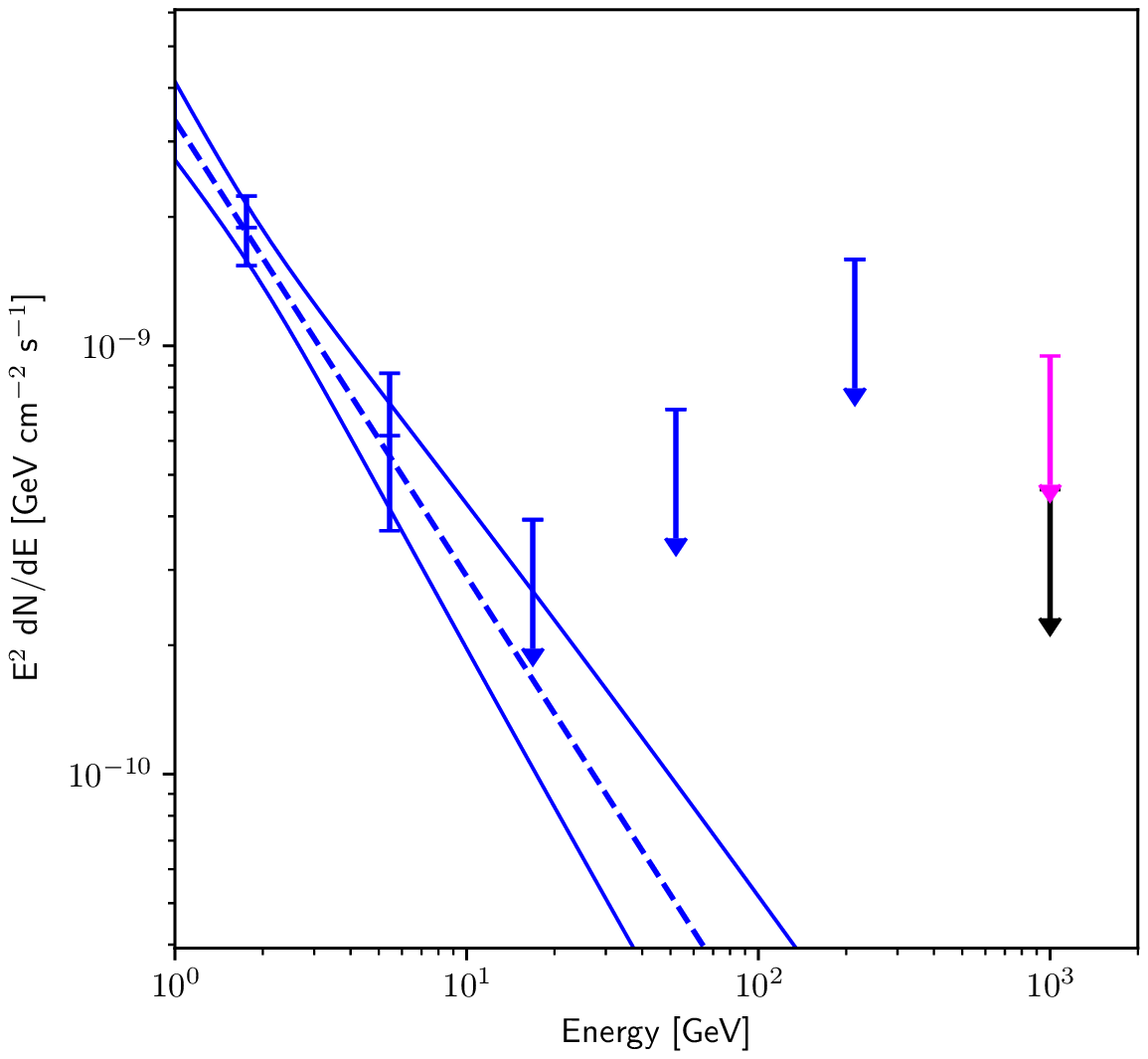}{0.21\textwidth}{(ap) FGL J2031.3+3857}
\fig{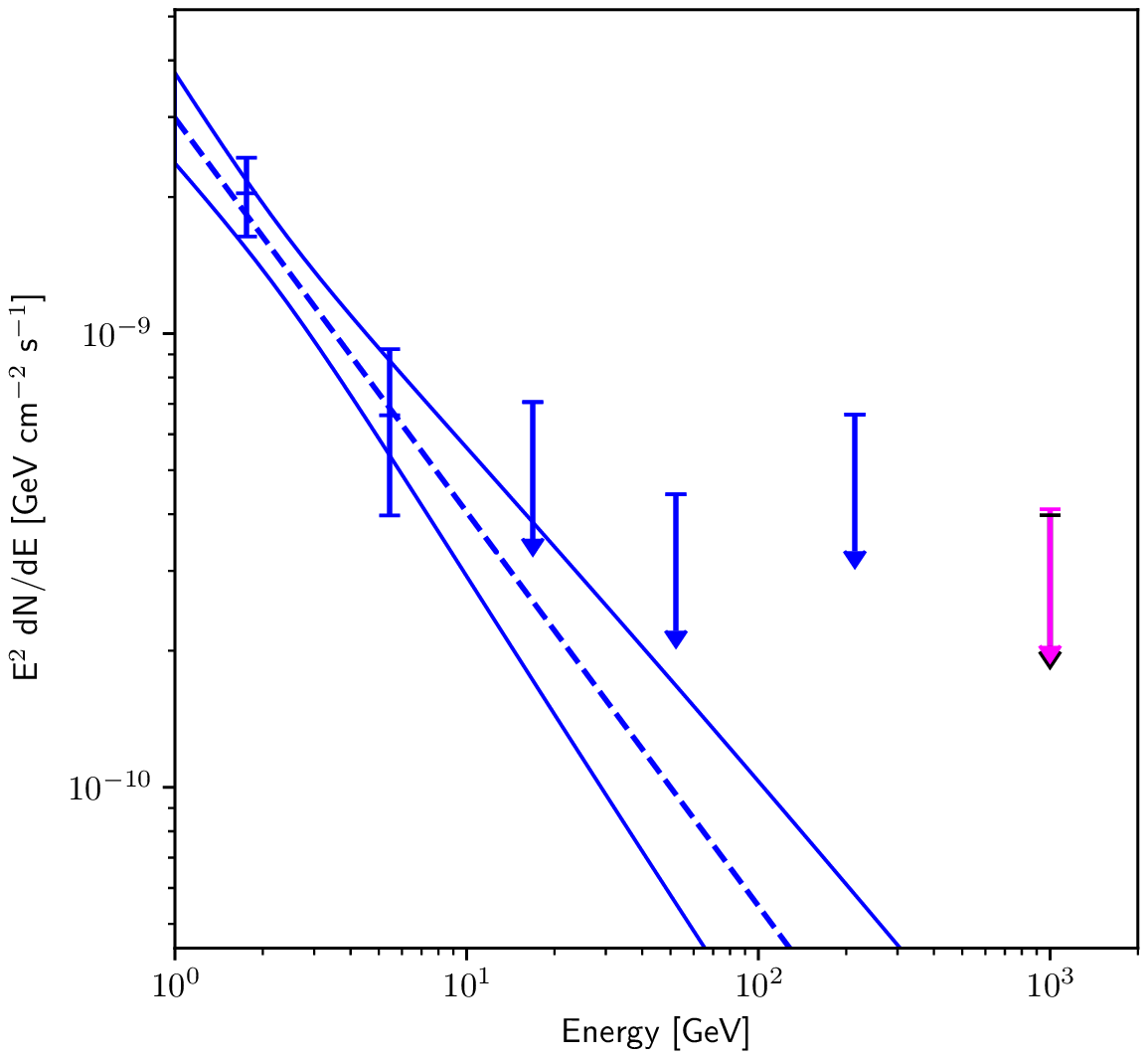}{0.21\textwidth}{(aq) FGL J2032.1+4058*}
\fig{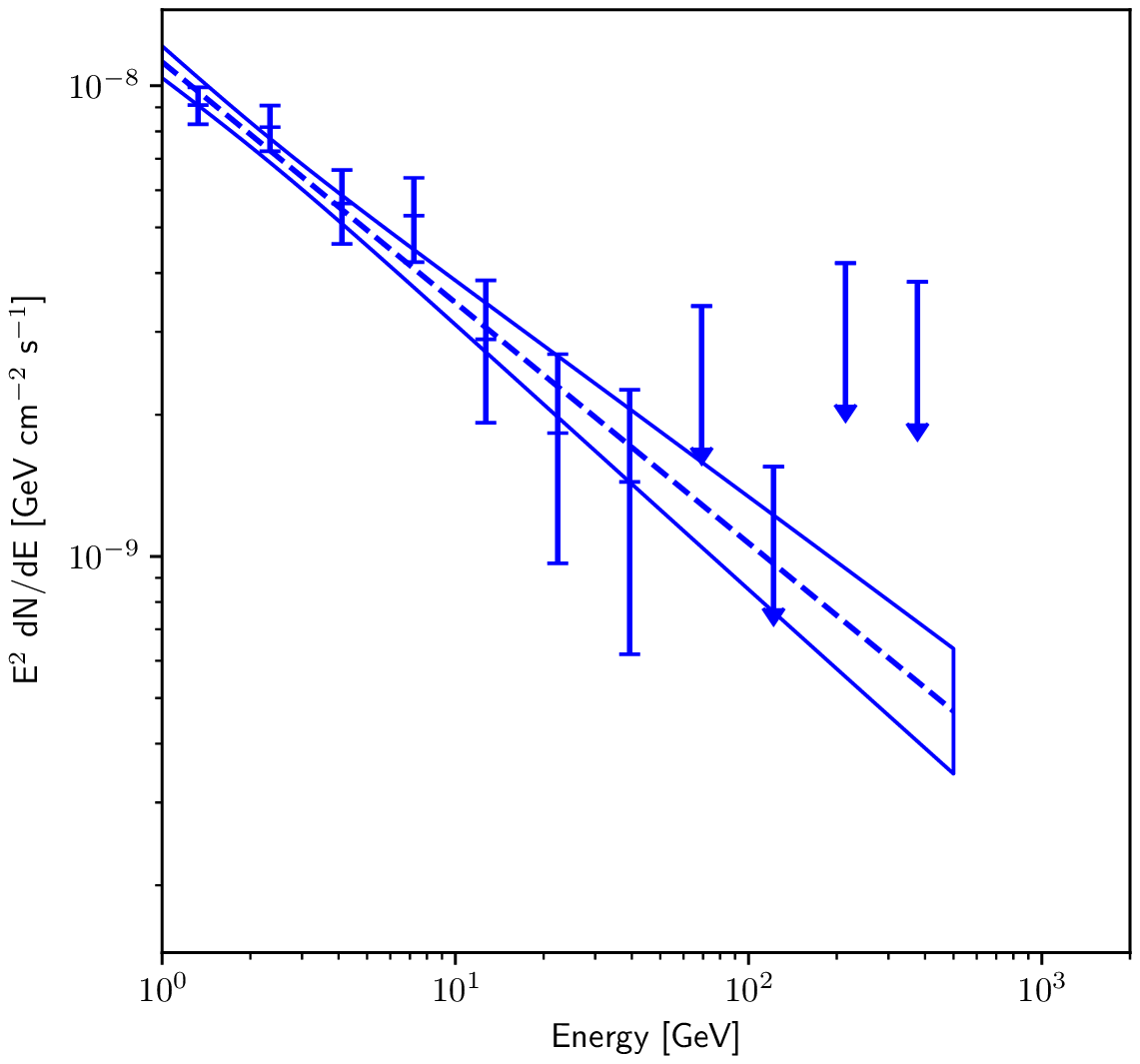}{0.21\textwidth}{(ar) FGL J2032.2+4128e*}
\fig{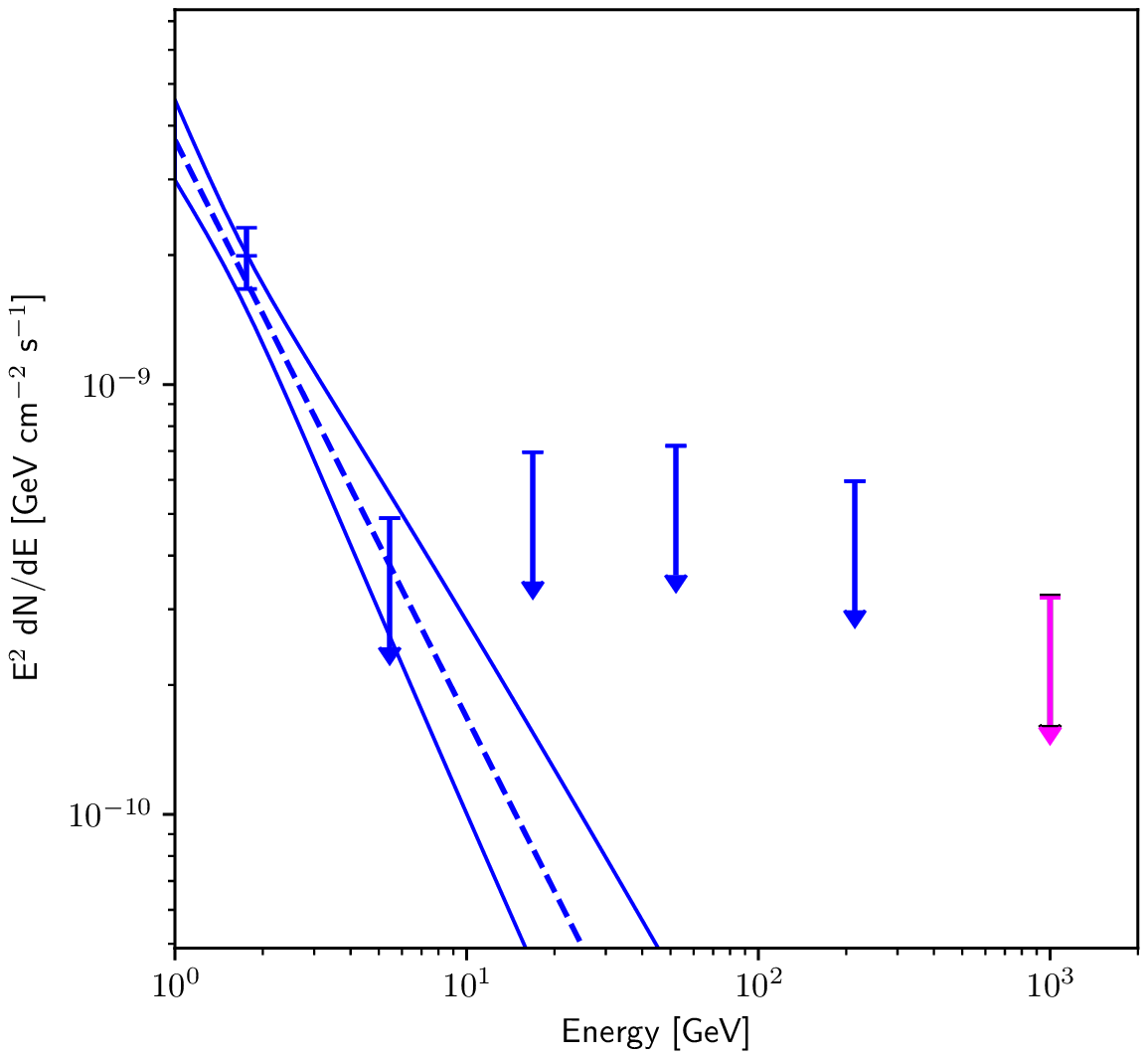}{0.21\textwidth}{(as) FGL J2032.7+4333}
}
\gridline{
\fig{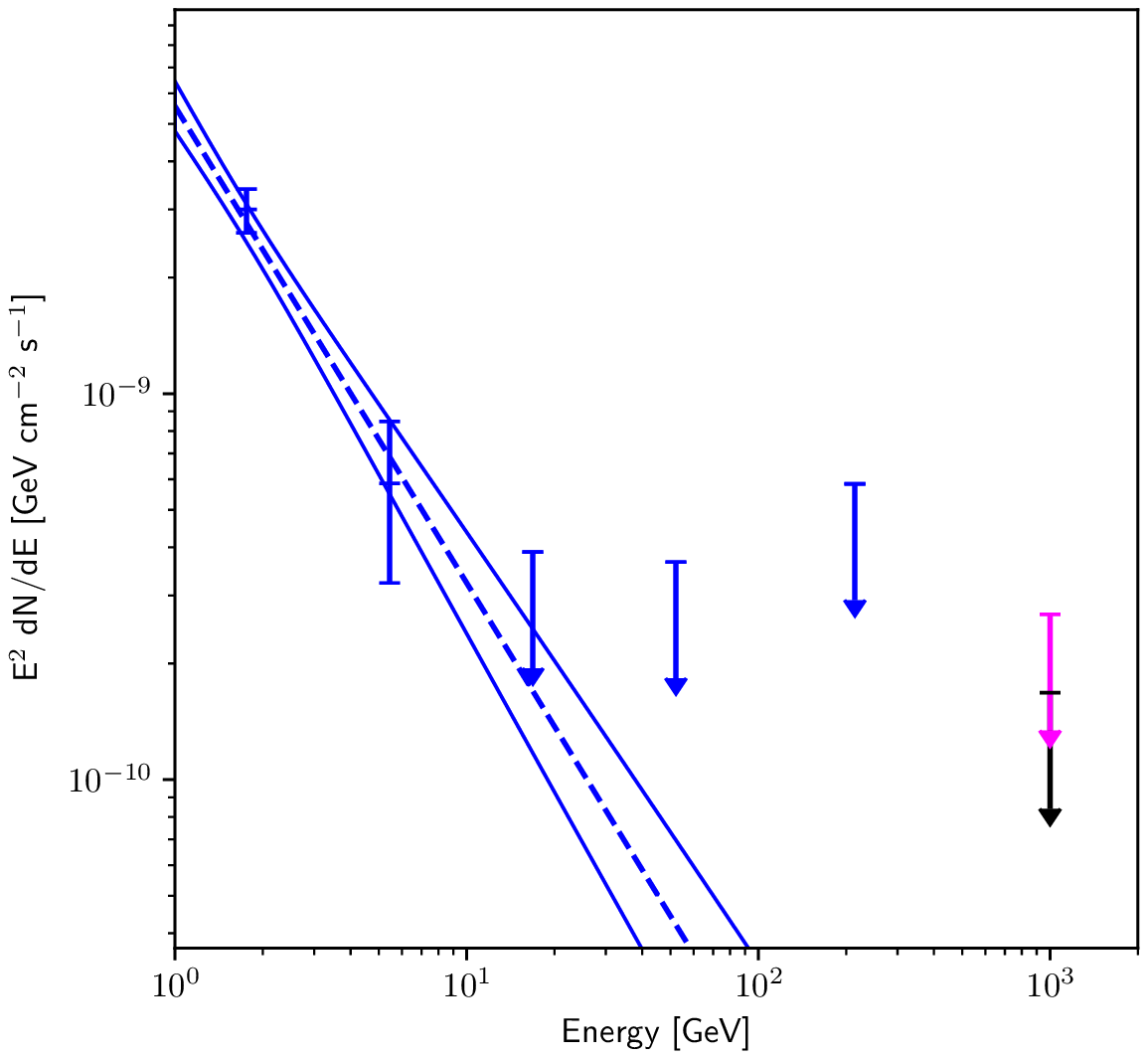}{0.21\textwidth}{(at) FGL J2032.9+3956*}
\fig{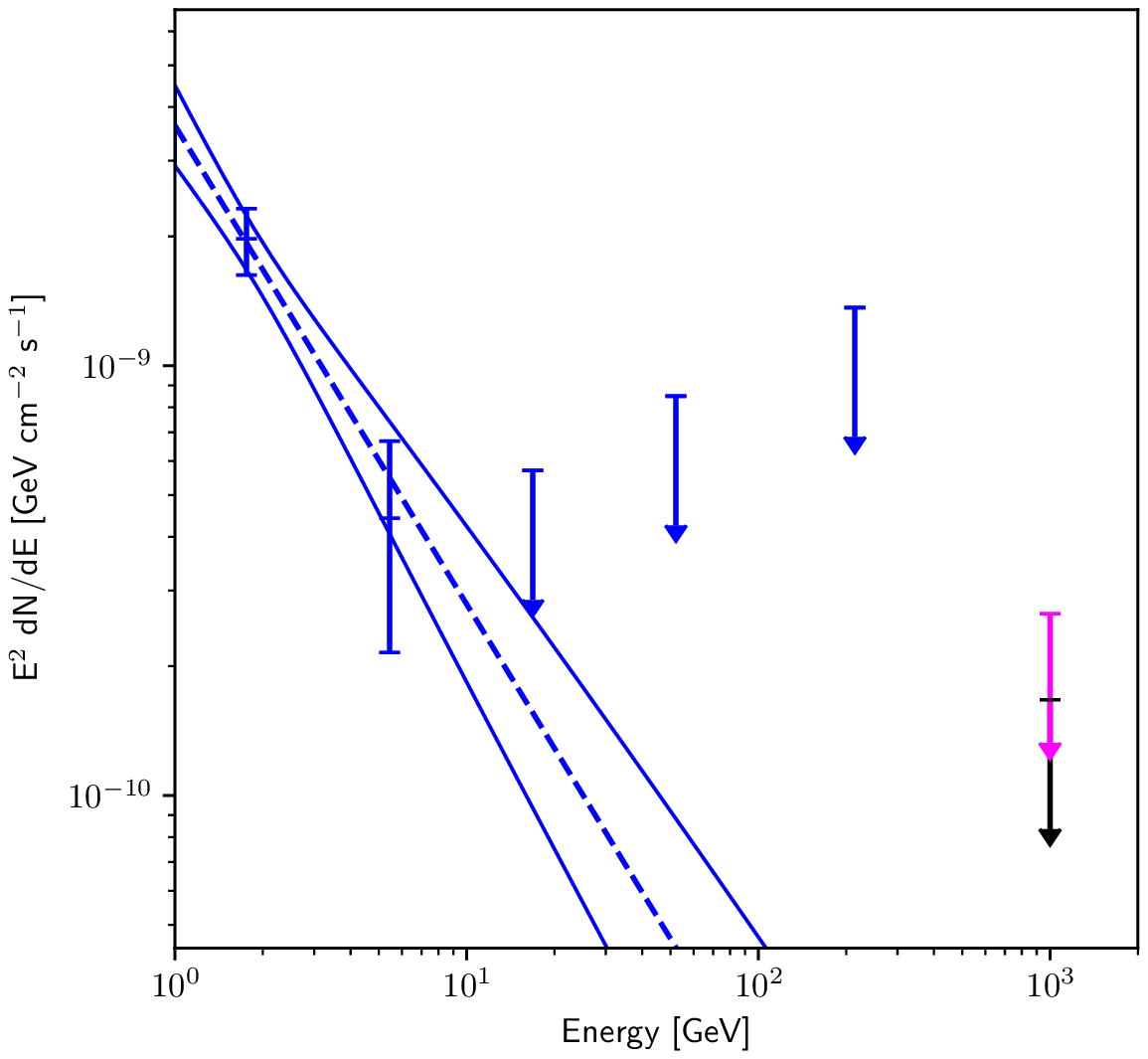}{0.21\textwidth}{(au) FGL J2034.3+4219*}
\fig{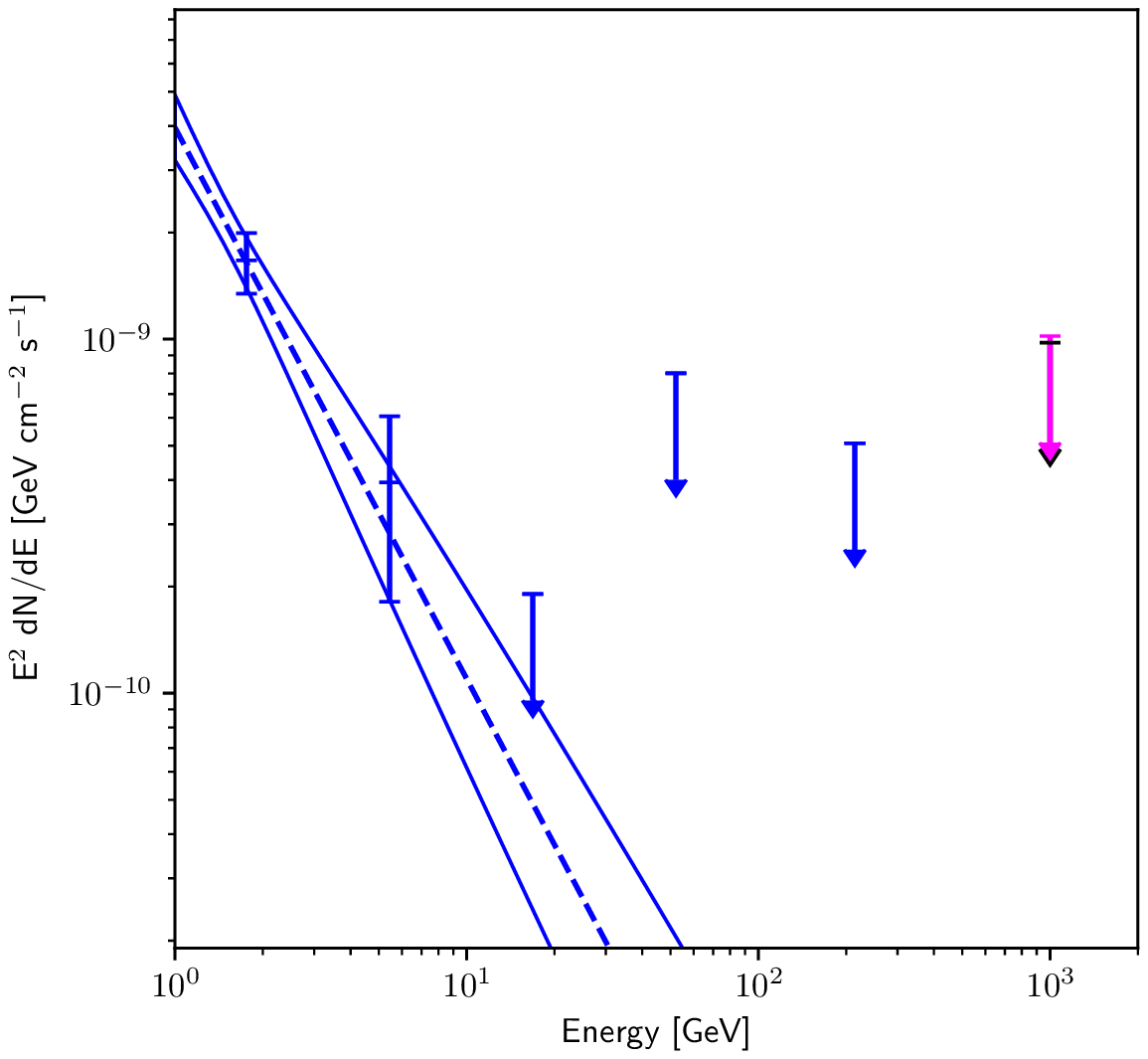}{0.21\textwidth}{(av) FGL J2036.9+4314}
\fig{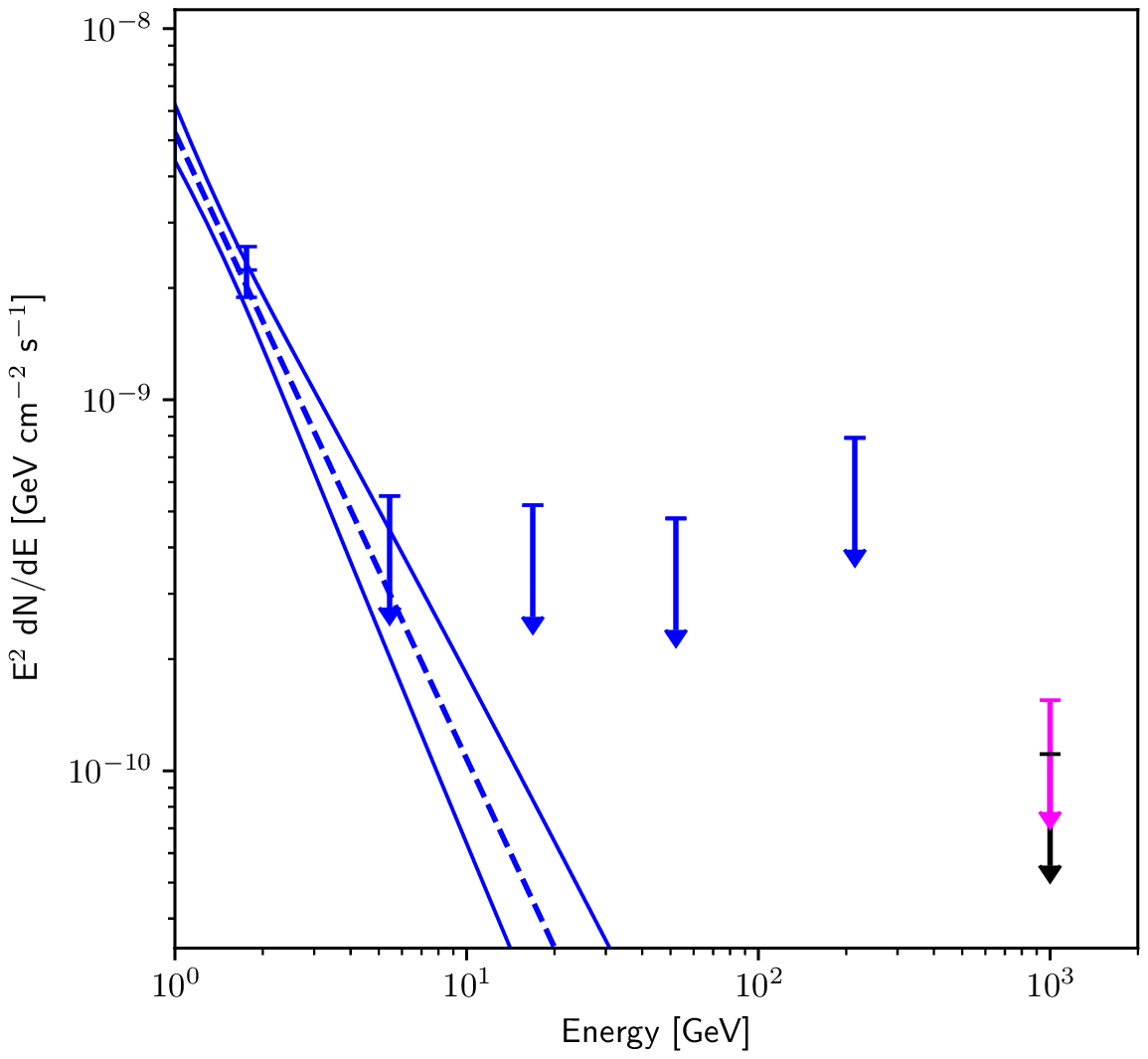}{0.21\textwidth}{(aw) FGL J2037.0+4005*}
}
\gridline{
\fig{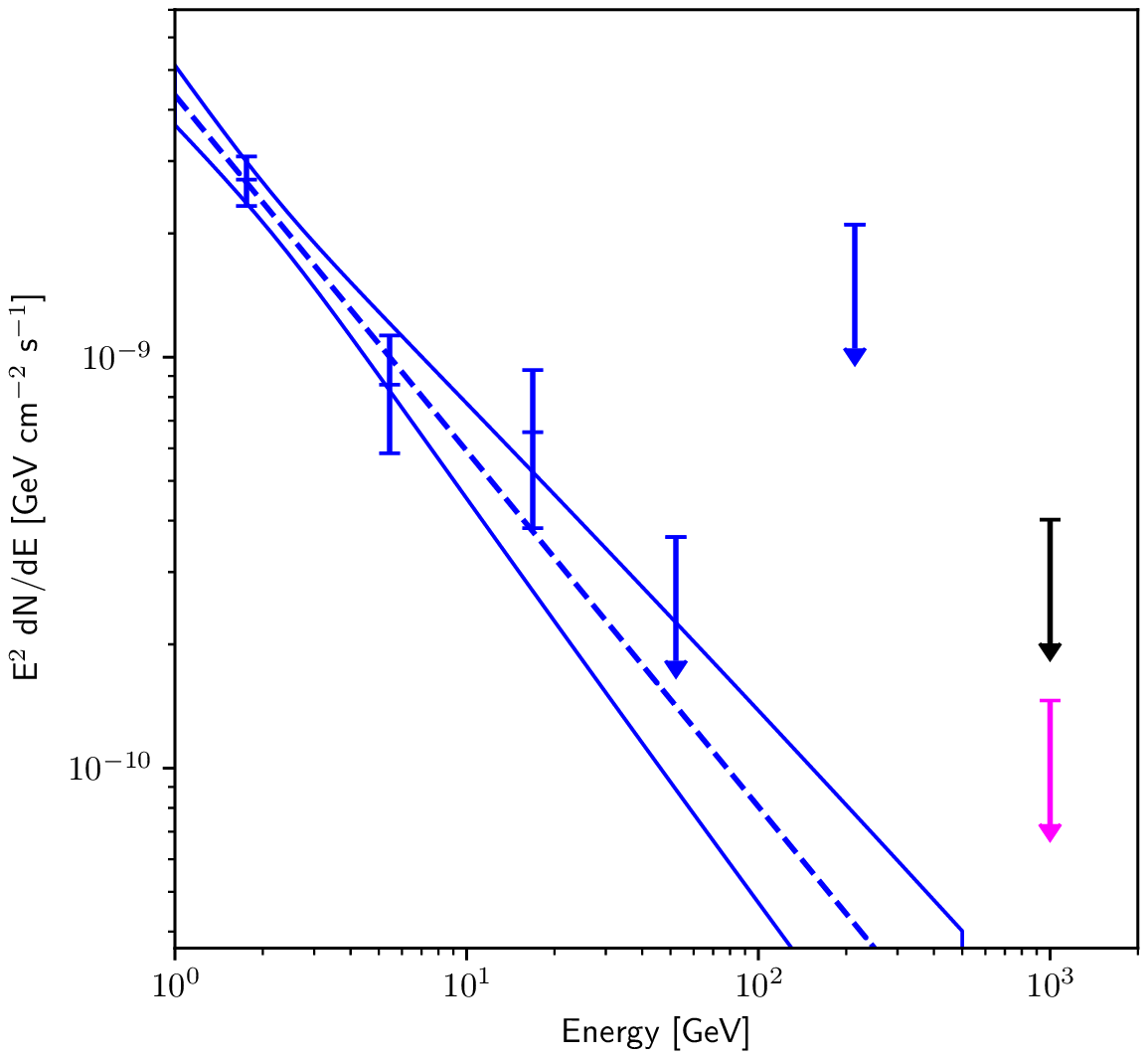}{0.21\textwidth}{(ax) FGL J2037.6+4152*}
\fig{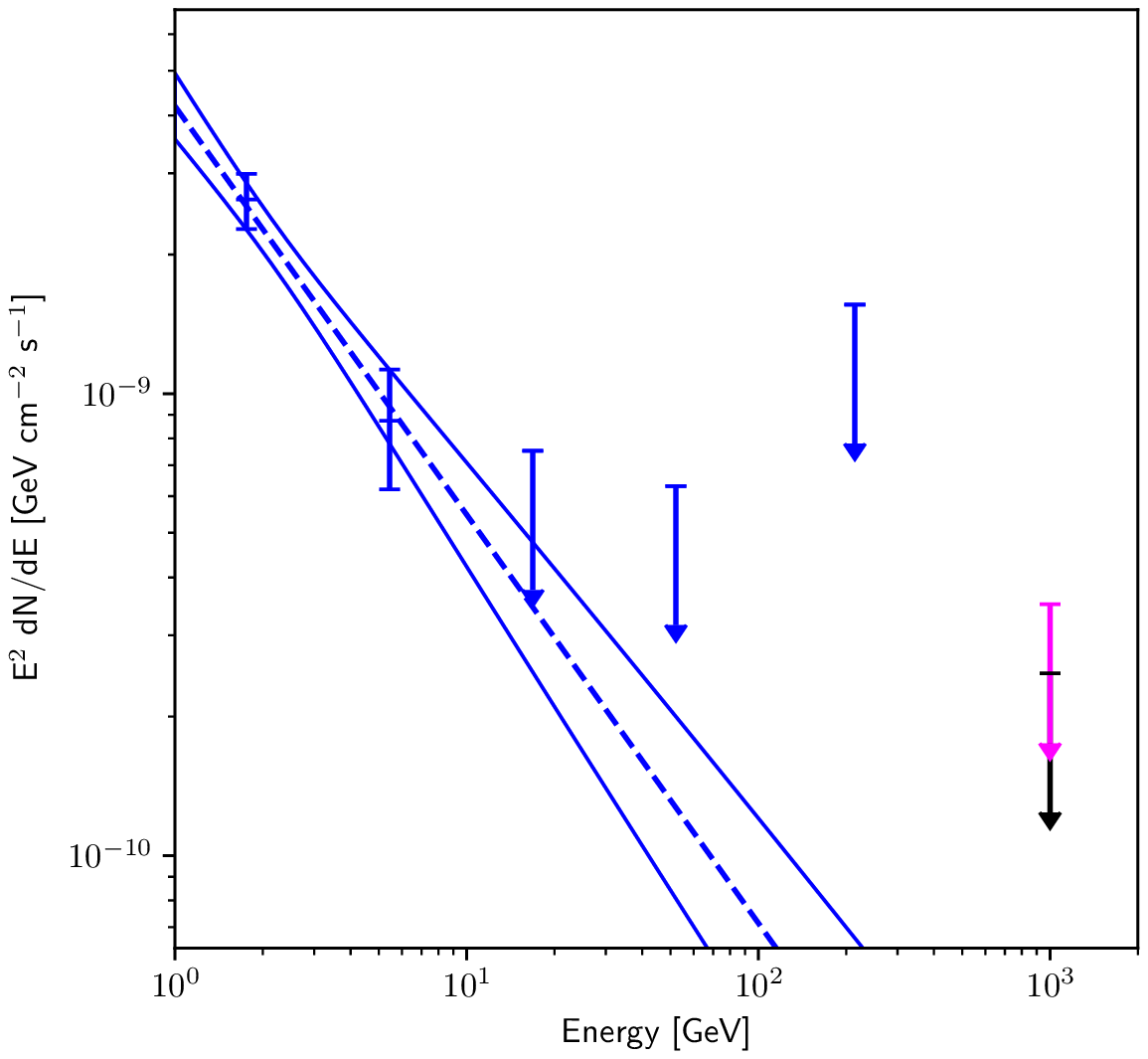}{0.21\textwidth}{(ay) FGL J2038.8+4235}
\fig{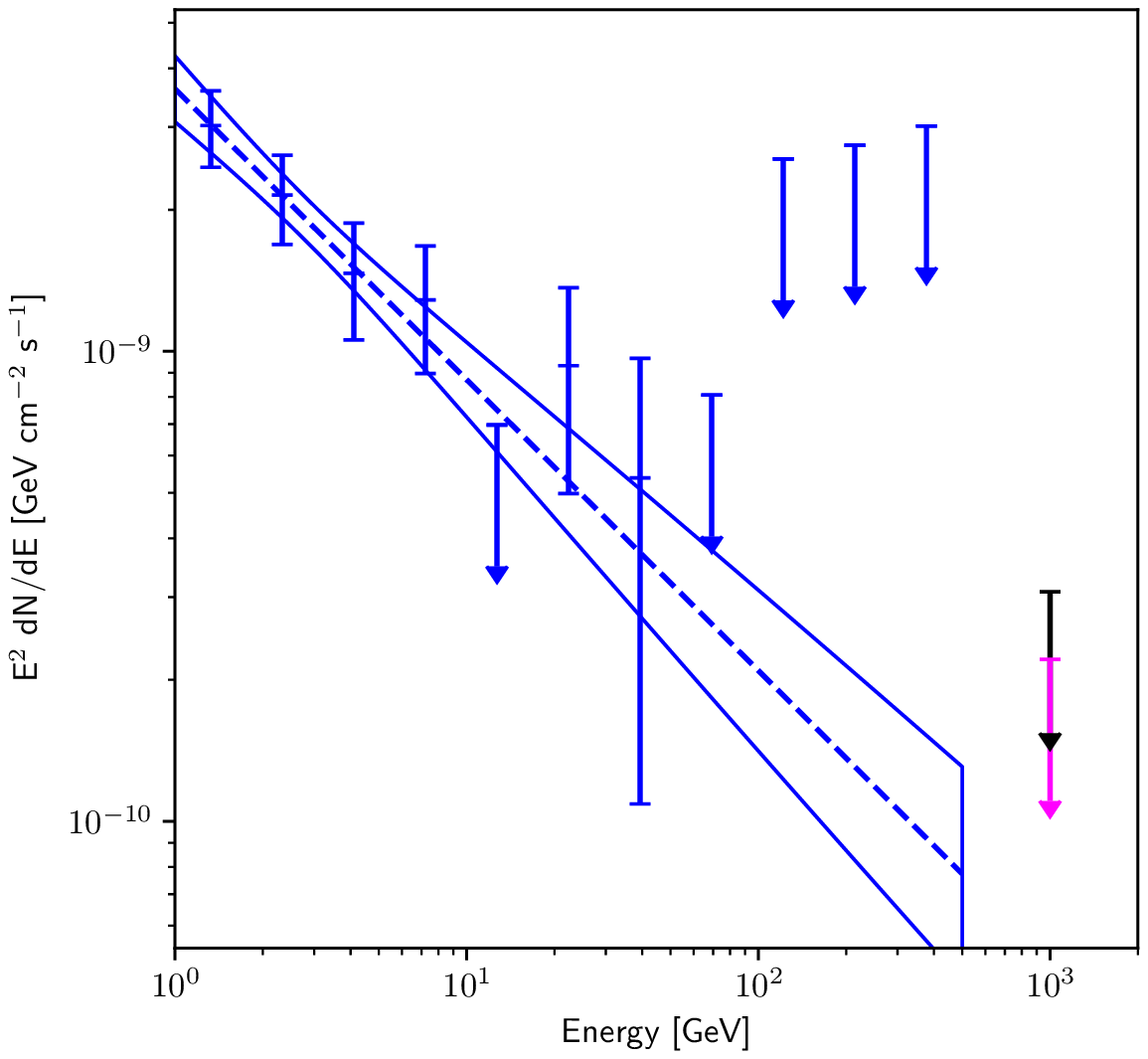}{0.21\textwidth}{(az) FGL J2040.1+4152}
\fig{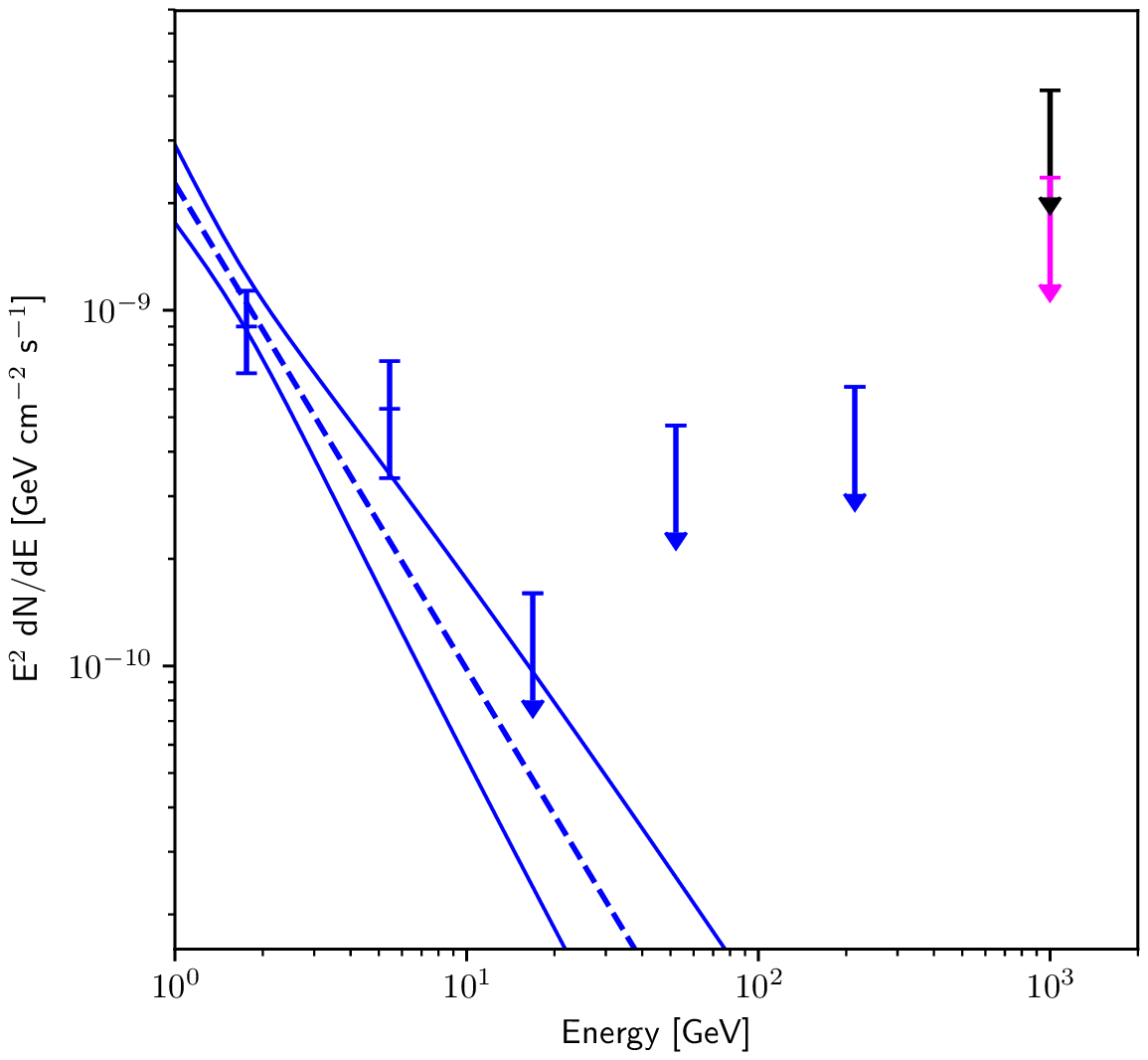}{0.21\textwidth}{(bb) FGL J2054.6+4130}
}
\caption{SEDs for all of the \LAT\ sources within the survey region (blue).
For sources that lie outside the \veritas\ source exclusion regions, \veritas\ 95\% confidence level differential upper limits are presented for the \point\ and \ext\ analysis in black and magenta respectively (see \autoref{sec:RestOfSurvey} for details). 
The red butterfly is the 3FGL catalog spectrum \citep{TheFermi-LAT:2015hja}, purple is from the 3FHL catalog \citep{2017ApJS..232...18A}.
The two extended sources 3FGL~J2021.0+4031e and 3FGL~J2028.6+4110e had their spectra fixed during the production of the 3FGL catalog and thus are plotted without butterflies.}
\label{fig:Fermi3FGLSpectra}
\end{figure}

\begin{longrotatetable}
\begin{deluxetable*}{cccPccccccc}
\tablecaption{$>$1 GeV \LAT\ analysis results for 3FGL sources that lie within the survey region. 
For each source, the Galactic coordinates for the source are listed together with the corresponding TS value at this location.
Associations as listed are from the 3FGL.  
Source class definitions: PSR = pulsar, FSRQ = flat spectrum radio quasar type of blazar, BCU = active galaxy of uncertain type, SNR = supernova remnant, SFR = star-forming region, SPP = special case, either SNR or PWN, U = unknown.  
Sources marked with a * next to their name lie in the field of the Cygnus Cocoon. 
The spectra were fit with a PL (\autoref{eq:PL}), LP (\autoref{eq:LP}), or a PLEC (\autoref{eq:PLEC}).
\label{tab:Fermi_Results}}
\tablehead{
\multirow{2}{*}{Source Name} & \multirow{2}{*}{$l$ [\arcdeg]} & 
\multirow{2}{*}{$b$ [\arcdeg]} & \multirow{2}{*}{Associations} & 
\multirow{2}{*}{Class} & \multirow{2}{*}{TS} & \colhead{Spectral} & 
\colhead{$N_0$} & \colhead{$E_0/E_b$} &  \multirow{2}{*}{$\gamma$} &  
\colhead{$\beta/E_c$} \\ 
\colhead{} & \colhead{} & \colhead{} & \colhead{} &
\colhead{} & \colhead{} & \colhead{Type} & 
\colhead{[GeV$^{-1}$cm$^{-2}$s$^{-1}$]} & \colhead{[GeV]} &
\colhead{} & \colhead{[GeV]} 
}
\startdata
 3FGL~J1951.6+2926 & 65.67 & 1.32 & None & SPP & 60.7 & PL & (6.29 $\pm$ 0.88)E-10 & 1.66 & 3.37 $\pm$ 0.27 & N/A \\
 3FGL~J1952.9+3253 & 68.78 & 2.83 & PSR~J1952+3252 1FHL~J1953.3+3251 & PSR & 10581.3 & PLEC & (7.52 $\pm$ 0.34)E-8 & 0.67 & 2.02 $\pm$ 0.08 & 5.56 $\pm$ 0.72 \\
 3FGL~J1958.6+2845 & 65.88 & -0.35 & LAT~PSR~J1958+2846 1FHL~J1958.6+2845 3EG~J1958+2909 & PSR & 4959.8 & PLEC & (2.71 $\pm$ 0.10)E-8 & 1.02 & 2.17 $\pm$ 0.11 & 5.60 $\pm$ 1.03 \\
 3FGL~J2004.4+3338 & 70.67 & 1.19 & 1FHL~J2004.4+3339 & U & 582.2 & PL & (5.21 $\pm$ 0.29)E-10 & 2.67 & 2.47 $\pm$ 0.07 & N/A \\
 3FGL~J2015.6+3709 & 74.89 & 1.2 & MG2~J201534+3710 VER~J2016+371 1FHL~J2015.8+3710 3EG~J2016+3657 & FSRQ & 1999.2 & LP & (5.46 $\pm$ 0.19)E-9 & 1.52 & 2.63 $\pm$ 0.07 & 0.02 $\pm$ 0.04 \\
 3FGL~J2017.9+3627 & 74.54 & 0.41 & MGRO~J2019+37 PSR~J2017+3625 & PSR & 1529.8 & LP & (5.38 $\pm$ 0.21)E-9 & 1.65 & 2.47 $\pm$ 0.11 & 0.69 $\pm$ 0.12 \\
 3FGL~J2018.5+3851 & 76.59 & 1.66 & TXS~2016+386 1FHL~J2018.3+3851 & BCU & 149.6 & PL & (8.29 $\pm$ 0.88)E-11 & 4.18 & 2.26 $\pm$ 0.10 & N/A \\
 3FGL~J2018.6+4213* & 79.4 & 3.53 & None & U & 34.8 & PL & (5.62 $\pm$ 1.12)E-11 & 3.66 & 2.31 $\pm$ 0.20 & N/A \\
 3FGL~J2021.0+4031e & 78.24 & 2.2 & Gamma~Cygni VER~J2019+407 1FHL~J2021.0+4031e 1AGL~J2022+4032 & SNR & 971.8 & PL & (3.04 $\pm$ 0.12)E-10 & 6.78 & 2.05 $\pm$ 0.03 & N/A \\
 3FGL~J2021.1+3651 & 75.23 & 0.11 & PSR~J2021+3651 MGRO~J2019+37 1FHL~J2021.0+3651 1AGL~J2021+3652 & PSR & 42295 & PLEC & (1.80 $\pm$ 0.03)E-7 & 0.84 & 2.03 $\pm$ 0.05 & 5.00 $\pm$ 0.35 \\
 3FGL~J2021.5+4026 & 78.23 & 2.08 & LAT~PSR~J2021+4026 1AGL~J2022+4032 & PSR & 57923.9 & PLEC & (5.09 $\pm$ 0.10)E-7 & 0.66 & 1.98 $\pm$ 0.05 & 3.32 $\pm$ 0.19 \\
 3FGL~J2022.2+3840* & 76.85 & 0.96 & None & SPP & 100.1 & PL & (1.66 $\pm$ 0.18)E-9 & 1.45 & 3.74 $\pm$ 0.27 & N/A \\
 3FGL~J2023.5+4126* & 79.25 & 2.34 & None & U & 60.4 & PL & (1.76 $\pm$ 0.26)E-10 & 2.84 & 2.55 $\pm$ 0.16 & N/A \\
 3FGL~J2025.2+3340 & 73.1 & -2.41 & B2~2023+33 & BCU & 206.3 & PL & (5.85 $\pm$ 0.49)E-10 & 2.03 & 2.86 $\pm$ 0.12 & N/A \\
 3FGL~J2028.5+4040c* & 79.19 & 1.13 & None & U & 52 & PL & (8.46 $\pm$ 1.25)E-10 & 1.68 & 3.34 $\pm$ 0.28 & N/A \\
 3FGL~J2028.6+4110e & 79.6 & 1.4 & Cygnus~Cocoon MGRO~J2031+41 1FHL~J2028.6+4110e & SFR & 3698.6 & PL & (6.70 $\pm$ 0.12)E-9 & 3.63 & 2.19 $\pm$ 0.02 & N/A \\
 3FGL~J2030.0+3642 & 76.13 & -1.43 & PSR~J2030+3641 & PSR & 1642.2 & PLEC & (9.87 $\pm$ 1.51)E-9 & 1.54 & 0.93 $\pm$ 0.28 & 1.82 $\pm$ 0.32 \\
 3FGL~J2030.8+4416 & 82.35 & 2.89 & LAT~PSR~J2030+4415 & PSR & 1237.8 & PLEC & (9.43 $\pm$ 2.61)E-9 & 1.6 & 1.73 $\pm$ 0.38 & 1.90 $\pm$ 0.58 \\
 3FGL~J2032.2+4126 & 80.22 & 1.02 & LAT~PSR~J2032+4127 TeV~J2032+4130 1FHL~J2032.1+4125 1AGL~J2032+4102 & PSR & 3245 & PLEC & (1.08 $\pm$ 0.05)E-8 & 1.56 & 1.03 $\pm$ 0.12 & 3.65 $\pm$ 0.38 \\
 3FGL~J2032.5+3921* & 78.57 & -0.27 & None & U & 38.6 & PL & (1.09 $\pm$ 0.20)E-10 & 3.13 & 2.57 $\pm$ 0.19 & N/A \\
 3FGL~J2032.5+4032* & 79.51 & 0.44 & 1AGL~J2032+4102 & U & 28.2 & PL & (9.08 $\pm$ 1.80)E-10 & 1.5 & 3.60 $\pm$ 0.39 & N/A \\
 3FGL~J2034.4+3833c* & 78.16 & -1.04 & None & U & 57.5 & LP & (2.28 $\pm$ 2.19)E-9 & 0.53 & 1.34 $\pm$ 1.49 & 1.78 $\pm$ 0.59 \\
 3FGL~J2034.6+4302* & 81.77 & 1.6 & None & U & 32.9 & PL & (4.20 $\pm$ 0.78)E-10 & 1.85 & 3.14 $\pm$ 0.30 & N/A \\
 3FGL~J2035.0+3634 & 76.63 & -2.32 & None & U & 432.3 & LP & (7.72 $\pm$ 0.63)E-10 & 2.13 & 1.43 $\pm$ 0.19 & 0.88 $\pm$ 0.16 \\
 3FGL~J2038.4+4212* & 81.53 & 0.54 & None & U & 49 & LP & (6.70 $\pm$ 1.63)E-10 & 1.65 & 0.73 $\pm$ 0.53 & 1.66 $\pm$ 0.48 \\
 3FGL~J2039.4+4111* & 80.83 & -0.21 & None & U & 89.5 & PL & (6.84 $\pm$ 0.80)E-10 & 1.87 & 2.96 $\pm$ 0.20 & N/A \\
 3FGL~J2042.4+4209* & 81.93 & -0.07 & None & U & 92.1 & PL & (8.47 $\pm$ 0.97)E-10 & 1.75 & 3.14 $\pm$ 0.20 & N/A \\
 \enddata
\tablecomments{The values for 3FGL J2015.6+3709 quoted in this table are from the best fit location of ($l$, $b$) = (74.89\arcdeg, 1.2\arcdeg) ((\ra, \dec) = (\hms{20}{15}{39.2}, \dms{37}{11}{31.1})), for details of this analysis see \autoref{sec:CTB}.}
\end{deluxetable*}
\end{longrotatetable}

\startlongtable
\begin{deluxetable*}{cccccccccc}
\tablecaption{$>$1 GeV \LAT\ analysis results of newly identified sources.  All sources were fit with a power law (\autoref{eq:PL}) and showed no significant evidence of curvature. 
Source class definitions are as in \autoref{tab:Fermi_Results}, sources marked with a * next to their name lie in the field of the Cygnus Cocoon. 
\label{tab:Fermi_NewPS_Results}}
\tablehead{
\multirow{2}{*}{Source Name} &\multirow{2}{*}{$l$ [$^\circ$]} & 
\multirow{2}{*}{$b$ [\arcdeg]} &
\multirow{2}{*}{Association} & \multirow{2}{*}{Class} & 
\multirow{2}{*}{TS}  & \colhead{$N_0$} & \colhead{$E_0$}  & 
\multirow{2}{*}{$\gamma$}  &\multirow{2}{*}{Notes} \\ 
\colhead{} & \colhead{} & \colhead{} & \colhead{} & \colhead{} & 
\colhead{} & \colhead{[GeV$^{-1}$cm$^{-2}$s$^{-1}$]} & 
\colhead{[GeV]} & \colhead{} & \colhead{}
}
\startdata
 FGL~J1949.0+3412 & 69.49 & 4.21 & None & U & 32.7 & (6.87 $\pm$ 1.40)E-11 & 2.62 & 2.54 $\pm$ 0.21 & \\
 FGL~J1955.0+3319 & 69.36 & 2.68 & None & U & 49.5 & (1.94 $\pm$ 0.31)E-10 & 2.3 & 2.78 $\pm$ 0.20 & \\
 FGL~J1958.6+3510 & 71.33 & 3.01 & Cygnus~X-1 & HMXB & 39.4 & (2.10 $\pm$ 0.45)E-11 & 4.78 & 2.18 $\pm$ 0.19 & \tablenotemark{a} \\
 FGL~J2005.7+3417 & 71.36 & 1.3 & 2HWC J2006+341 & U & 32.7 & (5.45 $\pm$ 1.02)E-10 & 1.61 & 3.22 $\pm$ 0.32 & \tablenotemark{b} \\
 FGL~J2006.3+3103 & 68.7 & -0.55 & PSR~J2006+3102 & PSR & 111 & (3.62 $\pm$ 0.40)E-10 & 2.16 & 2.80 $\pm$ 0.14 & \tablenotemark{c} \\
 FGL~J2009.9+3544 & 73.04 & 1.36 & HD~191612 & HMXB & 44.8 & (1.18 $\pm$ 0.24)E-11 & 6.75 & 2.03 $\pm$ 0.16 & \tablenotemark{d} \\
 FGL~J2013.3+3616 & 73.86 & 1.08 & G73.9+0.9 & SNR & 147 & (5.42 $\pm$ 0.52)E-10 & 2.06 & 2.94 $\pm$ 0.14 & \tablenotemark{e} \\
 FGL~J2017.3+3526 & 73.62 & -0.07 & None & U & 37 & (5.85 $\pm$ 1.16)E-11 & 3.47 & 2.39 $\pm$ 0.20 & \\
 FGL~J2018.1+4111* & 78.47 & 3.03 & None & U & 26.3 & (5.79 $\pm$ 1.19)E-10 & 1.62 & 3.48 $\pm$ 0.43 & \\
 FGL~J2022.6+3727 & 75.88 & 0.21 & None & U & 55 & (7.81 $\pm$ 1.12)E-10 & 1.74 & 3.31 $\pm$ 0.24 & \\
 FGL~J2024.4+3957* & 78.13 & 1.35 & None & U & 56.6 & (6.73 $\pm$ 0.96)E-10 & 1.89 & 3.24 $\pm$ 0.24 & \\
 FGL~J2025.9+3904 & 77.58 & 0.61 & None & U & 27.6 & (1.53 $\pm$ 0.34)E-9 & 1.26 & 3.63 $\pm$ 0.56 & \\
 FGL~J2029.4+3940* & 78.46 & 0.41 & None & U & 37.1 & (1.54 $\pm$ 0.27)E-9 & 1.34 & 4.10 $\pm$ 0.51 & \\
 FGL~J2031.3+3857 & 78.1 & -0.32 & None & U & 43.2 & (5.72 $\pm$ 0.93)E-10 & 1.78 & 3.06 $\pm$ 0.27 & \\
 FGL~J2032.1+4058* & 79.81 & 0.75 & Cygnus~X-3 & HMXB & 34.9 & (3.31 $\pm$ 0.60)E-10 & 2.15 & 2.87 $\pm$ 0.23 & \tablenotemark{f} \\
 FGL~J2032.2+4128e* & 80.24 & 1.04 & None & U & 321 & (1.44 $\pm$ 0.09)E-9 & 2.27 & 2.52 $\pm$ 0.07 & \\
 FGL~J2032.7+4333 & 81.96 & 2.19 & None & U & 45.7 & (7.81 $\pm$ 1.24)E-10 & 1.59 & 3.34 $\pm$ 0.36 & \\
 FGL~J2032.9+3956* & 79.08 & 0.03 & None & U & 80.2 & (8.82 $\pm$ 1.06)E-10 & 1.77 & 3.24 $\pm$ 0.19 & \\
 FGL~J2034.3+4219* & 81.14 & 1.23 & None & U & 49.2 & (4.91 $\pm$ 0.76)E-10 & 1.9 & 3.11 $\pm$ 0.30 & \\
 FGL~J2036.9+4314 & 82.16 & 1.39 & None & U & 40.8 & (9.24 $\pm$ 1.54)E-10 & 1.51 & 3.56 $\pm$ 0.41 & \\
 FGL~J2037.0+4005* & 79.67 & -0.53 & None & U & 60 & (1.19 $\pm$ 0.17)E-9 & 1.49 & 3.69 $\pm$ 0.35 & \\
 FGL~J2037.6+4152* & 81.15 & 0.47 & None & U & 87 & (4.81 $\pm$ 0.57)E-10 & 2.16 & 2.87 $\pm$ 0.18 & \\
 FGL~J2038.8+4235 & 81.85 & 0.72 & None & U & 87 & (2.65 $\pm$ 0.33)E-10 & 2.61 & 2.88 $\pm$ 0.18 & \\
 FGL~J2040.1+4152 & 81.43 & 0.1 & None & U & 110 & (2.62 $\pm$ 0.29)E-10 & 2.72 & 2.62 $\pm$ 0.13 & \\
 FGL~J2054.6+4130 & 82.85 & -2.23 & None & U & 31 & (4.89 $\pm$ 0.95)E-10 & 1.58 & 3.36 $\pm$ 0.43 & \\
\enddata
\tablenotetext{a}{FGL J1958.6+3510 is located close to Cygnus X-1 (separation = 3.6\arcmin ).
Cygnus X-1 has previously been identified in \citet{2016arXiv160505914Z} at (\ra,\dec) = (\hms{19}{58}{56.8}, \dms{+35}{11}{4.4}) as well as being identified in \citet{Zdziarski2016}.}
\tablenotetext{b}{FGL~J2005.7+3417 lies 0.15\arcdeg\ away from 2HWC J2006+341 which has a 1$\sigma$ positional uncertainty of 0.13\arcdeg.}
\tablenotetext{c}{FGL J2006.3+3103 lies coincident with PSR~J2006+3102 and was identified in \citet{2016arXiv160505914Z} as being the most likely counterpart due to the location and spectral shape.
No search for pulsations has been conducted and no evidence of spectral curvature was detected.}
\tablenotetext{d}{FGL J2009.9+3544 lies 0.075\arcdeg\ away from the source identified as n1 in \citet{Zdziarski2016} which they associated with HD 191612 (0.092\arcdeg\ from FGL J2009.9+3544), an O8 high mass binary detected by both \xmm\ and \rosat.}
\tablenotetext{e}{The addition of the new point source FGL J2013.3+3616 resulted in the removal of the 3FGL source 3FGL J2014.4+3606 (TS $<$ 25).
This source lies 12\arcmin\ away from the center of SNR G73.9+0.9 (which has an extension of 27\arcmin\ \citet{2014BASI...42...47G}).
This association has previously been reported in \citet{2016MNRAS.455.1451Z} and in \citet{2016arXiv160505914Z}.
It is included in the 1SC \citep{2016ApJS..224....8A} as a marginally classified candidate with a TS of 30 at a position of ($l$, $b$) = (73.86\arcdeg, 0.92\arcdeg) ((\ra, \dec) = (\hmsee{20}{13}{58}{10}{26}, \dmsee{36}{10}{48}{10}{22})}
\tablenotetext{f}{FGL J2032.1+4058 lies 4\arcmin\ away from Cygnus X-3, and close to the previously reported 1FGL and 2FGL catalog sources 1FGL J2032.4+4057 and 2FGL J2032.1+4049 which were both associated with Cygnus X-3.
It was not reported in the 3FGL due to flux variability \citep{TheFermi-LAT:2015hja}.}
\end{deluxetable*}

\section{Results for Known VHE Sources \& Other Objects of Interest}
\label{sec:KnownSources}

\subsection{\TeVJ}
\label{sec:TeVJ}
\subsubsection{Background}
In the vicinity of Cygnus OB2, the \GR\ source TeV J2032\allowbreak+4130 was the first unidentified VHE \GR\ source.
It was discovered with the \hegra\ IACTs \citep{Aharonian:2005ak} and has been further studied by the \whipple\ \citep{Konopelko:2007ab}, \magic\ \citep{Albert:2008ab} and \veritas\ \citep{aliu_observations_2014} IACTs. 
As the position of the source is coincident with the Cygnus OB2 star association and located north of the micro-quasar Cygnus X-3, it was suggested that these two sources could be the origin of the VHE \GR s \citep{Aharonian:2002ij}. 
More recently, it has been argued that the VHE source could be the PWN of the GeV pulsar PSR J2032\allowbreak+4127 \citep{aliu_observations_2014,Camilo:2009ab}. 
\citet{Lyne:2015oua} suggested that PSR J2032\allowbreak+4127 is part of a binary system with a 15M$_{\odot}$ Be star, though the authors argue that this binary might not be able to fully power the VHE emission. 
Observations of the region during the November 2017 periaston found correlated X- and \GR\ emission from the direction of the pulsar, confirming the binary nature of the system \citep{2017ATel10810....1V}, a paper is in preparation to present these results.

\TeVJ\ is located in the Cygnus Cocoon, a large excess of hard spectrum emission detected by \LAT\ that is associated with the star-forming region Cygnus X and has the potential to be detected at VHE energies \citep{Ackermann:2011}.  
The extensive air shower arrays \milagro\ and \argo\ have detected extended emission coming from the same region (MGRO J2031+41 \citep{Abdo:2007ad} and ARGO J2031+4157 \citep{Bartoli:2012tj}) at energies above  20 TeV and 1 TeV respectively, and these sources have been associated with the Cygnus Cocoon.
The Cygnus Cocoon overlaps with, but extends beyond, TeV J2032\allowbreak+4130 and the extent of this emission makes it unlikely to be detectable by \veritas\ using conventional techniques.
Localized sources of emission (of which TeV J2032\allowbreak+4130 could be an example) within the Cygnus Cocoon may be detectable and IACT observations may be able to identify the sources which are powering this emission.
Identification of the origin of the VHE emission from TeV J2032+4130 could help to clarify the origin of the emission from this complex region.

\subsubsection{Results}
In this analysis, \TeVJ\ was observed by \veritas\ at a peak (local) significance of 10.1$\sigma$ using the \ext\ integration radius.
As in \citet{aliu_observations_2014} the source was found to be extended and asymmetrical.  
The extension was fit with an asymmetrical Gaussian function and is given in \autoref{tab:VERITAS_Morphology}.

\begin{longrotatetable}
\begin{deluxetable*}{cccccccccc}
\tablecaption{
Source extension of the \veritas\ sources.  
The source extension was determined by fitting a sky map of the excess events with a two-dimensional Gaussian  distribution convolved with the \veritas\ PSF. 
The rotation angle is given east of Galactic North.
\label{tab:VERITAS_Morphology}}
\tabletypesize{\scriptsize}
\tablehead{
\multirow{2}{*}{Source} &
\multicolumn{2}{c}{Centroid} &
\colhead{$\sigma$ resp. $\sigma_{semi-major}$} &
\colhead{$\sigma_{semi-minor}$} &
\colhead{Rotation Angle} \\
\colhead{} &
\colhead{$l$, $b$} &
\colhead{\ra, \dec} &
\colhead{[\arcdeg]} &
\colhead{[\arcdeg]} &
\colhead{[\arcdeg]} 
}
\startdata
\TeVJ & \dee{80.25}{0.01}{0.01},\dee{1.20}{0.01}{0.01} & \hms{20}{31}{33},\dms{41}{34}{48} 
& \nee{0.19}{0.02}{0.01} &  \nee{0.08}{0.01}{0.03} & \nee{13}{4}{1} \\
\GCyg & \dee{78.30}{0.02}{0.01},\dee{2.55}{0.01}{0.01} & \hms{20}{20}{04.8},\dms{40}{45}{36}
& \nee{0.29}{0.02}{0.02} &  \nee{0.19}{0.01}{0.03} & \nee{176.7}{0.1}{2} \\
\cisne & \dee{74.97}{0.02}{0.01},\dee{0.35}{0.01}{0.01} & \hms{20}{19}{23},\dms{36}{46}{44}
& \nee{0.34}{0.02}{0.01} &  \nee{0.14}{0.01}{0.02} & \nee{127.0}{2.6}{0.1} \\
\cisneA & \dee{74.87}{0.01}{0.01}, \dee{0.42}{0.01}{0.01} & \hms{20}{18}{48}, \dms{36}{44}{24}
& \nee{0.18}{0.01}{0.04} &  \multicolumn{2}{c}{\textit{Symmetric}} \\
\cisneB & \dee{75.13}{0.01}{0.01}, \dee{0.19}{0.01}{0.01} & \hms{20}{20}{31},\dms{36}{49}{12}
& \nee{0.03}{0.01}{0.01}  &  \multicolumn{2}{c}{\textit{Symmetric}} \\
\CTB & \dee{74.94}{0.01}{0.01}, \dee{1.16}{0.01}{0.01} & \hms{20}{15}{57},\dms{37}{12}{31}
& \multicolumn{3}{c}{\textit{Point Source}} \\
\enddata
\end{deluxetable*}
\end{longrotatetable}

\begin{figure}[htb!]
\plotone{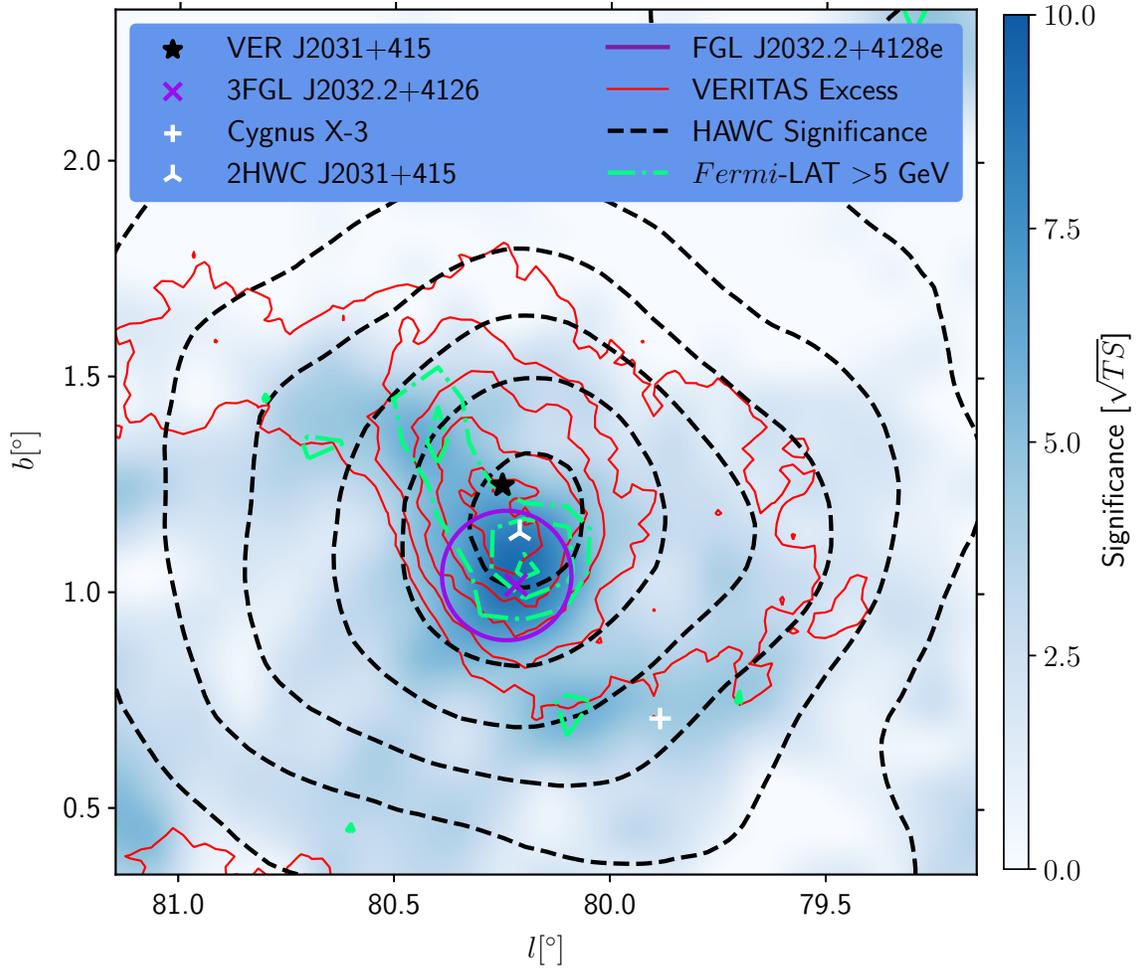}
\caption{Significance ($\sqrt{\mathrm{TS}}$) map of the \LAT\ observations in the region of \TeVJ\ with the pulsar PSR J2032+4127 (3FGL J2032.2+4126) fixed to the \on -pulse spectral parameters. 
The best fit extension of FGL~J2032.2+4128e, the source fit to this excess is shown in purple.
Overlaid in green dot-dash contours are the 3, 4 and 5 $\sqrt{\mathrm{TS}}$ contours from the analysis above 5~GeV.
Also shown are \veritas\ excess contours using the \ext\ integration region at levels of 50, 100, 150, 200, and 250 counts (red) and \hawc\ significance contours at 3, 4, 5, 6, and 7$\sigma$ (black dash).
The locations of \TeVJ\ (black star), 3FGL J2032.3+4126 (purple cross), 2HWC J2031+415 (white tri) and Cygnus X-3 (white plus) are also indicated in the figure.}
\label{fig:TeV2032FermiResid_VERConts}
\end{figure}

The \LAT\ emission in the region is dominated by the pulsar PSR J2032+4127 (3FGL J2032.2+4126) which cuts off sharply at a few GeV. 
To investigate the possibility of an additional source (for example a PWN) that is masked by the emission from PSR J2032+4127, a gated analysis was performed to remove the pulsed component using the method outlined in \autoref{sec:FermiAnalysis}, with the \off -pulse selected using phases 0.1 to 0.4 and 0.5 to 0.9.
This analysis showed an extended residual, as depicted in \autoref{fig:TeV2032FermiResid_VERConts}, with a centroid of ($l$, $b$) = (\dstat{80.24}{0.03}, \dstat{1.04}{0.03}) ((\ra, \dec) = (\hms{20}{32}{13}, \dms{41}{28}{39}).
This residual is fit by an extended symmetric Gaussian source of 68\% containment radius of $0.15^{\circ} {}^{+0.02^\circ}_{-0.03^\circ}$ centered on this location.
Thus, we name the source FGL~J2032.2+4128e.
This source has a TS of 321.1 and the TS of extension is 28.6.
A uniform disk source was also tested, the best fit had a 68\% containment radius of $0.15^{\circ} {}^{+0.02^\circ}_{-0.02^\circ}$ and a TS of extension of 24.7, the Gaussian source morphology was preferred and used for the rest of this work.

A cross check using an \off-pulse analysis identified a similar residual with a comparable morphology although, with the smaller statistics due to the combined effect of the phase cut and limited time period of the ephemeris (about a factor of 2 reduction in live time), the signal is weaker.
The similarity in the morphology between these two methods of removing the pulsar, and the offset and extension in the residual is supportive of this residual being due to a new source rather than an artifact due to errors in fitting to PSR J2032+4127.

An additional analysis was conducted above 5~GeV, this has the advantage that the PSF is reduced at these energies (to around 0.3\arcdeg\ at 5~GeV) which will improve the ability to separate the two sources. This also found moderately extended emission potentially associated with \TeVJ\ with a TS of 31.7 as shown in \autoref{fig:TeV2032FermiResid_VERConts}.  
A fit to this emission was conducted and it was found to peak at ($l$, $b$) = (\dstat{80.21}{0.04}, \dstat{1.10}{0.04}) ((\ra, \dec) = (\hms{20}{31}{54}, \dms{41}{29}{24}) with an extension of \nstat{0.14}{0.05} and a TS of extension of 10.6.

It is noted that an analysis of the \LAT\ data from the Cygnus Cocoon region is a very complex and, in this analysis, a simplistic model of the Cocoon emission was used (a single symmetric Gaussian as in the 3FGL).  
Further work with a more detailed morphology of the Cygnus Cocoon has the potential to provide significantly more insight here, but is beyond the scope of this work.
In particular there are significant challenges associated with disentangling the emission from the various objects and identifying whether an observed region of emission is associated with a peak in the emission from the Cygnus Cocoon or another source.
At present, VHE emission is either detected from a significantly extended region (\milagro\ and \argo) associated with the Cygnus Cocoon or a smaller, moderately extended region (\veritas, \magic, \hegra, \hawc\ and \whipple) associate with TeV J2032+4130.  
Understanding the relationship between these VHE observations will significantly improve our ability to understand this region in other wavelengths.  To do this will require further development of tools such as the ``3D Maximum Likelihood Method'' currently being developed \citep{2014NuPhS.256..136W,JoshThesis} which will allow for observations of significantly extended objects by \veritas.  
This will then allow the superior PSF of \veritas\ to be used to study the morphology of the Cygnus Cocoon, providing information that could be used to guide the HE analysis.
Until then it is not possible to definitively claim the detection of a HE counterpart to \TeVJ, nor to fully understand the relationship between the two objects. 
However, the presence of extend emission with similar morphology to those observed in \TeVJ\ is highly suggestive that there is HE emission associated with the source.

\begin{figure}[htb!]
\centering
\plotone{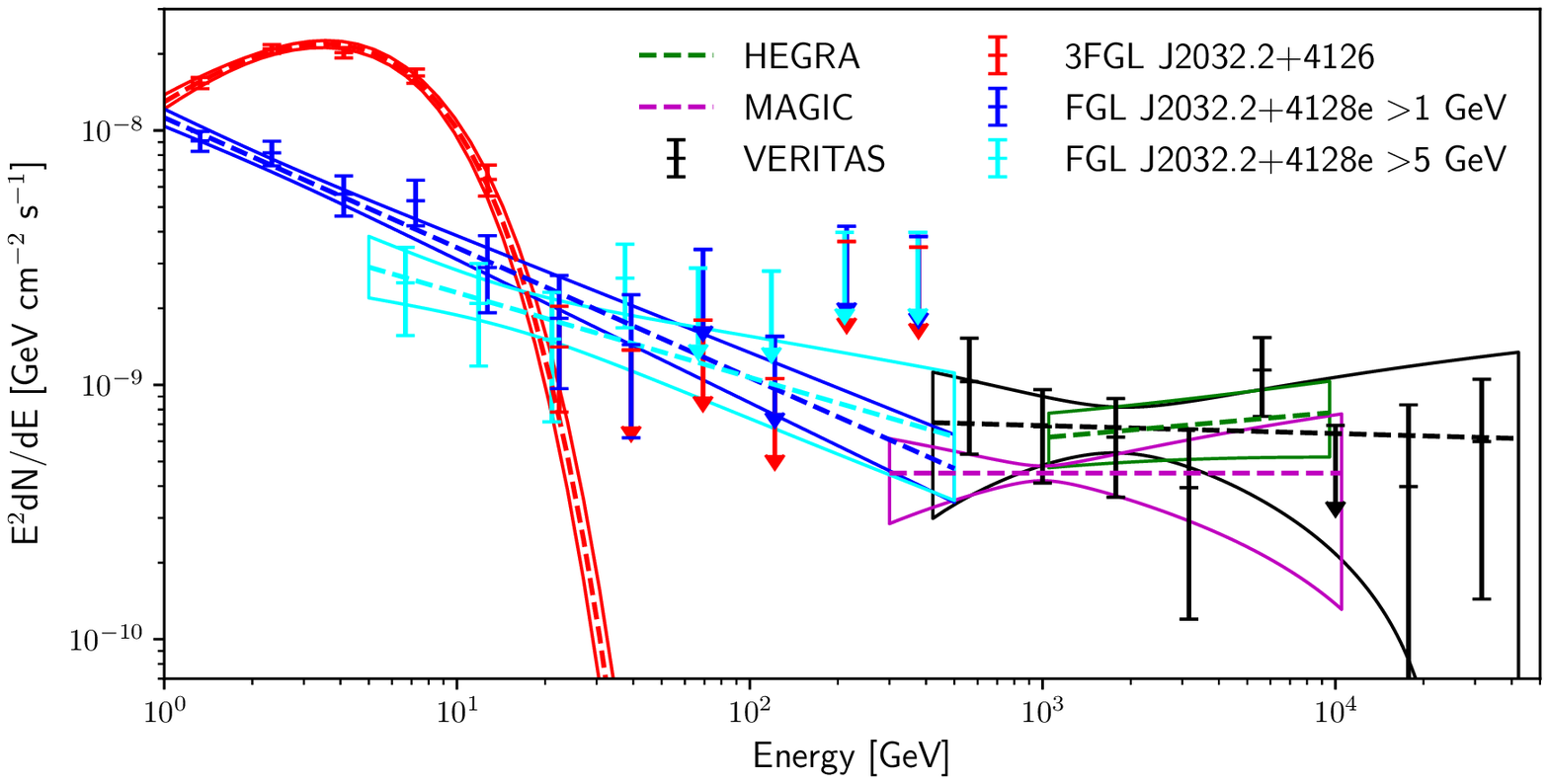}
\caption{SEDs from sources in the \TeVJ\ region.  Black is the \veritas\ \ext\ analysis, blue is FGL~J2032.2+4128e fit above 1~GeV, cyan above 5~GeV, the \LAT\ source that is located at the same position as \TeVJ\ and is potentially the low-energy extension of this emission, red is 3FGL~J2032.2+4126, the bright pulsar in the region.   Also plotted are the TeV J2032+4130 results from \hegra\ \citep{Aharonian:2005ak} and \magic\ \citep{Albert:2008ab}.  The butterflies show  statistical errors only. }
\label{fig:TeV2032Spectrum}
\end{figure}

The spectra of \TeVJ\ and FGL~J2032.2+4128e are shown in \autoref{fig:TeV2032Spectrum}. 
The \veritas\ spectrum is given in \autoref{tab:VERITAS_Spectra} and is consistent with the spectrum presented in \citet{aliu_observations_2014}.
The spectrum of HE emission can be described by a power law with index \nstat{2.52}{0.07} and a normalization of \fstat{1.44}{0.09}{-8} at 2.27 GeV.
Comparing the spectra of FGL~J2032.2+4128e and 3FGL~J2032.2+4126 shows that, below 1~GeV, the extrapolated flux from FGL~J2032.2+4128e would be stronger than 3FGL~J2032.2+4126.  
This is clearly not the case as it has not been detected in lower energy analyses.  
It is likely, therefore, that at low energies some of the emission from 3FGL~J2032.2+4126 and potentially the Cygnus Cocoon is being included in the measured flux from FGL~J2032.2+4128e.

A joint fit to the \veritas\ and \LAT\ data points for \TeVJ/FGL~J2032.2+4128e is well fit with a PL of index \nstat{2.39}{0.03} and normalization \fstat{3.61}{0.21}{-10} at 4.04~GeV, with a $\chi^2$ of 9.9 for 9 DoF.
However, it is clearly evident that the low energy \LAT\ spectral points are contaminated with emission from 3FGL~J2032.2+4126.  To reduce the impact of this and better match the regions from which the \veritas\ spectrum was extracted the analysis  above 5~GeV was used. 
This spectrum (index = \nstat{2.33}{0.22}, normalization = \fstat{8.80}{1.76}{-12} at 15.1~GeV,  \autoref{fig:TeV2032Spectrum}) shows good agreement with the \veritas\ spectrum for \TeVJ\ and together they are well fit with a PL of index \nstat{2.22}{0.06} and normalization \fstat{2.28}{0.33}{-15} at 221~GeV with a $\chi^2$ of 4.9 for 9 DoF. 

\begin{longrotatetable}
\begin{deluxetable*}{ccccccccccc}
\tablecaption{Observation results for the \veritas\ sources.  \on, \off, and $\alpha$ are from the RBM analysis.  The spectral analysis is conducted using the RR method (\autoref{sec:VERAnalysis}) and all sources were best fit with a power law (\autoref{eq:PL}).
\label{tab:VERITAS_Spectra}}
\tabletypesize{\scriptsize}
\tablehead{
\multirow{2}{*}{Source} &
\colhead{Integration} &
\multirow{2}{*}{\on} &
\multirow{2}{*}{\off} &
\multirow{2}{*}{$\alpha$} &
\colhead{Sig. } &
\colhead{Fit Range} &
\colhead{$N_0$} &
\colhead{$E_0$} &
\multirow{2}{*}{$\gamma$} &
\multirow{2}{*}{$\chi^2$/DoF}\\
\colhead{} &
\colhead{Region} &
\colhead{} &
\colhead{} &
\colhead{} &
\colhead{[$\sigma$]} &
\colhead{[GeV]} &
\colhead{[GeV$^{-1}$cm$^{-2}$s$^{-1}$]} &
\colhead{[GeV]} &
\colhead{} &
\colhead{}
} 
\startdata
\TeVJ & \ext & 871 & 5181 & 0.114 & 10.1& 422 -- 42200 & \pee{1.24}{0.24}{0.24}{-16} & 2300 & \nee{2.00}{0.19}{0.20}  & 3.02/5\\
\GCyg & \ext & 598 & 2767 & 0.151 & 7.6 & 750 -- 7500 & \pee{5.01}{0.93}{1.00}{-16} & 1500 & \nee{2.79}{0.39}{0.20} & 3.27/2 \\
\cisne &\ext & 951 & 4230 & 0.153 & 10.3 & 422 -- 42200 & \pee{1.02}{0.11}{0.20}{-16} & 3110 & \nee{1.98}{0.09}{0.20} & 8.73/6 \\
\cisneA &\point & 196 & 3692 & 0.028 & 7.8 & 750 -- 23700 & \pee{5.12}{0.94}{1.48}{-17} & 2710 & \nee{2.00}{0.21}{0.2} & 2.81/4 \\
\cisneB &\point & 191 & 3456 & 0.030 & 7.5 & 750 -- 23700 & \pee{3.00}{0.56}{0.60}{-17} & 3270 & \nee{1.71}{0.26}{0.2} &  4.71/4 \\
\CTB & \point & 157 & 2370 & 0.038 & 6.3 & 680 --  14700 & \pee{2.8}{1.2}{1.2}{-17} & 2510 & \nee{2.1}{0.8}{0.4} &  0.38/2  \\
\enddata
\end{deluxetable*}
\end{longrotatetable}

\subsubsection{Discussion}
\TeVJ\ has been detected by \hawc\ and is listed in the 2HWC catalog as 2HWC J2031+415.
The \veritas\ location lies just outside the \hawc\ 1$\sigma$ uncertainty at a distance of 0.11\arcdeg\ (though the \hawc\ position lies within the 1$\sigma$ extent of the \veritas\ emission).
HAWC sees evidence of extension with the source fit with both a point source and uniform disk of radius 0.7\arcdeg.
This extension is larger than that measured by \veritas\ and closer to the results from \milagro.
Using both extensions, \hawc\ detects a higher flux (\fstat{3.24}{0.32}{-17} and \fstat{6.16}{0.44}{-17} at 7000~GeV for the point and extended fit respectively) and softer spectral index (\nstat{-2.57}{0.07} and \nstat{-2.52}{0.05}) than reported in this work (and in previous observations by IACTs); this is likely due to contributions from the extended emission that is not detected by \veritas.

With similar morphologies and spectral properties, it is likely that both the HE and VHE emission share a common origin. 
Given the proximity of the emission from both sources to the pulsar PSR J2032+4127, a PWN origin of this emission is a strong possibility. 
However, \citet{aliu_observations_2014} comment that, taking into account the hard spectral index ($\sim$ -2) and due to the Klein-Nishina effect, a cutoff in the spectrum should be seen close to 10 TeV in order to be consistent with the PWN interpretation. 
Though it cannot be excluded using these data alone, there is no evidence of a spectral cutoff in this data which would be expected with a PWN origin for the emission.

\begin{figure*}[htbp]
\centering
	\fig{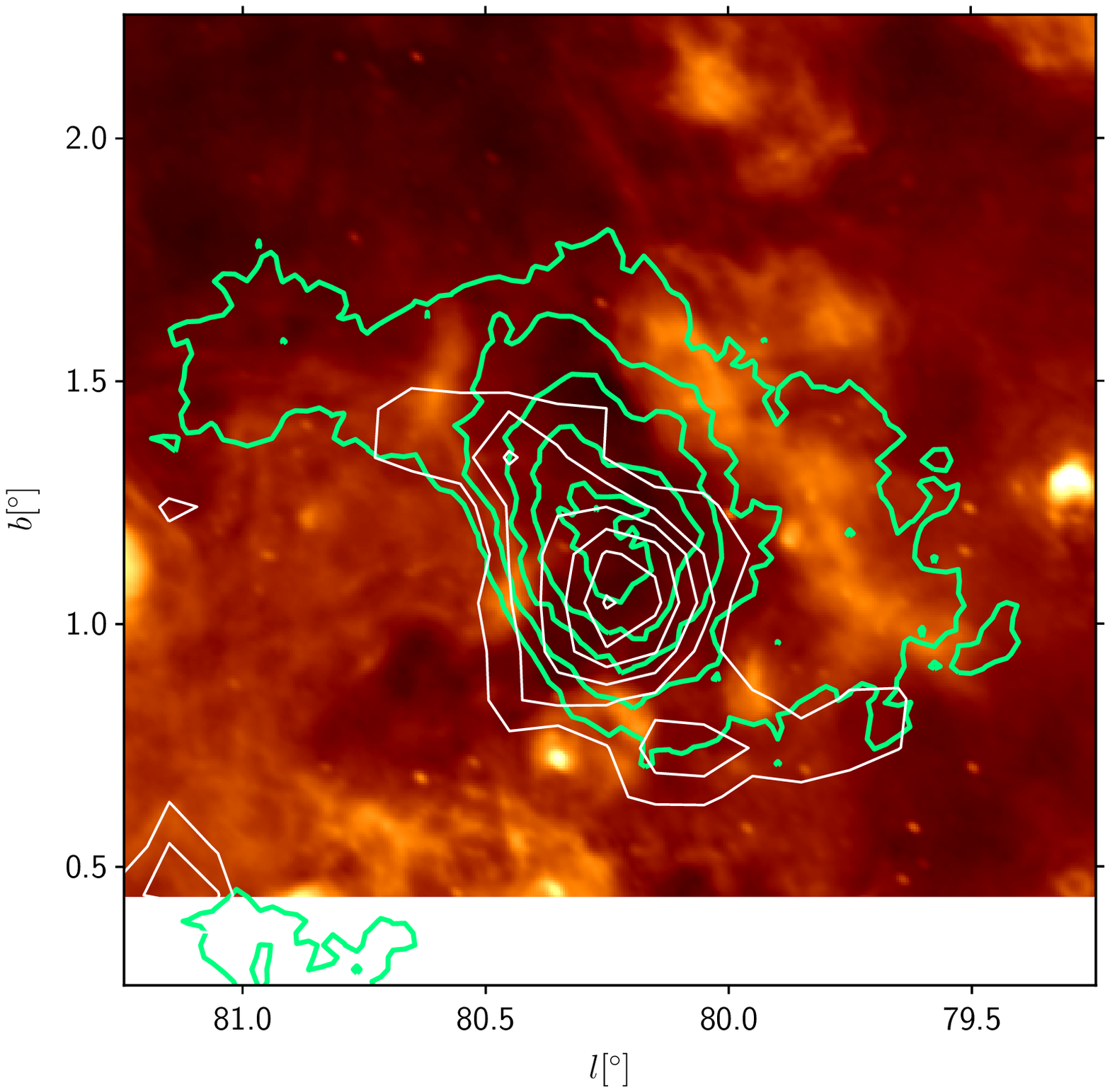}
	{0.48\textwidth}
	{(a) 1420 MHz image from the Canadian Galactic Plane Survey. \citep{Taylor:2003cp}}
	\fig{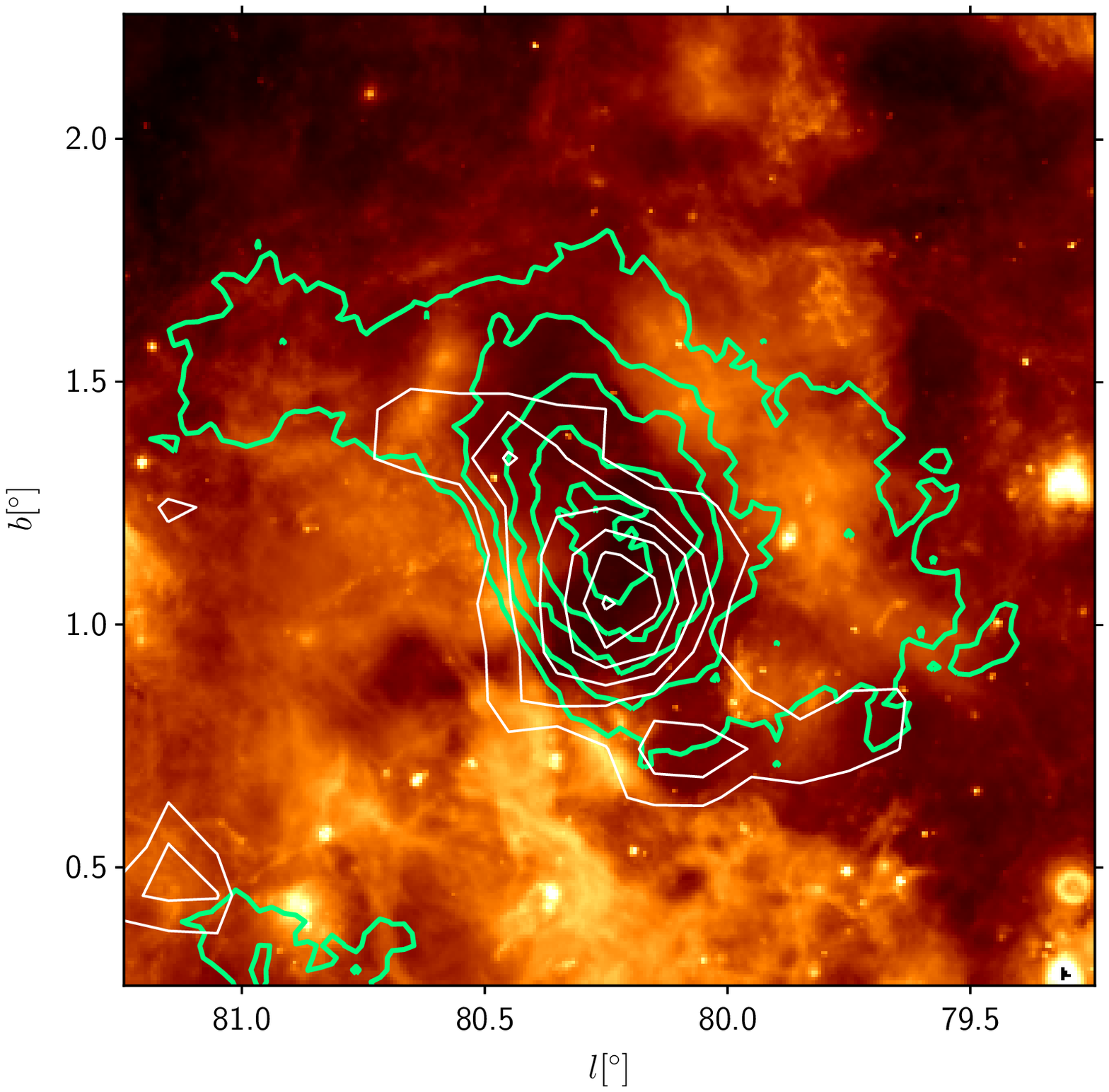}
	{0.48\textwidth}
	{(c) \spitzer\ MIPS 24~$\mu$m image from the MIPSGAL survey.}
\caption{The multiwavelength view of the region around \TeVJ .  
Overlaid are \veritas\ excess contours (green) produced using the \ext\ integration region at levels of 50, 100, 150, 200, and 250 counts and \LAT\ significance ($\sqrt{\mathrm{TS}}$, white) at 1$\sqrt{\mathrm{TS}}$ levels above 4$\sqrt{\mathrm{TS}}$.  
Both the \veritas\ and the \LAT\ emission are confined to a void in this mulitwavelength emission.
\label{fig:TeV2032_MWImages}}
\end{figure*}

Multiwavelength images of TeV J2032+4130 and its vicinity are shown in \autoref{fig:TeV2032_MWImages}. 
The infrared images from \spitzer\ are dominated by bright diffuse emission exhibiting complex structure caused by star-forming activity taking place in the Cygnus~X complex. 
In addition, as noted in \citet{aliu_observations_2014}, nearly all the VHE \GR\ emission happens to be confined within a void. 
One possible explanation proposed by \citet{Butt:2003xc} is that the large, mechanical power density from the stellar winds of the OB stars could be powering the emission.
However, even though the energy required to power the VHE emission is a fraction of the estimated wind energy, almost all of the massive stars are located outside of the VHE emission region.
It is also plausible that a supernova exploded within Cygnus OB2 and its remnant is expanding into the surrounding medium, creating the void and powering the \GR\ emission.

\subsection{\GCyg}
\label{sec:GammaCyg}
\subsubsection{Background}
The \GR\ emission in the vicinity of the SNR G78.2\allowbreak+2.1, located in the region of Gamma-Cygni, was first detected by the EGRET instrument on board the Compton Gamma Ray Observatory (CGRO) \citep{Thompson:1995ec}. 
There is a bright \LAT\ pulsar, PSR J2021\allowbreak+4026 (3FGL J2021.5+4026), at the center of the remnant \citep{Abdo:2010fs,Abdo:2010pc} as well as  extended \GR\ emission above 10 GeV (3FGL J2021.0+4031e) with an extension consistent with that seen in radio observations of G78.2+2.1 \citep{Lande:2012fe}. 
A VHE source, \GCyg , was detected by \veritas\ at a level of 7.5$\sigma$ on the northwestern rim of the radio shell of the SNR G78.2\allowbreak+2.1 \citep{aliu_discovery_2013}.  
\hawc\ has detected a source (2HWC J2020+403) that lies within G78.2+2.1, 0.43\arcdeg\ away from the location of \GCyg\ and close to PSR J2021\allowbreak+4026.
They do not report on any evidence of extension, and they report a power law spectrum for flux \fstat{1.85}{0.26}{-17} at 7000 GeV and spectral index (\nstat{-2.95}{0.10}).

3FGL J2021.0+4031e is generally modeled as a uniform disk of radius 0.63\arcdeg\ \citep{TheFermi-LAT:2015hja}, though
\citet{Ackermann:2011} found an improvement in their modeling of the region if, in addition to a disk source for the entire remnant, they included an elliptical source based on the morphology presented in \citet{weinstein_veritas_x}.
\citet{Fraija:2016shx} conducted an analysis above 4~GeV which showed a region of enhanced emission in the GeV energy range over and above the uniform disk model that coincides with the VHE emission from \GCyg.

The radio emission is brightest at the opposite side of the SNR from \GCyg, on the southern edge of the remnant.
\citet{Leahy01122013} discuss the radio and X-ray emission in detail and note that there are several possible sources of radio emission in the same line of sight as the southern edge of the remnant, which could give the appearance of the radio emission in that region being higher than it actually is.
In contrast, by examining \rosat\ X-ray observations they show that the diffuse emission is brightest at the northern and western edges.
It is noted that the northern emission may, in part, be due to the presence of a large X-ray shell associated with the early-type star V1685 Cyg which lies in the line of sight of G78.2+2.1 but at between half and one third of the distance (980 pc, cf. 1.7 to 2.6 kpc for G78.2+2.1 \citep{greenExt}).
G78.2+2.1 lies within the Cygnus Cocoon and could power at least some of the high energy emission seen from the region.

\subsubsection{Results}
In this analysis, \GCyg\ was observed by \veritas\ at a peak (local) significance of 7.6$\sigma$ using the \ext\ integration region.
The source was found to be extended and asymmetric, unlike the morphology reported in \cite{aliu_discovery_2013} where a symmetric fit was favored.
It was fit with an asymmetric Gaussian with the results given in \autoref{tab:VERITAS_Morphology}.
This difference in morphology likely reflects the improved PSF in this analysis in comparison to the discovery paper \citep{aliu_discovery_2013} resulting from the requirement of three telescopes in the reconstruction and a higher image intensity requirement, and the larger dataset.

\begin{figure}[htb!]
\plotone{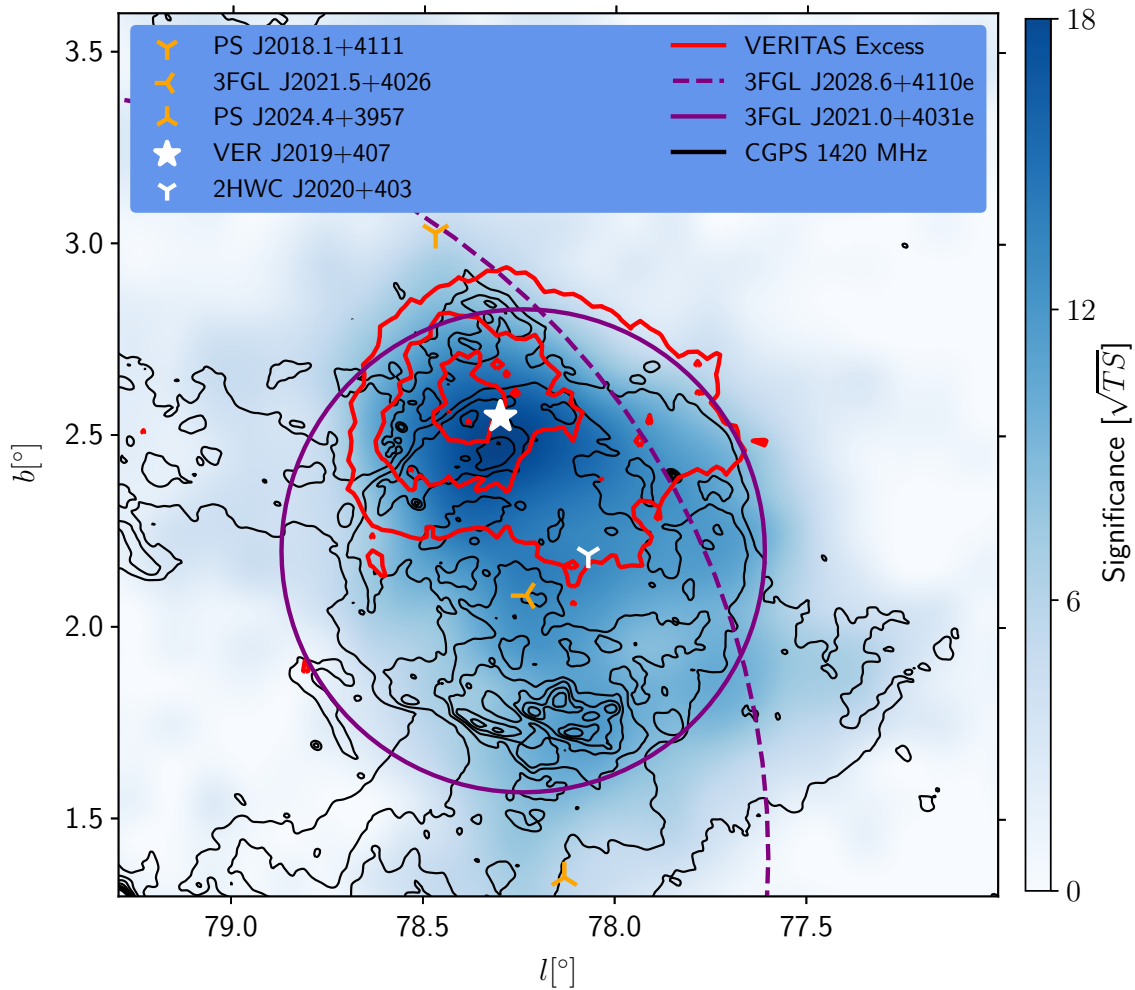}
\caption{\LAT\ $\sqrt{\mathrm{TS}}$ map of 3FGL J2021.0+4031e (SNR G78.2+2.1). 
Overlaid are VERITAS excess contours at levels of 50, 100, 150, 200, and 250 (red), and CGPS 1420~MHz (black) contours in 8 logarithmically spaced levels between brightness temperatures 20 and 150~K, \citep{Taylor:2003cp}.  
The locations of 2HWC J2020+403 (purple diamond) and the \LAT\ sources in the region (orange tris for point sources, circles showing the extent for extended sources) are indicated as well.
\label{fig:FermiGammaCyg}}
\end{figure}

The morphology of 3FGL J2021.0+4031e was examined by producing a TS map of the region with 3FGL J2021.0+4031e removed from the model of the region (\autoref{fig:FermiGammaCyg}).
The \LAT\ emission showed a disk-like structure, of similar size to the remnant detected in the CGPS 1420~MHz survey \citep{Taylor:2003cp}.
However, the HE \GR\ emission was not uniform across the disk, rather it was enhanced at the northern rim where the \veritas\ emission is detected.

The spectral fit to the \veritas\ data is given in \autoref{tab:VERITAS_Spectra}.  The updated analysis has a spectrum that is softer, but within the statistical errors of the earlier spectrum reported in \citet{aliu_discovery_2013}, with a normalization at 1000 GeV that is in agreement with the value in that work.

\begin{figure}[htb!]
\centering
\plotone{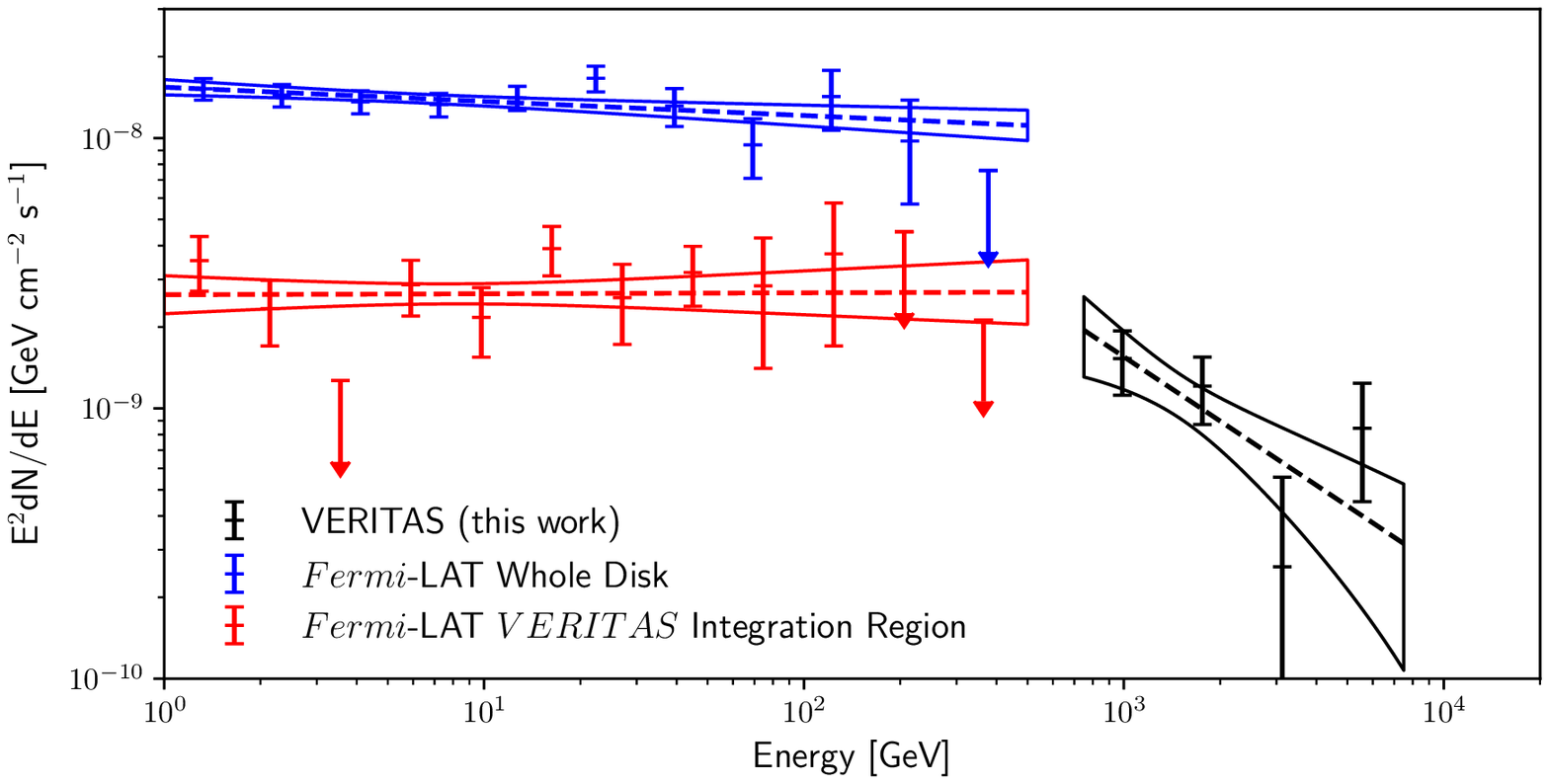}
\caption{SEDs from sources in the region of the G78.2+2.1.  
Black is the \GCyg\ spectrum from the \veritas\ \ext\ analysis.  
Also plotted are the \LAT\ SEDs from 3FGL J2021.0+4031e  (blue) and 3FGL J2021.0+4031eA, the region of 3FGL J2021.0+4031e that lies within the \veritas\ \ext\ integration region (red, 3FGL J2021.0+4031eB was also included in the analysis as an additional source whose spectrum is not shown for clarity).  The butterflies show statistical errors only.
\label{fig:GammaCygniSpectrum}}
\end{figure}

Two \LAT\ SEDs were produced for 3FGL J2021.0+4031e.
The first uses the full disk of radius 0.63\arcdeg\ as in the 3FGL catalog.
For the second, 3FGL J2021.0+4031e was broken up into two sources: the region inside the \veritas\ \ext\ integration region (a disk of radius 0.23\arcdeg\ centered on ($l$, $b$)  = (78.38, 2.56), 3FGL J2021.0+4031eA) the region from which the \veritas\ spectrum presented in \autoref{tab:VERITAS_Spectra} is extracted, and the remainder of the 0.63\arcdeg\ radius disk outside of the \veritas\ integration region, 3FGL J2021.0+4031eB.
This was done to investigate whether the emission seen by \veritas\ has a common origin with that observed with the \LAT.
Also, it was used to test whether the level of enhancement seen in the \LAT\ TS map of the remnant is reflected in a change in the spectral properties.
In both cases, the favored fit is a PL.
When treated as a single disk, 3FGL J2021.0+4031e is detected at a TS of 910, with spectral parameters of $\gamma$ = \nstat{2.02}{0.03}, \No\ = \fstat{2.50}{0.10}{-13} at \Eo\ = 6.78~GeV.
When divided into two sources, they are detected at test statistics of 186 and 474 for 3FGL J2021.0+4031eA and 3FGL J2021.0+4031eB respectively.
3FGL J2021.0+4031eA is fit with the parameters $\gamma$ = \nstat{2.00}{0.07}, \No\ = \fstat{5.77}{0.52}{-14} at \Eo\ = 6.78~GeV, whereas 3FGL J2021.0+4031eB is fit by $\gamma$ = \nstat{2.02}{0.04}, \No\ = \fstat{1.84}{0.10}{-13} at \Eo\ = 6.78~GeV.
\autoref{fig:GammaCygniSpectrum} shows the SEDs for 3FGL J2021.0+4031e and 3FGL J2021.0+4031eA.
3FGL J2021.0+4031eB is not shown for clarity.  
All have the same spectral indices within statistical errors, which suggests that they share a common origin of their emission.
A joint fit was conducted using the \veritas\ and the \LAT\ spectral points for 3FGL J2021.0+4031eA with the PL, BPL and LP spectral models.
The best fit is with a BPL  with parameters \No\ = \fstat{1.93}{0.50}{-14}, $E_b$ = 405 GeV, $\gamma_1$ = \nstat{1.97}{0.07} and $\gamma_2$ = \nstat{2.79}{0.22}, p-value = 0.55 cf. 0.33 for the LP and 0.11 for the PL.

\subsubsection{Discussion}
With a morphology that matches G78.2+2.1 the HE emission from 3FGL J2021.0+4031e is associated with the supernova remnant.  The agreement in the fluxes from 3FGL J2021.0+4031eA and \GCyg\ suggest that these two objects are associated, and thus that the \veritas\ emission is also associated with G78.2+2.1, with the current \veritas\ observations only able to detect the brightest part of the remnant.
The peak of this emission is located 0.5\arcdeg\ away from the pulsar (PSR J2021\allowbreak+4026) with no evidence of VHE emission from the region between \GCyg\ and the pulsar.  Though the pulsar is often offset from the center of the VHE emission, it generally lies within the emission region and/or is associated with another region of VHE emission as discussed in \citet{aliu_discovery_2013} and references therein.  The level of separation we have observed here, significantly more than the angular extent of the VHE emission, makes it unlikely that the emission is due to a PWN associated with PSR J2021\allowbreak+4026.

Comparing the joint spectral fit conducted in this work with that presented in \citet{Fraija:2016shx} shows a softer spectral index above and below the break (though within statistical errors), with a higher break energy. The softer spectral index in this work is in agreement with that predicted by standard test-particle shock acceleration, reducing the requirements for non-linear effects as discussed in that work.

The source detected by \hawc\ is offset from the \veritas\ emission, has a higher flux and a softer spectral index.  The origin of these differences is uncertain though it may be due to the higher energy threshold of the \hawc\ observations, though there is no evidence of this in the \veritas\ data.
The other possibility is that the significant diffuse emission in the region is affecting the analyses.
Further work is required to understand the emission in this region across all wavebands.
Deeper \veritas\ observations, in particular with the lower energy threshold following the 2012 camera upgrade, may enable the detection of emission from more of the remnant and the exploration of any energy dependent morphology which could explain the differences between the three instruments.

\subsection{\cisne}
\label{sec:CisneRes}
\subsubsection{Background}
\MGROcisne, one of the brightest sources observed in VHE \GR s, is located within the vicinity of the OB star association Cygnus OB1. 
This source was observed by the \milagro\ collaboration to have a flux level of 80\% of the Crab Nebula where it was fit with a exponentially cut-off power law (\autoref{eq:PLEC}) of parameters \No\ = $(7^{+5}_{-2})\times 10^{-17} \GeV^{-1}\cm^{-2}\s^{-1}$ at 10000 GeV, $\gamma$ = $2.0^{+0.5}_{-1.0}$ and $E_c$ = $29000^{+50000}_{-16000}$~GeV \citep{Abdo:2012bj} with no identified multiwavelength counterpart(s). 
\veritas\ has resolved \MGROcisne\ into two components, an extended source \cisne, which constitutes the bulk of the emission and a point source \CTB, which is coincident with the SNR CTB 87 \citep{aliu_spatially_2014} and discussed in \autoref{sec:CTB}. 

\cisne, with an extent of over 1\arcdeg\ along the major axis, is spatially close to several sources that could potentially power (at least part of) the VHE emission.  
These include the HII region \Sh, Wolf-Rayet stars,  the pulsars PSR J2021+3651 and PSR J2017+3625 (which are both detected in HE \GR s by the \LAT), the PWN G75.1+0.2 (powered by PSR J2021+3651), and the hard X-ray transient IGR J20188+3657 \citep{aliu_spatially_2014}.  

PSR J2021+3651 (3FGL J2021.1\allowbreak+3651) lies to the east of the \cisne\ region and is associated with the X-ray PWN G75.2+0.1, which stretches back towards the VHE emission region.
The possibility of these objects being linked to \MGROcisne\ is discussed in detail in \citet{2014ChPhC..38h5001H}.

\citet{Gotthelf:2016opn} report on \nustar\ observations of the western edge of the \cisne\ region.
They identified a number of hard X-ray sources in the region, three sources of particular interest were identified: 3XMM J201744.7\allowbreak+365045 (a bright point source which is most likely an active galactic nucleus behind the Galactic plane though it could be a pulsar), NuSTAR J201744.3\allowbreak+364812 (a nebula, most likely a galaxy cluster hidden behind the Galactic plane), and emission associated with a young massive stellar cluster enclosed by a radio spur at the eastern edge of \Sh.
Further observations, for example with \textit{Chandra} are required to determine whether these sources are related to the observed emission.

3FGL J2017.9+3627 lies to the southwest of \cisne\ and was recently identified as a \GR\ pulsar (PSR J2017+3625) by the Einstein@Home distributed computing project \citep{Pletsch:2014eh, Clark:2016eh}. 
The pulsar has a spin frequency of 6.00~Hz with a first frequency derivative of -4.89$\times 10^{-14}$~Hz s$^{-1}$, giving a characteristic age of 1943~kyrs and a spin-down power of 1.2$\times10^{34}$~erg s$^{-1}$ \citep{2016arXiv161101015C}.
3FGL J2017.9+3627 is coincident with a region of low energy VHE \GR\ emission (E~$<$~1~TeV)  \citep{aliu_spatially_2014}.
\citet{Gotthelf:2016opn} concluded that this pulsar could power the observed emission with an observed efficiency ($L_{TeV}/ \dot{E}) \sim $ 0.03, which is plausible for a $>10^5$ year-old pulsar.

\subsubsection{Results}
In this analysis, \cisne\ was observed at a peak (local) significance  of 10.3$\sigma$ using the \ext\ integration region.
Significantly extended and asymmetrical, the source is best fit with an asymmetric Gaussian (\autoref{tab:VERITAS_Morphology}) as in \citep{aliu_spatially_2014}, with the fit parameters in agreement with the results presented there.

In addition, a sky map produced using the \point\ integration region shows that the emission appears to be strongest in two regions, each of which is detected at a level greater than 7$\sigma_{local}$.
Between the two sources lies a valley in the significance where the significance drops below 4$\sigma_{local}$, giving a change in the significance that is greater than 3$\sigma_{local}$.
This suggests that the emission may be the result of two sources that were previously unresolved (though whether these are two independent sources, two enhancements of a single, extended source, or the result of statistical fluctuations is, as yet, uncertain).

\begin{figure}[htb!]
\centering
\plotone{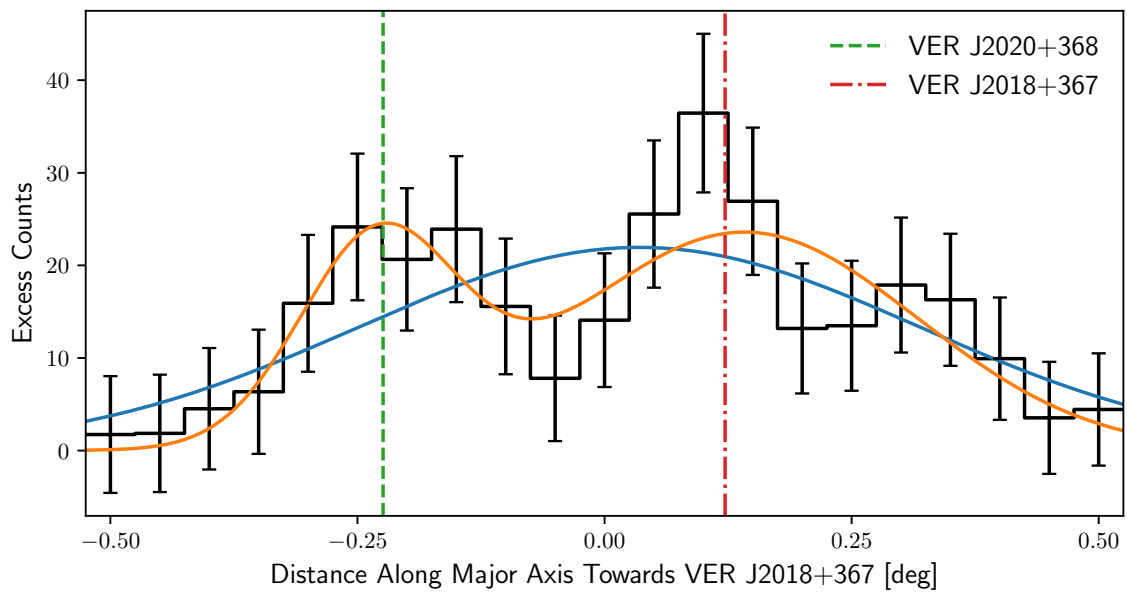}
\caption{A section along the major axis of \cisne\ towards \cisneA\ through the excess map using the \point\ integration radius (black steps).  
Each bin was produced using an aperture of width 0.05\arcdeg\ and height 0.2\arcdeg.
The data are fit with a single Gaussian (blue) and the sum of two Gaussians (orange). 
The locations of \cisneA\ (red dot-dashed) and \cisneB\ (green dashed) are also shown.}
\label{fig:CisneTranscept}
\end{figure}

To examine the extension of these two regions a joint fit was conducted using two symmetric Gaussian sources (\autoref{tab:VERITAS_Morphology}).  
This did not produce a significant improvement in the fit in comparison to a single, asymmetric Gaussian model ($\chi^2$/DoF = 6511/5992 for the two sources cf. 6600/5994 for the single, asymmetric Gaussian).
We also produced a section through the excess map along the major axis of the fit to \cisne\ (\autoref{fig:CisneTranscept}).
The data were binned using rectangles of width 0.05\arcdeg\ and height 0.2\arcdeg\ along a length of 1.0\arcdeg.  
This was fit with two models: a single Gaussian function and the sum of two Gaussian functions.
The single Gaussian function fits the data well with a $\chi^2$/DoF of 15.1/17 and the sum of two Gaussian functions mildly over fits the data with a $chi^2$/DoF of 8.2/14.
Neither of these tests provides statistically significant evidence that the emission is due to two independent sources, therefore we name the potential sources \cisneA\ and \cisneB\ where the * indicates that the two sources are still candidates and may be related.
Further observations are required to understand this morphology.

\begin{figure}[htb!]
\centering
\plotone{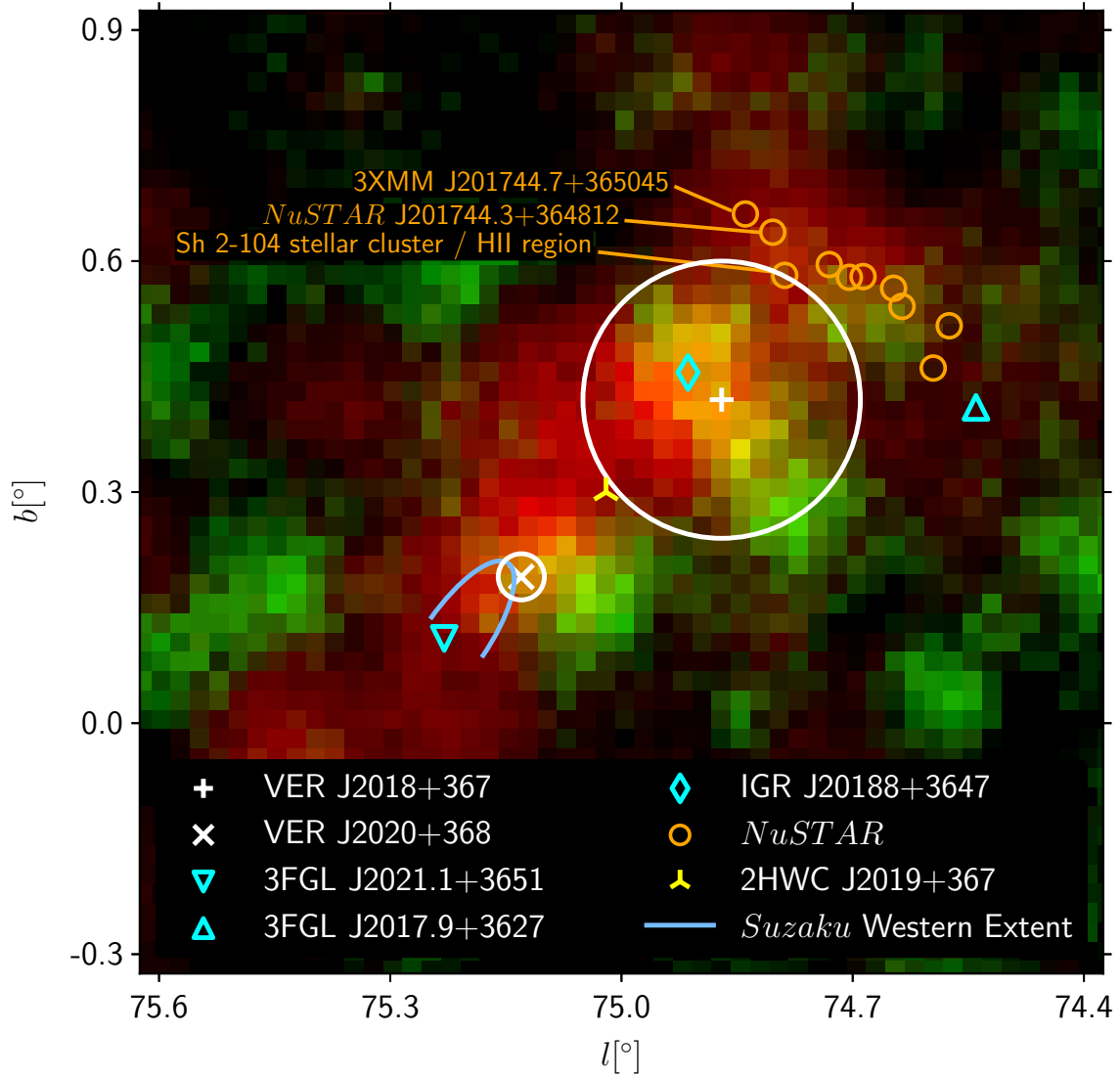}
\caption{Energy dependent excess map at $>$1000~GeV (red) and $<$1000~GeV (green) for the region around \cisne\ produced using the \point\ integration region, the transition from black to color is an excess $\geq$ 0 with a maximum excess $>$1000~GeV of 70 counts and a maximum excess $<$1000~GeV of 30 counts.  
Unlike \citet{aliu_spatially_2014} there is no evidence of a region of lower energy emission to the south of the extended emission from \cisne.  
This is either due to the higher energy threshold employed in this analysis reducing the sensitivity to low energy emission in this region or the earlier result was a statistical fluctuation which has diminished with the increased data.
\cisneA\ and \cisneB\ are shown with the 1$\sigma$ extents (solid white lines) and are located at regions which show enhanced emission below 1000 GeV.
Also shown are the locations of 2HWC J2019+367 (yellow, with the 1$\sigma$ error on its position), IGR J20188+3647 (cyan diamond), the two \LAT\ pulsars (3FGL J2021.1+3651 and 3FGL J2017.9+3627 (cyan triangles)), and the \nustar\ sources from \citet{Gotthelf:2016opn}.
The sky blue arc approximately traces the region with $>$ 50 photons s$^{-1}$ cm$^{-2}$ sr$^{-1}$ in the 2-10~keV energy range from \citet{2017arXiv170502733M}.
\label{fig:CisneRG}}
\end{figure}

\citet{aliu_spatially_2014} reported evidence of energy-dependent morphology with a low significance (3$\sigma$) region of lower energy (E $<$ 1000 GeV) emission to the southwest of the main emission region around (\ra, \dec) = (\hm{20}{18}, \dm{+36}{26}). 
This is not visible in this analysis with either the \point\ or \ext\ integration regions (\autoref{fig:CisneRG}).
The reason for this is either due to the higher energy threshold used in this analysis or that the original apparent energy dependence was a statistical fluctuation which has disappeared with additional data.
Further observations with a lower energy threshold following the 2012 camera upgrade may clarify this.
Interestingly, both \cisneA\ and \cisneB\ are also regions which show low-energy emission in \autoref{fig:CisneRG}. 
However, at the moment, it is not possible to say whether this is the cause of these excesses (regions of lower energy emission over a background of higher energy emission) or whether this is just an effect of having brighter emission from these regions across both energy ranges.

\begin{figure}[htb!]
\centering
\plotone{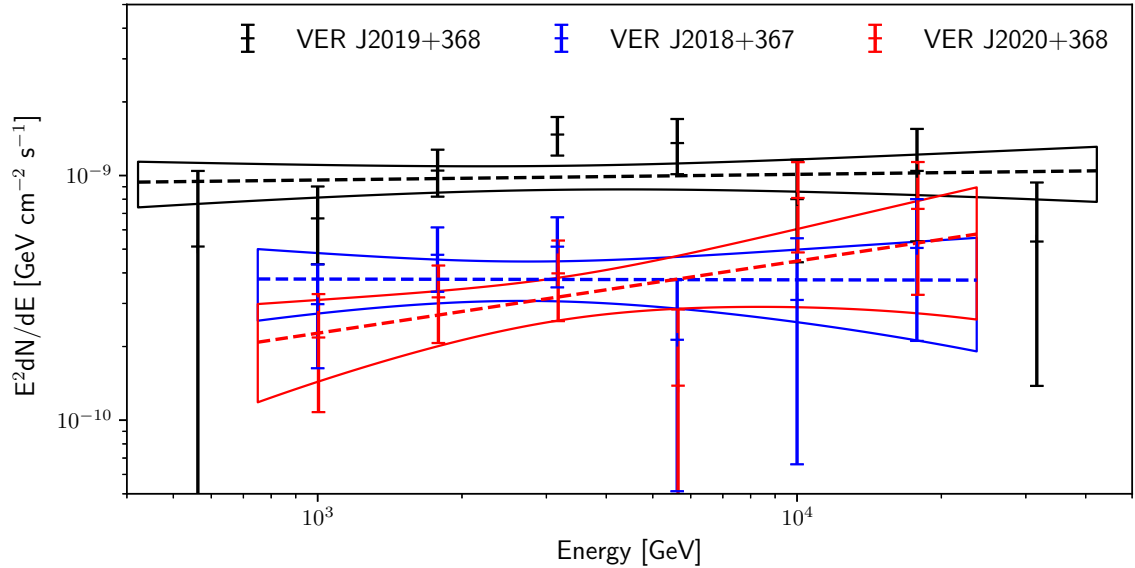}
\caption{Spectra of \veritas\ sources in the \MGROcisne\ region. 
The spectrum for \cisne\ was produced using the \ext\ integration region is shown in black along with the spectra for \cisneA\ (blue) and \cisneB\ (red) which were produced using the \point\ integration region. The butterflies show statistical errors only.
\label{fig:CisneSpectrum}}
\end{figure}

We reconstructed spectra for all three sources and the fit parameters are given in \autoref{tab:VERITAS_Spectra} and shown in \autoref{fig:CisneSpectrum}.  
In the case of \cisne\ there is some evidence of curvature.  
When fit with a LP spectrum the fit improved in comparison to the PL spectrum that we report in \autoref{tab:VERITAS_Spectra} but not significantly (F-test = 1.75$\sigma$). 
In comparison with the result presented in \citet{aliu_spatially_2014}, the spectrum has a significantly lower flux.  This is due to the smaller integration region used in this analysis (0.23\arcdeg\ radius rather than 0.5\arcdeg).

\subsubsection{Discussion}
2HWC J2019+367 lies between \cisneA\ and \cisneB, 0.07\arcdeg\ away from \cisne\ (within the 0.09\arcdeg\ statistical uncertainty of the \hawc\ location).  
As a point source (it was also detected with an extension of 0.7\arcdeg) it is fit with a softer spectral index (\nstat{-2.29}{0.06}) and a higher normalization (\fstat{3.02}{0.31}{-17} at 7000~GeV) than observed for \cisne\ in this work ($\gamma$ = \nee{1.98}{0.09}{0.20},  $N_0$ = \fee{1.03}{0.11}{0.2}{-16} at 3110~GeV).
It is noted that \CTB\ is not resolved in the 2HWC catalog, though there is evidence of excess emission in that direction in the maps presented in the associated paper.
Examining a \LAT\ TS map there is no evidence of emission coming from either of the two sources.

\subsubsection{Potential \cisneA\ Multiwavelength Counterparts}

3FGL J2017.9+3627 has recently been identified as a \GR\ pulsar (PSR J2017+3625)  by the Einstein@Home distributed computing project \citep{2016arXiv161101015C} and lies to the south west of \cisneA\ at a distance of 0.34\arcdeg. 
As pulsations have only recently been detected, no publicly available ephemeris is available.
This prevented us from using either of the analysis methods presented in \autoref{sec:FermiAnalysis} for determining whether a PWN is present but hidden by the strength of the pulsed emission.
Instead we performed an analysis above 10~GeV to test for any emission potentially associated with a PWN, none was detected and we place a 95\% integral flux upper limit on the point source emission of 6.2$\times 10^{-11}$~cm$^{-2}$s$^{-1}$ between 10 and 500~GeV.
We can compare the observed separation between PSR J2017+3625 divided by the extension of \cisneA\ with the values reported in \citet{2018A&A...612A...2H}.  The observed value of 1.9 is approximately a factor of 2 higher than all of the firm associations reported and above the rating criterion of 1.5 they applied to post-select associations.  However, with a characteristic age of 1943~kyr this is significantly older than all of the firm identification which they report and could explain the larger ratio.  Assuming a distance of 450~pc (from  \citet{Gotthelf:2016opn} based upon the gamma-ray efficiency of the pulsar) the TeV efficiency is 0.003, right at the lower edge of the scatter predicted by their varied model. In contrast, assuming a larger distance (\citet{Gotthelf:2016opn} place a lower limit on the distance using an empirical relationship between pulsars and their spin down power at 1~kpc) gives a TeV efficiency of greater than 0.01,  well withing the expected range .
Given the large separation and lack of evidence of an extended PWN in multiwavelength observations we are not able to associate the two objects.  However, based upon the observed properties of younger systems it is possible that \cisneA\ is a PWN associated with PSR J2017+3625.

In addition to PSR J2017+3625, other non-thermal sources are detected in the area which may power the observed VHE emission.
\citet{Gotthelf:2016opn} discuss the potential for a number of hard X-ray sources identified using \nustar\ data to power at least the western part of the emission from \cisne, the emission that is now associated with \cisneA .
We note that all of the identified sources lie to the west of \cisneA, and thus are unlikely to be the sole contributor to the emission.

Observations with \suzaku\ presented in \citet{2017arXiv170502733M} showed no evidence of X-ray emission in the 0.7~--~10~keV energy range in the region around \cisneA. 
The closest source, a point source that they suspect is a background active galactic nucleus, lies in the gap between \cisneA\ and \cisneB\ and is unlikely to be the dominant contributor to the observed, extended VHE emission.

\cisneA\ is spatially coincident with IGR J20188+3647, which remains a possible source for the emission from this region.
Based upon the fast rise time of the emission from IGR J20188+3647 ($\sim$ 10 minutes), followed by the slower decay ($\sim$ 50 minutes) and the spectral parameters, \citet{2009arXiv0902.0245S} consider this to be a candidate supergiant fast X-ray transient.  
They concluded that though it is spatially coincident with \MGROcisne , due to the extended and diffuse nature of the VHE emission region it is unlikely to be associated with it.
However, the resolution of \cisne\ into two sources has significantly reduced this constraint.

\subsubsection{Potential \cisneB\ Multiwavelength Counterparts}

3FGL J2021.1\allowbreak+3651 (PSR J2021+3651) is a \GR\ pulsar that lies 0.11\arcdeg\ away from \cisneB.
The spectrum cuts off sharply at a few tens of GeV ($E_{c}$ = 5.14 \plm\ 0.36~GeV) and thus it is not expected to contribute to the observed VHE emission, but it could be masking emission associated with a PWN.
Using the methods outlined in  \autoref{sec:FermiAnalysis} an examination was conducted for a possible counterpart using an \off-pulse phase of 0.7 to 1. 
The \texttt{BRIDGE} pulse phase from 0.2 to 0.4 was included in the \on-pulse phase.
No evidence of a source was present in either of the analyses.
The lack of detection of a \LAT\ PWN does not rule out the presence of such an object given the analysis challenges associated with such a bright pulsar, nor does it prevent the known PWN visible in other wavelengths from being responsible for at least in part for the VHE emission.

\xmm\ clearly shows a bright point source spatially coincident with the \LAT\ detected pulsar PSR J2021.1\allowbreak+3651, with associated extended emission likely from a PWN (G75.2+0.1) powered by the pulsar.
The PWN is also visible in VLA 20~cm data.
\citet{aliu_spatially_2014} discuss the potential for G75.2+0.1 to contribute to the VHE emission in detail.
The authors note that the bulk of \veritas\ emission lies to the west of these objects, though the pulsar is moving in an eastwards direction with G75.2+0.1 stretching back towards \cisneB.  
\citet{2017arXiv170502733M} reported on \suzaku\ observations that have allowed for a more detailed study of this westerly emission from G75.2+0.1 in the 0.7~--~2 and 2~--~10~keV bands.
They show significant emission that extends to the west of PSR J2021+3651 back towards \cisneB\ though, as in the earlier observations, the X-ray emission they observe does not cover the region of highest VHE emission.
Comparing the X-ray and VHE emission they explore the relationship between the earlier results on \cisne\ presented by \veritas\ in \citet{aliu_spatially_2014} and the X-ray data they studied.
The results presented here show a similar spectral index for \cisneB\ as for the results reported for \cisne\ in \citet{aliu_spatially_2014} (\nee{-1.71}{0.26}{0.2} cf. \nee{-1.75}{0.08}{0.2}) though the 1-10~TeV flux is approximately a factor of six lower (7.3$\times$10$^{-10}$~GeV cm$^{-2}$ s$^{-1}$ rather than 4.2$\times$10$^{-9}$~GeV cm$^{-2}$ s$^{-1}$).
This is due to the smaller integration region size used (0.1\arcdeg\ cf. 0.5\arcdeg) but with an area that is 25 times smaller the flux per unit solid angle is higher. 
The significant decrease in the TeV flux results in an increase in the ratio $U_{mag}/U_{ph}$ from about one as presented in \citet{2017arXiv170502733M} to about six, with the smaller integration region better matching the size of the region from which the X-ray flux was extracted.
This change results in a larger average magnetic $B$ field of about 18~$\mu$G. 
This implies a lower energy population of electrons powering the observed X-ray emission (by a factor of about $\sqrt{6}$) bringing them down to about the same energy as those powering the \veritas\ emission.  
Thus, the difference in the spectral indices between the X-ray and \GR\ observations cannot be explained by a break in the electron spectrum.  
The similarities between the populations also make it difficult to explain the origins of the observed morphology differences between the two wavebands.  
This higher magnetic field is closer to that detected in HESS~J1825-137 \citep{2009PASJ...61S.189U} though \cisneB\ has a much smaller difference between the X-ray and \GR\ spectral indices.
In summary, it is likely that \cisneB\ is associated with G75.2+0.1 (the PWN of PSR J2021+3651), though more detailed observations are required to fully understand the system, in particular to understand the relationship between the observed properties of the X- and \GR\ emission and the high inferred magnetic field.

\subsection{\CTB}
\label{sec:CTB}
\subsubsection{Background}
In addition to the extended emission from \cisne, \citet{aliu_spatially_2014} resolved from \MGROcisne\  a point source of emission coincident with the SNR CTB 87 (\CTB). 
There is a bright source detected by \LAT\ which lies close to the position of \CTB\ in the 1FGL, 2FGL and 3FGL catalogs (1FGL J2015.7\allowbreak+3708, 2FGL J2015.6\allowbreak+3709 and 3FGL J2015.6\allowbreak+3709 respectively) and also in the higher energy 1FHL, 2FHL and 3FHL catalogs (1FHL J2015.8+3710, 2FHL J2016.2+3713, 3FHL J2015.9+3712) and in the 1st Fermi LAT Supernova Remnant Catalog (1SC, \citealt{2016ApJS..224....8A}).
The location of this source is slightly different in each of the three energy ranges.
In the 1FGL/2FGL/3FGL catalogs the source is located close to QSO J2015+371, a BL Lacertae object at z = 0.859 \citep{2013ApJ...764..135S}, whereas in the 1FHL/2FHL/3FHL catalogs it lies closer to CTB 87. 
At the lower energies of the 1/2/3 FGL catalog analysis, there is evidence of flux variability, which has, along with the location, lead to the source being associated with the blazar QSO J2015+371.
The higher energies see reduced evidence of variability and also see a hardening of the spectral index from -2.53\plm0.04 in the 3FGL to -1.74\plm0.46 in the 2FHL.
Combined with the change in the source location this has led to it being associated with CTB 87 in those catalogs.
This suggests that there could be two HE \GR\ sources which lie close together and are not being resolved in the \LAT\ data.

\subsubsection{Results}
In this analysis, \CTB\ was observed using the ring background technique and the \point\ integration region at a level of 6.3$\sigma_{local}$.
Previously reported as a point source \citep{aliu_spatially_2014} we found no improvement when fitting with a single symmetric Gaussian.
Thus we still consider it to be a point source.
To identify its best fit location we modeled it as a point source and the best fit location is given in \autoref{tab:VERITAS_Morphology}.

\begin{figure}[htb!]
\centering
\plotone{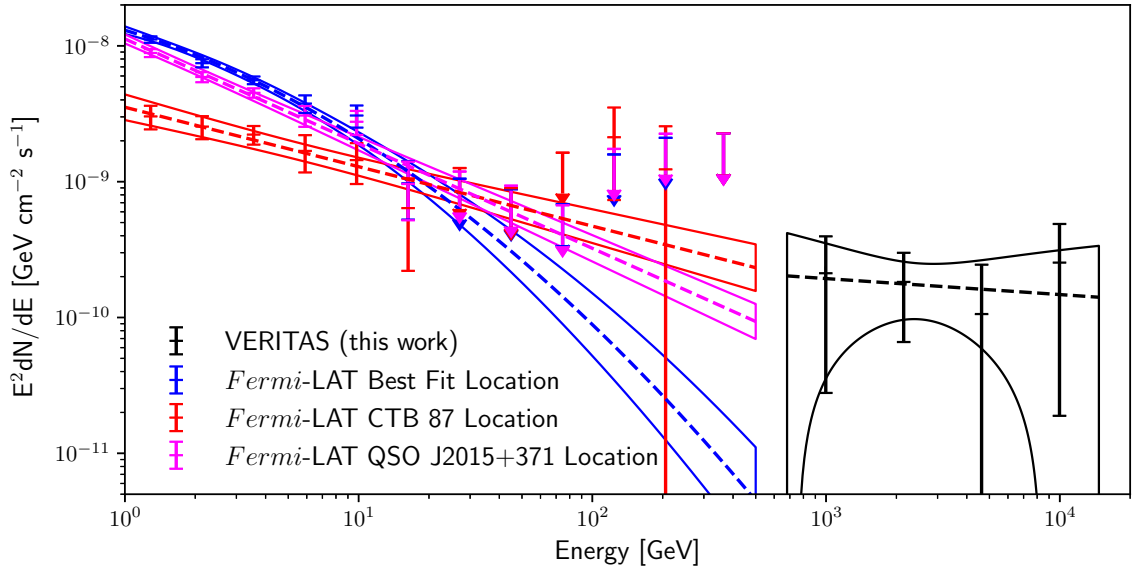}
\caption{Spectra of sources in the CTB 87 region.  
The \veritas\ spectrum for \CTB\ (black) is plotted along with the \LAT\ spectra derived from a single source located at the position of 3FGL J2015.6\allowbreak+3709 (blue).
The spectra from an analysis where this was replaced with two sources, one located at the radio position of CTB 87 (red) and one located at the radio position of QSO J2015+371 (magenta), are also shown.  The butterflies show statistical errors only.}
\label{fig:CTB87Spectrum}
\end{figure}

A spectral fit was conducted using all of the data taken with four telescopes operational and with a pointing offset of less than two degrees (a looser cut on pointing than used for the other sources, since the majority of the observations that had the source in the FoV were targeted at \cisne\ and thus have a larger pointing offset).
A SED was produced using three logarithmically spaced bins per decade in energy and the \point\ integration region (\autoref{fig:CTB87Spectrum}, \autoref{tab:VERITAS_Spectra}).
The updated spectrum is consistent (within statistical errors) with that reported in \citet{aliu_spatially_2014}.

In the \LAT\ analysis, when the source is fit at the location of 3FGL J2015.6\allowbreak+3709 (($l$, $b$) = (74.87\arcdeg, 1.186\arcdeg)) an additional source is detected at ($l$, $b$) = (74.99\arcdeg, 1.16\arcdeg) (see \autoref{tab:Fermi_ctb87_results} for details).
The location of the \LAT\ source is known to vary depending upon the energy threshold so this additional source is likely due to a change in the source location associated with the higher energy threshold in this analysis in comparison to the 3FGL.
The \fermipy\ method \emph{localize} was run on 3FGL J2015.6\allowbreak+3709  without the additional point source included in the model.  
This resulted in an updated source position of ($l$, $b$) = (\dstat{74.893}{0.006}, \dstat{1.200}{0.004}).
This position lies 0.03\arcdeg\ from the catalog location of 3FGL J2015.6\allowbreak+3709 and is shown relative to the \LAT\ catalog positions and overlaid on an \xmm\ counts map with  \veritas\ excess and CGPS 1420 MHz contours in \autoref{fig:CTB87MultiWavelength}.
With the updated source location, no second source is detected.  
For the updated location, the \fermipy\ method \emph{extension} was run to determine if there was any evidence of spatial extension, none was detected and a 95\% confidence level upper limit of extension was found of 0.07\arcdeg\ with a 2D symmetrical Gaussian as the source template. 
Given the location of the source it is likely dominated by emission from QSO J2015+371 and the SED (\autoref{fig:CTB87Spectrum}) is consistent with that from the 3FGL J2015.6\allowbreak+3709 as presented in the 3FGL (shown in \autoref{fig:Fermi3FGLSpectra}).

To test the relative contributions of the two associations, instead of using a single source to model the emission, two power law sources were used and placed at the radio locations of CTB 87 and QSO J2015+371 respectively (the positions are given in \autoref{tab:Fermi_ctb87_results}).
When this model was fit to the data, the two sources had test statistics of 102 and 1087 for CTB 87 and QSO J2015+371, respectively.
It is not possible to definitively state that the \LAT\ data supports the presence of two sources that are currently unresolved, but there is evidence to suggest that this is the case.

The spectra of the two sources are noticeably different, with the source located at the site of CTB 87 being weaker and harder.
Plotting these spectra with the spectrum from the updated 3FGL location, and with the \veritas\ spectrum of \CTB\ (\autoref{fig:CTB87Spectrum}), shows good agreement between the spectrum of the \LAT\ source located at CTB 87 and the spectrum of \CTB, whereas the source located at the location of QSO J2015+371 would require a spectral hardening to fit the \veritas\ results.
Conducting a joint fit to both the spectral points from the \LAT\ emission at the CTB 87 position and \veritas\ emission from \CTB\ with a PL gives the parameters \No\ = \fstat{6.67}{0.60}{-11} at \Eo\ = 5.2 GeV and spectral index \nstat{-2.39}{0.05} with a $\chi^2$ of 4.4 and 9 DoF.

\begin{deluxetable*}{cccccNNQcOO}
\tablecaption{Results of an examination of the HE emission in the region of \CTB. All errors are statistical only.
\label{tab:Fermi_ctb87_results}}
\tabletypesize{\scriptsize}
\tablehead{
\multicolumn{2}{c}{\multirow{2}{*}{Model}} & 
\multirow{2}{*}{TS} & \colhead{\ra} & \colhead{\dec} &
\colhead{$l$} & \colhead{$b$} & \colhead{\No} &
\colhead{\Eo / $E_b$} & \multirow{2}{*}{$\gamma$} & \multirow{2}{*}{$\beta$}  \\ 
\colhead{} & \colhead{} & \colhead{} & \colhead{[\arcdeg]} &
\colhead{[\arcdeg]} & \colhead{[\arcdeg]} & \colhead{[\arcdeg]} &
\colhead{[GeV$^{-1}$cm$^{-2}$s$^{-1}$]} &
\colhead{[GeV]} & \colhead{} & \colhead{}
} 
\startdata
Single Source & 3FGL Location & 1888  & 303.91 & 37.16 & 74.86 & 1.19 & (4.43 \plm\ 0.18)E-12 & 1520 & 2.62 \plm\ 0.09 & 0.12 \plm\ 0.06 \\ \hline
Single Source & Best Fit Location & 1964 & 303.91 & 37.19 & 74.893 \plm\ 0.006 & 1.200 \plm\ 0.004 & (4.86 \plm\ 0.17)E-12 & 1520 & 2.63 \plm\ 0.07 & 0.02 \plm\ 0.04 \\ \hline
\multirow{3}{*}{Two Sources}  & 3FGL J2015.6+3709 & 1363& 303.91 & 37.16 & 74.86 & 1.19 & (4.43 \plm\ 0.18)E-12 & 1520 & 2.62 \plm\ 0.09 & 0.12 \plm\ 0.06 \\
& \textit{New Source} & 59.8 & 304.02 & 37.25 & 74.99 & 1.16 & (1.19 \plm\ 0.34)E-12 & 1000 &  2.11 \plm\ 0.12 & N/A \\ \hline
\multirow{3}{*}{Two Sources} & CTB 87 & 102 &  304.01 & 37.21 & 74.95 & 1.15 & (3.53 \plm\ 0.86)E-12 & 1000 & 2.44 \plm\ 0.10 &  N/A \\ 
 & QSO J2015+371 &  1087 & 303.87 & 37.18 & 74.87 & 1.22 & (1.13 \plm\ 0.09)E-11 & 1000 & 2.77 \plm\ 0.06   & N/A\\ 
\enddata
\end{deluxetable*}

\subsubsection{Discussion}
Combined, the location of \CTB\ and the spectrum of the \LAT\ emission, when fit as two sources, suggests that the \fermi\ emission from the direction of CTB 87 and \CTB\ are the same source and that they are associated with the SNR CTB 87. 
However, a contribution from QSO J2015+371 to the VHE emission cannot be ruled out.

\begin{figure}[htb!]
\centering
\plotone{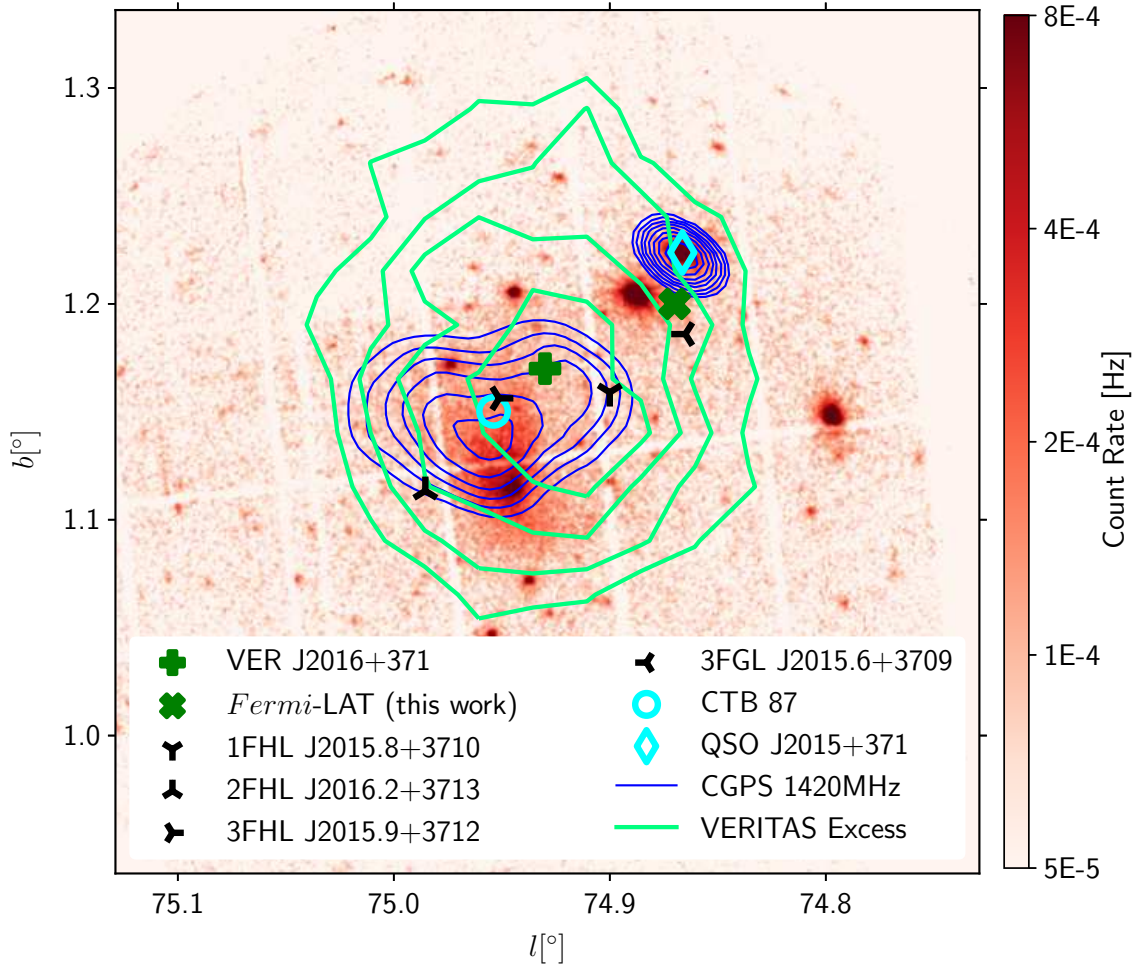}
\caption{\xmm\ counts (red color scale, observation id. 0744640101, no background subtraction nor correction for variation in exposure across the FoV has been applied).
\veritas\ excess with the \point\ integration region is shown in levels of ten counts starting at 30 (bright green) along with CGPS 1420~MHz contours (eight logarithmically spaced levels between brightness temperatures 20 and 150~K  \citep{Taylor:2003cp}, dark blue).  
The locations of \CTB\ (green plus), the \LAT\ position from this work (green cross) and the positions from the different catalogs (black tri symbols) and the radio locations of CTB 87 (cyan o) and QSO J2015+371 (cyan diamond) are shown for comparison.}
\label{fig:CTB87MultiWavelength}
\end{figure}

Examining the region in other wavelengths highlights the two sources identified above along with a number of other, weaker, non-thermal sources.
\autoref{fig:CTB87MultiWavelength} shows an \xmm\ counts map of the region (observation id. 0744640101) overlaid with contours from the CGPS 1420MHz survey.
In both of these observations, CTB 87 and the blazar QSO J2015+371 are clearly visible, with the \veritas\ emission centered on a location that lies within the radio shell of CTB 87, supporting the theory that the VHE \GR\ emission is from that source.
In contrast, the lower energy \LAT\ emission is located between the two sources and outside of any of the X-ray emission regions.
Chandra observations of the region clearly show a bright point source (CXOU J201609.2+371110) that lies within the X-ray emission from CTB 87 at ($l$, $b$) = (74.9438, +01.1140), though pulsations have yet to be detected, it has been suggested that this is the location of the pulsar powering the PWN.
The implications of this, and the location and morphology of the X-ray emission, are discussed in detail in \citet{2013ApJ...774...33M}, concluding that CTB 87 is an evolved ($\sim$5~--~28 kyr), crushed PWN with the extended radio emission and likely a `relic' PWN, as in Vela-X and G327.1-1.1, both of which are detected in VHE \GR s.
\citet{2016MNRAS.460.3563S} conducted a multiwavelength study of the object, where they attributed excess emission in the MeV-GeV range to an additional Maxwellian component.  In contrast, the single power law fit presented here does not require this component. However, the emission above 1~TeV still requires the BPL electron distribution as presented in Figure 4 of that work.  This BPL electron population can be understood in the context of the multiwavelength data with the electron distribution before the break responsible for the radio emission, and the population after the break responsible for the X-ray emission.  Due to the relatively poor angular resolution of \veritas\ in comparison to the radio and X-ray observations it is not possible to observe any differences in the VHE spectrum from these two locations.

\section{Source Population In Comparison to \hess\ Galactic Plane Survey}
It is interesting to compare the populations of HE and VHE sources in the Cygnus region to the region covered by the \hess\ Galactic plane survey.
In the region surveyed by \hess\ (250\arcdeg\ $<$ $l$ $<$ 65\arcdeg, -3.5\arcdeg\ $<$ $b$ $<$ 3.5\arcdeg, 2700 hours of quality-selected data), they detected 78 sources \citep{2018A&A...612A...1H}.
The average sensitivity for their survey was approximately 1.5\% Crab Nebula fluxfor point sources, though it varied across the region.  In comparison the average sensitivity in this work is about 3\% of the Crab, two times higher, though again it shows significant variation with position.
In the \LAT\ 3FGL catalog there are 339 total sources in the region covered by \hess\ and 37 in the region covered by this work.
We would therefore expect to detect 37/339$\times$78 = 8 VHE sources in the \veritas\ survey region, roughly twice the number that we detect (depending on whether \cisneA\ and \cisneB\ are considered as independent sources or as hot spots in \cisne ).
Performing similar calculations with the 2FHL and 3FHL catalogs gives 3/40$\times$78 = 6 and 13/119$\times$78 = 9 VHE sources within the \veritas\ region respectively.
All three predictions agree well with the number of sources observed given the difference in sensitivity between the two surveys and suggest that there is likely a correlation between the numbers of HE and VHE sources in a region.
It should, however, be noted that this is a very simplistic calculation and does not take into account the distances to the sources, the possibility of source confusion, the differences in the diffuse Galactic background or the relative depth of the surveys.

\section{Conclusions}
\label{sec:8}
The results of the \veritas\ survey of the Cygnus region covering a 15\arcdeg\ by 5\arcdeg\ area centered ($l$, $b$) = (74.5\arcdeg, 1.5\arcdeg) conducted between 2007 April and 2008 December with a total observing time of 135 h (120 h live time) were presented. 
We also included targeted and follow-up observations of 174 h (151 h live time) made by \veritas\ between 2008 November and 2012 June, for a total observing time of about 309 h (271 h live time).
A Pass 8 \LAT\ analysis of the same region using over 7 years (2008 August - 2016 January) of data above 1 GeV was also conducted.

No new sources were detected in the \veritas\ analysis and the updated results of the four previously detected \veritas\ sources show that:
\begin{description}
\item[\GCyg] The morphology of the HE \GR\ emission from 3FGL J2021.0+4031e is not a uniform disk as in the 3FGL and peaks at the north eastern rim of the SNR, in the same region as the \veritas\ emission from \GCyg.    
A \LAT\ spectrum extracted from the region coincident with the region from which the \veritas\ spectrum is extracted was found to agree well with the \veritas\ data and a joint fit is well described by a broken power law showing a common origin.

\item[\TeVJ] An extended (68\% containment radius = $0.15\arcdeg^{+0.02\arcdeg}_{-0.03\arcdeg}$) \LAT\ potential counterpart to \TeVJ\ (TeV J2032+4130) is detected at a test statistic of 321.
The spectral points for \TeVJ\ and the \LAT\ counterpart are jointly fit by a single power law.

\item[\cisne] Two source candidates were identified in \cisne, \cisneA\ and \cisneB.  
Both sources are detected at a level greater than 7$\sigma_{local}$ and the spectra are fit with a single power law.

\item[\CTB] When the \LAT\ emission is fit as two point sources, one located at the radio location of CTB 87, one at the radio location of QSO J2015+371, a single power law fits the data from the CTB 87 located \LAT\ source and the emission detected by \veritas. 
\end{description}

In addition to the detected sources, \veritas\ upper limits have been produced for 71 locations at an average level of 2.25\% (2.87\%) of the Crab Nebula flux for a 0.1\arcdeg\ (0.23\arcdeg) integration region.
These locations have, on average, a positive significance of 0.33 (0.18), a 2.9$\sigma$ (1.7$\sigma$) deviation from the expected average significance.

The Cygnus region has  significant potential for follow-up observations by \veritas\ (especially following the summer 2012 camera upgrade and the development of advanced analysis methods aimed at extended sources) and for future work with \cta\ \citep{2013APh....43..317D} and \hawc.

\section*{Acknowledgments}
\noindent This research is supported by grants from the U.S. Department of Energy Office of Science, the U.S. National Science Foundation and the Smithsonian Institution, and by NSERC in Canada. We acknowledge the excellent work of the technical support staff at the Fred Lawrence Whipple Observatory and at the collaborating institutions in the construction and operation of the instrument. The \veritas\ Collaboration is grateful to Trevor Weekes for his seminal contributions and leadership in the field of VHE \GR\ astrophysics, which made this study possible.

This research made use of NASA's Astrophysics Data System; data and/or software provided by the High Energy Astrophysics Science Archive Research Center (HEASARC), which is a service of the Astrophysics Science Division at NASA/GSFC and the High Energy Astrophysics Division of the Smithsonian Astrophysical Observatory; data from the Canadian Galactic Plane Survey, a Canadian project with international partners, supported by the Natural Sciences and Engineering Research Council; the SIMBAD database, operated at CDS, Strasbourg, France. This research is based on observations obtained with \xmm , an ESA science mission with instruments and contributions directly funded by ESA Member States and NASA and observations made with the \spitzer\ Space Telescope, which is operated by the Jet Propulsion Laboratory, California Institute of Technology under a contract with NASA.

We wish to thank the anonymous referee for their constructive comments on this paper.

\facilities{VERITAS, Fermi}
\software{Astropy \citep{2013A&A...558A..33A}, Fermipy \citep{2017arXiv170709551W}, matplotlib \citep{Hunter:2007}, SciPy \citep{jones_scipy_2001}, ROOT \citep{1997NIMPA.389...81B}}

\appendix
\section{Calculation of \veritas\ Spectral Errors}
\label{sec:SpecDef}
For the \veritas\ spectra presented in this work, statistical errors on the spectra were computed and plotted to indicate the 1$\sigma$ confidence range of the fitted models (a ``butterfly'' plot).
For a power law with parameters \No\ \plm\  $\Delta$\No, and spectral index $-\gamma$ \plm\ $\Delta\gamma$, and covariance $cov\,(N_0 , \gamma)$, the contour is defined by

\begin{align}
\frac{\Delta F^2}{F^2} =  \left(\frac{\Delta N_0}{N_0}\right)^2 + \Delta \gamma^2\ln^2 \left( \frac{E}{E_0} \right) - 2 \frac{  cov\,(N_0 , \gamma) }{N_0} \ln \left( \frac{E}{E_0} \right) \label{eq:PLerr}
\end{align}

where the narrowest point in the butterfly occurs at the decorrelation energy ($E_d$)

\begin{align}
E_d = E_0  \exp\left(\frac{cov\,(N_0, \gamma)}{(N_0  \Delta \gamma^2)}\right) . \label{eq:Edec}
\end{align}
The normalization (\No) is quoted at $E_d$ to minimize the covariance between the errors.

The butterfly for the log parabola is defined by
\begin{align}
\begin{split}
\frac{\Delta F^2}{F^2} = &\left(\frac{\Delta N_0}{N_0}\right)^2  
+ \Delta\gamma^2 \ln^2\left(\frac{E}{E_0}\right) 
+ \Delta\beta^2 \ln^4\left(\frac{E}{E_0}\right)  \\
& - 2  \frac{  cov\,(N_0 , \gamma) }{N_0} \ln\left(\frac{E}{E_0}\right) 
- 2  \frac{  cov\,(N_0 , \beta) }{N_0} \ln^2\left(\frac{E}{E_0}\right)
+ 2  cov\,(\gamma , \beta)  \ln^3 \left(\frac{E}{E_0}\right) .
 \label{eq:LPerr}
\end{split}
\end{align}
For LP spectral fits \Eo\ was fixed to $E_d$ calculated in the PL fit (which was calculated prior to the LP fit).

When fitting the BPL, there was a large correlation between $E_b$ and $N_0$, which severely limited the quality of the fit, especially when $\delta \gamma$ ($= |\gamma_1 - \gamma_2|$) was small. 
To overcome this, a preliminary fit was conducted with $E_b$ free (along with $N_0$, $\gamma_1$ and $\gamma_2$).
$E_b$ was then fixed to this value and the fit was conducted again, which results in a much better constrained fit.
The butterfly is given by

\begin{align}
\frac{\Delta F^2}{F^2} =
  \begin{cases}
    &\left(\frac{\Delta N_0}{N_0}\right)^2 
    + \Delta\gamma_1^2 \ln^2\left(\frac{E}{E_b}\right) 
    - 2  \frac{  cov\,(N_0 , \gamma_1) }{N_0} \ln\left(\frac{E}{E_b}\right) 
    \quad \text{if } E \leqslant E_b\\
    &\left(\frac{\Delta N_0}{N_0}\right)^2 
    + \Delta\gamma_2^2 \ln^2\left(\frac{E}{E_b}\right) 
	- 2  \frac{  cov\,(N_0 , \gamma_2) }{N_0} \ln\left(\frac{E}{E_b}\right)    
    \quad \text{if } E >E_b . \\
    \end{cases} \label{eq:BPLerr} 
\end{align}

For the ECPL the butterfly is given by

\begin{align}
\begin{split}
\frac{\Delta F^2}{F^2} = &\left(\frac{\Delta N_0}{N_0}\right)^2  
+ \Delta\gamma^2 \ln^2\left(\frac{E}{E_0}\right) 
+ \left( \frac{E \Delta E_c}{E_c^2} \right)^2 \\
& - 2  \frac{  cov\,(N_0 , \gamma) }{N_0} \ln\left(\frac{E}{E_0}\right) 
+ 2  \frac{  cov\,(N_0 , E_c) E}{N_0 E_c^2}
- 2  \frac{cov\,(\gamma , E_c) E}{E_c^2} \ln \left(\frac{E}{E_0}\right) .
 \label{eq:ECPLerr}
\end{split}
\end{align}

As for the LP fit, an initial PL fit was conducted to determine \Eo.

\section{Trials Factor Estimation}
\label{sec:trials}
A significant number of statistical trials ($X_T$) are associated with blind searches for \GR\ sources of unknown location in a survey such as this. 
These trials increase the chance of an observed positive significance at any location being the result of statistical fluctuations of the background and need to be accounted for when determining the significance of any observed deviation from the null hypothesis (no \GR\ sources in the region, assuming only a uniform cosmic-ray background).
Since the grid points are correlated, both in the \on -region and the \off -region, the number of trials will be less than the number of grid points.
To estimate the number of trials in this analysis, Monte Carlo simulations were conducted to estimate the chance of measuring a given significance in the absence of a signal.
Background-only sky maps were generated by throwing the expected number of background ``events'' with a probability distribution constructed from the acceptance map shown in \autoref{fig:Exposure}.
An RBM analysis was then conducted for each simulated sky map and the significance for each point was calculated.
The significance of an \emph{a priori} chosen random point ($\sigma_{local}$) and the maximum significance of any point ($\sigma_{global}$) were recorded.  
In total, about $1.5 \times 10^6$ simulated sky maps were produced.

The trials factor ($X_T$) was  found by considering that searching the significance sky map for a signal constitutes a simple success/failure test at each location on the sky with a probability distribution described by the Binomial probability function.
In this case, a ``success'' was classed as the measured signal being due to a random statistical fluctuation of the background. 
When dealing with the probability of observing more than $n$ successes in $N$ attempts, the binomial expression given above was summed and expanded
\begin{align}
P_{(n \geqslant 1)} &= \sum^{N}_{n=1} \frac{N!}{n! (N - n)!} p^n (1-p)^{(N-n)} \\
 &  = 1-(1-p)^N  \label{eq:CumBin}
\end{align}

Where $P_{(n \geqslant 1)}$ is the probability of observing one or more values with individual probabilities $p$. 
Hence, for larger and larger values of $N$ (i.e. more trials), the likeliness of observing this success increases. 
This equation was then re-arranged to get an expression for the $X_T \, (\equiv N$) which relates the local ($P_{local} \, ( \equiv  p$) which was calculated using the distribution of $\sigma_{local}$ ) and global ($P_{global} \, (\equiv P_{(n\geq1)})$ which was calculated using the distribution of $\sigma_{global}$) probabilities.

\begin{align}
X_T = \frac{\ln(1 - P_{global})}{\ln(1-P_{local})} \label{eq:Trials Factor}
\end{align}

\begin{figure}[htb!]
\centering
\plotone{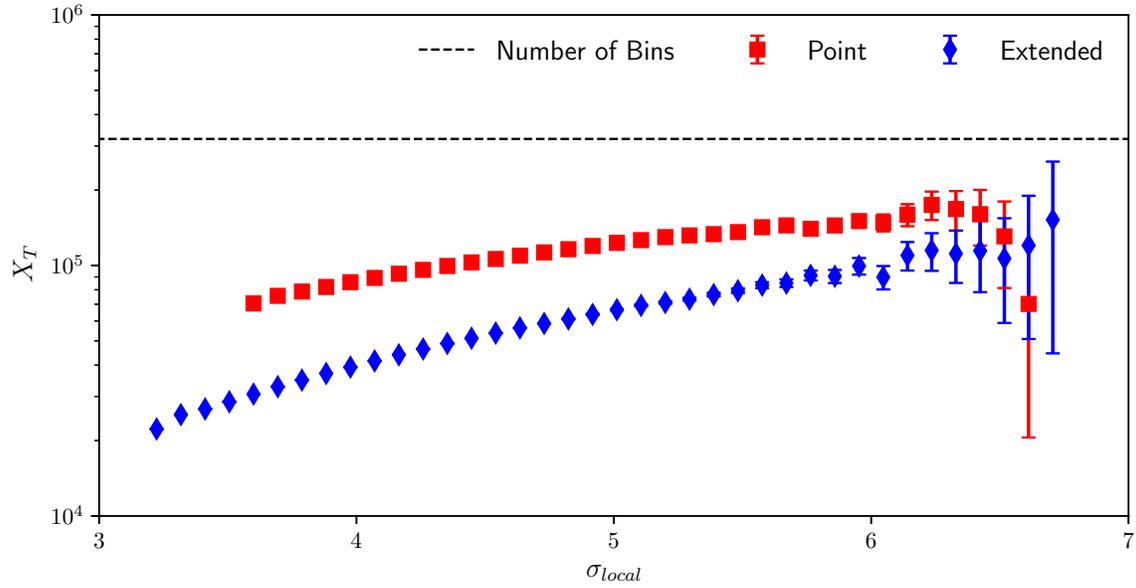}
\caption{Trial factor, $X_{T}$, as a function of local significance, $\sigma_{local}$. 
The black dashed line presents the total number of bins in the sky map of the whole Cygnus region observed with \veritas . At low $\sigma$, $P(\sigma_{global} > \sigma) $ approaches 1, thus \autoref{eq:Trials Factor} has no solution.}
\label{fig:TrialSims_ErrorTrials}
\end{figure}

\begin{figure}[htb!]
\centering
\plotone{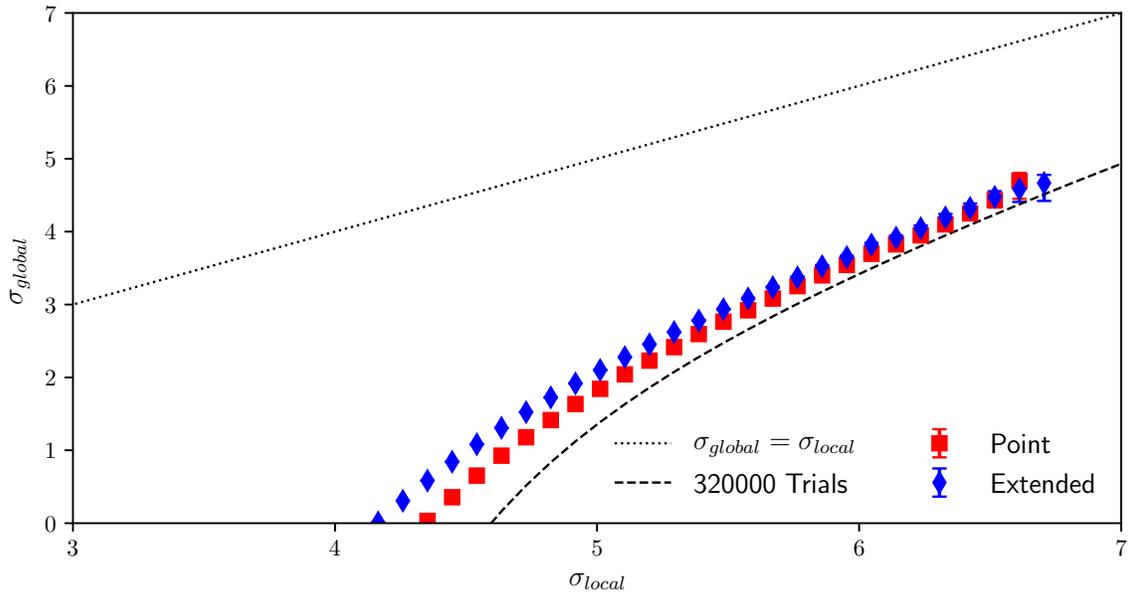}
\caption{Global ($\sigma_{global}$) vs. local ($\sigma_{local}$) significance for the sky survey region.  Below $\sigma_{local} \approx 4.5$ the global significance is zero, i.e. a fluctuation of this magnitude is expected every observation.}
\label{fig:TrialSims_SigRel}
\end{figure}

\autoref{fig:TrialSims_ErrorTrials} shows the dependence of $X_T$ on $\sigma_{local}$. 
$X_T$ increases with increasing $\sigma_{local}$ but it is always significantly smaller than the number of bins in the sky map, reflecting the correlation in both the signal (due to the size of the \on -region relative to the grid spacing) and background (due to the size of the \off -region).
The \ext\ integration region has a smaller $X_T$ than the \point\ integration region, reflecting the higher level of correlation between the \on -regions.
\autoref{fig:TrialSims_SigRel} shows how the global significance varies as a function of local significance. 
Below a  $\sigma_{local}$ of about 4.5 the $\sigma_{global}$ is 0 since a fluctuation of this magnitude is expected in every observation.
Achieving a $\sigma_{global}$ that is greater than 5 requires a $\sigma_{local}$ that is greater than 7.

\section{3FGL Sources Not Detected In This Work}
\label{sec:3FGLremoved}
\begin{deluxetable*}{ccccccc}[htb!]
\tablecaption{The catalog values for the 3FGL sources in the survey region ((83.5\arcdeg\ $>$ $l$ $>$ 65.5\arcdeg) and  (5.5\arcdeg\ $>$ $b$ $>$ -2.5\arcdeg)) that were not significantly detected in this analysis (TS $<$ 25) and removed. * denotes that the source lies in the field of the Cygnus Cocoon.
\label{tab:3FGL_not_this}}
\tabletypesize{\scriptsize}
\tablehead{
\colhead{Name} & \colhead{Spectrum Type} & \colhead{3FGL Sig. [$\sigma$]} & 
\colhead{\No\ [GeV$^{-1}$cm$^{-2}$s$^{-1}$]} &
\colhead{\Eo / $E_b$ [GeV]} & \colhead{$\gamma$} & \colhead{$\beta$}
} 
\startdata
 3FGL J1958.6+3844 & PowerLaw & 4.84 & (\nstat{1.63}{0.31})E-9 & 0.8 & \nstat{2.64}{0.13} & N/A \\
 3FGL J2011.1+4203 & PowerLaw & 4.31 & (\nstat{1.23}{0.25})E{-9} & 0.95 & \nstat{2.54}{0.12} & N/A \\
 3FGL J2014.4+3606 & PowerLaw & 4.48 & (\nstat{4.49}{0.94})E{-10} & 1.69 & \nstat{2.40}{0.11} & N/A \\
 3FGL J2024.6+3747 & LogParabola & 7.04 & (\nstat{6.30}{0.87})E{-9} & 0.87 & \nstat{2.39}{0.58} & \nstat{0.49}{0.22} \\
 3FGL J2026.8+4003 & LogParabola & 9.03 & (\nstat{2.00}{0.21})E{-8} & 0.7 & \nstat{2.61}{0.19} & 1 \\
 3FGL J2033.3+4348* & PowerLaw & 4.39 & (\nstat{6.57}{1.52})E{-11} & 3.84 & \nstat{2.24}{0.14} & N/A \\
 3FGL J2036.8+4234* & PowerLaw & 5.38 & (\nstat{1.52}{0.29})E{-10} & 3.21 & \nstat{2.25}{0.10} & N/A \\
 3FGL J2037.4+4132* & PowerLaw & 6.45 & (\nstat{9.67}{1.66})E{-11} & 4.17 & \nstat{2.18}{0.11} & N/A \\
 3FGL J2043.1+4350  & LogParabola & 8.41 & (\nstat{3.32}{0.38})E{-8} & 0.47 & \nstat{2.46}{0.19} & 1 \\
\enddata
\end{deluxetable*}

\section{Results Tables and Figures}

\begin{figure}[htb!]
\centering
\gridline{
	\fig{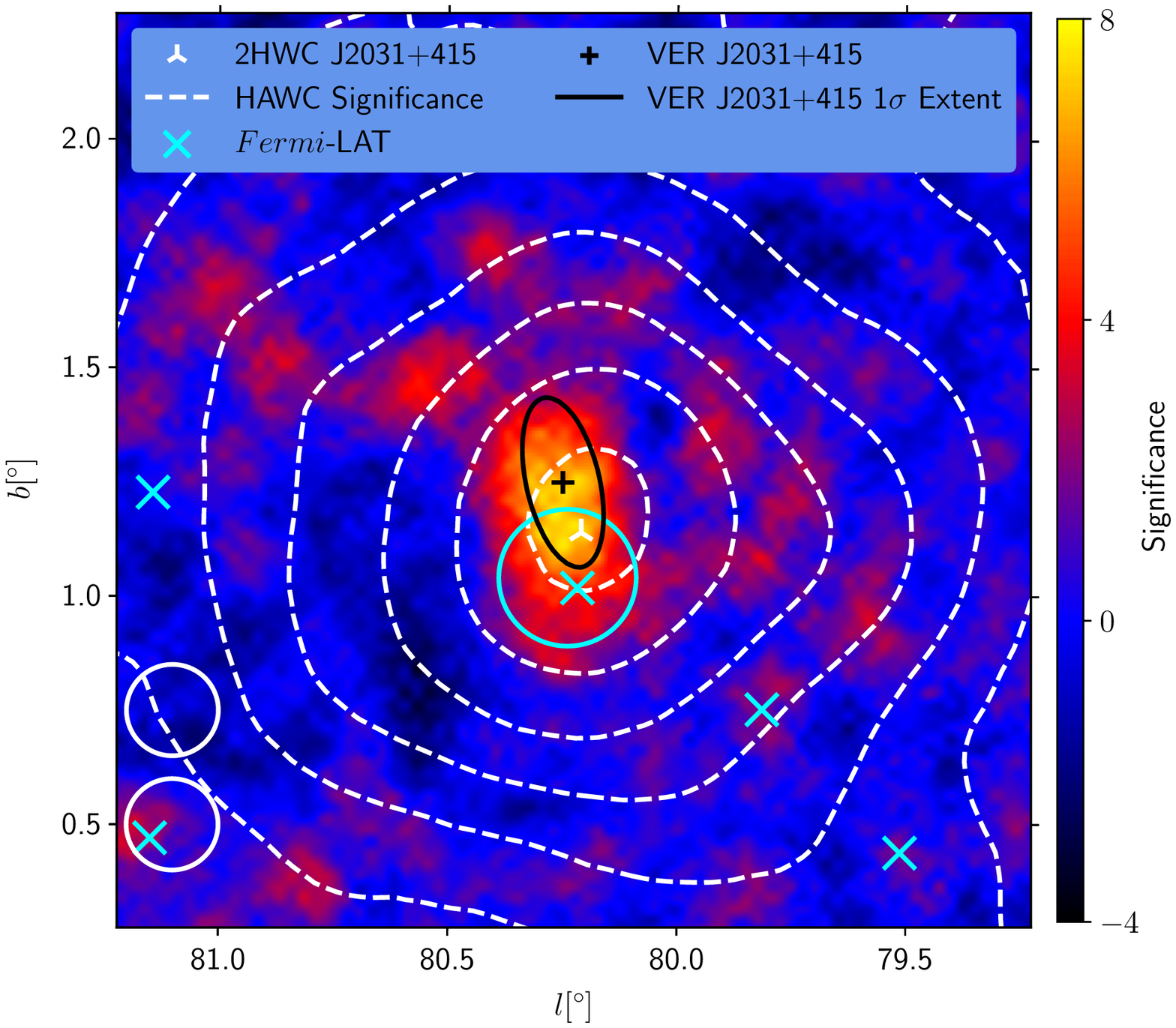}
	{0.48\textwidth}
	{(a) \point\ integration region (radius = 0.1\arcdeg)}
	\fig{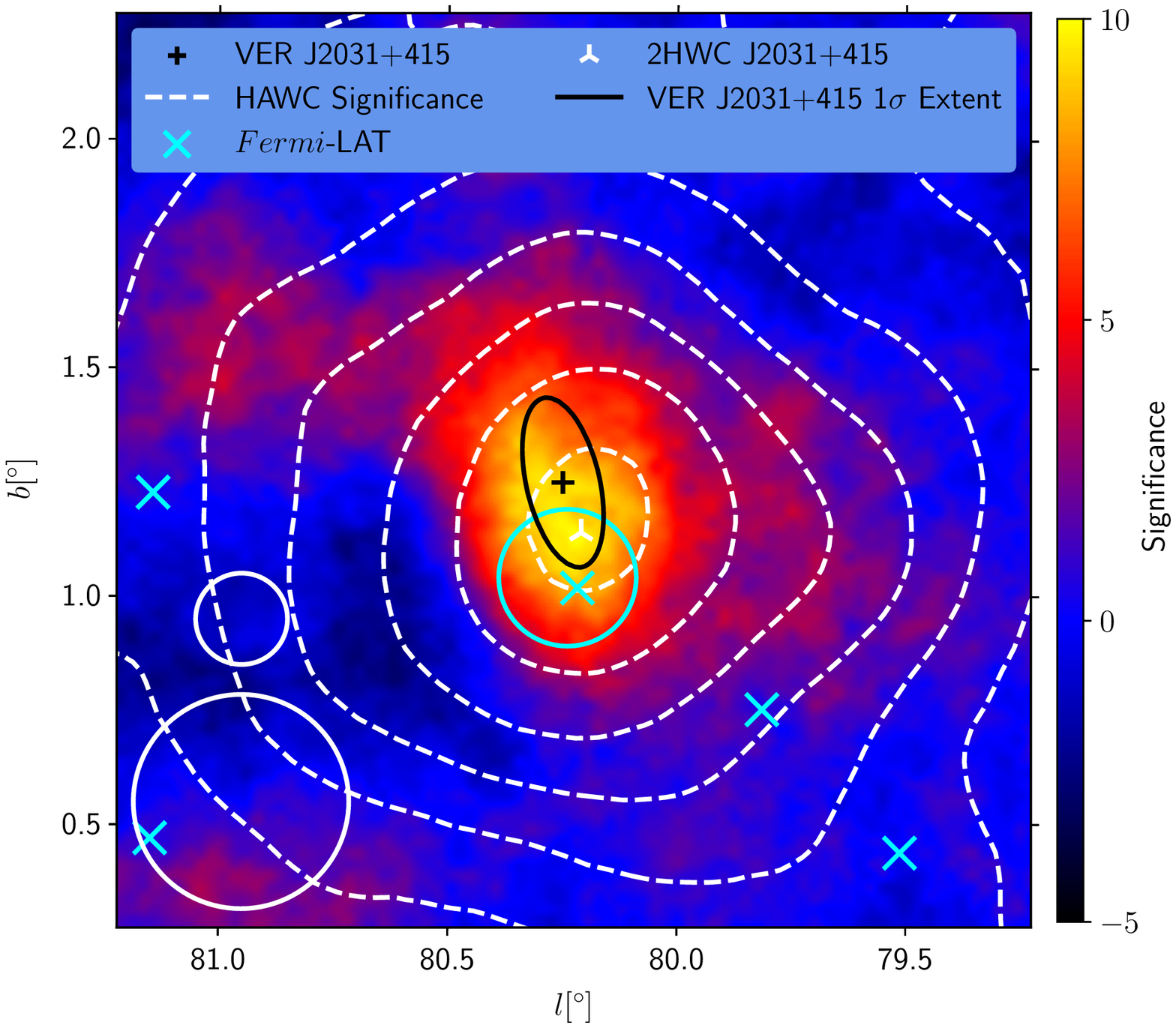}
	{0.48\textwidth}
	{(b) \ext\ integration region (radius = 0.23\arcdeg)}}
\caption{Significance maps of the region around \TeVJ.  
Overlaid are the 1$\sigma$ ellipses for the source extension and its centroid (black) along with the location of 2HWC J2031+415 (white tri), the positions of the \LAT\ sources (cyan) and \hawc\ significance contours (at 2$\sigma$ levels starting at 4$\sigma $, white dash).  
The upper white circle in the lower left is the PSF, whilst the lower circle is the integration region.
\label{fig:VERSkymapsTeV2032}}
\end{figure}

\begin{figure}[htb!]
\centering
\gridline{
	\fig{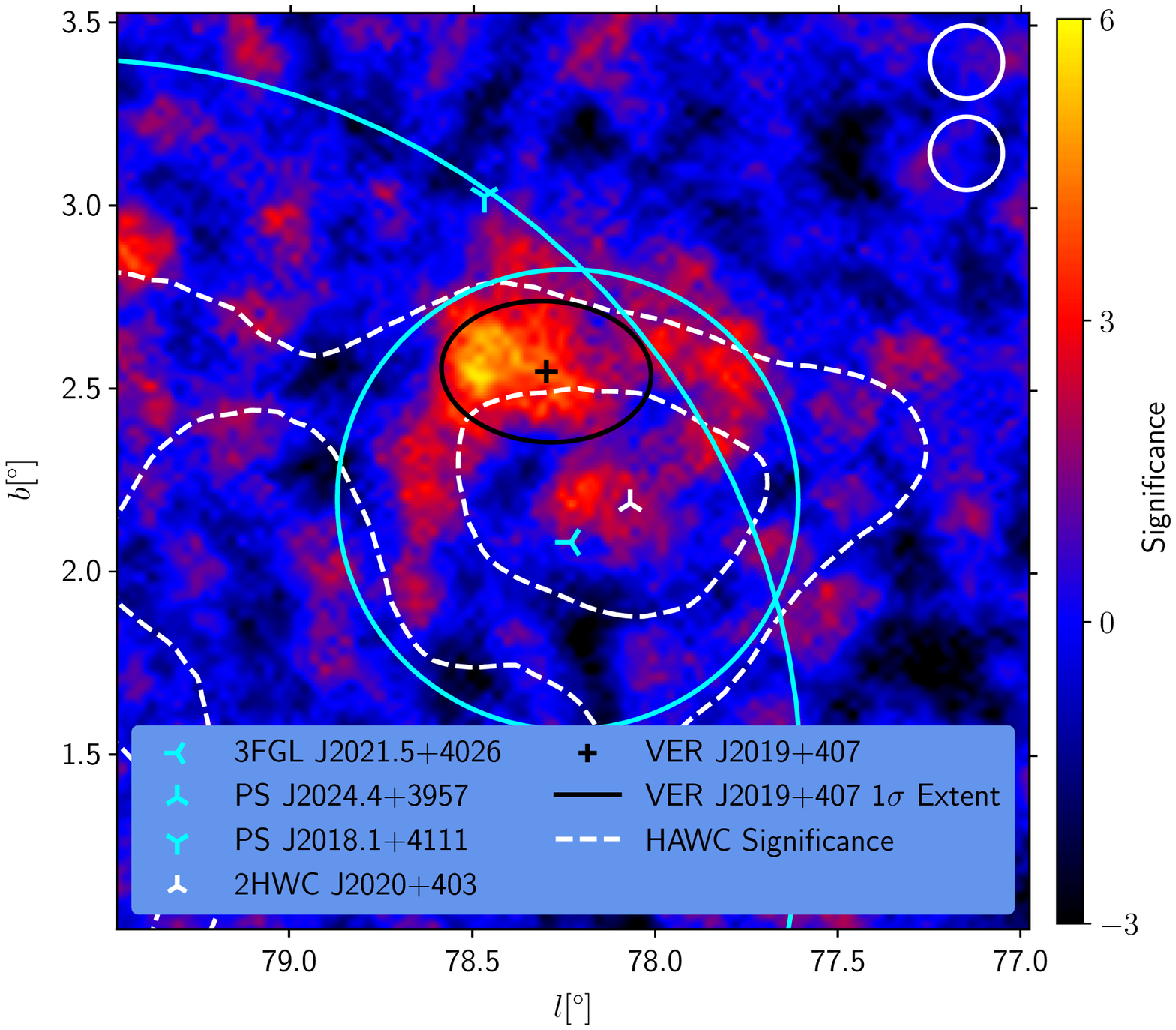}
	{0.48\textwidth}
	{(a) \point\ integration region (radius = 0.1\arcdeg)}
	\fig{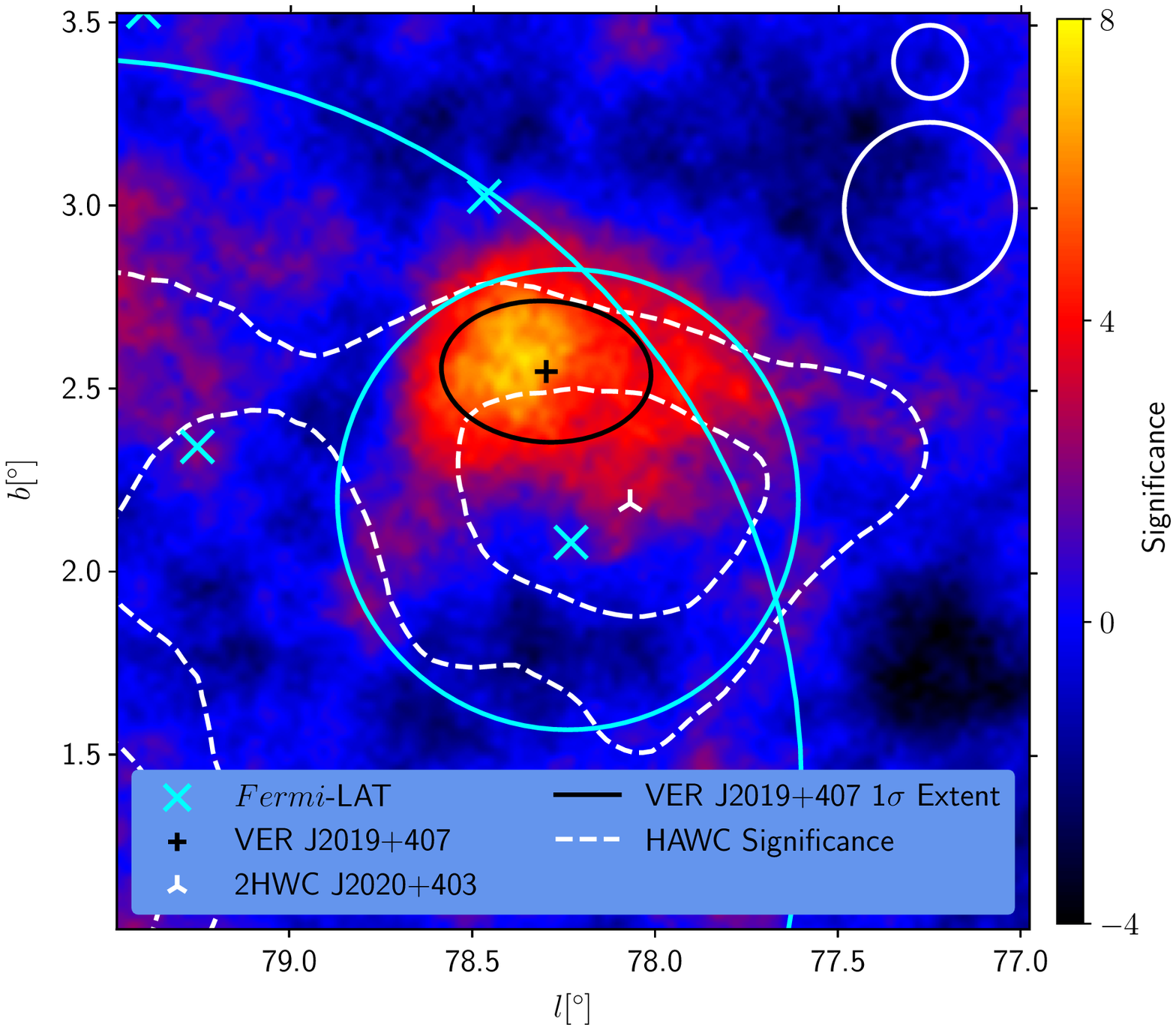}
	{0.48\textwidth}
	{(b) \ext\ integration region (radius = 0.23\arcdeg)}}
\caption{Significance maps of the region around \GCyg.  
Overlaid are the 1$\sigma$ ellipses for the source extension and its centroid (black) along with the location of 2HWC J2019+407 (white tri), the positions of the \LAT\ sources (cyan) and \hawc\ significance contours (at 2$\sigma$ levels starting at 4$\sigma $, white dash).  
The smaller cyan circle shows the extent of 3FGL J2021.0+4031e (G78.2+2.1) and the arc of a larger cyan circle the extent of 3FGL J2028.6-4110e (the Cygnus Cocoon).
The upper white circle in the upper right is the PSF, whilst the lower circle is the integration region.
\label{fig:VERSkymapsGCyg}}
\end{figure}

\begin{figure}[htb!]
\centering
\gridline{
	\fig{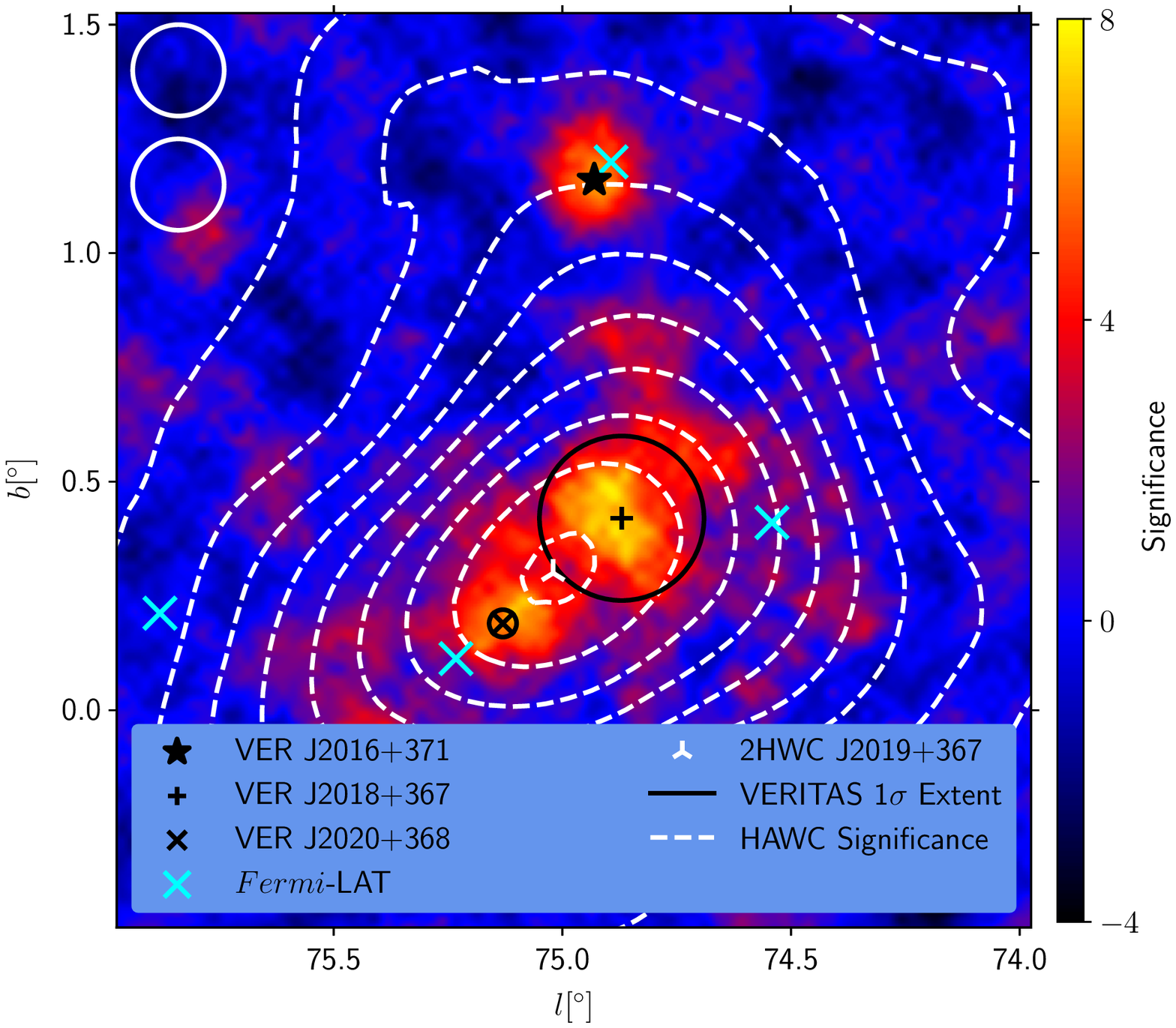}
	{0.48\textwidth}
	{(a) \point\ integration region (radius = 0.1\arcdeg)}
	\fig{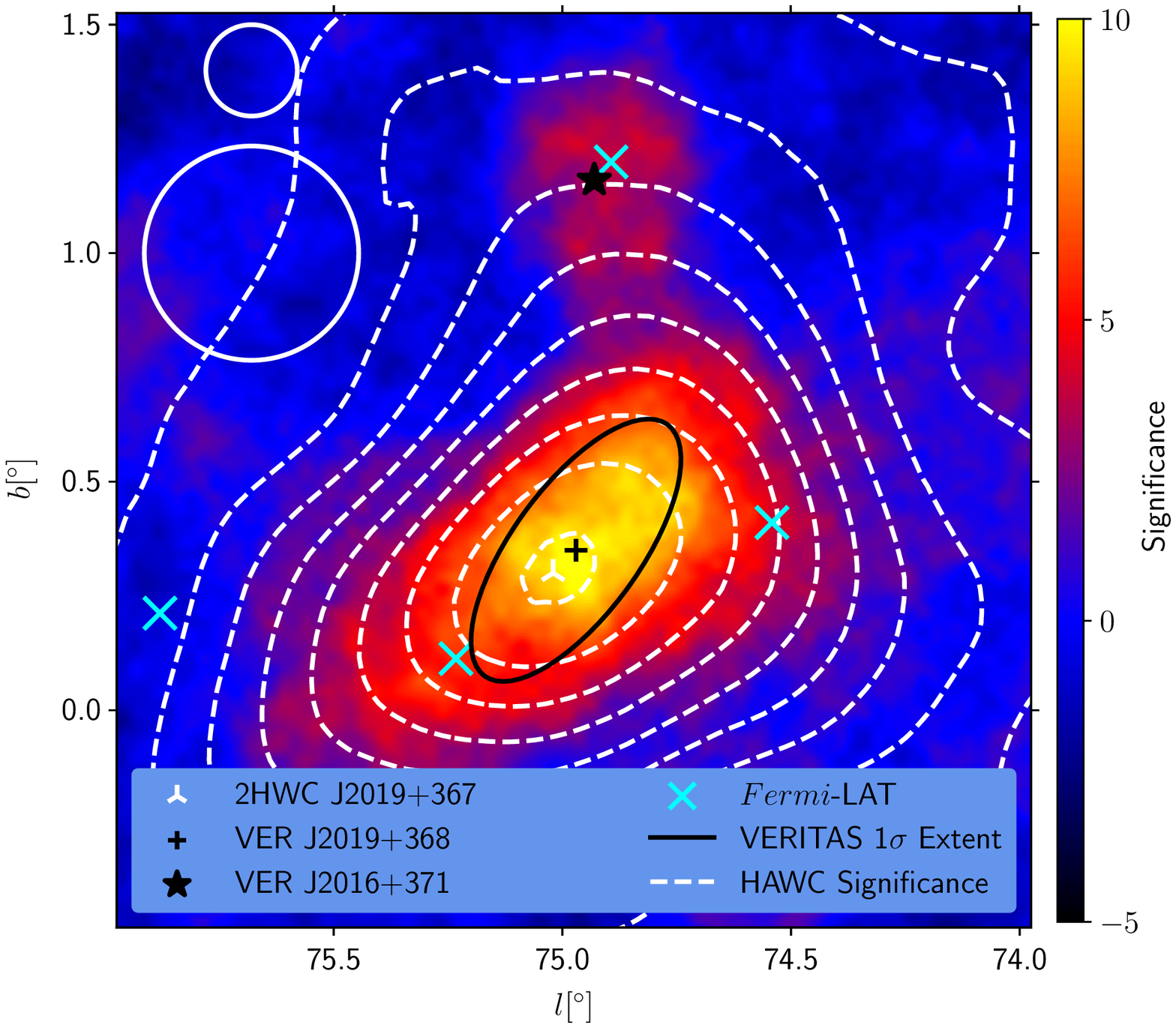}
	{0.48\textwidth}
	{(b) \ext\ integration region (radius = 0.23\arcdeg)}}
\caption{Significance maps of the region around \cisneA, \cisneB\ and \CTB.  
Overlaid are the 1$\sigma$ ellipses for the source extension and  their centroids (black) along with the location of 2HWC J2019+367 (white tri), the positions of the \LAT\ sources (cyan) and \hawc\ significance contours (at 2$\sigma$ levels starting at 4$\sigma $, white dash).  
The upper white circle in the upper left is the PSF, whilst the lower circle is the integration region.
\label{fig:VERSkymapsCisne}}
\end{figure}

\begin{deluxetable*}{cccc}[htb!]
\tablecaption{\veritas\ spectral points for \TeVJ\ fit with a power law (\autoref{eq:PL}).}
\tablehead{
\colhead{Energy Low} & \colhead{Energy High} & \colhead{Flux} & 
\colhead{Flux Error}  \\ 
\colhead{$[\GeV]$} & \colhead{$[\GeV]$} & \colhead{[$\GeV^{-1} \cm^{-2} \s^{-1}$]} 
& \colhead{[$\GeV^{-1} \cm^{-2} \s^{-1}$]} 
}
\startdata
 422 & 750 & 3.26E-15 & 1.57E-15  \\
 750 & 1330 & 6.85E-16 & 2.73E-16  \\
 1330 & 2370 & 1.97E-16 & 8.26E-17  \\
 2370 & 4220 & 3.95E-17 & 2.75E-17  \\
 4220 & 7500 & 3.62E-17 & 1.24E-17  \\
 7500 & 13300 & 6.92e-18 & \textit{Upper Limit} \\
 13300 & 23700 & 1.26E-18 & 1.38E-18 \\
 23700 & 42200 & 5.99E-19 & 4.55E-19 \\
\enddata
\end{deluxetable*}

\begin{deluxetable*}{cccc}[htb!]
\tablecaption{\veritas\ spectral points for \GCyg\ fit with a power law (\autoref{eq:PL}).}
\tablehead{
\colhead{Energy Low} & \colhead{Energy High} & \colhead{Flux} & 
\colhead{Flux Error}  \\ 
\colhead{$[\GeV]$} & \colhead{$[\GeV]$} & \colhead{[$\GeV^{-1} \cm^{-2} \s^{-1}$]} 
& \colhead{[$\GeV^{-1} \cm^{-2} \s^{-1}$]} }
\startdata
 750 & 1330 & 1.56E-15 & 4.17E-16  \\
 1330 & 2370 & 3.91E-16 & 1.09E-16  \\
 2370 & 4220 & 2.65E-17 & 3.03E-17  \\
 4220 & 7500 & 2.73E-17 & 1.27E-17 \\
\enddata
\end{deluxetable*}

\begin{deluxetable*}{cccc}[htb!]
\tablecaption{\veritas\ spectral points for \cisne\ fit with a power law (\autoref{eq:PL}).}
\tablehead{
\colhead{Energy Low} & \colhead{Energy High} & \colhead{Flux} & 
\colhead{Flux Error}  \\ 
\colhead{$[\GeV]$} & \colhead{$[\GeV]$} & \colhead{[$\GeV^{-1} \cm^{-2} \s^{-1}$]} 
& \colhead{[$\GeV^{-1} \cm^{-2} \s^{-1}$]} 
}
\startdata
 422 & 750 & 1.62E-15 & 1.68E-15  \\
 750 & 1330 & 6.68E-16 & 2.34E-16  \\
 1330 & 2370 & 3.31E-16 & 7.2E-17  \\
 2370 & 4220 & 1.47E-16 & 2.64E-17  \\
 4220 & 7500 & 4.29E-17 & 1.09E-17  \\
 7500 & 13300 & 8.01E-18 & 3.59E-18  \\
 13300 & 23700 & 3.3E-18 & 1.6E-18  \\
 23700 & 42200 & 5.37E-19 & 3.99E-19  \\
\enddata
\end{deluxetable*}

\begin{deluxetable*}{cccc}
\tablecaption{\veritas\ spectral points for \cisneA\ fit with a power law (\autoref{eq:PL}).}
\tablehead{
\colhead{Energy Low} & \colhead{Energy High} & \colhead{Flux} & 
\colhead{Flux Error}   \\ 
\colhead{$[\GeV]$} & \colhead{$[\GeV]$} & \colhead{[$\GeV^{-1} \cm^{-2} \s^{-1}$]} 
& \colhead{[$\GeV^{-1} \cm^{-2} \s^{-1}$]} 
}
\startdata
 750 & 1330 & 2.98E-16 & 1.35E-16  \\
 1330 & 2370 & 1.5E-16 & 4.41E-17  \\
 2370 & 4220 & 5.12E-17 & 1.63E-17  \\
 4220 & 7500 & 6.74E-18 & 5.12E-18  \\
 7500 & 13300 & 3.1E-18 & 2.44E-18  \\
 13300 & 23700 & 1.6E-18 & 9.33E-19  \\
 \enddata
\end{deluxetable*}

\begin{deluxetable*}{cccc}[htb!]
\tablecaption{\veritas\ spectral points for \cisneB\ fit with a power law (\autoref{eq:PL}).}
\tablehead{
\colhead{Energy Low} & \colhead{Energy High} & \colhead{Flux} & 
\colhead{Flux Error}   \\ 
\colhead{$[\GeV]$} & \colhead{$[\GeV]$} & \colhead{[$\GeV^{-1} \cm^{-2} \s^{-1}$]} 
& \colhead{[$\GeV^{-1} \cm^{-2} \s^{-1}$]}
}
\startdata
 750 & 1330 & 2.16E-16 & 1.09E-16  \\
 1330 & 2370 & 9.96E-17 & 3.48E-17  \\
 2370 & 4220 & 3.95E-17 & 1.43E-17  \\
 4220 & 7500 & 4.34E-18 & 4.55E-18 \\
 7500 & 13300 & 8.03E-18 & 3.22E-18  \\
 13300 & 23700 & 2.29E-18 & 1.27E-18  \\
\enddata
\end{deluxetable*}

\begin{deluxetable*}{cccc}[htb!]
\tablecaption{\veritas\ spectral points for \CTB\ fit with a power law (\autoref{eq:PL}).}
\tablehead{
\colhead{Energy Low} & \colhead{Energy High} & \colhead{Flux} & 
\colhead{Flux Error}   \\ 
\colhead{$[\GeV]$} & \colhead{$[\GeV]$} & \colhead{[$\GeV^{-1} \cm^{-2} \s^{-1}$]} 
& \colhead{[$\GeV^{-1} \cm^{-2} \s^{-1}$]} 
}
\startdata
 681 & 1470 & 2.13E-16 & 1.85E-16  \\
 1470 & 3160 & 3.96E-17 & 2.53E-17  \\
 3160 & 6810 & 4.95E-18 & 6.48E-18  \\
 6810 & 14700 & 2.55E-18 & 2.36E-18  \\
\enddata
\end{deluxetable*}

\bibliography{bibliography}


\end{document}